\documentclass[12pt, hyper]{JHEP3}  

\usepackage{amsmath,amssymb,amsfonts,amsxtra, mathrsfs, makeidx,graphics,graphicx,amsthm,epsfig}
\usepackage[all]{xy}

\pdfoutput=1

\newcommand{\be}{\begin{equation}}
\newcommand{\ee}{\end{equation}}
\newcommand{\beq}{\begin{equation}}
\newcommand{\eeq}{\end{equation}}
\newcommand{\ba}{\begin{array}}
\newcommand{\ea}{\end{array}}
\newcommand{\bi}{\begin{itemize}}
\newcommand{\ei}{\end{itemize}}
\newcommand{\bea}{\begin{eqnarray}}
\newcommand{\eea}{\end{eqnarray}}
\newcommand{\ben}{\begin{enumerate}}
\newcommand{\een}{\end{enumerate}}
\newcommand{\bean}{\begin{eqnarray*}}
\newcommand{\eean}{\end{eqnarray*}}
\newcommand{\eref}[1]{(\ref{#1})}

\newcommand{\tref}[1]{Table~\ref{#1}}
\newcommand{\fref}[1]{Figure~\ref{#1}}
\newcommand{\nn}{\nonumber}

\newcommand{\tr}{\mathop{\rm Tr}}

\newcommand{\BC}{\mathbb{C}}

\newcommand{\BP}{\mathbb{P}}
\newcommand{\BZ}{\mathbb{Z}}

\newcommand{\cB}{{\cal B}}
\newcommand{\cC}{{\cal C}}
\newcommand{\cD}{{\cal D}}
\newcommand{\cE}{{\cal E}}
\newcommand{\cF}{{\cal F}}
\newcommand{\cP}{{\cal P}}

\newcommand{\cT}{{\cal T}}
\newcommand{\cO}{{\cal O}}

\newcommand{\comment}[1]{}

\newcommand{\CMm}{{\cal M}^{\mathrm{mes}}}
\newcommand{\gm}{ g^{\mathrm{mes}}}

\newcommand{\BF}{\mathbb{F}}

\newcommand{\perm}{\mathrm{perm}}

\newcommand{\tmat}[1]{{\tiny \left(\begin{matrix} #1 \end{matrix}\right)}}

\newcommand{\ud}{\mathrm{d}}

\newcommand{\PL}{\mathrm{PL}}

\newcommand{\firr}[1]{{}^{{\rm Irr}}\!{\cal F}^{\flat}_{#1}}

 \title{M2-Branes and Fano 3-folds}
\author{John Davey, Amihay Hanany, Noppadol Mekareeya and Giuseppe Torri\\
Theoretical Physics Group, The Blackett Laboratory \\
Imperial College London, Prince Consort Road\\ 
London,  SW7 2AZ,  UK \\
Email: {\tt j.davey07, a.hanany, n.mekareeya07, giuseppe.torri08@imperial.ac.uk}}
\abstract{A class of supersymmetric gauge theories arising from M2-branes probing Calabi-Yau 4-folds which are cones over smooth toric Fano 3-folds is investigated. For each model, the toric data of the mesonic moduli space is derived using the forward algorithm.  The generators of the mesonic moduli space are determined using Hilbert series. The spectrum of scaling dimensions for chiral operators is computed.}

\preprint{Imperial/TP/11/AH/03}

\begin{document}

\section{Introduction}
Supersymmetric Chern-Simons (CS) theories are believed to be excellent candidates for describing the world-volume theory on a stack of M2-branes on various back-grounds. The first step in this direction was taken by Bagger-Lambert \cite{BL} and Gustavsson \cite{gus}, who succeeded in writing an $\mathcal{N}=8$ supersymmetric classical action for CS gauge fields coupled to matter with gauge group $SO(4)$. This model was then reformulated as an $SU(2)\times SU(2)$ gauge theory with Chern-Simons levels $k$ and $-k$ \cite{VanRaamsdonk:2008ft}. Subsequently, a similar $U(N)\times U(N)$ gauge theory with CS levels $k$ and $-k$ was proposed by Aharony, Bergman, Jafferis and Maldacena (ABJM) \cite{Aharony:2008ug} as a model describing $N$ M2-branes in the $\BC^4/\BZ_k$ background (see \cite{Klebanov:2009sg} for a recent review). 

In the vastness of supersymmetric Chern-Simons theories, those that can be represented by a \textbf{quiver} certainly constitute an interesting subclass. Quivers are graphs that encode relevant pieces of information about the supersymmetric Lagrangian of a theory and, in the context of M2-branes in various backgrounds, they are believed to describe the $(2+1)$-dimensional CS theories that live on the M2-branes' world-volume \cite{Martelli:2008si, Ueda:2008hx, Hanany:2008cd, Hanany:2008fj,Franco:2008um, taxonomy, phase, Hanany:2009vx, Aganagic:2009zk,Davey:2009qx, Hewlett:2009bx}.

Certain quiver CS theories can be represented by a \textbf{brane tiling}\footnote{The reader is reminded that although every brane tiling gives rise to a supersymmetric quiver gauge theory, the converse is not necessarily true.}. Although tilings were first introduced to describe $(3+1)$-dimensional supersymmetric gauge theories \cite{Hanany:2005ve, Franco:2005rj, Franco:2005sm, Feng:2005gw, Franco:2006gc, Broomhead:2008an}, the idea can be used to study $(2+1)$-dimensional CS theories as well \cite{Hanany:2008cd, Hanany:2008fj, phase, Davey:2009qx, Davey:2009bp, Taki:2009wf, Davey:2009et}. Given the brane tiling of a CS theory, with the use of the \textbf{forward algorithm} it is possible to investigate the classical vacuum moduli space, its generators and the relations among these generators\footnote{For a detailed explanation of the forward algorithm for CS theories see \cite{phase, Davey:2009qx,Davey:2009et}.}. In particular, brane tilings are very useful for computing the R-charges of the chiral fields. Since for a generic $(2 + 1)$-dimensional supersymmetric gauge theory the known $a$-maximization method for D3 branes \cite{Gubser:1998vd, Butti:2005vn} fails, the computation of these R-charges could be a difficult task, especially when the global symmetry of the theory contains many abelian factors. However, the forward algorithm provides a simple method of finding the R-charges of the chiral fields through volume minimisation \cite{Hanany:2008fj}.

Applied to $(2+1)$-dimensional CS theories, brane tilings have allowed the investigation of very interesting phenomena such as \textit{toric duality} \cite {Franco:2008um,taxonomy,phase,Franco:2009sp,Amariti:2009rb}, which corresponds to the situation where two or more CS theories with different quivers have the same CY 4-fold as their mesonic moduli space. Brane tilings have also provided a simple way of relating various theories via the \textit{Higgs mechanism} \cite{Davey:2009qx, Taki:2009wf, Benishti:2009ky}.

This paper focuses on supersymmetric CS theories on M2-branes probing a specific class of CY 4-folds, namely the complex cones over smooth toric Fano 3-folds. Fano varieties in lower dimensions 
are well known in the string theory literature as, for example, they have played an important role in the study of supersymmetric gauge theories living on D3-branes probing a CY 3-fold given by the complex cone over a the smooth toric Fano 2-folds. It is known that there  are only 5 such fano 2-folds: the zeroth Hirzebruch surface $\BF_0$ and the del Pezzo surfaces $dP_{n=0,1,2,3}$\footnote{The other del Pezzo surfaces are Fano varieties but are not toric.}. The study of supersymmetric gauge theories corresponding to these CY 3-folds led to the discovery of the first examples of toric duality for $(3+1)$-dimensional gauge theories \cite{Feng:2000mi, Feng:2001xr, Feng:2002zw, Feng:2002fv, Feng:2001bn, Franco:2003ea, Franco:2003ja, Franco:2002mu, Forcella:2008ng}. The del Pezzo surfaces have also been studied in a more phenomenological context \cite{Verlinde:2005jr, Malyshev:2007yb, Conlon:2008wa, Blumenhagen:2008zz, Blumenhagen:2009gk, Blumenhagen:2009yv, Balasubramanian:2009tv}.

One of the appealing features of the toric Fano varieties is that for every complex dimension there always exist a finite number of smooth toric fanos. It is known \cite {database, toricfano4} that for the three dimensional case there are 18 smooth toric fano 3-folds, each of which can be used to construct a toric CY 4-fold. Below some CS gauge theories that have these CY 4-folds as their mesonic moduli spaces are investigated. It is our hope that the gauge theories that are introduced shall receive the same amount of attention as those living on D3-branes on cones over Fano 2-folds.

\section{The Fano Varieties}

Fano varieties are defined as projective $d$-dimensional algebraic varieties whose anti-canonical sheaf is ample. They are characterized by a positive curvature, so that they can be used to construct Calabi-Yau $(d+1)$-folds by taking a complex cone over them. In the following, smooth Fano varieties are investigated, even though many examples of non-smooth cases are well-known. The reason for this is that for every complex dimension there always exists a finite number of smooth Fano varieties \cite{KMM1, KMM2}, thus giving the hope for a complete classification of them.

In one complex dimension the only Fano variety is $\BP^1$, which can also be thought of as the real 2-sphere. 

It is a classical result that in 2 complex dimensions there are exactly 10 Fano varieties, up to deformations: the zeroth Hirzebruch surface, $\BF_0$, and the 9 del Pezzo surfaces $dP_{n=0,\ldots,8}$. The former is simply a product of two projective lines, $\BP^1\times \BP^1$, whereas the latter is a family of 9 CY 3-folds obtained from $\BP^2$ by blowing up $0\leq n \leq 8$ points in general position. Of these 10 Fano varieties, only $\BF_0$ and $dP_{n=0,1,2,3}$ are toric.

The first important results towards a classification of Fano 3-folds were obtained by Iskovskih \cite{isk1, isk2}, and a complete classification was given by Mori and Mukai \cite{MM1, MM2, MM3} (see also \cite{murre, cutkosky}). They found 88 varieties up to deformations of which 18 are toric \cite{database, toricfano3, toricfano4}. For higher dimensions a complete classification of smooth Fano varieties is still an open problem \cite{toricfano4, Kreuzer:2007zy, Oebro07b}.

\subsection{The smooth toric Fano three-folds}

It is of interest to construct CY 4-folds which can be probed with M2-branes and so cones over three dimensional Fano varieties are considered. Furthermore, since tiling technology is used, we restrict our attention to those smooth Fanos are toric. As is mentioned in the preceding section, there are only 18 such varieties. A detailed presentation of these varieties and their main geometric features is given in Table \ref{t:fanotable}.

\begin{table}[h!]
\[
\hspace{-1cm}
\begin{array}{|c|c|c|c|c|c|}
\hline
 & Sym & \mbox{Toric Data} & \mbox{Geometry} & \mbox{Id of \cite{database}}
 & (b_2,g) \\
\hline \hline
\BP^3 & U(4) & 
\tmat{1 &-1 & 0 & 0 & 0 \\
      0 & 1 &-1 & 0 & 0 \\
      0 & 0 & 1 &-1 & 0} & \BP^3 & 4 & (1,33) 
	\\
	\hline 
\cB_4 & [3,2,1] &
\tmat{ 1 &-1 & 0 & 0 & 0 & 0 \\
       0 & 1 &-1 & 0 & 0 & 0 \\
       0 & 0 & 0 & 1 & -1 & 0 } & \BP^2 \times \BP^1  &  24 &
 (2,28) 
	\\
	\hline
\cB_1 & [3,1^2] &
\tmat{
 1 &-1 & 0 & 0 & 0 & 0 \\
 0 & 1 &-1 & 0 & 0 & 0 \\
 0 & 0 & 2 &-1 & 1 & 0
} & \BP(\cO_{\BP^2} \oplus \cO_{\BP^2}(2)) &  35 &
 (2,32) 
	\\
	\hline
\cB_2 &  [3,1^2] &
\tmat{
 1 &-1 & 0 & 0 & 0 & 0 \\
 0 & 1 &-1 & 0 & 0 & 0 \\
 0 & 0 & 1 &-1 & 1 & 0
} &  \BP(\cO_{\BP^2} \oplus \cO_{\BP^2}(1)) & 36 &
 (2,29) 
	\\
	\hline
\cC_3 & [2^3,1] &
\tmat{
 1 &-1 & 0 & 0 & 0 & 0 & 0 \\
 0 & 0 & 1 &-1 & 0 & 0 & 0 \\
 0 & 0 & 0 & 0 & 1 &-1 & 0 } & \BP^1 \times \BP^1 \times \BP^1 & 62 &
 (3,25) 
	\\
	\hline
\cC_4 & [2^2,1^2] &
\tmat{
 1 &-1 & 0 & 0 & 0 & 0 & 0 \\
 0 & 0 & 1 &-1 & 0 & 0 & 0 \\
 0 & 0 & 1 & 1 &-1 & 0 & 0 } & dP_1 \times \BP^1 & 123 &
 (3,25) 
	\\
	\hline
\cC_5 & [2^2,1^2] &
\tmat{ 1 &-1 & 0 &  0 & 0 & 0 & 0 \\
       0 & 0 & 1 & -1 & 0 & 0 & 0 \\
       0 & 1 & 0 & -1 & 1 &-1 & 0} & 
\BP(\cO_{\BP^1 \times \BP^1} \oplus \cO_{\BP^1 \times \BP^1}(1,-1) ) & 68 &
 (3,23) 
	\\
	\hline
\cB_3 & [2^2,1^2] &
\tmat{
 1 &-1 & 0 & 0 & 0 & 0 \\
 0 & 0 & 1 &-1 & 0 & 0 \\
 0 & 1 & 0 &-1 &-1 & 0
} & \BP(\cO_{\BP^1} \oplus \cO_{\BP^1} \oplus \cO_{\BP^1}(1)) & 37 &
 (2,28) 
	\\
	\hline
\cC_1 & [2^2,1^2] & 
\tmat{
 1 &-1 & 0 &  0 & 0 & 0 & 0 \\
 0 & 0 & 1 & -1 & 0 & 0 & 0 \\
 0 & 1 & 0 &  1 &-1 & 1 & 0 } & 
\BP(\cO_{\BP^1 \times \BP^1} \oplus \cO_{\BP^1 \times \BP^1}(1,1) ) & 105 &
 (3,27) 
	\\
	\hline
\cC_2 & [2,1^3] &
\tmat{
 1 &-1 & 0 & 0 & 0 & 0 & 0 \\
 0 & 1 &-1 &-1 & 0 & 0 & 0 \\
 0 & 0 & 1 & 2 &-1 & 1 & 0} &
\BP( \cO_{dP_1} \oplus \cO_{dP_1}(\ell)), \quad 
\left.\ell^2\right|_{dP_1} = 1 &  136 &
 (3,26) 
	\\
	\hline
\cD_1 & [2,1^3] &
\tmat{
 1 &-1 & 0 & 0 & 0 & 0 & 0 \\
 0 & 1 & 1 &-1 & 0 & 0 & 0 \\
 0 & 0 & 0 & 1 &-1 & 1 & 0} & 
\BP^1\mbox{-blowup of }\cB_2 & 131 &
 (3,26) 
	\\
	\hline
\cD_2 & [2,1^3] &
\tmat{
 1 &-1 & 0 & 0 & 0 & 0 & 0 \\
 0 & 1 &-1 &-1 & 0 & 0 & 0 \\
 0 & 0 & 0 & 1 & 1 &-1 & 0} & \BP^1\mbox{-blowup of }\cB_4 & 139 &
 (3,24) 
	\\
	\hline
\cE_1 & [2,1^3] & 
\tmat{1 &-1 & 0 & 0 & 0 & 0 & 0 & 0 \\
      0 & 1 & 1 & 1 &-1 & 0 & 0 & 0 \\
      0 & 0 & 0 & 1 &-1 &-1 & 1 & 0} &
dP_2 \mbox{ bundle over } \BP^1 & 218 &
 (4,24) 
	\\
	\hline
\cE_2 & [2,1^3] &
\tmat{
 1 &-1 & 0 & 0 & 0 & 0 & 0 & 0 \\
 0 & 0 & 1 &-1 &-1 & 0 & 0 & 0 \\
 0 & 1 & 0 & 0 & 1 &-1 & 1 & 0} &
dP_2 \mbox{ bundle over } \BP^1 & 275 &
 (4,23) 
	\\
	\hline
\cE_3 & [2,1^3] &
\tmat{
 1 &-1 & 0 & 0 & 0 & 0 & 0 & 0 \\
 0 & 0 & 1 &-1 & 1 & 0 & 0 & 0 \\
 0 & 0 & 0 & 0 & 1 & 1 &-1 & 0} & 
dP_2 \times \BP^1 & 266 &
 (4,22) 
	\\
	\hline
\cE_4 & [2,1^3] &
\tmat{
 1 &-1 & 0 & 0 & 0 & 0 & 0 & 0 \\
 0 & 1 & 1 &-1 &-1 & 0 & 0 & 0 \\
 0 & 0 & 0 & 0 & 1 & 1 &-1 & 0} &
dP_2 \mbox{ bundle over } \BP^1 & 271 &
 (4,21) 
	\\
	\hline
\cF_2 & [2,1^3] & 
\tmat{
 1 &-1 & 0 & 0 & 0 & 0 & 0 & 0 & 0 \\
 0 & 1 & 1 &-1 & 1 &-1 & 0 & 0 & 0 \\
 0 & 0 & 0 & 0 & 1 &-1 &-1 & 1 & 0} & 
dP_3 \mbox{ bundle over } \BP^1 & 369 &
 (5,19) 
\\
\hline
\cF_1 & [2,1^3] &
\tmat{
 1 &-1 & 0 & 0 & 0 & 0 & 0 & 0 & 0 \\
 0 & 0 &-1 & 1 &-1 & 1 & 0 & 0 & 0 \\
 0 & 0 & 0 & 0 & 1 &-1 &-1 & 1 & 0} & 
dP_3 \times \BP^1 & 324 &
 (5,19) 
	\\
	\hline
\end{array}
\]
\caption{The 18 smooth toric Fano 3-folds and some important geometric data.}
\label{t:fanotable}
\end{table}

A detailed explanation of \tref{t:fanotable} is in order. 

The first column contains the `name' of the Fano 3-fold, according to the nomenclature defined in \tref{t:fanonames}. In particular, the Fano varieties can be divided into subclasses based upon the number of external points in the toric diagram \cite{toricfano4}\footnote{The reason why both $\cC_i$ and $\cD_i$ are used to denote varieties having 6 external points has to do with the structure of the toric diagram. The reader is referred to \cite{toricfano4} for a detailed discussion of this point.}.

\begin{table}[h!]
\[
\hspace{-1cm}
\begin{array}{|c||c|c|c|c|c|c|}
\hline
\mbox{Number of external points} &  \;\;4\;\;  &  \;\;5\;\;  &  \;\;6\;\;  &  \;\;6\;\;  &  \;\;7\;\;  &  \;\;8\;\;\\
\hline
\mbox{Number of varieties}       &  \;\;1\;\;  &  \;\;4\;\;  &  \;\;5\;\;  &  \;\;2\;\;  &  \;\;4\;\;  &  \;\;2\;\; \\
\hline
\mbox{Nomenclature}              &  \;\;\BP^3\;\;  &  \;\;\cB_i\;\;  &  \;\;\cC_i\;\;  &  \;\;\cD_i\;\;  &  \;\;\cE_i\;\;  &  \;\;\cF_i\;\; \\
\hline
\end{array}
\]
\caption{The number of smooth toric Fano three-folds for each number of external points in the toric diagram.}
\label{t:fanonames}
\end{table}
This subdivision provides a natural nomenclature for the Fano varieties to which we shall adhere in the following sections. 

The second column of \tref{t:fanotable} encodes information about the symmetry of the CY 4-fold constructed by taking a complex cone over the Fano variety considered. In order to make the table compact, the following notation is used:
\bea
[3^{k_3}, 2^{k_2}, 1^{k_1}] = SU(3)^{k_3}\times SU(2)^{k_2}\times U(1)^{k_1},
\eea
and, since the symmetry group of the CY must be of rank 4, we have:
\bea
2k_3 + k_2 + k_1 = 4.
\eea

The order of the rows of \tref{t:fanotable} is determined by the amount of symmetry of the corresponding CY, with the rule that manifolds with the greatest number of non-abelian factors of highest rank come first. 

The third column contains the $G_t$ matrices that represent the toric diagram\footnote{True up to multiplicities.} of the CY 4-folds corresponding to the Fano varieties. In particular, the entries of each column of the matrices are the coordinates in a three-dimensional lattice of a point in the toric diagram of a specific CY. 

Note that the point $\left(0,0,0\right)$ is always internal. This situation is similar to the cases where del Pezzo surfaces are considered: their toric diagram contains precisely one internal point. This property of the toric diagrams is not a mere coincidence, as it corresponds to the condition that the variety is Fano.

An interesting property of the $G_t$ (and $G_K$) matrix of a model is that it is always possible to perform a series of elementary row operations such that the resulting matrix contains the simple roots of the non-abelian symmetries of the mesonic moduli space of the considered model. For example, the mesonic symmetry of the real cone over $M^{1,1,1}$ is $SU(3)\times SU(2)\times U(1)$ and the $G_t$ matrix of this model can be written as:
\bea
G_t = \left(
\begin{array}{cccccc}
 1 & -1 &  0 &  0 &  0 & 0 \\ 
 0 &  1 & -1 &  0 &  0 & 0 \\
 0 &  0 &  0 &  1 & -1 & 0 
\end{array}
\right)~.
\eea
The first two rows of this matrix clearly contain the simple roots of $SU(3)$, whereas the third row contains the simple root of $SU(2)$. Therefore it can be seen that this example is consistent with the fact that the non-abelian mesonic symmetry of a given model is encoded in the coordinates of the toric diagram.

Finally, the last column of Table \ref{t:fanotable} contains two topological invariants that characterize the Fano 3-folds, the second Betti number and the genus.\\
The former, which is denoted by $b_2$, can be related to the toric diagram by
\bea
b_2 = E - 3,
\eea
where $E$ is the number of external points. The second Betti number can also be related to the number of baryonic symmetries of the CS gauge theory. In fact, each conserved baryonic charge in the theory corresponds to a 2-cycle in the Sasaki-Einstein 7-fold. By Poincar\'e duality, in this 7-fold there are as many 2-cycles as 5-cycles. Therefore, since the number of 5-cycles equals the number of external points in the toric diagram subtracted by 4, it can be seen that a certain CS gauge theory has $E-4$ baryonic symmetries, or, in terms of the second Betti number, $b_2-1$.

The genus, which is denoted by $g$, is another important quantity used to characterize a manifold. It is of interest here because it can give information about the number of generators of the CS gauge theories living on an M2-branes probing the CY 4-folds considered in this paper.
In particular, as a consequence of its defining property, a Fano variety can always be embedded in a projective space. It can be shown that this embedding is of degree $d = c_1(X)^3$ into $\BP^{g+1}$, with $d = 2g-2$. The $g+2$ homogeneous coordinates of the ambient space are given precisely by the gauge invariant operators that generate the vacuum moduli space of the CS gauge theory.

A word of caution is necessary here. In the mathematical literature the usual approach is to embed a Fano variety in a projective space where all the coordinates have the same weight under multiplication by a scalar. This is known as the \textit{canonical embedding}. If we think of the homogeneous coordinates of the ambient space as gauge invariant operators, then the standard embedding corresponds to the situation where all these operators have the same R-charge. This situation is equivalent to the UV limit of the gauge theory. The physically interesting properties are in the IR, of course, where R-charges and scaling dimensions vary according to the dynamics of the theory. This translates to having gauge invariant operators with different R-charges that corresponds to embedding the Fano variety in a weighted projective space.

It is important to observe that although in principle one is free to choose how to embed the Fano, the requirement that the volume of the Sasaki-Einstein 7-fold (a \textit{real} cone constructed over the Fano 3-fold) is minimised forces the choice of a specific embedding. In other words, the IR dynamics of the theory chooses a very specific embedding of the Fano 3-fold.

Since the genus of the Fano 3-fold is related to the number of generators of the mesonic moduli space, it is expected that the Hilbert series can be written with explicit $g$ dependence \cite{pleth}. In fact, by looking at all the examples reported in this paper, it is easy to see that the Hilbert series of the mesonic moduli space can be written as\footnote{For the sake of simplicity it is assumed that $X$ is embedded in a projective space in a canonical way.}:
\begin{equation}
\gm(t;~X) = \frac{1 + (g-2)t + (g-2)t^2 + t^3}{(1-t)^4} = 
\sum_{n=0}^\infty \frac{t^n}{6}(2n+1)((g-1)n^2+(g-1)n+6)
\ , 
\end{equation}
where $X$ is a Fano 3-fold of genus $g$ \cite{Hanany:2009vx}.

\subsection{The R-charges}
As is mentioned above, one of the advantages of using Hilbert series is that they provide an easy way of finding the R-charges of the perfect matchings and quiver fields of CS theories.

The mesonic moduli space of a given theory is a Calabi-Yau manifold $X$ having a Sasaki-Einstein manifold $H$ as its base. The former can be obtained by the symplectic quotient $\BC^c//(\BC^{*})^{(c-4)}$, where $c$ is the number of perfect matchings of the model. If the fugacity $s_{\alpha}$ is assigned to the perfect matching $p_\alpha$, then the Hilbert series of the mesonic moduli space can be computed by the usual Molien-Weyl formula \cite{Hanany:2008cd, Hanany:2008fj, phase, Davey:2009qx, pleth}:
\bea
\gm(s_\alpha; X) = \oint \prod^{c-G-2}_{i=1} \frac{\ud z_i}{2\pi i z_i} \oint \prod^{G-2}_{j=1} \frac{\ud b_j}{2 \pi i b_j}\; \frac{1}{\prod^{c}_{\alpha=1}\left(1- s_\alpha Z_\alpha(z_i, b_j)\right)},
\eea
where $G$ is the number of gauge groups of the theory and $Z_\alpha(z_i, b_j)$ represents the monomial weight of the perfect matching $p_\alpha$ under the fugacities $z_i$, coming from linear relations among the perfect matchings, and $b_j$, arising from the D-terms. 

In principle, the Hilbert series resulting from this computation is a function of $c$ variables. However, since the mesonic moduli space is a four-dimensional toric manifold, it is possible to express the Hilbert series as a function of only four fugacities\footnote{When the R-charges are irrational, additional fugacities are needed in order to have a Taylor expansion of the Hilbert series that involves only integer powers of all the fugacities.} $t_i$, each corresponding to a global mesonic $U(1)$ symmetry, some of which can be subgroups of non-abelian mesonic symmetry groups. 

Furthermore, if the fugacities are written in terms of the chemical potentials, $t_i = e^{-\mu R_i}$, the Hilbert series has the property that:
\bea
\lim_{\mu \rightarrow 0} \mu^4 \gm(t_i; X) = { {\rm Vol}(R_i)}{}
\eea
where ${\rm Vol}(R_i)$ is the volume of the family of Sasaki-Einstein manifolds having $\vec{R} = (R_1, R_2, R_3, R_4)$ as their Reeb vector. At the superconformal point, the R-symmetry corresponds precisely to the Reeb vector that provides a metric on the cone. This vector can be determined by minimising the function ${\rm Vol}(R_i)$. Note that the requirement that the top holomorphic form on the Calabi-Yau has R-charge 2, reduces the number of variables involved in the minimisation process by one. 

In a generic situation the mesonic symmetry can contains non-abelian factors and their corresponding chemical potentials minimize of the volume of the base when they are equal to 0. Accordingly, the number of variables involved in the minimisation problem equals the number of abelian factors in the mesonic symmetry minus one.

At this point, the crucial observation is that every perfect matching can be associated to a point in the toric diagram, which, in turn, can be 
associated to a divisor of the toric Calabi-Yau $X$. 

The R-charge of any perfect matching is simply the volume of the corresponding divisor, normalized by the minimised volume of the base of the manifold. Letting $D_\beta$ be the divisor corresponding to the perfect matching $p_\beta$, its associated Hilbert series can be computed by the following formula:
\bea
g(s_\alpha; D_\beta) = \oint \prod^{c-G-2}_{i=1} \frac{\ud z_i}{2\pi i z_i} \oint \prod^{G-2}_{j=1} \frac{\ud b_j}{2 \pi i b_j}\; \frac{\left(s_\beta Z_\beta(z_i, b_j)\right)^{-1}}{\prod^{c}_{\alpha=1}\left(1- s_\alpha Z_\alpha(z_i, b_j)\right)},
\eea
The normalized volume of the divisor $D_\beta$ or, equivalently, the R-charge of the perfect matching $p_\beta$ is given by the following formula:
\bea
\lim_{\mu\rightarrow 0}\frac{1}{\mu} \left[ \frac{g(e^{-\mu \hat{R}_i}; D_\beta)}{\gm(e^{-\mu \hat{R}_i}; X)}- 1 \right] = R(p_\beta)~,
\eea
where $\hat{R}_i$ are the components of the Reeb vector whose values minimise the volume of the base of the Calabi-Yau.

Another method to compute the R-charges of the perfect matchings of a given model is based directly on its toric data \cite{Hanany:2008fj}. 

Let's consider an external point in the toric diagram, $v_\alpha$, associated to the perfect matching $p_\alpha$. Let's then consider the clockwise ordered sequence of points $w_\beta$, with $\beta=1,\ldots,m_\alpha$ of the toric diagram that are connected to $v_\alpha$. Since the toric diagram is to be considered as living in $\BZ^4$, $v_\alpha$ and $w_\beta$ can be thought of as vectors defined on the four-dimensional lattice. Therefore, $F_\alpha$ can be defined to be:
\bea
F_\alpha = \sum^{m_\alpha-1}_{\beta=2} \frac{(v_\alpha,w_{\beta-1},w_\beta,w_{1})(v_\alpha,w_\beta,w_1,w_{m_\alpha})}{(v_\alpha,R,w_\beta,w_{\beta+1})(v_\alpha,R,w_{\beta-1},w_\beta)(v_\alpha,R,w_1,w_{m_\alpha})},
\eea
where $\left(v_1, v_2, v_3, v_4\right)$ corresponds to the determinant of the $4\times 4$ matrix constructed by the juxtaposition of the four vectors $v_1, v_2, v_3$ and $v_4$. 

The sum $\sum^{p}_{\alpha=1} F_\alpha$ now plays the role of the volume functional. It explicitly depends on the Reeb vector $R$ and it has to be minimised with respect to the four components of this vector $R_i$. However, the Calabi-Yau condition ensures that, by performing linear transformations on the toric diagram, it is always possible to set the fourth coordinate of each point $v_\alpha$ to 1. This allows us to set $R_4=4$, which reduces by one the number of variables involved in the minimisation process.

The R-charge of the perfect matching $p_\alpha$ associated to the divisor $D_\alpha$ can be computed by:

\bea
R_\alpha = \frac{2F_\alpha}{\sum^{p}_{\beta=1} F_\beta}.
\eea

Now that some useful tools have been introduced, let us investigate 14 of the Fano 3-folds in detail. For each Fano 3-fold a tiling and a set of Chern-Simons levels are given leading to a gauge theory with the corresponding CY 4-fold as its mesonic moduli space. The forward algorithm is applied to each tiling and each model's mesonic and baryonic symmetries are investigated. Volume minimisation is used to find the R-charges of the quiver fields. The refined hilbert series of the master space and mesonic moduli space are then computed and the generators of the mesonic moduli space are given along with the lattice of generators.

\section{$\cB_4$ (Toric Fano 24): $\BP^2 \times \BP^1$ (The $M^{1,1,1}$ Theory)}
\begin{figure}[ht]
\includegraphics[totalheight=5.5cm]{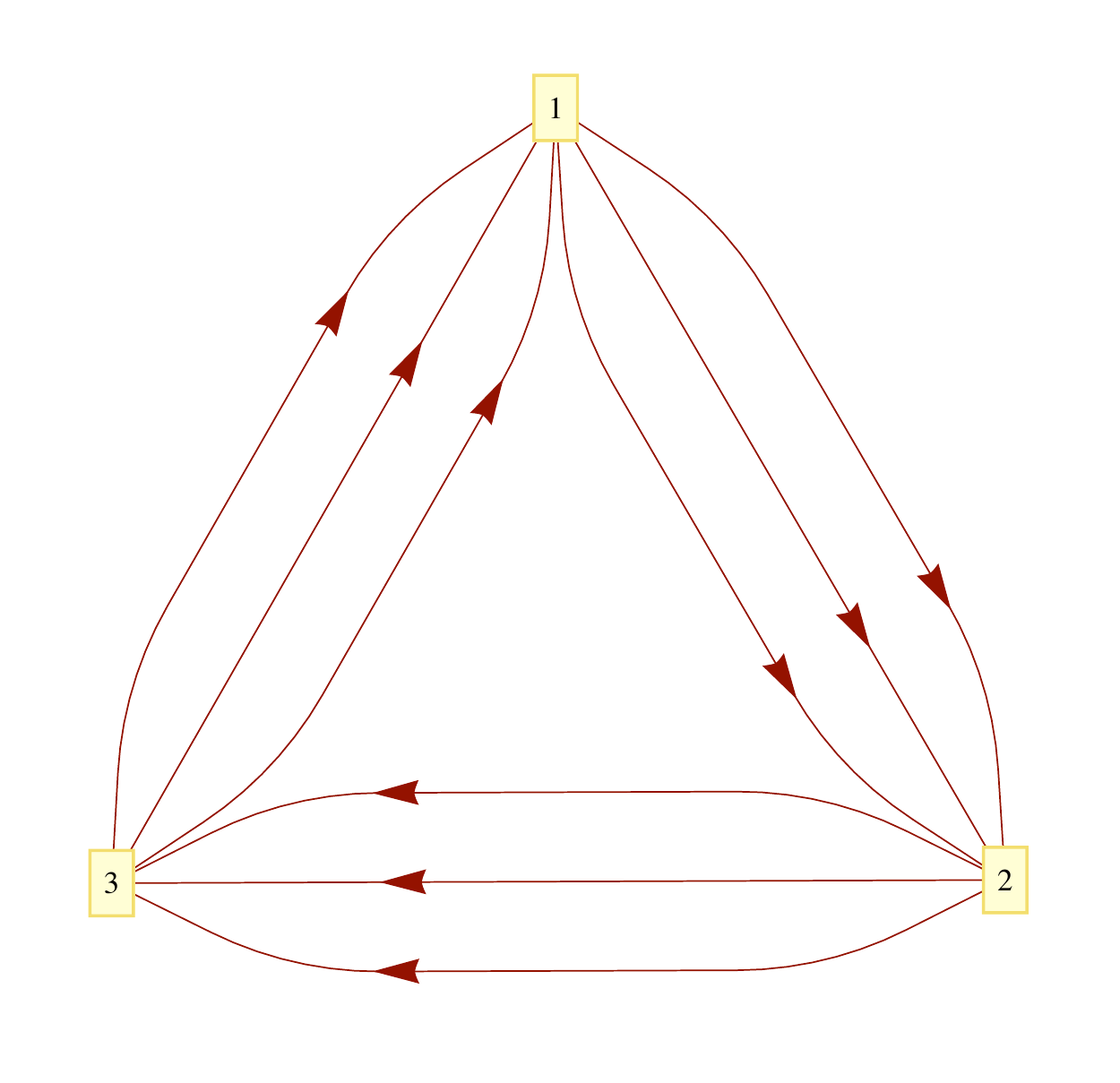}
\hskip 2cm
\includegraphics[totalheight=5.5cm]{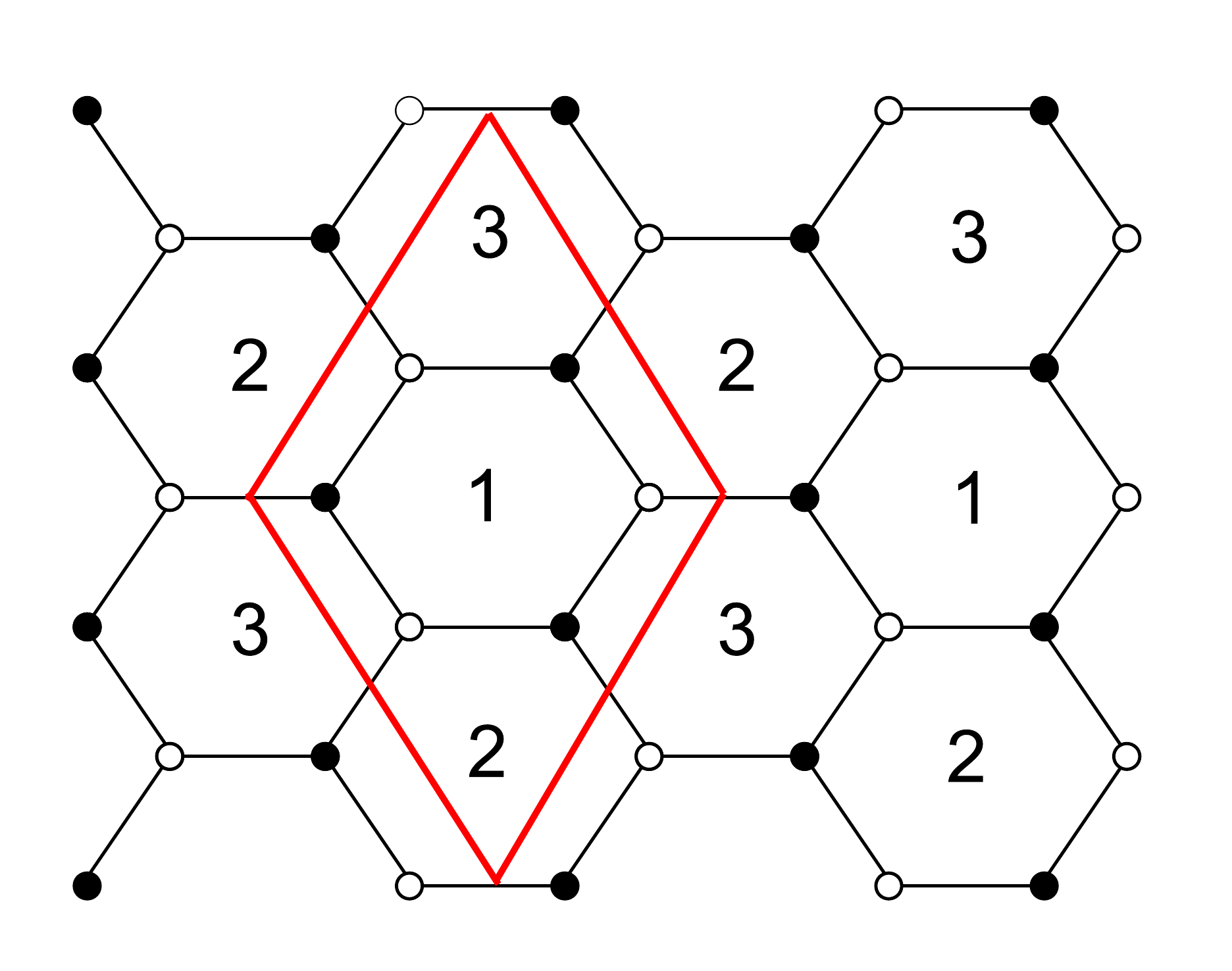}
\caption{(i) Quiver diagram of the $M^{1,1,1}$  theory.\ (ii) Tiling of the $M^{1,1,1}$  theory.}
  \label{f:m111}
\end{figure}

In this section, the theory living on M2-branes placed at the tip of the cone over $\cB_4$, also known as $M^{1,1,1}$, is discussed. 
The $M^{1,1,1}$ theory \cite{Hanany:2008cd, Martelli:2008si, Hanany:2008fj,phase, Hanany:2009vx,Davey:2009qx,Davey:2009bp,Davey:2009et, Petrini:2009ur, Fabbri:1999hw} has 3 gauge groups and 9 chiral multiplets, which are denoted as $X_{12}^i, X_{23}^i, X_{31}^i$ (with $i=1,2,3$). The quiver diagram and tiling are given in \fref{f:m111}.  
Note that in $3+1$ dimensions, this tiling corresponds to the gauge theory living on D3-branes probing the cone over the $dP_0$ surface.
The superpotential is given by
\bea
W= \tr \left( \epsilon_{ijk} X^i_{12} X^j_{23} X^k_{31} \right)~. 
\eea
The CS levels are $\vec{k} = (1, -2, 1) $.

\comment{
\begin{figure}[ht]
\begin{center}
\includegraphics[totalheight=8cm]{fdM111.pdf}
 \caption{The fundamental domain of tiling for the $M^{1,1,1}$ theory.}
  \label{f:fdm111}
  \end{center}
\end{figure}
}

\paragraph{The Kasteleyn matrix.} The Chern-Simons levels $k_a$ for gauge groups can be written in terms of the integers $n_i$ that correspond to Chern-Simons variables for fields. The two variables are related by the incidence matrix $k_a = \sum_i d_{ai} n_i$ \cite{phase}. In this case $n^i_{jk}$ are related to the levels $k_a$ by
\bea
\begin{array}{rcl}
\text{Gauge group 1~:} \qquad k_1   &=   n^{1}_{12} + n^{2}_{12} + n^{3}_{12} - n^{1}_{31} - n^{2}_{31} - n^{3}_{31}~, \nn \\
\text{Gauge group 2~:} \qquad k_2   &=   n^{1}_{23} + n^{2}_{23} + n^{3}_{23} - n^{1}_{12} - n^{2}_{12} - n^{3}_{12}~, \nn \\
\text{Gauge group 3~:} \qquad k_3   &=   n^{1}_{31} + n^{2}_{31} + n^{3}_{31} - n^{1}_{23} - n^{2}_{23} - n^{3}_{23}~.  
\label{e:kafano24}
\end{array}
\eea  
and a particular choice of $n^i_{jk}$ can be
\bea
n^{1}_{12}=-n^{1}_{23}= 1,\qquad n^i_{jk}=0 \;\;\text{otherwise}~.
\eea
The Kasteleyn matrix is calculated as follows. Since the fundamental domain contains 3 pairs of black and white nodes, the Kasteleyn matrix is a $3 \times 3$ matrix: 
\bea
K =   \left(
\begin{array}{c|ccc}
& w_1 & w_2 & w_3\\
\hline
b_1 & z^{n^{1}_{31}} &  z^{n^{3}_{12}} & y z^{n^{2}_{23}} \\
b_2 &  \frac{1}{x} z^{n^{3}_{23}} & z^{n^{2}_{31}} & z^{n^{1}_{12}} \\
b_3 &  z^{n^{2}_{12}} &  \frac{x}{y} z^{n^{1}_{23}} & z^{n^{3}_{31}} \end{array}
\right) ~.
\label{e:kastfano24}
\eea
The permanent of the Kasteleyn matrix is given by
\bea
\mathrm{perm}(K) &=&   x y^{-1} z^{(n^{1}_{12} + n^{1}_{23} + n^{1}_{31})}
+ y z^{(n^{2}_{12} + n^{2}_{23} + n^{2}_{31})}
+ x^{-1} z^{(n^{3}_{12} + n^{3}_{23} + n^{3}_{31})} \nn \\
&+& z^{(n^{1}_{12} + n^{2}_{12} + n^{3}_{12})}
+ z^{(n^{1}_{23} + n^{2}_{23} + n^{3}_{23})}
+ z^{(n^{1}_{31} + n^{2}_{31} + n^{3}_{31})}\nn \\&=&
 x y^{-1}+ y+ x^{-1}+z + z^{-1}+1\;\nn \\
&& \text{(for $n^{1}_{12} = - n^{1}_{23} = 1,~ n^i_{jk}=0 \;
\text{otherwise}$)~.}
\label{e:charpolyfano24}
\eea

\paragraph{The perfect matchings.} From (\ref{e:charpolyfano24}), the perfect matchings can be taken to be
\bea 
&& p_1 = \left\{X^1_{12}, X^1_{23}, X^1_{31}\right\}, \;\; p_2 = \left\{X^2_{12}, X^2_{23}, X^2_{31}\right\}, \;\; p_3 = \left\{X^3_{12}, X^3_{23}, X^3_{31}\right\}, \nn \\  
&&  r_1 = \left\{X^1_{12}, X^2_{12}, X^3_{12}\right\}, \;\; r_2 = \left\{X^1_{23}, X^2_{23}, X^3_{23}\right\}, \;\; v_1 = \left\{X^1_{31}, X^2_{31}, X^3_{31}\right\}\ . \qquad
\eea
From (\ref{e:charpolyfano24}), it can be seen that the perfect matchings $p_1,~p_2,~p_3,~r_1,~r_2$ correspond to external points in the toric diagram, whereas $v_1$ corresponds to the internal point.
The chiral fields can be written in terms of perfect matchings as follows:
\bea
&& X^1_{12} = p_1 r_1 , \quad X^1_{23} = p_1 r_2 , \quad X^1_{31} = p_1 v_1 \nn \\
&& X^2_{12} = p_2 r_1 , \quad X^2_{23} = p_2 r_2 , \quad X^2_{31} = p_2 v_1 \nn \\
&& X^3_{12} = p_3 r_1 , \quad X^3_{23} = p_3 r_2 , \quad X^3_{31} = p_3 v_1 ~.
\eea
These pieces of information can be collected in the perfect matching matrix:
\beq
P=\left(\begin{array} {c|cccccc}
  \;& p_1 & p_2 & p_3 & r_1 & r_2 & v_1\\
  \hline 
  X^{1}_{12}& 1&0&0&1&0&0\\
  X^{1}_{23}& 1&0&0&0&1&0\\
  X^{1}_{31}& 1&0&0&0&0&1\\
  X^{2}_{12}& 0&1&0&1&0&0\\
  X^{2}_{23}& 0&1&0&0&1&0\\
  X^{2}_{31}& 0&1&0&0&0&1\\
  X^{3}_{12}& 0&0&1&1&0&0\\
  X^{3}_{23}& 0&0&1&0&1&0\\
  X^{3}_{31}& 0&0&1&0&0&1\\
  \end{array}
\right).
\eeq
The null space of $P$ is 1 dimensional and is spanned by the vector that can be written in the row of the following charge matrix:
\be
Q_F =   \left(
\begin{array}{cccccc}
1,&1,&1,&-1,&-1,&-1
\end{array}
\right)~.  \label{e:qffano24}
\ee
Hence, among the perfect matchings there is a relation, which is given by:
\bea
p_1 + p_2 + p_3 - r_1 - r_2 - v_1 = 0.
\label{e:relpmfano24}
\eea

\paragraph{The toric diagram.} The toric diagram of this model is constructed using two methods:
\begin{itemize}
\item {\bf The charge matrices.}
Because the number of gauge groups of this model is $G = 3$, there is $G-2 =1$ baryonic charge coming from the D-terms. The baryonic charges of the perfect matchings are collected in the $Q_D$ matrix:
\be
Q_D =   \left(
\begin{array}{cccccc}
0, & 0, & 0, & 1,& 1,& -2
\end{array}
\right). \label{e:qdfano24}
\ee
\eref{e:qffano24} and \eref{e:qdfano24} are combined in the total charge matrix, $Q_t$, that contains all the charges of the perfect matchings that need to be integrated over to compute the Hilbert series of the mesonic moduli space:
\be
Q_t = { \Blue Q_F \choose \Green Q_D \Black } =   \left( 
\begin{array}{cccccc} \Blue
1 & 1& 1& -1& -1& -1 \\ \Green
0 & 0& 0&  1&  1& -2 \Black
\end{array}
\right).
\label{e:qtfano24}
\ee
The $G_t$ matrix is obtained and, after the removal of the first row, a matrix whose columns represent the coordinates of the toric diagram is generated:
\bea
G'_t = \left(
\begin{array}{cccccc}
 1 & -1 &  0 &  0 &  0 & 0 \\ 
 0 &  1 & -1 &  0 &  0 & 0 \\
 0 &  0 &  0 &  1 & -1 & 0 
\end{array}
\right)~. \label{e:toricdiafano24}
\eea
The toric diagram is presented in Figure \ref{f:tdtoricfano24}.  Note that the 4 blue points form the toric diagram of $\BP^2$, and the 2 black points together with the blue internal point form the toric diagram of $\BP^1$.  This indicates that this theory arises from the cone over $\BP^2 \times \BP^1$.

\begin{figure}[ht]
\begin{center}
  \includegraphics[totalheight=3.5cm]{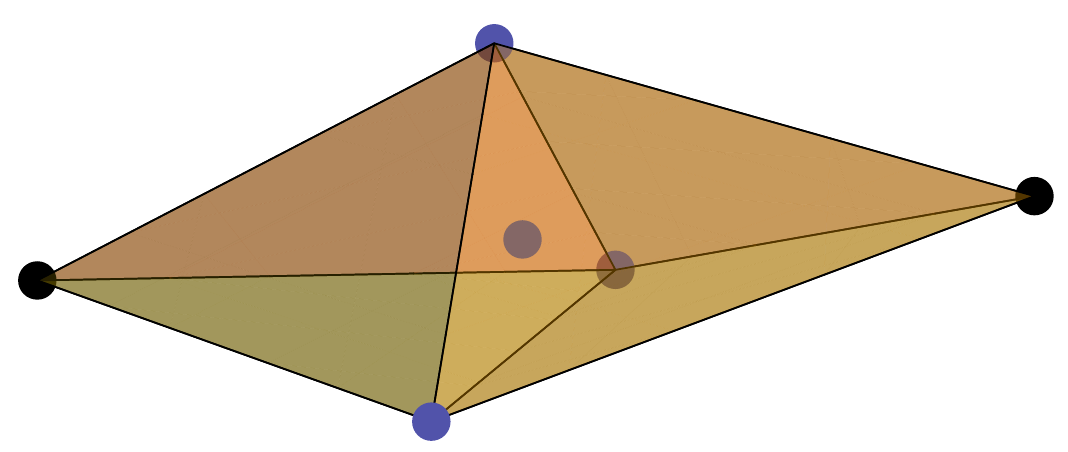}
 \caption{The toric diagram of the $M^{1,1,1}$ theory.}
  \label{f:tdtoricfano24}
\end{center}
\end{figure}

\item {\bf The Kasteleyn matrix.} The powers of $x, y, z$ in each term of \eref{e:charpolyfano24} give the coordinates of each point in the toric diagram. These points are collected in the columns of the following $G_K$ matrix: 
\bea
G_K = \left(
\begin{array}{cccccc}
  1 & 0 & -1 & 0 &  0 & 0 \\
 -1 & 1 &  0 & 0 &  0 & 0\\
  0 & 0 &  0 & 1 & -1 & 0
\end{array}
\right)~.
\eea
Note that the toric diagrams constructed from $G_K$ and from $G'_t$ are the same up to a transformation {\footnotesize $\cT = \left(
\begin{array}{ccc}
  1 & 1 & 0 \\
 -1 & 0 & 0 \\
  0 & 0 & 1
\end{array}
\right) \in GL(3, \BZ)$},
where $G_K = \cT \cdot G'_t$.
\end{itemize}

\paragraph{The baryonic charges.} Since the toric diagram of this model has 5 external points, the number of baryonic symmetries is precisely $5-4 = 1$, which shall be denoted as $U(1)_{B}$. From the discussion on the charge matrices above, it is understood that the baryonic charge of the perfect matchings comes from the row of the $Q_D$ matrix.

\paragraph{The global symmetry.} It is observed from \eref{e:qtfano24} that $Q_t$ has a pair and a `triplet' of repeated columns. Since the total rank of the mesonic symmetries is 4, the mesonic symmetry of this model is $SU(3)\times SU(2)\times U(1)$.  This can also be seen from the $G'_t$ matrix \eref{e:toricdiafano24} by noticing that that the first two rows contain weights of $SU(3)$ and the third row contains the weight of $SU(2)$. 
Since there is precisely one factor of $U(1)$, this can be unambiguously identified with the R-symmetry of the theory. 
The global symmetry of this theory is a product of mesonic and baryonic symmetries: $SU(3)\times SU(2)\times U(1)_R \times U(1)_B$.

The R-charge of each perfect matching can be determined as follows.  

\paragraph{R-charges of the perfect matchings.}  
In order to compute the R-charge of the perfect matching $p_1$ the refined Hilbert series of the mesonic moduli space must be found.
Since the non-abelian weights do not play any role in the volume minimisation, they are set to unity.  Also, since the R-charge of the internal perfect matching $v_1$ is zero, the corresponding fugacity can be set to unity.  We denote by $s_1$ the R-charge fugacity of $p_1,p_2,p_3$, and by $s_2$ the R-charge fugacity of $r_1, r_2$. From the $Q_t$ matrix (\ref{e:qtfano24}), the Hilbert series of the mesonic moduli space $\CMm= \BC^6//Q_t$ is given by:
\bea
\gm (s_1,s_2; \cB_4) &=& \oint \limits_{|z| =1} {\frac{\ud z}{2 \pi i z }}\oint \limits_{|b| =1} {\frac{\ud b}{2 \pi i b }} \frac{1}{\left(1- s_1 z\right)^3\left(1-\frac{s_2 b}{ z }\right)^2\left(1- \frac{1}{z b^2}\right)}\nn \\
&=& \frac{1 + 26 s^3_1 s^2_2 + 26 s_1^6 s_2^4 +  s^9_1 s^6_2}{(1- s^3_1 s^2_2)^4},~
\label{e:HStrfano24}
\eea
where $z$ is the fugacity associated with the $Q_F$ charges, $b$ is the fugacity associated with the $Q_D$ charges. The computation shows that the result of the integration depends only on a specific combination of the $s_{\alpha}$'s, namely $s_1^3 s^2_2$. 
This is not surprising, since there exists only one $U(1)$ in the mesonic symmetry (and non-abelian fugacities have been set to unity).
Therefore, a new fugacity $t$ can be defined such that:
\bea
t^{18} = s^3_1 s^2_2~,
\eea
where the power 18 is introduced for convenience.  The Hilbert series of the mesonic moduli space can be rewritten in terms of $t$ as:
\bea
\gm (t; \cB_4) = \frac{1 + 26 t^{18} + 26 t^{36} +  t^{54}}{(1-t^{18})^4}~.
\label{e:HStrtfano24}
\eea
Each term in the superpotential is actually the product of all the external perfect matchings and, therefore, it scales like $t^{18}$.
Since the R-charge of the superpotential is 2, the R-charge associated with $t$ is $1/9$. 
In other words, let us write:
\bea
t = e^{- \mu/9}~,
\eea 
where $\mu$ is the chemical potential of the R-charge associated with $t$.

Next, let us compute the Hilbert series of the divisor corresponding to $p_1$ (which is referred to as $D_1$). This would be the integral over the baryonic fugacities of the Hilbert series of the space of perfect matchings multiplied by the inverse of the fugacity relative to $p_1$:
\bea
g (D_1; s_1, s_2; \cB_4) &=& \oint \limits_{|z| =1} {\frac{\ud z}{2 \pi i z }}\oint \limits_{|b| =1} {\frac{\ud b}{2 \pi i b }} \frac{(s_1 z)^{-1}}{\left(1- s_1 z\right)^3\left(1-\frac{s_2 b}{z}\right)^2\left(1- \frac{1}{z b^2}\right)},\nn \\
\label{e:HSD1fano24}
\eea
where again the non-abelian fugacities have been set to unity as they do not play a role in the computation of volumes. As before, the result of the integration depends only on the product of $s_\alpha$'s and, therefore, it can be rewritten in terms of $t$:
\bea
g (D_1; t ; \cB_4) &=& \frac{3(1 + 11t^{18} + 6t^{36})}{(1-t^{18})^4}~.
\eea
Thus, the R-charge of the perfect matching $p_1$ is given by:
\bea
\lim_{\mu\rightarrow0}\frac{1}{\mu} \left[ \frac{g(D_1; e^{- \mu/9 }; \cB_4) }{\gm(e^{-\mu/9}; \cB_4)}- 1 \right]= \frac{4}{9}~.
\eea
The computations for the other perfect matchings can be done in a similar way.  The results, as well as the charges under the other global symmetries,  are presented in Table \ref{t:chargefano24}:
\begin{table}[h]
 \begin{center}  
  \begin{tabular}{|c||c|c|c|c|c|c|}
  \hline
  \;& $SU(3)$&$SU(2)$&$U(1)_R$&$U(1)_B$&fugacity\\
  \hline  \hline 
  $p_1$&$(1,0) $&$0$&4/9&0&$t^4 y_1  $\\
  \hline
  $p_2$&$(-1,1)$&$0$&4/9&0&$t^4 y_2 / y_1 $\\
  \hline
  $p_3$&$(0,-1)$&$0$&4/9&0&$ t^4 / y_2 $\\
  \hline
  $r_1$&$(0,0)$&1&1/3&$1$&$ t^3 x  b$ \\
  \hline
  $r_2$&$(0,0)$&$-1$&1/3&$1$&$t^3 b/x$\\
  \hline
  $v_1$&$(0,0)$&0&0&$-2$&$  1/b^2$\\
  \hline
  \end{tabular}
  \end{center}
\caption{Charges of the perfect matchings under the global symmetry of the $M^{1,1,1}$ theory. Here $t$ is the fugacity of the R-charge (in units of $1/9$), $y_1,y_2$ are the weights of the $SU(3)$ symmetry, $x$ is the weight of the $SU(2)$ symmetry and $b$ is the fugacity of the $U(1)_B$ symmetry.  The notation $(a,b)$ is used to represent a weight of $SU(3)$.}
\label{t:chargefano24}
\end{table}

\begin{table}[h]
 \begin{center}  
  \begin{tabular}{|c||c|}
  \hline
  \; Quiver fields &R-charge\\
  \hline  \hline 
  $X^{i}_{12}$ &  $7/9$\\
  \hline
  $X^{i}_{23}$ &  $7/9$\\
  \hline
  $X^{i}_{31}$ &  $4/9$\\
  \hline
  \end{tabular}
  \end{center}
\caption{R-charges of the quiver fields for the $M^{1,1,1}$ theory.}
\label{t:Rgenfano24}
\end{table}

\paragraph{The Hilbert series.} The coherent component of the Master space is generated by the perfect matchings which are subject to the relation (\ref{e:relpmfano24}):
\bea
\firr{\cB_4} = \BC^6//Q_F~.  \label{firrfano24}
\eea
It follows that the Hilbert series of the coherent component of the Master space of this model can be obtained by integrating the Hilbert series of the space of the perfect matchings over the fugacity $z$:
{\small
\bea
g^{\firr{}} (t, x, y_1, y_2, b; \cB_4) &=& \oint \limits_{|z| =1} {\frac{\ud z}{2 \pi i z }} \frac{1}{\left(1- t^4 y_1 z\right)\left(1-\frac{t^4 y_2 z}{y_1}\right)\left(1-\frac{t^4 z}{y_2}\right)}\nn \\
&\times& \frac{1}{\left(1-\frac{t^3 x b}{ z }\right)\left(1-\frac{t^3 b}{x z}\right)\left(1- \frac{1}{z b^2}\right)}~. \nn \\
\label{e:HSmasterfano24}
\eea}
The unrefined version of the result of the integration can be written as:
{\small
\bea
g^{\firr{}} (t, 1, 1, 1, 1; \cB_4) &=& \frac{1 - 6 t^{11} - 3 t^{14} + 2 t^{15} + 12 t^{18} + 2 t^{21} - 3 t^{22} - 
 6 t^{25} + t^{36}}{\left(1-t^4\right)^3\left(1-t^7\right)^6}~, \nn \\
\eea}
which allows us to see that the coherent component of the Master space is a 5 dimensional Calabi-Yau space.
Integrating (\ref{e:HSmasterfano24}) over the baryonic charge $b$ gives the Hilbert series of the mesonic moduli space:
{\small \bea
\gm (t,x,y_1,y_2; \cB_4) &=& \oint_{|b|=1} \frac{\ud b}{2\pi i b}\;\; g^{\firr{}} (t, x, y_1, y_2, b; \cB_4) \nn \\
&=&\frac{P(t,x,y_1,y_2; \cB_4)}{\left(1-\frac{t^{18} y^3_1}{x^2}\right)\left(1-t^{18} x^2 y^3_1\right)\left(1-\frac{t^{18} x^2}{y^3_2}\right)\left(1-\frac{t^{18}}{x^2 y^3_2}\right)}\nn \\
&\times& \frac{1}{\left(1-\frac{t^{18} y^2_2}{x^2 y^3_1}\right)\left(1-\frac{t^{18} x^2 y^3_2}{y^3_1}\right)} \nn \\ 
&=& \sum^{\infty}_{n=0}\left[3n,0;2n\right]t^{18n}~,
\label{e:HSmesfano24}
\eea}
where $P(t,x,y_1,y_2; \cB_4)$ is a polynomial of degree 90.
 

The generators of the mesonic moduli space can be determined from the plethystic logarithm of (\ref{e:HSmesfano24}):
\bea
\PL[\gm (t,x,y_1,y_2, \cB_4)] = \left[3,0;2\right]t^{18} -O(t^{36})~.  \label{plm111}
\eea
The 30 generators can be written in terms of perfect matchings as:
\bea
p_i ~p_j ~p_k ~r_l ~r_m ~v_1~,  \label{genm111}
\eea
where $i,j,k=1,2,3$ and $l,m=1,2$. 
As a check, let us note that $p_i p_j p_k$ has $\frac{3 \times 4 \times 5}{3!} = 10$ independent components and $r_l r_m$ has $\frac{2 \times 3}{2!}  = 3$ independent components, which implies that there are indeed 30 generators.  

All of the generators of the mesonic moduli space have R-charge equal to 2.

\begin{figure}[ht]
\begin{center}
\includegraphics[totalheight=6cm]{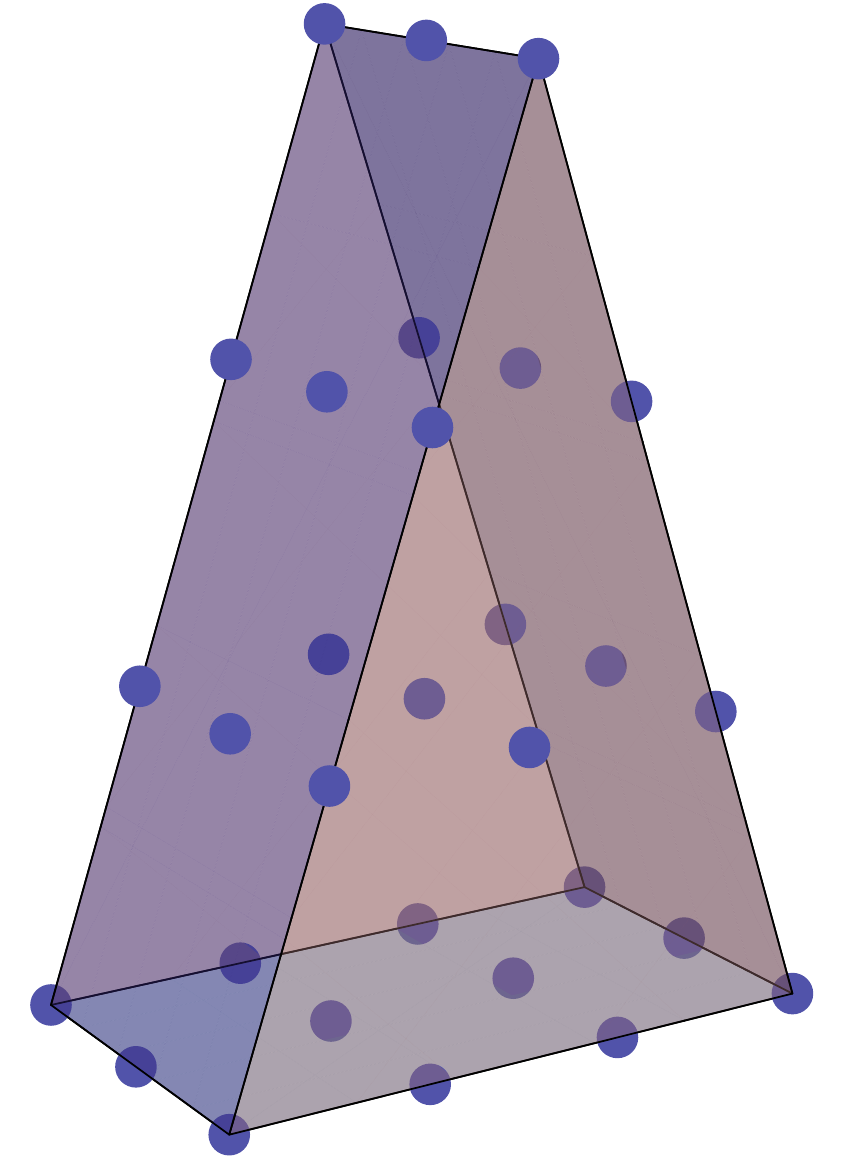}
\caption{The lattice of generators of the $M^{1,1,1}$ theory.}
  \label{f:lattm111}
\end{center}
\end{figure}

\paragraph{The lattice of generators.} The generators \eref{genm111} can be drawn in a lattice (\fref{f:lattm111}) by plotting the powers of 
the weights of the characters in \eref{plm111}.
Note that the lattice of generators is the dual of the toric diagram (nodes are dual to faces and edges are dual to edges): the toric diagram has 5 nodes (external points of the polytope), 9 edges and 6 faces, whereas the generators form a convex polytope that has 6 nodes (corners of the polytope), 9 edges and 5 faces. 

\comment{
\begin{table}[h]
 \begin{center}  
  \begin{tabular}{|c||c|}
  \hline
  \; Generators &$U(1)_R$\\
  \hline  \hline 
  $p_i ~p_j ~p_k ~r_l ~r_m ~v_1$ & 2 \\
  \hline
  \end{tabular}
  \end{center}
\caption{R-charges of the generators of the mesonic moduli space for the $M^{1,1,1}$ theory.}
\label{t:Rgenfano24}
\end{table}}


\section{$\cC_3$ (Toric Fano 62): $\BP^1 \times \BP^1 \times \BP^1$ (The $Q^{1,1,1}/\BZ_2$ Theory)}
This theory was introduced in \cite{Hanany:2008cd, Hanany:2008fj} as a modified $\BF_0$ theory. In the following subsections, two phases of this theory are examined in detail \cite{phase,Davey:2009qx,Davey:2009bp,Davey:2009et,Franco:2009sp,Amariti:2009rb}.

\subsection{Phase I of The $Q^{1,1,1}/\BZ_2$ Theory}
\begin{figure}[ht]
\begin{center}
  \vskip 0cm
  \hskip -7cm
  \includegraphics[totalheight=5.7cm]{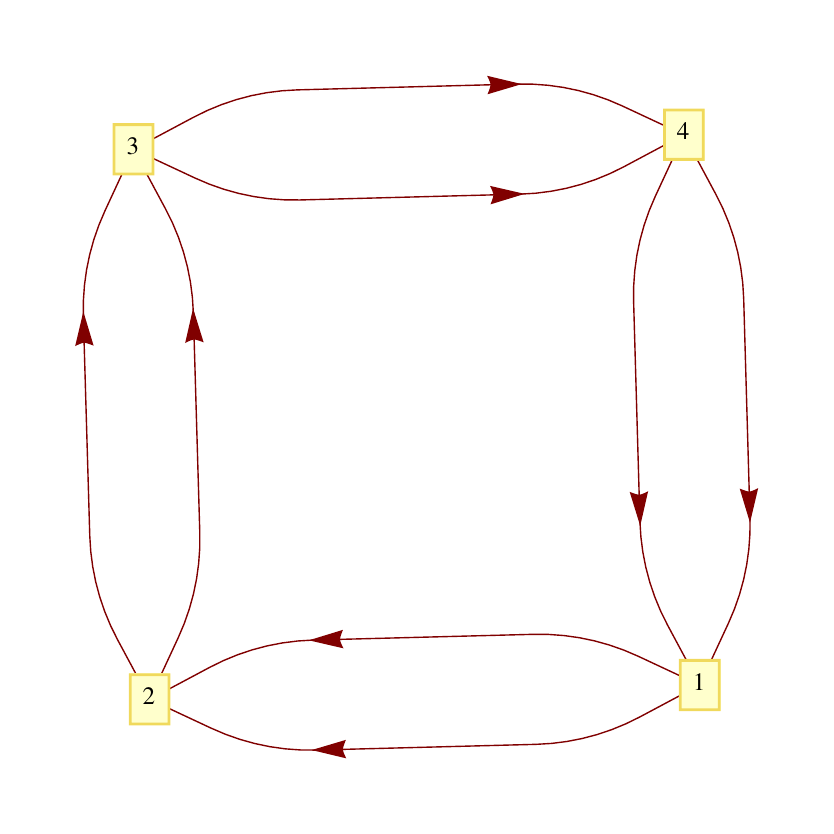}
    \vskip -5.5cm
  \hskip 8cm
  \includegraphics[totalheight=5cm]{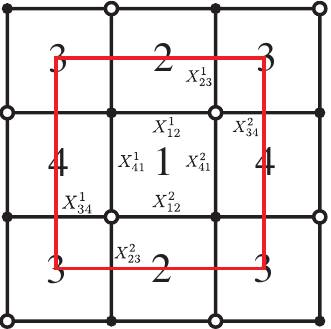}
 \caption{(i) Quiver for Phase I of $Q^{1,1,1}/\BZ_2$. \ (ii) Tiling for Phase I of $Q^{1,1,1}/\BZ_2$.}
  \label{f:phase1f0}
\end{center}
\end{figure}

This theory has 4 gauge groups and has bi-fundamental fields $X_{12}^i$, $X_{23}^i$, $X_{34}^i$ and $X_{41}^i$ (with $i=1,2$). The superpotential is given by
\bea
W = \epsilon_{ij} \epsilon_{pq} \tr(X_{12}^i X_{23}^p X_{34}^j X_{41}^q)~. \label{suppotph1q111z2}
\eea
The quiver diagram and tiling are shown in Figure \ref{f:phase1f0}.  
The fields are assigned to the edges in the tiling according to \fref{f:phase1f0} (ii).
Note that, in 3+1 dimensions, this quiver and this tiling correspond to Phase I of the $\BF_0$ theory \cite{ Forcella:2008ng, master, Butti:2007jv}.
The CS levels are chosen to be $\vec{k} = (1,-1,-1,1)$.

\comment{
\begin{figure}
\begin{center}
   \includegraphics[totalheight=8.0cm]{fdph1f0.pdf}
 \caption{The fundamental domain of tiling for Phase I of $Q^{1,1,1}/\BZ_2$.}
  \label{f:fdph1f0}
\end{center}
\end{figure}
}

\paragraph{A discrete symmetry of $\BF_0$.}  Since the toric diagram of the $\BF_0$ theory is a square (with internal points at the origin), the  $\BF_0$ theory is expected to possess the dihedral symmetry $D_4$.  Indeed, the $D_4$ symmetry appears in the brane tiling and hence at the level of the Lagrangian.  This can be shown as follows.

\begin{figure}[h]
\begin{center}
   \includegraphics[totalheight=5.0cm]{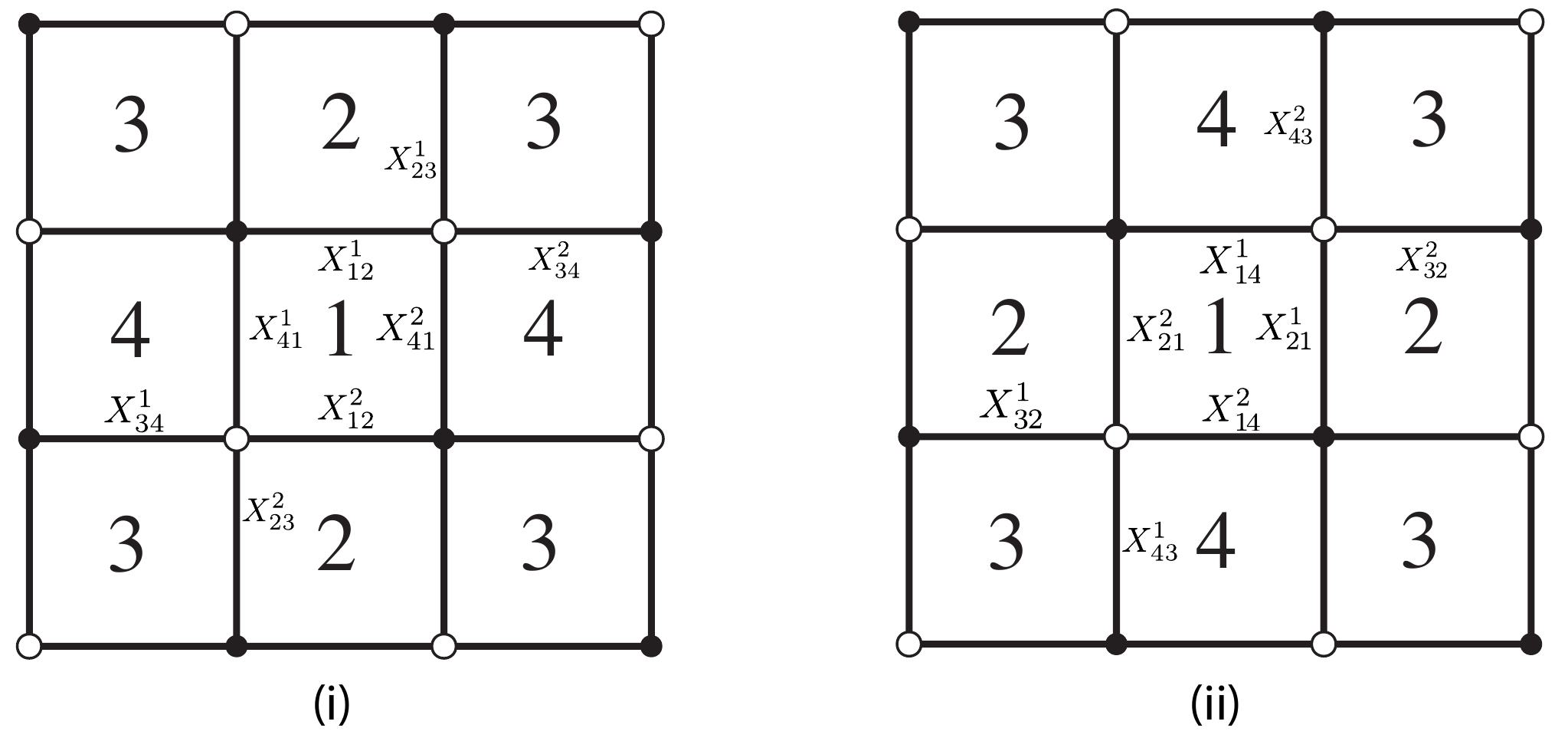}
 \caption{(i) The original tiling of Phase I of $\BF_0$.  (ii) The tiling after a 90 degree rotation around tile 1.  This tiling also gives the moduli space of $\BF_0$. Note that it is related by the original tiling by the charge conjugation (which corresponds to swapping white and black nodes).}
  \label{f:D4rot}
\end{center}
\end{figure}

\begin{figure}[h]
\begin{center}
   \includegraphics[totalheight=5.0cm]{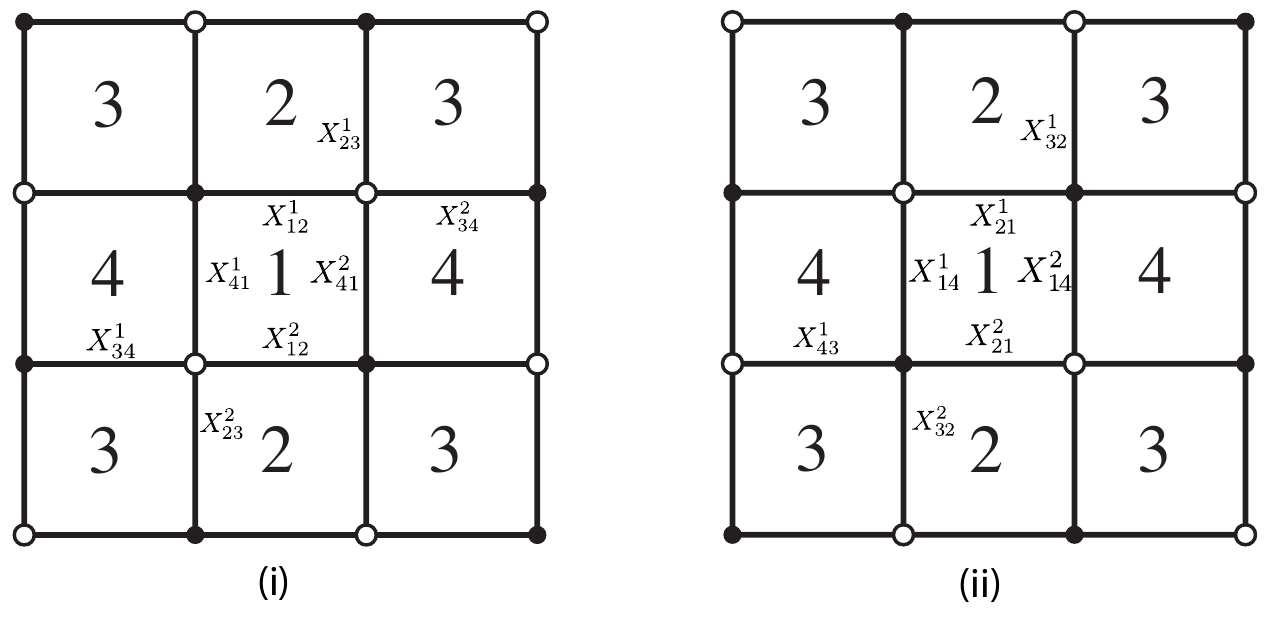}
 \caption{(i) The original tiling of Phase I of $\BF_0$.  (ii) The tiling after the rotation with respect to the vertical line passing through the centre. This tiling also gives the moduli space of $\BF_0$. Note that it is related by the original tiling by the charge conjugation (which corresponds to swapping white and black nodes).}
  \label{f:D4refl}
\end{center}
\end{figure}

The $D_4$ symmetry is generated by a rotation of the tiling by 90 degrees and also a reflection around an axis of symmetry. The symmetry of the tiling can be checked by explicitly considering the actions of these two generators on the tiling (Figures \ref{f:D4rot} and \ref{f:D4refl}).  As a result of these actions, even though some of the labels of the gauge groups and the fields get permuted, the moduli space corresponding to these resulting brane tilings is still $\BF_0$.  Hence, Phase I of $\BF_0$ indeed possess the $D_4$ symmetry.  (Note that a similar argument can also be applied to Phase II of $\BF_0$.)


\paragraph{The Kasteleyn matrix.} The CS levels can be written in terms of the integers $n^i_{jk}$ as:
\bea \label{e:csfano62phI}
\begin{array}{rcl}
\text{Gauge group 1~:} \qquad k_1 &=  n^{1}_{12} + n^{2}_{12} - n^{1}_{41} - n^{2}_{41} ~, \\
\text{Gauge group 2~:} \qquad k_2 &=  n^{1}_{23} + n^{2}_{23} - n^{1}_{12} - n^{2}_{12} ~, \\
\text{Gauge group 3~:} \qquad k_3 &=  n^{1}_{34} + n^{2}_{34} - n^{1}_{23} - n^{2}_{23} ~, \\
\text{Gauge group 4~:} \qquad k_4 &=  n^{1}_{41} + n^{2}_{41} - n^{1}_{34} - n^{2}_{34} ~.
\end{array}
\eea  

\noindent The Kasteleyn matrix can be computed from the tiling. The fundamental domain contains two black nodes and two white nodes and, therefore, the Kasteleyn matrix is a $2\times 2$ matrix:
\be
K =   \left(
\begin{array}{c|cc}
& w_1 & w_2 \\
\hline
b_1 & z^{n^{2}_{12}} +  \frac{1}{x} z^{n^{1}_{34}} &\ z^{n^{2}_{41}} +  \frac{1}{y} z^{n^{1}_{23}}   \\
b_2 & z^{n^{1}_{41}} +  y z^{n^{2}_{23}} &\ z^{n^{1}_{12}} + x z^{n^{2}_{34}}  
\end{array}
\right) ~. \nn
\ee
The permanent of this matrix is given by
\bea 
\perm~K &=&  x z^{(n^{2}_{12} + n^{2}_{34})} +  x^{-1} z^{(n^{1}_{12} + n^{1}_{34})} + y z^{(n^{2}_{23} + n^{2}_{41})} + y^{-1} z^{(n^{1}_{23} + n^{1}_{41})}\nn \\
&&+ z^{(n^{1}_{12} + n^{2}_{12})} + z^{(n^{1}_{34} + n^{2}_{34})} +  z^{(n^{1}_{23} + n^{2}_{23})} + z^{(n^{1}_{41} + n^{2}_{41})} \nn \\ 
&=&  x  +  x^{-1} +  y  + y^{-1} + z+ z^{-1} +  2 \nn \\
&&\qquad \text{(for $n^{2}_{12} = - n^{2}_{34} = 1,~ n^i_{jk}=0 \; \text{otherwise}$)} ~. 
\label{permKph1f0}
\eea

\paragraph{The perfect matchings.} From \eref{permKph1f0}, the perfect matchings can be written as collections of fields as follows:
\bea 
&& p_1 = \{X^{2}_{12}, X^{2}_{34} \}, \;\; p_2 = \{ X^{1}_{12},  X^1_{34} \}, \;\; q_1=  \{ X^{2}_{23}, X^2_{41} \}, \;\; q_2 = \{ X^1_{23}, X^1_{41} \}, \nn \\  
&& r_1 = \{ X^1_{12},  X^2_{12} \},\;\; r_2 = \{X^{1}_{34}, X^{2}_{34} \}, \;\; 
v_1 = \{  X^1_{23},  X^2_{23} \}, \; \; v_2 = \{ X^{1}_{41}, X^2_{41}  \} \ . \qquad
\eea
The perfect matchings $p_i, q_i, r_i$ correspond to external points in the toric diagram, whereas the perfect matchings $v_i$ correspond to the internal point at the origin.
In turn the fields can be written in terms of perfect matchings:
\bea
&& X^2_{12} = p_1 r_1 , \quad X^1_{12} = p_2 r_1 , \quad X^2_{34} = p_1 r_2, \quad X^1_{34} = p_2 r_2, \nn \\
&& X^2_{23} = q_1 v_1, \quad X^1_{23} = q_2 v_1, \quad X^2_{41} = q_1 v_2, \quad X^1_{41} = q_2 v_2~.
\eea
This is summarized in the perfect matching matrix:
\beq
P=\left(\begin{array} {c|cccccccc}
  \;& p_1 & p_2 & q_1 & q_2 & r_1 & r_2 & v_1 & v_2\\
  \hline 
  X^{2}_{12}& 1&0&0&0&1&0&0&0\\
  X^{1}_{12}& 0&1&0&0&1&0&0&0\\
  X^{2}_{34}& 1&0&0&0&0&1&0&0\\
  X^{1}_{34}& 0&1&0&0&0&1&0&0\\
  X^{2}_{23}& 0&0&1&0&0&0&1&0\\
  X^{1}_{23}& 0&0&0&1&0&0&1&0\\
  X^{2}_{41}& 0&0&1&0&0&0&0&1\\
  X^{1}_{41}& 0&0&0&1&0&0&0&1
  \end{array}
\right).
\eeq
Basis vectors of the null space of $P$ are given in the rows of the charge matrix:
\be
Q_F =   \left(
\begin{array}{cccccccc}
 1 & 1 &  0 &  0 & -1 & -1 &  0 &  0 \\
 0 & 0 &  1 &  1 &  0 &  0 & -1 & -1
\end{array}
\right)~.  \label{qfph1q111z2}
\ee
Hence, the relations between the perfect matchings are given by
\bea
p_1+p_2-r_1-r_2 &=& 0~, \nn \\ 
q_1+q_2-v_1-v_2 &=& 0~. \label{relf0I}
\eea
Since the coherent component $\firr{}$ of the Master space is generated by the perfect matchings (subject to the relation \eref{relf0I}), it follows that 
\bea
\firr{} = \BC^8//Q_F~.  \label{firrS4}
\eea

\paragraph{The toric diagram.} The two methods of constructing the toric diagram are demonstrated. 
\begin{itemize}
\item{\bf The charge matrices.}  Since the number of gauge groups is $G=4$, there are $G-2 = 2$ baryonic symmetries coming from the D-terms.  The charges of the perfect matchings are collected in the $Q_D$ matrix:
\bea
Q_D = \left(
\begin{array}{cccccccc}
 0 & 0 &  0 &  0 &  1 &  1 & -2 &  0 \\
 0 & 0 &  0 &  0 &  0 &  0 &  1 & -1
 \end{array}
\right)~.  \label{qdph1q111z2}
\eea
From \eref{qfph1q111z2} and \eref{qdph1q111z2}, the total charge matrix is given by
\be
Q_t = { \Blue Q_F \choose \Green Q_D \Black } =   \left( 
\begin{array}{cccccccc} \Blue
 1 & 1 &  0 &  0 & -1 & -1 &  0 &  0 \\
 0 & 0 &  1 &  1 &  0 &  0 & -1 & -1 \\ \Green
 0 & 0 &  0 &  0 &  1 &  1 & -2 &  0 \\
 0 & 0 &  0 &  0 &  0 &  0 &  1 & -1
 \Black
\end{array}
\right) 
\label{qtph1q111z2}
\ee
The kernel of $Q_t$ gives the matrix $G_t$ and, after removing the first row, its columns give the coordinates of points in the toric diagram:  
\bea
G'_t = \left(
\begin{array}{cccccccc}
 1 &-1 & 0 & 0 & 0 & 0 & 0 & 0 \\
 0 & 0 & 1 &-1 & 0 & 0 & 0 & 0 \\
 0 & 0 & 0 & 0 & 1 &-1 & 0 & 0
\end{array}
\right)~.
\eea
The toric diagram is shown in Figure \ref{f:torq111z2}.  Note that the 3 blue points form the toric diagram of $\BP^1$, and so do the yellow points (together with the internal point) and the black points (together with the internal point).  The mesonic moduli space can be identified as the cone over $\BP^1 \times \BP^1 \times \BP^1$.
\begin{figure}[h]
\begin{center}
  \includegraphics[totalheight=5cm]{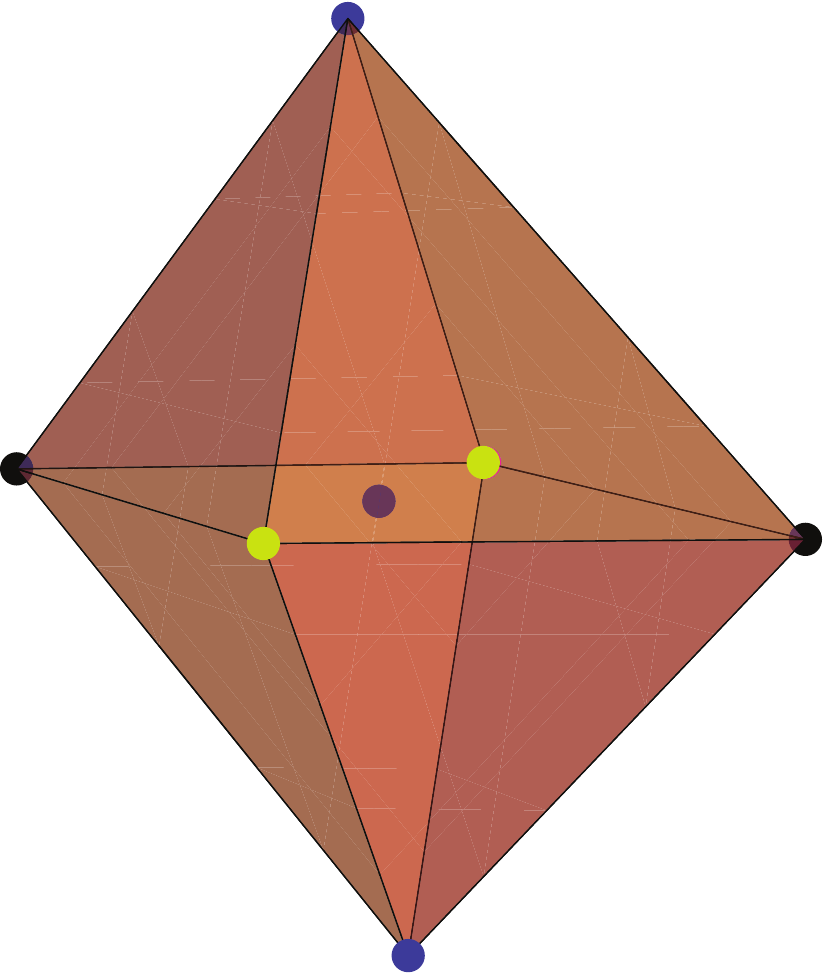}
 \caption{The toric diagram of $Q^{1,1,1}/\BZ_2$}
  \label{f:torq111z2}
\end{center}
\end{figure}

\item{\bf The Kasteleyn matrix.}  The powers of $x, y, z$ in each term of \eref{permKph1f0} give the coordinates of each point in the toric diagram.  These points are collected in the columns of the following $G_K$ matrix:
\bea
G_K = \left(
 \begin{array}{cccccccc}
 1 &-1 & 0 & 0 & 0 & 0 & 0 & 0 \\
 0 & 0 & 1 &-1 & 0 & 0 & 0 & 0 \\
 0 & 0 & 0 & 0 & 1 &-1 & 0 & 0
\end{array}
\right) = G'_t~.
\eea
Thus, the toric diagrams constructed from these two methods are indeed identical.
\end{itemize}

\paragraph{The baryonic charges.}  Since the toric diagram has 6 external points, this model has precisely $6-4 = 2$ baryonic symmetries, which are denoted by $U(1)_{B_1}$ and $U(1)_{B_2}$.  As discussed above, these symmetries arise from the D-terms.  The baryonic charges of the perfect matchings are given by the rows of the $Q_D$ matrix. 

\paragraph{The global symmetry.}  Since the $Q_t$ matrix has 3 pairs of repeated columns, it follows that the mesonic symmetry of this model is $SU(2)^3 \times U(1)_R$. This mesonic symmetry can also be seen from the $G_K$ (or $G'_t$) matrix by noticing that the three rows contain weights of $SU(2)$.
Since $v_1$ and $v_2$ are the perfect matchings corresponding to the internal point of the toric diagram, a zero R-charge is assigned to them.  
The remaining 6 external perfect matchings are completely symmetric and the requirement of R-charge 2 to the superpotential divides 2 equally among them, resulting in R-charge of 1/3 per each.

The global symmetry of the theory is a product of mesonic and baryonic symmetries: $SU(2)^3 \times U(1)_R \times U(1)_{B_1} \times U(1)_{B_2}$.
In Table \ref{chargeph1f0}, a consistent way of assigning charges to the perfect matchings under these global symmetries is presented.
\begin{table}[h]
 \begin{center}  
  \begin{tabular}{|c||c|c|c|c|c|c|c|}
  \hline
  \;& $SU(2)_{1}$&$SU(2)_{2}$&$SU(2)_{3}$&$U(1)_R$&$U(1)_{B_1}$&$U(1)_{B_2}$& fugacity\\
  \hline \hline  
 
  $p_1$& $1$&$0$&$0$ &$1/3$&$ 0$&$0$& $t x_1$ \\
  \hline
  $p_2$& $-1$&$0$&$0$&$1/3$&$ 0$&$0$ & $t / x_1$ \\
  \hline  
  $q_1$& $0$&$1$&$0$ &$1/3$&$ 0$&$0$  & $t x_2$\\
  \hline
  $q_2$& $0$&$-1$&$0$&$1/3$&$ 0$&$0$& $ t/ x_2$\\
  \hline
  $r_1$& $0$&$0$&$1$& $1/3$&$ 1$&$0$& $ t x_3  b_1$\\
  \hline
  $r_2$& $0$&$0$&$-1$&$1/3$&$ 1$&$0$& $ t b_1/ x_3 $\\
  \hline
  $v_1$& $0$&$0$&$0$   &$0$&$-2$&$1$&  $ b_2/ b^2_1$ \\
  \hline
  $v_2$& $0$&$0$&$0$   &$0$&$ 0$&$-1$ & $1 / b_2$ \\
  \hline
  $\Blue v_3$& $0$&$0$&$0$&$0$&$0$&$0$ & $1$ \\ 
  \hline
  \end{tabular}
  \end{center} \Black
  \caption{Charges under the global symmetry of the $Q^{1,1,1}/\BZ_2$ theory. Here $t$ is the fugacity of R-charge, $x_1,x_2,x_3$ are weights of $SU(2)_{1}, SU(2)_{2}, SU(2)_3$, and $b_1, b_2$ are baryonic fugacities of $U(1)_{B_1}, U(1)_{B_2}$. Note that the perfect matching $v_3$ (represented in blue) does not exist in Phase I but exists in Phase II.}
\label{chargeph1f0}
\end{table}

\begin{table}[h]
 \begin{center}  
  \begin{tabular}{|c||c|}
  \hline
  \; Quiver fields &R-charge\\
  \hline  \hline 
  $X^{i}_{12}, X^i_{34}$ &  $2/3$\\
  \hline
  $X^{i}_{23}, X^{i}_{41}$ &  $1/3$\\
  \hline
  \end{tabular}
  \end{center}
 \caption{R-charges of the quiver fields of $Q^{1,1,1}/\BZ_2$, Phase I.}
 \label{t:Rgenph1f0}
 \end{table}

\paragraph{The Hilbert series.} From \eref{firrS4}, the Hilbert series of the coherent component of the Master space is computed by integrating the Hilbert series of $\BC^8$ over the fugacities $z_1$ and $z_2$ associated with the $Q_F$ charges:
\bea
g^{\firr{}}_1 (t,x_1,x_2,x_3,b_1,b_2; \cC^{(I)}_3) &=& \frac{1}{(2\pi i)^2} \oint \limits_{|z_1| =1} {\frac{\ud z_1}{z_1 }} \oint \limits_{|z_2| =1} {\frac{\ud z_2}{z_2}}  \frac{1}{\left(1- t x_1 z_1  \right)\left(1- \frac{t z_1}{x_1}\right)\left(1- t x_2 z_2 \right)} \nn \\
&\times& \frac{1}{\left(1-\frac{t z_2}{x_2}\right)\left(1- \frac{t x_3 b_1}{z_1} \right)\left(1-\frac{t b_1}{x_3 z_1}\right)\left(1- \frac{b_2}{b^2_1 z_2} \right)\left(1- \frac{1}{z_2 b_2} \right)} \nn \\
&=& \frac{\left(1-\frac{t^2}{b_1^2}\right)}{\left(1-\frac{t x_2}{b_2}\right)\left(1-\frac{t}{x_2 b_2}\right)\left(1-\frac{t x_2 b_2}{b_1^2}\right)\left(1- \frac{t b_2}{x_2 b_1^2}\right)\left(1-\frac{t^2 b_1}{x_1 x_3}\right)} \nn \\
&\times& \frac{\left(1-t^4 b_1^2\right)}{\left(1-\frac{t^2 b_1 x_3}{x_1}\right)\left(1-\frac{t^2 b_1 x_1}{x_3}\right)\left(1-t^2 b_1 x_1 x_3\right)}~. \label{hsfirrs4}
\eea
The unrefined Hilbert series of the Master space can be written as:
\bea
g^{\firr{}}_1 (t,1,1,1,1,1; \cC^{(I)}_3) = \frac{1-t^2}{\left(1-t\right)^4}\times \frac{1-t^4}{\left(1-t^2\right)^4}~. \label{hsfph1f0}
\eea
This space can be seen to be product of two conifolds.  The Hilbert series of the mesonic moduli space can be obtained by integrating \eref{hsfirrs4} over the two baryonic fugacities $b_1$ and $b_2$:
\bea
\gm_1 (t,x_1,x_2,x_3; \cC^{(I)}_3) &=& \frac{1}{(2\pi i)^2} \oint \limits_{|b_1| =1} {\frac{db_1}{b_1}} \oint \limits_{|b_2| =1} {\frac{db_2}{b_2}} g^{\firr{}}_1 (t,x_1,x_2,x_3,b_1,b_2; \cC^{(I)}_3)\nn \\
&=& \frac{P(t,x_1,x_2,x_3; \cC^{(I)}_3)}{\left(1 - t^6 x_1^2 x_2^2 x_3^2 \right)\left(1 - \frac{t^6 x_1^2 x_2^2}{x_3^2}\right)\left(1 - \frac{t^6 x_1^2 x_3^2}{x_2^2}\right)\left(1 - \frac{t^6 x_2^2 x_3^2}{x_1^2}\right)} \nn \\
&\times& \frac{1}{\left(1 - \frac{t^6 x_1^2}{x_2^2 x_3^2}\right)\left(1 - \frac{t^6 x_2^2}{x_1^2 x_3^2}\right)\left(1 - \frac{t^6 x_3^2}{x_1^2 x_2^2}\right)\left(1 - \frac{t^6}{x_1^2 x_2^2 x_3^2}\right)}\nn \\
&=& \sum^{\infty}_{n=0}[2n;2n;2n]t^{6n}~.
\label{meshsph1fo}
\eea
where $P(t,x_1,x_2,x_3; \cC^{(I)}_3)$ is a polynomial of degree $42$ in $t$ which is not presented here. The unrefined Hilbert series of the mesonic moduli space can be written as:
\bea
\gm_1 (t,1,1,1; \cC^{(I)}_3) = \frac{1+23t^6+23t^{12}+t^{18}}{\left(1-t^6\right)^4}~.
\eea
This indicates that the mesonic moduli space is a Calabi--Yau 4-fold, as expected.
The plethystic logarithm of the mesonic Hilbert series is given by 
\bea
\PL[\gm_1 (t,x_1,x_2,x_3;  \cC^{(I)}_3)] &=& [2;2;2] t^6 - O(t^{12})~.
\label{PLf0}
\eea

\paragraph{The generators.} Each of the generators can be written as a product of the perfect matchings:
\bea
p_i ~p_j ~q_k ~q_l ~r_m ~r_n ~v_1 ~v_2~,  \label{genph1f0}
\eea
where the indexes $i, j, k,  l, m, n$ run from 1 to 2.  
Since, for example, $p_i p_j$ has 3 independent components, $p_1 p_1, ~ p_1 p_2$ and $p_2 p_2$, it follows that there are indeed 27 independent generators.  All generators have R-charges equal to 2.

\begin{figure}[ht]
\begin{center}
  \includegraphics[totalheight=5.0cm]{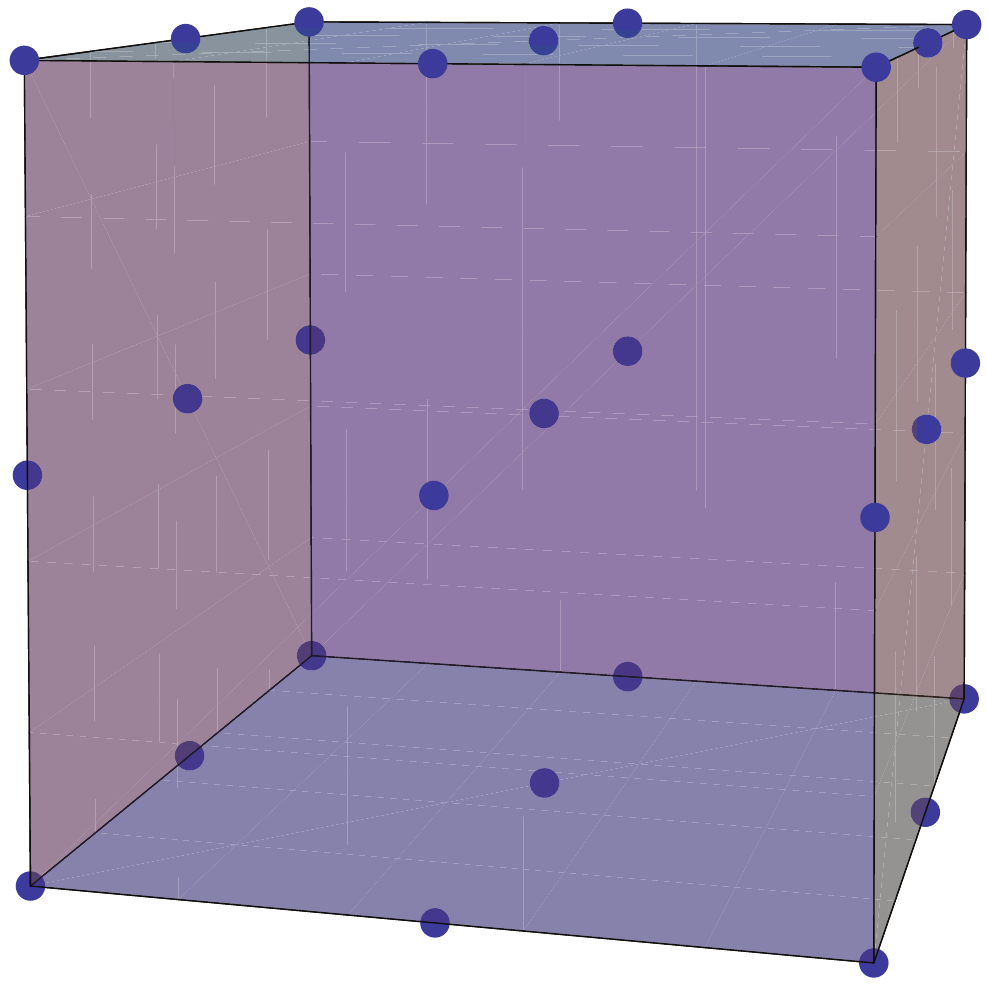}
 \caption{The lattice of generators of the $Q^{1,1,1}/\BZ_2$ theory.}
  \label{f:latq111z2}
\end{center}
\end{figure}

The generators can be drawn in a lattice (Figure \ref{f:latq111z2}) by plotting the powers of the weights of the characters in \eref{PLf0}.  Note that the lattice of generators is the dual of the toric diagram (nodes are dual to faces and edges are dual to edges): the toric diagram has  6 nodes (external points), 12 edges and 8 faces, whereas the generators form a convex polytope that has 8 nodes (corners of the cube), 12 edges and 6 faces.

\comment{
\begin{table}[h]
 \begin{center}  
  \begin{tabular}{|c||c|}
  \hline
  \; Generators &$U(1)_R$\\
  \hline  \hline 
  $p_i ~p_j ~q_k ~q_l ~r_m ~r_n ~v_1~ v_2$ & 2 \\
  \hline
  \end{tabular}
  \end{center}
\caption{R-charges of the generators of the mesonic moduli space for Phase I of $Q^{1,1,1}/\BZ_2$.}
\label{t:Rgenfano62Ph1}
\end{table}}

\paragraph{A discrete symmetry of $Q^{1,1,1}/\BZ_2$.}  As can be seen from \fref{f:torq111z2} and \fref{f:latq111z2},  the toric diagram is a regular octahedron and the lattice of generators is a cube.  The symmetry group of these polyhedra is $S_4 \times \BZ_2$.  This symmetry can also be seen in the gauge theory as follows.  The gauge groups can be relabelled by permuting the labels 1, 2, 3 and 4. To each permutation, there are two \emph{distinct}\footnote{Given a theory with the CS levels $\vec{k}$, the CS levels $-\vec{k}$ correspond to the same theory.  We do not count $\vec{k}$ and $-\vec{k}$ as distinct.} choices of CS levels which give $Q^{1,1,1}/\BZ_2$; for example, for the labels chosen in \fref{f:phase1f0}, the two distinct CS levels that give $Q^{1,1,1}/\BZ_2$ are $\vec{k} = (1,-1,-1,1)$ and $\vec{k} = (1,1,-1,-1)$.  Hence, the $Q^{1,1,1}/\BZ_2$ theory indeed possesses a discrete symmetry $S_4 \times \BZ_2$. (Note that a similar argument can also be applied to Phase II of $Q^{1,1,1}/\BZ_2$.)

\subsection{Phase II of The $Q^{1,1,1}/\BZ_2$ Theory}
\begin{figure}[ht]
\begin{center}
\vskip 0.5cm
  \hskip -6.5cm
  \includegraphics[totalheight=4cm]{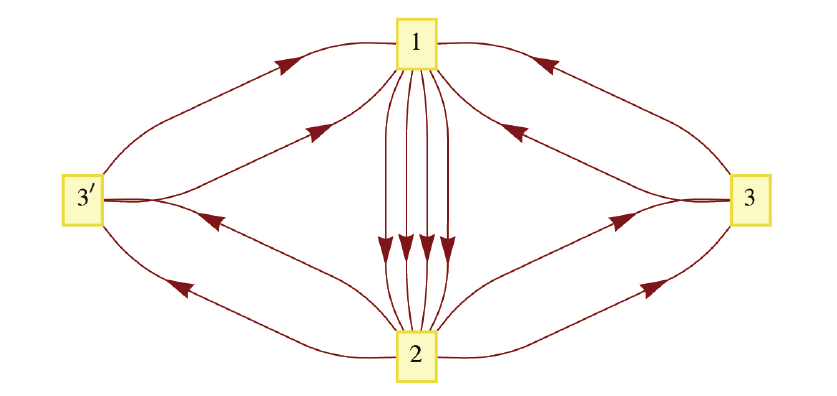}
    \vskip -5cm
  \hskip 8.9cm
  \includegraphics[totalheight=6cm]{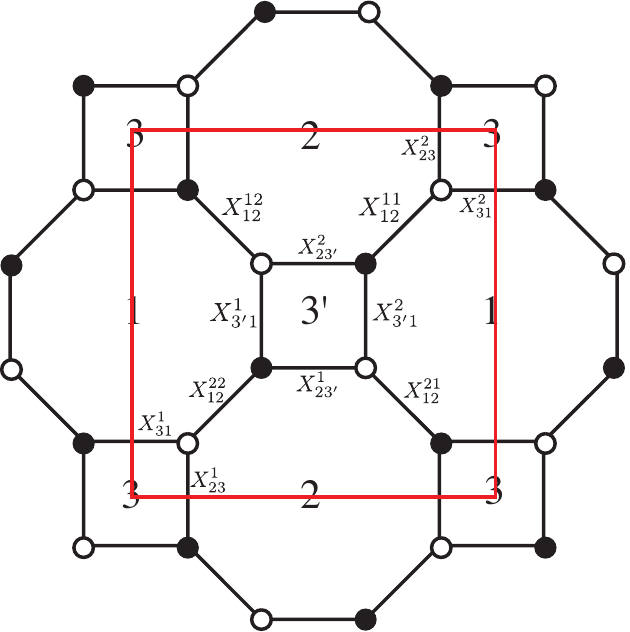}
 \caption{(i) Quiver for Phase II of $Q^{1,1,1}/\BZ_2$.\quad (ii) Tiling for Phase II of $Q^{1,1,1}/\BZ_2$.}
  \label{f:phase2f0}
\end{center}
\end{figure}

This model, first studied in \cite{Hanany:2008fj}, has four gauge groups and bi-fundamental fields $X_{12}^{ij}$, $X_{23}^i$, $X_{23'}^i$, $X_{31}^i$ and $X^{i}_{3'1}$ (with $i,j=1,2$).  From the features of this quiver gauge theory, this phase is also known as a \emph{three-block model} \cite{Benvenuti:2004dw}. The superpotential is
\bea
W &=& \epsilon_{ij}\epsilon_{kl} \tr(X^{ik}_{12}X^{l}_{23}X^{j}_{31}) - \epsilon_{ij}\epsilon_{kl} \tr(X^{ki}_{12}X^{l}_{23'} X^{j}_{3'1})~.
\label{e:spotfano62phII}
\eea
The quiver diagram and tiling of this phase of the theory are given in Figure \ref{f:phase2f0}.  Note that in 3+1 dimensions this tiling corresponds to Phase II of the $\BF_0$ theory \cite{Forcella:2008ng, master}.  
The CS levels can be chosen to be $\vec{k} = (-1,-1,1,1)$.
\\

\comment{
\begin{figure}[h]
\begin{center}
   \includegraphics[totalheight=6.5cm]{fdph2f0.pdf}
 \caption{The fundamental domain of tiling for Phase II of $Q^{1,1,1}/\BZ_2$.}
  \label{f:fdph2f0}
\end{center}
\end{figure}
}

\paragraph{The Kasteleyn matrix.}  The Chern-Simons levels can be written in terms of the integers $n^i_{kl}$ and $n^{ij}_{kl}$ as:
\bea
\begin{array}{ll}
\text{Gauge group 1~:} \qquad k_1    &=   n^{11}_{12} + n^{12}_{12} + n^{21}_{12} + n^{22}_{12} - n^{1}_{31} - n^{2}_{31} - n^{1}_{3'1} - n^{2}_{3'1}  ~,  \\
\text{Gauge group 2~:} \qquad k_2    &=   n^{1}_{23} + n^{2}_{23} + n^{1}_{23'} + n^{2}_{23'} - n^{11}_{12} - n^{12}_{12} - n^{21}_{12} - n^{22}_{12} ~,  \\
\text{Gauge group 3~:} \qquad k_3    &=   n^{1}_{31} + n^{2}_{31} - n^{1}_{23} - n^{2}_{23} ~,  \\
\text{Gauge group $3'$~:}\qquad k_{3'} &=   n^{1}_{3'1} + n^{2}_{3'1} - n^{1}_{23'} - n^{2}_{23'}  ~.
\label{e:csfano62phII}
\end{array}
\eea  
and are chosen to be
\bea
 n^{2}_{3'1} = 1, \quad n^{2}_{23} = -1,\quad n^i_{jk}=0 \; ~ \text{otherwise}~.
\eea

\noindent The Kasteleyn matrix can now be computed.  Since the fundamental domain contains 4 black nodes and 4 white nodes, the Kasteleyn matrix is a $4\times 4$ matrix:\be
K =   \left(
\begin{array}{c|cccc}
 \;& w_1&w_2&w_3&w_4 \\
  \hline 
b_1 &  \frac{1}{y} z^{n^{2}_{23}} &\ x z^{n^{1}_{31}} &\ 0 &\  {z^{n^{21}_{12}}} \\
b_2 &  \frac{1}{x} z^{n^{2}_{31}} &\ y z^{n^{1}_{23}} &\  z^{n^{12}_{12}} &\ 0  \\
b_3 &  0 &\   z^{n^{22}_{12}} &\  z^{n^{1}_{3'1}} &\  z^{n^{1}_{23'}}  \\
b_4 &  z^{n^{11}_{12}} &\ 0 &\  z^{n^{2}_{23'}} &\  z^{n^{2}_{3'1}}
\end{array}
\right) ~.
\ee
The permanent of this matrix is given by:
\bea \label{permKph2f0}
\perm~K &=&  x z^{(n^{1}_{23'} + n^{1}_{31} + n^{11}_{12} + n^{12}_{12})} +  x^{-1} z^{(n^{2}_{23'} + n^{2}_{31} + n^{21}_{12} + n^{22}_{12})} + y z^{(n^{1}_{3'1} + n^{1}_{23} + n^{11}_{12} + n^{21}_{12})}\nn \\
&&+  y^{-1} z^{(n^{2}_{3'1} + n^{2}_{23} + n^{12}_{12} + n^{22}_{12})} + z^{(n^{1}_{31} + n^{2}_{31} + n^{1}_{3'1} + n^{2}_{3'1})} +  z^{(n^{1}_{23'} + n^{2}_{23'} + n^{2}_{23} + n^{1}_{23})}  \nn \\
&&+  z^{(n^{11}_{12} + n^{21}_{12} + n^{12}_{12} + n^{22}_{12})}+  z^{(n^{1}_{3'1} + n^{2}_{3'1} + n^{1}_{23} + n^{2}_{23})} + z^{(n^{1}_{31} + n^{2}_{31} + n^{2}_{23'} + n^{1}_{23'} )} \nn \\
&=& x + x^{-1} + y + y^{-1} + z + z^{-1}  + 3~. \nn \\
&& \qquad \text{(for $n^{2}_{3'1} = 1, \quad n^{2}_{23} = -1,\quad n^i_{jk}=0 \; ~ \text{otherwise}$)} ~.
\eea

\paragraph{The perfect matchings.} The relationship between quiver fields and perfect matchings is summarised in the $P$ matrix below: 
\beq
P=\left(\begin{array} {c|ccccccccc}
  \;& p_1 & p_2 &q_1&q_2&r_1&r_2&v_1&v_2&v_3\\
  \hline
  X^{1}_{31}   & 1 & 0 & 0 & 0 & 1 & 0 & 0 & 0 & 1 \\
  X^{2}_{31}   & 0 & 1 & 0 & 0 & 1 & 0 & 0 & 0 & 1 \\
  X^{1}_{23'}  & 1 & 0 & 0 & 0 & 0 & 1 & 0 & 0 & 1 \\
  X^{2}_{23'}  & 0 & 1 & 0 & 0 & 0 & 1 & 0 & 0 & 1 \\
  X^{1}_{3'1}  & 0 & 0 & 1 & 0 & 1 & 0 & 0 & 1 & 0 \\
  X^{2}_{3'1}  & 0 & 0 & 0 & 1 & 1 & 0 & 0 & 1 & 0 \\
  X^{1}_{23}   & 0 & 0 & 1 & 0 & 0 & 1 & 0 & 1 & 0 \\
  X^{2}_{23}   & 0 & 0 & 0 & 1 & 0 & 1 & 0 & 1 & 0 \\
  X^{11}_{12}  & 1 & 0 & 1 & 0 & 0 & 0 & 1 & 0 & 0 \\
  X^{12}_{12}  & 1 & 0 & 0 & 1 & 0 & 0 & 1 & 0 & 0 \\
  X^{21}_{12}  & 0 & 1 & 1 & 0 & 0 & 0 & 1 & 0 & 0 \\
  X^{22}_{12}  & 0 & 1 & 0 & 1 & 0 & 0 & 1 & 0 & 0 \\
  \end{array}
\right)~.
\eeq
From (\ref{permKph2f0}), The perfect matchings $p_i, q_i, r_i$ correspond to the external points in the toric diagram, whereas the perfect matchings $v_i$ correspond to the internal point at the origin.
Basis vectors of the null space of $P$ are given in the rows of the charge matrix:
\be
Q_F =   \left(
\begin{array}{ccccccccc}
 1 & 1 & 0 & 0 & 0 & 0 & -1 &  0 & -1\\
 0 & 0 & 1 & 1 & 0 & 0 & -1 & -1 &  0 \\
 0 & 0 & 0 & 0 & 1 & 1 &  0 & -1 & -1
 \end{array}
\right)~. \label{qfph2f0}
\ee
Hence, the relations between the perfect matchings are given by
\bea
p_1 + p_2 - v_1 - v_3 &=& 0~, \nn\\
q_1 + q_2 - v_1 - v_2 &=& 0~, \nn \\ \
r_1 + r_2 - v_2 - v_3 &=& 0~. \label{relf0ii}
\eea
Since the coherent component of the Master space is generated by the perfect matchings (subject to the relations \eref{relf0ii}), it follows that 
\bea
\firr{} = \BC^9//Q_F~.  \label{firrph2f0}
\eea

\paragraph{The toric diagram.} Two methods of constructing the toric diagram are demonstrated. 
\begin{itemize}
\item{\bf The charge matrices.}    Since the number of gauge groups is $G=4$, there are $G-2 = 2$ baryonic symmetries coming from the D-terms.  These charges are collected in the $Q_D$ matrix:
\bea
Q_D = \left(
\begin{array}{ccccccccc}
 0 & 0 & 0 & 0 & 1 & 1 &-2 &  0 & 0 \\
 0 & 0 & 0 & 0 & 0 & 0 & 1 & -1 & 0
\end{array}
\right)~. \label{qdph2f0}
\eea
From \eref{qfph2f0} and \eref{qdph2f0}, the total charge matrix is given by
\be
Q_t = { \Blue Q_F \choose \Green Q_D \Black } =   \left( 
\begin{array}{ccccccccc} \Blue
 1 & 1 & 0 & 0 & 0 & 0 & -1 &  0 & -1\\
 0 & 0 & 1 & 1 & 0 & 0 & -1 & -1 &  0 \\
 0 & 0 & 0 & 0 & 1 & 1 &  0 & -1 & -1\\ \Green
 0 & 0 & 0 & 0 & 1 & 1 &-2 &  0 & 0 \\
 0 & 0 & 0 & 0 & 0 & 0 & 1 & -1 & 0
 \Black
\end{array}
\right) 
\label{qtph2q111z2}
\ee
The matrix $G_t$ is obtained and, after removing the first row, the columns give the coordinates of points in the toric diagram:  
\bea
G'_t = \left(
\begin{array}{ccccccccc}
 1 &-1 & 0 & 0 & 0 & 0 & 0 & 0 & 0 \\
 0 & 0 & 1 &-1 & 0 & 0 & 0 & 0 & 0 \\
 0 & 0 & 0 & 0 & 1 &-1 & 0 & 0 & 0
\end{array}
\right)~.
\eea
The toric diagram is given in Figure \ref{f:torq111z2}, with three degenerate internal points at the center.  Comparing Figure \ref{f:torq111z2} with the 2d toric diagram of Phase II of $\BF_0$ theory \cite{master, Butti:2007jv}, it can be seen that the CS levels split two of the five points at the center of the 2d toric diagram along the vertical axis into the two tips and the rest remains at the center of the octahedron.

\item {\bf The Kasteleyn matrix.} The powers of $x, y, z$ in each term of the permanent of the Kasteleyn matrix give the coordinates of each point in the toric diagram.  These points can be collected in the columns of the following $G_K$ matrix:
\bea
G_K = \left(
\begin{array}{ccccccccc}
 1 &-1 & 0 & 0 & 0 & 0 & 0 & 0 & 0 \\
 0 & 0 & 1 &-1 & 0 & 0 & 0 & 0 & 0 \\
 0 & 0 & 0 & 0 & 1 &-1 & 0 & 0 & 0
\end{array}
\right) = G'_t~.
\eea
The rows of this matrix contain the powers of the weights of three $SU(2)$ groups, which implies that the mesonic symmetry of this model is $SU(2)^3 \times U(1)_R$.
\end{itemize}

\paragraph{The baryonic charges.}
Since the toric diagram has 6 external points, this model has precisely $6-4 = 2$ baryonic symmetries which shall be denoted by $U(1)_{B_1}, U(1)_{B_2}$.  From the above discussion, they can be seen to have arisen from the D-terms.  Therefore, the baryonic charges of the perfect matchings are given by the rows of the $Q_D$ matrix. 

\paragraph{The global symmetry.} From the $Q_t$ matrix, the symmetry of the mesonic moduli space can be seen to be $SU(2)^3\times U(1)_R$. 
Since $v_1, v_2$ and $v_3$ are the perfect matchings corresponding to the internal point in the toric diagram, they are each assigned zero R-charge.  
The remaining 6 external perfect matchings are completely symmetric and so the requirement that each superpotential term has an R-charge of 2 implies each of these external perfect matchings have an R-charge of 1/3. 
The global symmetry of the theory is a product of mesonic and baryonic symmetries: $SU(2)^3 \times U(1)_R \times U(1)_{B_1} \times U(1)_{B_2}$.
In Table \ref{chargeph1f0}, a consistent global charge assignment for the perfect matchings is given.

\begin{table}[h]
 \begin{center}  
  \begin{tabular}{|c||c|}
  \hline
  \; Quiver fields &R-charge\\
  \hline  \hline 
  $X^i_{23}, X^i_{31}$ &  $2/3$\\
  \hline
  $X^i_{23'}, X^i_{3'1}$ &  $2/3$\\
  \hline
  $X^{ij}_{12}$ &  $2/3$\\
  \hline
  \end{tabular}
  \end{center}
 \caption{R-charges of the quiver fields of $Q^{1,1,1}/\BZ_2$, Phase II.}
 \label{t:Rgenph2f0}
 \end{table}

\paragraph{The Hilbert series.} From \eref{firrph2f0}, the Hilbert series of the coherent component of the Master space is computed by integrating the Hilbert series of $\BC^{9}$ over the fugacities $z_1, z_2, z_3$ associated with the $Q_F$ charges:
\bea
g^{\firr{}}_1 (t,x_1,x_2,x_3,b_1,b_2; \cC^{(II)}_3) &=& \frac{1}{(2 \pi i)^3} \oint \limits_{|z_1|=1} {\frac{dz_1}{z_1}} \oint \limits_{|z_2|=1} {\frac{dz_2}{z_2}}\oint \limits_{|z_3|=1} { \frac{dz_3}{z_3}}  \frac{1}{\left(1- t x_1 z_1\right)\left(1-\frac{t z_1}{x_1}\right)}\nn \\
&\times& \frac{1}{\left(1- t x_2 z_2\right) \left(1-\frac{t z_2}{x_2}\right)\left(1 - t x_3 z_3 b_1 \right)\left(1-\frac{t z_3 b_1}{x_3}\right)} \nn \\
&\times& \frac{1}{\left(1-\frac{b_2}{b^2_1 z_1 z_2}\right)\left(1-\frac{1}{b_2 z_2 z_3}\right)\left(1- \frac{1}{z_1 z_3} \right)}~.
\eea
The unrefined Hilbert series of the Master space can be written as:
\bea
g^{\firr{}}_1 (t,1,1,1,1,1;\cC^{(II)}_3) &=& \frac{1+6t^2+6t^4+t^6}{\left(1-t^2\right)^6}~. \label{hsfph2f0}
\eea
Integrating the Hilbert series of the Master space over the baryonic fugacities gives the Hilbert series of the mesonic moduli space:
\bea
\gm_1 (t,x_1,x_2,x_3; \cC^{(II)}_3) &=& \frac{1}{(2\pi i)^2} \oint \limits_{|b_1|=1} {\frac{db_1}{b_1}}\oint \limits_{|b_2|=1} {\frac{db_2}{b_2}} g^{\firr{}}_1 (t,x_1,x_2,x_3,b_1,b_2; \cC^{(II)}_3) \nn \\
&=& \frac{P(t,x_1,x_2,x_3;\cC^{(II)}_3)}{\left(1 - t^6 x_1^2 x_2^2 x_3^2 \right)\left(1 - \frac{t^6 x_1^2 x_2^2}{x_3^2}\right)\left(1 - \frac{t^6 x_1^2 x_3^2}{x_2^2}\right)\left(1 - \frac{t^6 x_2^2 x_3^2}{x_1^2}\right)} \nn \\
&\times&  \frac{1}{\left(1 - \frac{t^6 x_1^2}{x_2^2 x_3^2}\right)\left(1 - \frac{t^6 x_2^2}{x_1^2 x_3^2}\right)\left(1 - \frac{t^6 x_3^2}{x_1^2 x_2^2}\right)\left(1 - \frac{t^6}{x_1^2 x_2^2 x_3^2}\right)} \nn \\
&=& \sum^{\infty}_{n=0}[2n;2n;2n]t^{6n}~. \label{meshsph2f0}
\eea
where $P(t,x_1,x_2,x_3;\cC^{(II)}_3)$ is a polynomial of order $42$ in $t$ mentioned in \eref{meshsph1fo}. 
This is precisely identical to the Hilbert series \eref{meshsph1fo} of the mesonic moduli space of Phase I .   

\paragraph{The generators.} Each of the generators can be written as a product of perfect matchings:
\bea
p_i ~p_j ~q_k ~q_l ~r_m ~r_n ~v_1~ v_2 ~v_3~,  \label{genph2f0}
\eea
where the indexes $i, j, k,  l, m, n$ run from 1 to 2.  Note that the generators of this model are identical to those of Phase I, apart from a factor of the internal perfect matching $v_3$. All generators of this model have R-charges equal to 2.

\comment{
\begin{table}[h]
 \begin{center}  
  \begin{tabular}{|c||c|}
  \hline
  \; Generators &$U(1)_R$\\
  \hline  \hline 
  $p_i ~p_j ~q_k ~q_l ~r_m ~r_n ~v_1~ v_2~ v_3$ & 2 \\
  \hline
  \end{tabular}
  \end{center}
\caption{R-charges of the generators of the mesonic moduli space for Phase II of $Q^{1,1,1}/\BZ_2$.}
\label{t:Rgenfano62Ph2}
\end{table}}

\section{$\cC_4$ (Toric Fano 123): $dP_1 \times \BP^1$}
This model has 4 gauge groups and chiral fields $X_{14}$, $X_{12}$, $X_{32}$, $X^i_{43}$, $X^j_{24}$ and $X^j_{31}$ (with $i=1,2,3$ and $j=1,2$).
The quiver diagram and the tiling are presented in Figure \ref{f:tqfano123}.
Note that in $3+1$ dimensions this tiling corresponds to the gauge theory on D3-branes probing a cone over the $dP_1$ surface.
The superpotential can be read off from the tiling and can be written as:
\bea
W = \tr \left[ \epsilon_{ij} \left( X_{14}X^{i}_{43}X^{j}_{31} + X_{32}X^{i}_{24}X^{j}_{43} - X_{12}X^{i}_{24}X^{3}_{43}X^{j}_{31}\right) \right]~.
\label{e:sptoric123}
\eea

The CS levels are chosen to be $\vec{k} = (1,1,-1,-1)$

\begin{figure}[ht]
\begin{center}
  \hskip -1cm
  \includegraphics[totalheight=5cm]{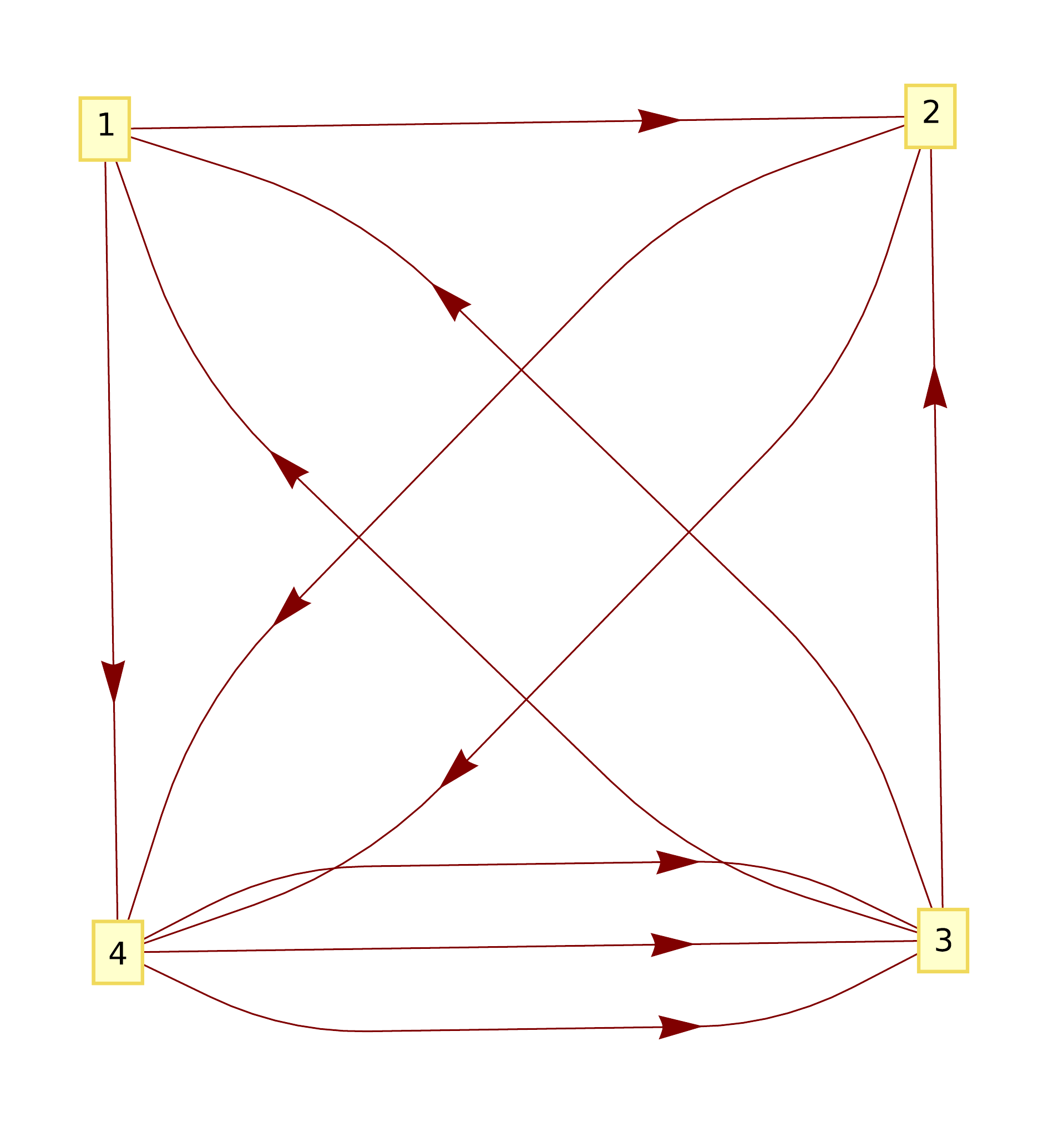}
  \hskip 1cm
  \includegraphics[totalheight=5cm]{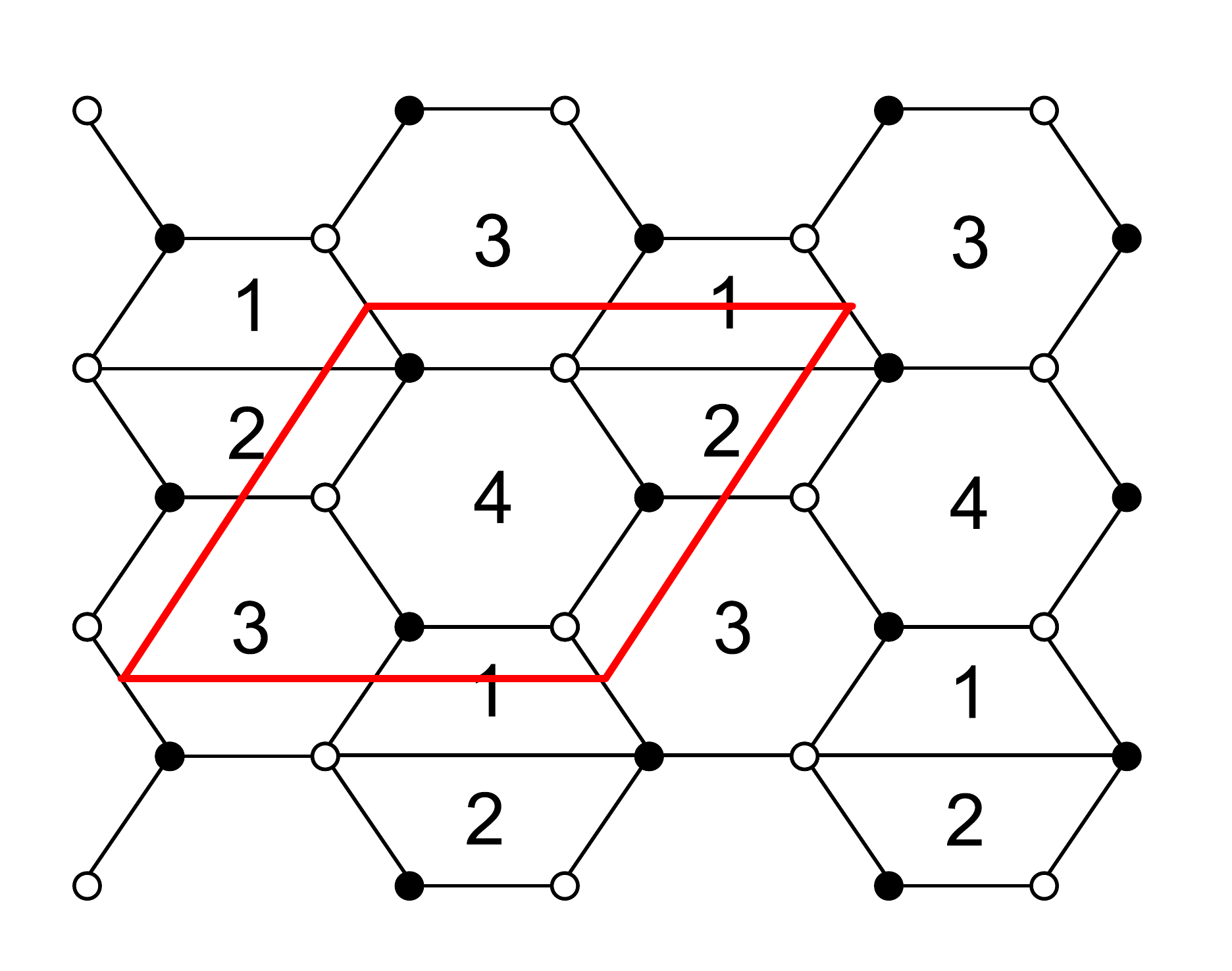}
 \caption{(i) Quiver diagram of the $dP_1 \times \BP^1$ theory.\ (ii) Tiling of the $dP_1 \times \BP^1$ theory.}
  \label{f:tqfano123}
\end{center}
\end{figure}

\comment{
\begin{figure}[ht]
\begin{center}
   \includegraphics[totalheight=6cm]{fdc4.pdf}
 \caption{The fundamental domain of the tiling for the $dP_1 \times \BP^1$ theory.}
  \label{f:fdtoricfano123}
\end{center}
\end{figure}}

\paragraph{The toric diagram.} Two methods of computing the toric diagram for this model are demonstrated

\begin{itemize}
\item {\bf The Kasteleyn matrix.} The Chern-Simons levels can be parametrized as follows:
\bea
\begin{array}{ll}
\text{Gauge group 1~:} \qquad k_1  &=   n_{12} + n_{14} - n^{1}_{31} - n^{2}_{31}~, \nn \\
\text{Gauge group 2~:} \qquad k_2  &=   n^{1}_{24} + n^{2}_{24} - n_{12} - n_{32}~, \nn \\
\text{Gauge group 3~:} \qquad k_3  &=   n_{32} + n^{1}_{31} + n^{2}_{31} - n^{1}_{43} - n^{2}_{43} - n^{3}_{43} ~, \nn \\
\text{Gauge group 4~:} \qquad k_4  &=   n^{1}_{43} + n^{2}_{43} + n^{3}_{43} - n^{1}_{24} - n^{2}_{24} - n_{14}~. \nn \\

\label{e:kafano123}
\end{array}
\eea
Let us choose:
\bea
n^1_{24} = - n^1_{31} = 1,\quad n^i_{jk}=0 \; \text{otherwise}~.
\eea
The Kasteleyn matrix $K$ can be computed for this model. The fundamental domain contains three black and three white nodes, hence $K$ is a $3\times 3$ matrix:
\bea
K =   \left(
\begin{array}{c|ccc}
& b_1 & b_2 & b_3\\
\hline
w_1 &  z^{n_{14}} &  z^{n^1_{43}} &  \frac{x}{y} z^{n^2_{31}} \\
w_2 &  y z^{n^1_{31}} &  z^{n^2_{24}} &  z^{n^3_{43}} +  x z^{n_{12}} \\
w_3 &  z^{n^2_{43}} &  \frac{1}{x} z^{n_{32}} &  z^{n^1_{24}} \end{array}
\right) ~.
\label{e:kastfano123}
\eea
The permanent of this matrix is given by:
\bea
\perm~K &=& 
   z^{(n^1_{24} + n^{2}_{24} + n_{14})} + z^{(n^{1}_{31} + n^2_{31} + n_{32})} +  y z^{(n^{1}_{31} + n^{1}_{24} + n^{1}_{43})} \nn \\
&+&  x y^{-1} z^{(n^{2}_{31}+ n^{2}_{24} + n^{2}_{43})} +  x z^{(n^1_{43} + n^{2}_{43} + n_{12})}  +  x^{-1} z^{(n^{3}_{43} + n_{14} + n_{32})} \nn \\
&+&  z^{(n^1_{43} + n^{2}_{43} + n^{3}_{43})} +  z^{(n_{12} + n_{14}+ n_{32})}\nn \\
&=& z  + z^{-1} + y + x y^{-1} + x + x^{-1} + 2\nn\\ 
&& \text{(for $n^1_{24} = - n^1_{31} = 1,\quad n^i_{jk}=0\; ~ \text{otherwise}$)} ~.
\label{e:permKfano123}
\eea
The coordinates of the toric diagram are collected in the columns of the following matrix:
\bea
\left(
\begin{array}{cccccccc}
  0 & 0 & 0 &  1 & 1 & -1 & 0 & 0\\
  0 & 0 & 1 & -1 & 0 &  0 & 0 & 0\\
  1 &-1 & 0 &  0 & 0 &  0 & 0 & 0
\end{array}
\right)~.
\eea
Multiplying on the left by {\footnotesize $\left( \begin{array}{ccc} 0&0&1\\0&1&0\\1&0&0 \end{array} \right) \in GL(3, \BZ)$}, the $G_K$ matrix can be computed to be:
\bea
G_K = \left(
\begin{array}{cccccccc}
  1 & -1 &  0 &  0 & 0 &  0 & 0 & 0\\
  0 &  0 &  1 & -1 & 0 &  0 & 0 & 0\\
  0 &  0 &  0 &  1 & 1 & -1 & 0 & 0
\end{array}
\right)~.
\eea
Observe that the first two rows contain the weights of two $SU(2)$'s. This implies that the mesonic symmetry for this model is $SU(2)^2\times U(1)^2$.
The permanent of the Kasteleyn matrix can be used to write the perfect matchings in terms of the chiral fields of the model:
\bea 
&&   p_1 = \left\{X^1_{24}, X^2_{24},X_{14}\right\},
\;\; p_2 = \left\{X^1_{31}, X^2_{31},X_{32}\right\},
\;\; q_1 = \left\{X^1_{31}, X^1_{24},X^1_{43}\right\},\nn \\
&&   q_2 = \left\{X^2_{31}, X^2_{24},X^2_{43}\right\},
\;\; r_1 = \left\{X^1_{43}, X^2_{43},X_{12}\right\},
\;\; r_2 = \left\{X^3_{43}, X_{32},X_{14}\right\},\nn \\
&&   v_1 = \left\{X^1_{43}, X^2_{43},X^3_{43}\right\},
\;\; v_2 = \left\{X_{12}, X_{32},X_{14}\right\}\ . \qquad
\eea
It can be seen from the Kasteleyn matrix that $v_1$ and $v_2$ correspond to the internal point in the toric diagram, whereas the others correspond to external points.
The chiral fields can be parametrised in terms of perfect matchings:
\bea
&& X^1_{24} = p_1 q_1, \quad X^2_{24} = p_1 q_2, \quad  X^1_{31} = p_2 q_1, \quad  X^2_{31} = p_2 q_2,\nn \\
&& X^1_{43} = q_1 r_1 v_1, \quad X^2_{43} = q_2 r_1 v_1,\quad X^3_{43} = r_2 v_1,\nn \\
&& X_{12} = r_1 v_2, \quad  X_{14} = p_1 r_2 v_2, \quad X_{32} = p_2 r_2 v_2~.
\eea
All of this information can be collected in the perfect matching matrix:
\beq
P=\left(\begin{array} {c|cccccccc}
  \;& p_1 & p_2 & q_1 & q_2 & r_1 & r_2 & v_1 & v_2\\
  \hline 
  X^{1}_{24} & 1&0&1&0&0&0&0&0\\
  X^{2}_{24} & 1&0&0&1&0&0&0&0\\
  X^{1}_{31} & 0&1&1&0&0&0&0&0\\
  X^{2}_{31} & 0&1&0&1&0&0&0&0\\
  X^{1}_{43} & 0&0&1&0&1&0&1&0\\
  X^{2}_{43} & 0&0&0&1&1&0&1&0\\
  X^{3}_{43} & 0&0&0&0&0&1&1&0\\
  X_{12}     & 0&0&0&0&1&0&0&1\\
  X_{14}     & 1&0&0&0&0&1&0&1\\
  X_{32}     & 0&1&0&0&0&1&0&1\\
  \end{array}
\right).
\eeq
The null space of the $P$ matrix is spanned by two vectors that can be cast in the rows of the following matrix
\be
Q_F =   \left(
\begin{array}{cccccccc}
1 & 1 & -1 & -1 &  0 & -1 & 1 & 0 \\
0 & 0 &  0 &  0 &  1 &  1 &-1 &-1 
\end{array}
\right)~.  \label{e:qffano123}
\ee
Hence, among the perfect matchings there are two relations, which are given by:
\bea
p_1 + p_2 - q_1 - q_2 - r_2 + v_1 &=& 0 \nn \\
r_1 + r_2 - v_1 - v_2 &=& 0~.
\label{e:relpmfano123}
\eea

\item {\bf The charge matrices.}
The number of gauge groups of this model is $G = 4$, hence there are $G-2 =2$ baryonic symmetries coming from the D-terms. The charges of the perfect matchings under this baryonic symmetries can be collected in the rows of the $Q_D$ matrix:
\be
Q_D =   \left(
\begin{array}{cccccccc}
 1 & 1 &  0 &  0 &  0 & 0 & 0 & -2\\
 0 & 0 &  0 &  0 &  0 & 0 & 1 & -1
\end{array}
\right) \label{e:qdfano123}
\ee
$Q_F$ and $Q_D$ can be combined to obtain the total charge matrix $Q_t$:
\be
Q_t = { \Blue Q_F \choose \Green Q_D \Black } =   \left( 
\begin{array}{cccccccc} \Blue
 1 & 1 & -1 & -1 &  0 &-1 & 1 & 0 \\
 0 & 0 &  0 &  0 &  1 & 1 &-1 &-1 \\\Green
 1 & 1 &  0 &  0 &  0 & 0 & 0 &-2\\
 0 & 0 &  0 &  0 &  0 & 0 & 1 &-1\Black
\end{array}
\right) 
\label{e:qtfano123}
\ee

The null space of this matrix gives the $G_t$ matrix from which, eliminating the first row, the $G'_t$ matrix is obtained, whose rows give the coordinates of the toric diagram:
\bea
G'_t = \left(
\begin{array}{cccccccc}
  1 & -1 &  0 &  0 & 0 &  0 & 0 & 0\\
  0 &  0 &  1 & -1 & 0 &  0 & 0 & 0\\
  0 &  0 &  0 &  1 & 1 & -1 & 0 & 0
\end{array}
\right) = G'_K~. 
\label{e:gtfano123}
\eea
The toric diagram constructed from (\ref{e:gtfano123}) is presented in Figure \ref{f:tdtoricfano123}.
Note that the three points aligning along the vertical line represent the toric diagram of $\BP^1$, and the other 4 points (as well as the internal point) form that of $dP_1$.  Thus, the mesonic moduli space of this theory is $dP_1 \times \BP^1$. 

\begin{figure}[ht]
\begin{center}
  \includegraphics[totalheight=4.0cm]{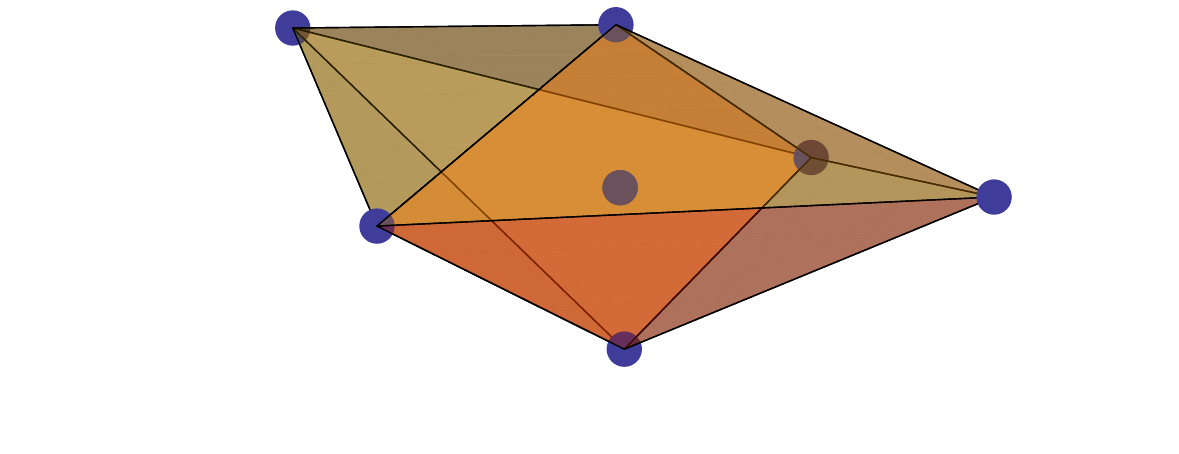}
 \caption{The toric diagram of $dP_1 \times \BP^1$.}
  \label{f:tdtoricfano123}
\end{center}
\end{figure}
\end{itemize}

\paragraph{The baryonic charges.} The toric diagram of this model is characterized by 6 external points. Thus, there are 2 baryonic symmetries which shall be denoted as $U(1)_{B_1}$ and $U(1)_{B_2}$. The charges of the perfect matchings under these two symmetries are collected in the rows of the $Q_D$ matrix presented in (\ref{e:qdfano123}).

\paragraph{The global symmetry.} The two pairs of repeated columns in the $Q_t$ matrix confirm that the mesonic symmetry of this model is $SU(2)^2\times U(1)^2$, where one of the abelian factors can be identified with the R-symmetry. The perfect matchings $p_1$ and $p_2$ transform as a doublet under the first $SU(2)$, whereas the perfect matchings $q_1$ and $q_2$ transform as a doublet under the second $SU(2)$. 

Since the internal perfect matchings $v_1$ and $v_2$ have zero R-charge, the corresponding fugacities are set to unity. Also, since the non-abelian symmetries do not have a role in the volume minimisation, the fugacities $x_1$ and $x_2$ can be set to 1.

Let us denote the fugacity for R-charges of the external perfect matchings $p_1$ and $p_2$ as $s_1$, that of $q_1$ and $q_2$ as $s_2$, and those of $r_1$ and $r_2$ by $s_3$ and $s_4$ respectively.
Given the charges of the perfect matchings written in the rows of (\ref{e:qtfano123}), the Hilbert series of the mesonic moduli space can be computed as:
{\small
\bea
\gm (t_\alpha; \cC_4) &=& \oint \limits_{|z_1|=1}{\frac{\ud z_1}{2\pi i z_1}}\oint \limits_{|z_2|=1}{\frac{\ud z_2}{2\pi i z_2}}\oint \limits_{|b_1|=1}{\frac{\ud b_1}{2\pi i b_1}}\oint \limits_{|b_2|=1}{\frac{\ud b_2}{2\pi i b_2}}
\frac{1}{\left(1-s_1 z_1 b_1\right)^2\left(1-\frac{s_2}{z_1}\right)^2}\nn \\
&\times& \frac{1}{\left(1- s_3 z_2\right)\left(1- \frac{s_4 z_2}{z_1}\right)\left(1- \frac{z_1 b_2}{z_2}\right)\left(1-\frac{1}{ b_1^2 b_2 z_2}\right)} \nn \\
&=& \frac{P\left(s_\alpha; \cC_4\right)}{\left(1-s^2_1 s^3_2 s^2_3 \right)^3\left(1-s^2_1 s_2 s^2_4 \right)^3}~,
\label{e:hsvolfano123}
\eea}
where $P\left(s_\alpha; \cC_4\right)$ is a polynomial that is not reported here. 

Since there are two abelian factors in the mesonic symmetry, the Hilbert series actually depends only on the following two combinations of $s_{\alpha}$'s:
\bea
t^2_1 = s^2_1 s^3_2 s^2_3~, \quad t^2_2 = s^2_1 s_2 s^2_4~.
\label{TC4}
\eea
Then, (\ref{e:hsvolfano123}) can be rewritten as
{\small
\bea
\gm (t_1, t_2; \cC_4) &=& \frac{1}{\left(1-t^2_1\right)^3\left(1-t^2_2\right)^3} \times (1 + 9 t_1^2 + 2 t_1^4 + 9 t_1 t_2 + 3 t_1^3 t_2 + 3 t_2^2 - 20 t_1^2 t_2^2-\nn \\
&& 7 t_1^4 t_2^2 - 7 t_1 t_2^3 - 20 t_1^3 t_2^3 + 3 t_1^5 t_2^3 + 3 t_1^2 t_2^4 + 9 t_1^4 t_2^4 + 2 t_1 t_2^5 + 9 t_1^3 t_2^5+ t_1^5 t_2^5) ~. \nn \\
\eea}
Let $R_1$ and $R_2$ be the R-charges corresponding respectively to  $t_1$ and $t_2$.  Since the R-charge of the superpotential is 2, it follows that the Calabi-Yau condition can be written as $R_1+R_2=2$.  Therefore, the volume of $dP_1 \times \BP^1$ is
\bea
\lim_{\mu \rightarrow 0} \;\mu^4 \gm (e^{- \mu R_1}, e^{- \mu (2-R_1)}; \cC_4) = \frac{R^2_1 - 4 R_1 + 6}{R^3_1\left(R_1-2\right)^3}~.
\eea
This function can be easily shown to have a minimum at:
\bea
R_1 = \frac{1}{6}\left(11-\frac{35}{(18\sqrt{359}-271)^{1/3}} + (18\sqrt{359}-271)^{1/3}\right) \approx 1.105
\eea
As discussed in the introduction, the R-charge of each perfect matching can be determined by computing the normalized volume of the corresponding divisor. For example, for the perfect matching $p_1$:
{\small
\bea
g (D_1; s_\alpha; \cC_4) &=& \oint \limits_{|z_1|=1}{\frac{\ud z_1}{2\pi i z_1}}\oint \limits_{|z_2|=1}{\frac{\ud z_2}{2\pi i z_2}}\oint \limits_{|b_1|=1}{\frac{\ud b_1}{2\pi i b_1}}\oint \limits_{|b_2|=1}{\frac{\ud b_2}{2\pi i b_2}}
\frac{(s_1 z_1 b_1)^{-1}}{\left(1-s_1 z_1 b_1\right)^2\left(1-\frac{s_2}{z_1}\right)^2}\nn \\
&\times& \frac{1}{\left(1- s_3 z_2\right)\left(1- \frac{s_4 z_2}{z_1}\right)\left(1- \frac{z_1 b_2}{z_2}\right)\left(1-\frac{1}{ b_1^2 b_2 z_2}\right)}~.
\eea}
This can be written in terms of $t_1$ and $t_2$ as follows:
{\small
\bea
g (D_1; t_1, t_2; \cC_4) &=& \frac{2}{\left(1-t^2_1\right)^3\left(1-t^2_2\right)^3} \times   (1 + 5 t_1^2 + 6 t_1 t_2 + t_2^2 - 12 t_1^2 t_2^2 - t_1^4 t_2^2 - 6 t_1 t_2^3-\nn \\
&& 8 t_1^3 t_2^3 + 2 t_1^5 t_2^3 + 3 t_1^2 t_2^4 + 3 t_1^4 t_2^4 + 2 t_1 t_2^5 +  4 t_1^3 t_2^5)~. \qquad \qquad
\eea}
Thus, the R-charge of the perfect matching $p_1$ is given by
\bea
\lim_{\mu\rightarrow0}\frac{1}{\mu} \left[ \frac{g(D_1; e^{- \mu R_1 }, e^{- \mu R_2 }; \cC_4) }{\gm(e^{-\mu R_1}, e^{- \mu R_2 };\cC_4)}- 1 \right] \approx 0.334~.
\eea
The computations for the other perfect matchings can be done in a similar way.  The results are shown in Table \ref{t:chargefano123}.
The assignment of charges under the remaining abelian symmetry can be done by requiring that the superpotential is not charged under it and that the charge vectors are linearly independent. The assignments are shown in Table \ref{t:chargefano123}.
\begin{table}[h!]
 \begin{center}  
  \begin{tabular}{|c||c|c|c|c|c|c|c|}
  \hline
  \;& $SU(2)_1$&$SU(2)_2$&$U(1)_q$&$U(1)_R$&$U(1)_{B_1}$&$U(1)_{B_2}$&fugacity\\
  \hline\hline  
   
  $p_1$&$  1$&$  0$&$ 1$&$0.335$&$ 1$&$ 0$ & $s_1 x_1 q b_1$\\
  \hline
  
  $p_2$&$ -1$&$  0$&$ 1$&$0.335$&$ 1$&$ 0$ & $s_1  q b_1/x_1$\\
  \hline  
  
  $q_1$&$  0$&$  1$&$ 1$&$0.353$&$ 0$&$ 0$ & $s_2 x_2 q$\\
  \hline
  
  $q_2$&$  0$&$ -1$&$ 1$&$0.353$&$ 0$&$ 0$ & $s_2 q /x_2  $\\
  \hline
   
  $r_1$&$  0$&$  0$&$-2$&$0.241$&$ 0$&$ 0$ & $s_3 / q^2$\\
  \hline
      
  $r_2$&$  0$&$  0$&$-2$&$0.383$&$ 0$&$0$ & $s_4 / q^2 $\\
  \hline
  
  $v_1$&$  0$&$  0$&$ 0$&$    0$&$ 0$&$ 1$ & $b_2 $\\
  \hline

  $v_2$&$  0$&$  0$&$ 0$&$    0$&$-2$&$-1$ & $1/(b^2_1  b_2)$\\
  \hline
 
     \end{tabular}
  \end{center}
\caption{Charges of the perfect matchings under the global symmetry of the $\cC_4$ theory. Here $s_i$ are the fugacities of the R-charges, $x_1, x_2$ are the weight of the $SU(2)$ symmetries, $q, b_1$ and $b_2$ are, respectively, the charges under the mesonic abelian symmetries $U(1)_q$ and under the two baryonic $U(1)_{B_1}$ and $U(1)_{B_2}$.}
\label{t:chargefano123}
\end{table}

\begin{table}[h]
 \begin{center}  
  \begin{tabular}{|c||c|}
  \hline
  \; Quiver fields &R-charge\\
  \hline  \hline 
  $X^i_{31}, X^i_{24}$ &  $0.688$\\
  \hline
  $X^1_{43}, X^2_{43}$ &  $0.594$\\
  \hline
  $X_{32}, X_{14}$ &  $0.718$\\
  \hline
  $X^3_{43}$ &  $0.383$\\
  \hline
  $X_{12}$ &  $0.241$\\
  \hline
  \end{tabular}
  \end{center}
\caption{R-charges of the quiver fields of $dP_1 \times \BP^1$.}
\label{t:Rgenchfano123}
\end{table}

\paragraph{The Hilbert series.} Restoring the non-abelian fugacities and integrating over $z_1$ and $z_2$, the fully refined Hilbert series for the Master space can be obtained:
{\small \bea
g^{\firr{}}  (s_{\alpha},q,x_1,x_2, b_1, b_2; \cC_4) &=& \oint \limits_{|z_1| =1} {\frac{\ud z_1}{2 \pi i z_1}} \oint \limits_{|z_2| =1} {\frac{\ud z_2}{2 \pi i z_2}} \frac{1}{\left(1-s_1 x_1 q b_1 z_1\right)\left(1-\frac{s_1 q b_1 z_1}{x_1}\right)\left(1-\frac{s_2 x_2 q}{z_1}\right)}\nn\\
&\times& \frac{1}{\left(1-\frac{s_2 q}{x_2 z_1}\right)\left(1-\frac{s_3 z_2}{q^2}\right)\left(1-\frac{s_4 z_2}{q^2 z_1}\right)\left(1- \frac{b_2 z_1}{z_2}\right)\left(1-\frac{1}{b^2_1 b_2 z_2}\right)} \nn \\
&=& \frac{\mathcal{P}\left(s_{\alpha},q,x_1,x_2, b_1, b_2; \cC_4\right)}{\left(1- s_1 s_2 x_1 x_2 q^2 b_1 \right)\left(1-\frac{s_1 s_2 x_1 q^2 b_1}{x_2}\right)\left(1-\frac{s_1 s_2 x_2 q^2 b_1}{x_1}\right)\left(1-\frac{s_1 s_2 q^2 b_1}{x_1 x_2}\right)}\nn \\
&\times& \frac{1}{\left(1-\frac{s_1 s_4 x_1}{q b_1 b_2}\right)\left(1-\frac{s_1 s_4}{x_1 q b_1 b_2}\right)\left(1-\frac{s_2 s_3 x_2 b_2}{q}\right)\left(1-\frac{s_2 s_3 b_2}{x_2 q}\right)}\nn \\
&\times& \frac{1}{\left(1-\frac{s_4 b_2}{q^2}\right)\left(1-\frac{s_3}{q^2 b^2_1 b_2}\right)}~,
\label{e:HSmasterfano123}
\eea}
where $\mathcal{P}\left(s_{\alpha},q,x_1,x_2, b_1, b_2; \cC_4\right)$ is a polynomial that is not reported here.

The Hilbert series of the mesonic moduli space can be obtained by integrating (\ref{e:HSmasterfano123}) over the two baryonic fugacities $b_1$ and $b_2$:
{\small
\bea
\gm (s_{\alpha}, x_1, x_2, q; \cC_4) &=& \oint \limits_{|b_1| =1} {\frac{\ud b_1}{2 \pi i b_1}} \oint \limits_{|b_2| =1} {\frac{\ud z_2}{2 \pi i b_2}}g^{\firr{}}  (s_{\alpha},x_1,x_2,q, b_1, b_2; \cC_4)\nn \\
&=& \frac{P\left(s_{\alpha}, x_1, x_2, q; \cC_4\right)}
{\left(1- s^2_1 s^3_2 s_3^2 x^3_1 x^2_2 q\right)\left(1-\frac{s^2_1 s^3_2 s_3^2 x^3_1 q}{x^2_2}\right)\left(1-\frac{s^2_1 s^3_2 s^2_3 x^2_2 q}{x^3_1}\right)\left(1-\frac{s^2_1 s^3_2 s^2_3 q}{x^3_1 x^2_2}\right)}\nn \\
&\times& \frac{1}{\left(1-\frac{s^2_1 s_2 s^2_4 x_1 x^2_2}{q}\right)\left(1-\frac{s^2_1 s_2 s^2_4 x^2_2}{x_1 q}\right)\left(1-\frac{s^2_1 s_2 s^2_4 x_1}{x^2_2 q}\right)\left(1-\frac{s^2_1 s_2 s^2_4}{x_1 x^2_2 q}\right)}~,\nn \\
\label{e:HSmesfano123}
\eea}
where $P\left(s_{\alpha}, x_1, x_2, q; \cC_4\right)$ is a polynomial that is not reported here.

\begin{figure}[ht]
\begin{center} 
\includegraphics[totalheight = 4cm]{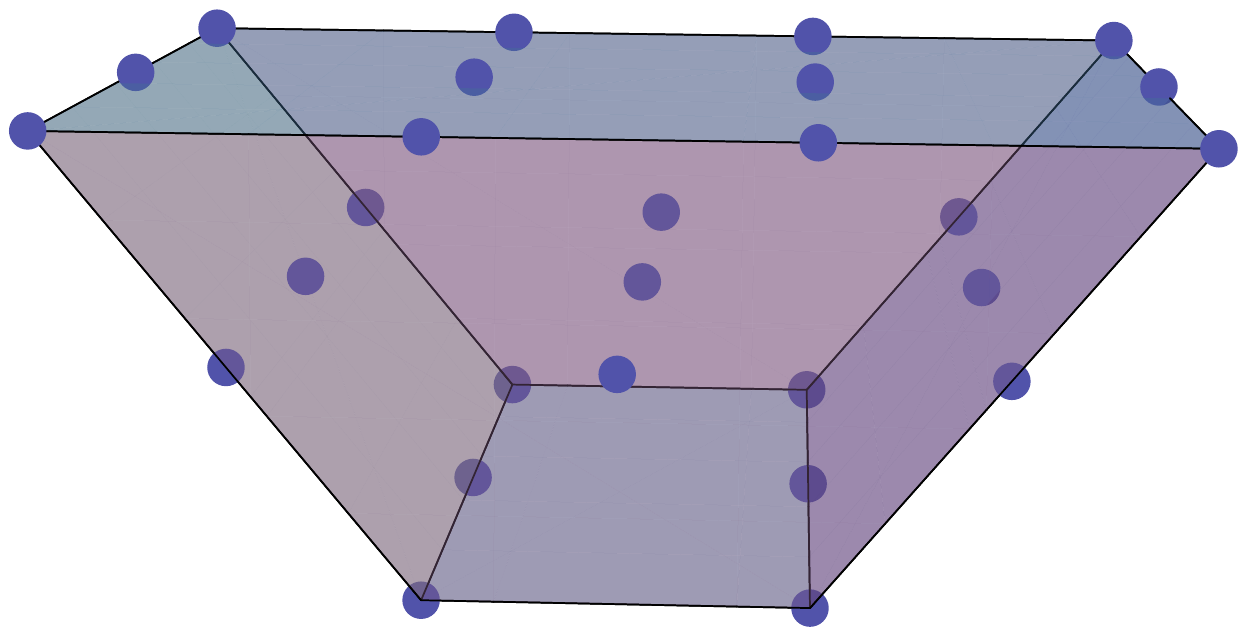}
\caption{The lattice of generators of the $\cC_4$ theory.}
  \label{f:latc4}
  \end{center}
\end{figure}

The plethystic logarithm of the mesonic Hilbert series is
\bea
\PL [\gm (s_{\alpha}, x_1, x_2, q; \cC_4)] 
&=& [2;3]q t_1^2 + [2;2] t_1 t_2 + [2;1] \frac{t_2^2}{q} - O(t_1^3) O(t_2^3)~,
\eea
where \eref{TC4} has been used to obtain the last line.
Therefore, the generators of the mesonic moduli space are
\bea
 p_i p_j q_k q_l q_m r^2_1 v_1 v_2,\qquad p_i p_j q_k q_l r_1 r_2 v_1 v_2,\qquad p_i p_j q_k r^2_2 v_1 v_2
\eea
with $i,j,k,l,m=1,2$. The R-charges of the generators of the mesonic moduli space are presented in Table \ref{t:Rgenfano123}.  The lattice of generators is drawn in \fref{f:latc4}.

\begin{table}[h]
 \begin{center}  
  \begin{tabular}{|c||c|}
  \hline
  \; Generators &$U(1)_R$\\
  \hline  \hline 
  $p_i p_j q_k q_l q_m r^2_1 v_1 v_2$ & 2.211 \\
  \hline
  $p_i p_j q_k q_l r_1 r_2 v_1 v_2$ & 2 \\
  \hline
  $p_i p_j q_k r^2_2 v_1 v_2$ & 1.789 \\
  \hline
  \end{tabular}
  \end{center}
\caption{R-charges of the generators of the mesonic moduli space for the $\cC_4$ Model.}
\label{t:Rgenfano123}
\end{table}

\section{$\cC_5$ (Toric Fano 68): $\BP \left(\mathcal{O}_{\BP^1 \times \BP^1} \oplus \mathcal{O}_{\BP^1 \times \BP^1}(1,-1) \right)$}
Below two phases of the $\cC_5$ theory are analysed in detail. The quiver diagram and tiling of each phase are identical to those of $Q^{1,1,1}/\BZ_{2}$, but the CS levels are different.

\subsection{Phase I of The $\cC_5$ Theory}
The quiver diagram and tiling of this model is given in \fref{f:phase1f0}.  The superpotential are given by \eref{suppotph1q111z2} and the CS levels are  $\vec{k} = (1,-2,1,0)$.

\paragraph{The toric diagram.} Two methods of computing the toric diagram are presented.

\begin{itemize}
\item {\bf The Kasteleyn matrix.} 
Although they have different values, the Chern-Simons levels can be parametrized in terms of the integers $n^i_{jk}$ in the same way as shown in (\ref{e:csfano62phI}). In particular, for this model the following choice is made:
\bea
n^{1}_{12} = - n^{2}_{23} = 1,\quad n^i_{jk}=0 \; \text{otherwise}~.
\eea
The Kasteleyn matrix $K$ can be computed for this model. The fundamental domain contains two black nodes and two white nodes, which implies that $K$ is a $2\times 2$ matrix\footnote{Although the tiling of this model is identical to that of the first phase of $Q^{1,1,1}/\BZ_2$, a different weight assignment is used in the Kasteleyn matrix. This choice will make the non-abelian factors of the global symmetry more apparent in the $G_K$ matrix.}:
\be
K= \left(
\begin{array}{c|cc}
& w_1 & w_2 \\
\hline
b_1 &  z^{n^{2}_{12}} +  x z^{n^{1}_{34}} &\  z^{n^{2}_{41}} +  y z^{n^{1}_{23}}   \\
b_2 &  z^{n^{1}_{41}} +  \frac{1}{y} z^{n^{2}_{23}} &\  z^{n^{1}_{12}} + \frac{1}{x} z^{n^{2}_{34}}  
\end{array}
\right) ~. \label{e:kastfano68ph1}
\ee
The permanent of the Kasteleyn matrix can be written as:
\bea
\mathrm{perm}~K 
&=&  x z^{(n^{1}_{12} + n^{1}_{34})} +  x^{-1} z^{(n^{2}_{12} + n^{2}_{34})} +   y z^{(n^{1}_{23} + n^{1}_{41})} +  y^{-1} z^{(n^{2}_{23} + n^{2}_{41})}\nn \\
&+& z^{( n^{1}_{12} + n^{2}_{12})}+  z^{( n^{1}_{23} + n^{2}_{23} )} + z^{(n^{1}_{34} + n^{2}_{34})} + z^{( n^{1}_{41} + n^{2}_{41})}\nn \\
&=&   x + x^{-1} z + y + y^{-1} z^{-1} +  z + z^{-1} + 2\nn \\
&& \text{(for $n^{2}_{12} = - n^{2}_{23} = 1,\quad n^i_{jk}=0\; \text{otherwise}$)} ~. 
\label{e:permKfano68ph1}
\eea
The coordinates of the toric diagram are collected in the columns of the following matrix:
\bea
G_K = \left(
\begin{array}{cccccccc}
   1 & -1 &  0 &  0 & 0 &  0 & 0 & 0 \\
   0 &  0 &  1 & -1 & 0 &  0 & 0 & 0 \\
   0 &  1 &  0 & -1 & 1 & -1 & 0 & 0
\end{array}
\right)~.
\eea
Note that the first two rows of the $G_K$ matrix contain the weights of two $SU(2)$ groups; this implies that the non-abelian part of the mesonic symmetry is $SU(2)\times SU(2)$.  From (\ref{e:permKfano68ph1}), the perfect matchings are
\bea 
&&   p_1 = \left\{X^1_{12}, X^1_{34}\right\}, 
\;\; p_2 = \left\{X^2_{12}, X^2_{34}\right\},
\;\; q_1 = \left\{X^1_{23}, X^1_{41}\right\},
\;\; q_2 = \left\{X^2_{23}, X^2_{41}\right\},\nn \\
&&   r_1 = \left\{X^1_{12}, X^2_{12}\right\},
\;\; r_2 = \left\{X^1_{23}, X^2_{23}\right\},
\;\; v_1 = \left\{X^1_{34}, X^2_{34}\right\},
\;\; v_2 = \left\{X^1_{41}, X^2_{41}\right\}\ . \qquad
\eea
It can be seen from $G_K$ that the perfect matchings $p_1, p_2, q_1, q_2, r_1, r_2$ correspond to external points in the toric diagram, while $v_1$ and $v_2$ correspond to the internal point.
The chiral fields can be written in terms of perfect matchings as follows:
\bea
&& X^1_{12} = p_1 r_1, \quad X^2_{12} = p_2 r_1, \quad  X^1_{23} = q_1 r_2, \quad X^2_{23} = q_2 r_2,\nn \\
&& X^1_{34} = p_1 v_1, \quad X^2_{34} = p_2 v_1, \quad  X^1_{41} = q_1 v_2, \quad X^2_{41} = q_2 v_2~.
\eea
All of these pieces of information can be summarised in the perfect matching matrix:
\beq
P=\left(\begin{array} {c|cccccccc}
  \;& p_1 & p_2 & q_1 & q_2 & r_1 & r_2 & v_1 & v_2\\
  \hline 
  X^{1}_{12}& 1&0&0&0&1&0&0&0\\
  X^{2}_{12}& 0&1&0&0&1&0&0&0\\
  X^{1}_{23}& 0&0&1&0&0&1&0&0\\
  X^{2}_{23}& 0&0&0&1&0&1&0&0\\
  X^{1}_{34}& 1&0&0&0&0&0&1&0\\
  X^{2}_{34}& 0&1&0&0&0&0&1&0\\
  X^{1}_{41}& 0&0&1&0&0&0&0&1\\
  X^{2}_{41}& 0&0&0&1&0&0&0&1
  \end{array}
\right).
\eeq
The null space of $P$ is two-dimensional and is spanned by two vectors that can be written as the rows of the following charge matrix:
\be
Q_F =   \left(
\begin{array}{cccccccc}
1 & 1 & 0 & 0 & -1 &  0 & -1 &  0\\
0 & 0 & 1 & 1 &  0 & -1 &  0 & -1
\end{array}
\right)~.  \label{e:qffano68ph1}
\ee
Hence, among the perfect matchings there are two relations, given by:
\bea
p_1 + p_2 - r_1 - v_1 = 0\\
q_1 + q_2 - r_2 - v_2 = 0
\label{e:relpmfano68ph1}
\eea

\item {\bf The charge matrices.}
Because the number of gauge groups of this model is $G = 4$, there are $G-2 =2$ baryonic symmetries coming from the D-terms. The charges of the perfect matchings can be collected in the $Q_D$ matrix:
\be
Q_D =   \left(
\begin{array}{cccccccc}
 0 & 0 & 0 & 0 & 1 & 1 & 0 &-2\\
 0 & 0 & 0 & 0 & 0 & 0 & 1 &-1
\end{array}
\right) \label{e:qdfano68ph1}
\ee
(\ref{e:qffano68ph1}) and (\ref{e:qdfano68ph1}) can be combined in a single matrix, $Q_t$, that contain all charges of the perfect matchings that need to be integrated over in order to compute the Hilbert series of the mesonic moduli space:

\be
Q_t = { \Blue Q_F \choose \Green Q_D \Black } =   \left( 
\begin{array}{cccccccc} \Blue
 1 & 1 & 0 & 0 & -1 &  0 & -1 &  0\\
 0 & 0 & 1 & 1 &  0 & -1 &  0 & -1 \\ \Green
 0 & 0 & 0 & 0 &  1 &  1 &  0 & -2\\
 0 & 0 & 0 & 0 &  0 &  0 &  1 & -1 \Black
\end{array}
\right) 
\label{e:qtfano68ph1}
\ee
The $G_t$ matrix can be computed from the null space of the matrix written in (\ref{e:qtfano68ph1}), which, after the removal of the first row, contains the coordinates of the toric diagram in its columns:
\bea
G'_t = \left(
\begin{array}{cccccccc}
   1 & -1 &  0 &  0 & 0 &  0 & 0 & 0 \\
   0 &  0 &  1 & -1 & 0 &  0 & 0 & 0 \\
   0 &  1 &  0 & -1 & 1 & -1 & 0 & 0
\end{array}
\right) = G_K~. \label{e:gtfano68ph1}
\eea
The toric diagram constructed from (\ref{e:gtfano68ph1}) is presented in Figure \ref{f:tdtoricfano68ph1}:
\begin{figure}[ht]
\begin{center}
  \includegraphics[totalheight=3.0cm]{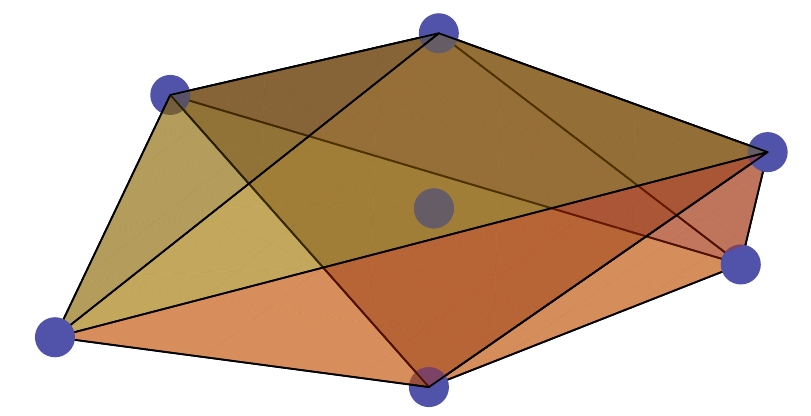}
 \caption{The toric diagram of the $\cC_5$ theory.}
  \label{f:tdtoricfano68ph1}
\end{center}
\end{figure}
\end{itemize}

\paragraph{The baryonic charges.} The toric diagram of this phase of the theory has 6 external points and, therefore, the number of baryonic symmetries is $6-4=2$, which will be called $U(1)_{B_1}$ and $U(1)_{B_2}$. The charges of the perfect matchings under these two baryonic symmetries can be read off from the $Q_D$ matrix written above.

\paragraph{The global symmetry.} The two pairs of repeated columns confirm that the global symmetry contains a non-abelian factor $SU(2)^2$. In turn, this implies that the mesonic symmetry is $SU(2)^2\times U(1)^2$, where one of the abelian factors can be identified with the R-symmetry.

The perfect matchings $p_1$ and $p_2$ transform as a doublet under the first $SU(2)$, whereas $q_1$ and $q_2$ are a doublet under the second $SU(2)$. The perfect matchings $v_1$ and $v_2$ correspond to the internal point of the toric diagram, accordingly, carry zero R-charge. Note that the presence of two abelian symmetries requires us to solve a volume minimisation problem to determine the correct R-charges in the IR. The procedure is similar to the one that is discussed in the previous sections.

Let $s_1$ be the fugacity for the R-charge of $p_1$ and $p_2$, $s_2$ that of $q_1$ and $q_2$, and $s_3, s_4$ those of  $r_1, r_2$ respectively.  The mesonic Hilbert series can be written as:
\bea
\gm (s_{\alpha}; \cC^{(I)}_5) &=& \oint \limits_{|z_1| =1} {\frac{\ud z_1}{2\pi i z_1}}\oint \limits_{|z_2| =1} {\frac{\ud z_2}{2\pi i z_2}} \oint \limits_{|b_1| =1} {\frac{\ud b_1}{2\pi i b_1 }}\oint \limits_{|b_2| =1} {\frac{\ud b_2}{2\pi i b_2 }} \frac{1}{\left(1 - s_1 z_1\right)^2 \left(1- s_2 z_2\right)^2}\nn \\
&\times& \frac{1}{\left(1- \frac{s_3 b_1}{z_1} \right)\left(1 - \frac{s_4 b_1}{z_2}\right)\left(1 - \frac{b_2}{z_1}\right)\left(1 - \frac{1}{b^2_1 b_2 z_2}\right)}~.
\eea
Since there are two factors of $U(1)$ in the mesonic symmetry, the mesonic Hilbert series depends on two combinations of $s_\alpha$'s. Defining:
\bea
t_1^2 = s_1^3 s_2 s_3^2 ~, \quad t_2^2 = s_1 s_2^3 s_4^2 ~,
\eea
the mesonic Hilbert series can be written as
{\small
\bea
\gm (t_1, t_2; \cC^{(I)}_5) &=& \frac{1}{\left(1-t_1^2 \right)^3 \left(1-t_2^2 \right)^3 }\times (1+5 t_1^2+9 t_1 t_2-3 t_1^3 t_2+5 t_2^2-14 t_1^2 t_2^2-3 t_1^4 t_2^2 -\nn \\
&& 3 t_1 t_2^3  -14 t_1^3 t_2^3+5 t_1^5 t_2^3-3 t_1^2 t_2^4+9 t_1^4 t_2^4+5 t_1^3 t_2^5+t_1^5 t_2^5)~,\nn \\
\eea}
If $R_1$ and $R_2$ are the R-charges corresponding to $t_1$ and $t_2$, the requirement that the superpotential, which has fugacity $t_1 t_2$, has R-charge 2 implies:
\bea
R_1 + R_2 = 2~. \label{ph1c5cy}
\eea
The volume of $\cC_5$ is given by
\bea
\lim_{\mu \rightarrow 0} \mu^4 \gm (e^{-\mu R_1}, e^{-\mu(2-R_1)};  \cC_5)= \frac{12 - 2 R_1 + R_1^2}{4 (2 - R_1)^3 R_1^3}~,
\eea
This has a minimum at $R_1 = 1$, at which the relation \eref{ph1c5cy} implies that $R_2=1$.  

The R-charges of the external perfect matchings are given by the following formula:
\bea
\lim_{\mu\rightarrow0}\frac{1}{\mu} \left[ \frac{g(D_\alpha; e^{- \mu R_1}, e^{- \mu R_2 }; \cC_5) }{\gm(e^{-\mu R_1}, e^{- \mu R_2 };\cC_5)}- 1 \right]~,
\eea
where $g(D_\alpha; e^{- \mu R_1}, e^{- \mu R_2 }; \cC_5)$ is the Molien-Weyl integral with the insertion of the inverse of the weight corresponding to the divisor $D_\alpha$. The results of computations are shown in Table \ref{t:chargefano68ph1}.

The charge assignment for $U(1)_q$ is subject to the condition that the superpotential and the internal perfect matchings carry zero charges, and that the charge vectors are all linearly independent.  All of the results are collected in Table \ref{t:chargefano68ph1}.
\begin{table}[h!]
 \begin{center}  
  \begin{tabular}{|c||c|c|c|c|c|c|c|}
  \hline
  \;& $SU(2)_1$&$SU(2)_2$&$U(1)_q$&$U(1)_R$&$U(1)_{B_1}$&$U(1)_{B_2}$&fugacity\\
  \hline\hline  
   
  $p_1$&$  1$&$  0$&$ 1$&$4/11$&$ 0$&$ 0$ & $t^4 x_1 q $\\
  \hline
  
  $p_2$&$ -1$&$  0$&$ 1$&$4/11$&$ 0$&$ 0$ & $t^4 q / x_1 $\\
  \hline  
  
  $q_1$&$  0$&$  1$&$-1$&$4/11$&$ 0$&$ 0$ & $t^4 x_2 / q $\\
  \hline
  
  $q_2$&$  0$&$ -1$&$-1$&$4/11$&$ 0$&$ 0$ & $t^4/ (x_2 q) $\\
  \hline
   
  $r_1$&$  0$&$  0$&$ 0$&$3/11$&$ 1$&$ 0$ & $t^3 b_1 $\\
  \hline
  
  $r_2$&$  0$&$  0$&$ 0$&$3/11$&$ 1$&$ 0$ & $t^3 b_1 $\\
  \hline
  
  $v_1$&$  0$&$  0$&$ 0$&$   0$&$ 0$&$ 1$ & $ b_2 $\\
  \hline

  $v_2$&$  0$&$  0$&$ 0$&$   0$&$-2$&$-1$ & $1 / (b_1^2 b_2) $\\
  \hline
  
  $\Blue v_3$&$  0$&$  0$&$ 0$&$   0$&$ 0$&$ 0$ & $1 $\\
  \hline
   \end{tabular}
  \end{center} \Black
\caption{Charges of the perfect matchings under the global symmetry of both phases of the $\cC_5$ theory. Here $t$ is the fugacity of the R-charge (in units of $1/11$), $x_1,x_2$ are the weights of the $SU(2)$ symmetry, $q, b_1$ and $b_2$ are, respectively, the charges under the mesonic abelian symmetry $U(1)_q$ and of the two baryonic symmetries $U(1)_{B_1}$ and $U(1)_{B_2}$. Note that the perfect matching $v_3$ (represented in blue) does not exist in Phase I but exists in Phase II.}
\label{t:chargefano68ph1}
\end{table}

\begin{table}[h]
 \begin{center}  
  \begin{tabular}{|c||c|}
  \hline
  \; Quiver fields &R-charge\\
  \hline  \hline 
  $X^i_{12}, X^i_{23}$ &  $7/11$\\
  \hline
  $X^i_{34}, X^i_{41}$ &  $4/11$\\
  \hline
  \end{tabular}
  \end{center}
\caption{R-charges of the quiver fields of $\cC_5$, Phase I.}
\label{t:Rgenfano68phI}
\end{table}

\begin{figure}[ht]
\begin{center}
 \includegraphics[totalheight=4cm]{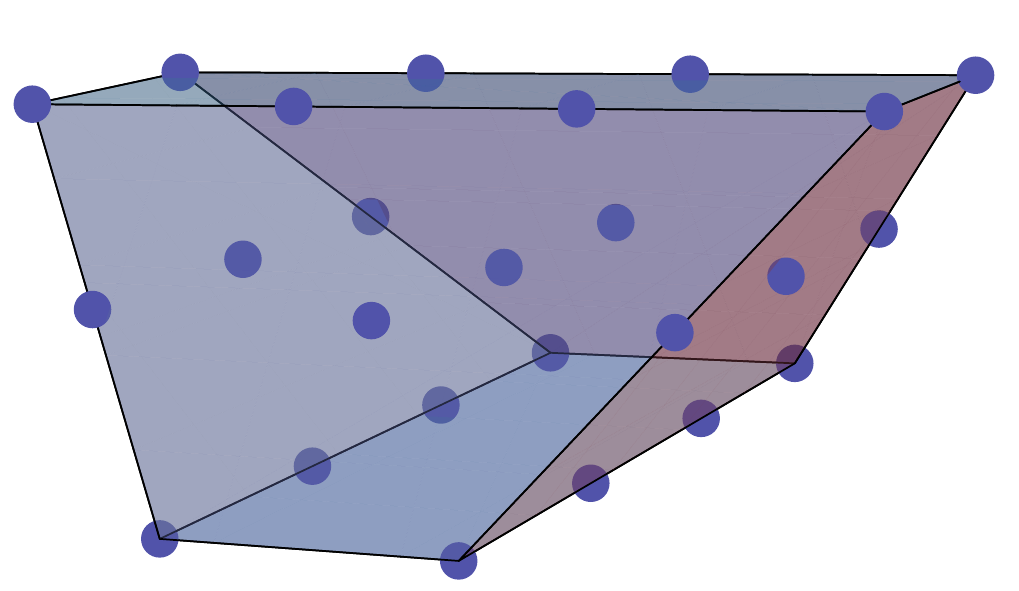}
\caption{The lattice of generators of the $\cC_5$ theory.}
  \label{f:latc5}
\end{center}
\end{figure}

\paragraph{The Hilbert series.} The Hilbert series of the Master space can be obtained by integrating that of the space of perfect matchings over the two fugacities $z_1$ and $z_2$:
\bea
g^{\firr{}} (t, x_1, x_2, q, b_1, b_2; \cC^{(I)}_5) &=& \oint \limits_{|z_1| =1} {\frac{\ud z_1}{2 \pi i z_1}} \oint \limits_{|z_2| =1} {\frac{\ud z_2}{2 \pi i z_2}} \frac{1}{\left(1- t^4 x_1 q z_1\right)\left(1- \frac{t^4 q z_1}{x_1} \right)\left(1- \frac{t^4 x_2 z_2}{q}\right)}\nn \\
&\times& \frac{1}{\left(1- \frac{t^4 z_2}{x_2 q}\right)\left(1- \frac{t^3 b_1}{z_1}\right)\left(1- \frac{t^3 b_1}{z_2}\right)\left(1-\frac{b_2}{z_1}\right)\left(1- \frac{1}{b^2_1 b_2 z_2}\right)}\nn \\
&=& \frac{\left(1- t^{11} q^2 b_1 b_2\right)}{\left(1-t^7 x_1 q b_1\right)\left(1-\frac{t^7 q b_1}{x_1}\right)\left(1- t^4 x_1 q b_2\right)\left(1- \frac{t^4 q b_2}{x_1}\right)}\nn \\
&\times& \frac{\left(1-\frac{t^{11}}{q^2 b_1 b_2}\right)}{\left(1-\frac{t^7 x_2 b_1}{q}\right)\left(1-\frac{t^7 b_1}{x_2 q}\right)\left(1-\frac{t^4 x_2}{q b^2_1 b_2}\right)\left(1-\frac{t^4}{x_2 q b^2_1 b_2}\right)}~.
\label{e:HSmasterfano68ph1}
\eea
The unrefined Hilbert series of the Master space can be written as:
\bea
g^{\firr{}} (t,1,1,1,1,1; \cC^{(I)}_5) &=& \left[ \frac{1-t^{11}}{\left(1-t^4\right)^2\left(1-t^7\right)^2} \right]^2  ~.
\eea
The Hilbert series indicates that the Master space of this theory is the product of two conifolds, as for Phase I of $Q^{1,1,1}/\BZ_2$.
The mesonic Hilbert series is obtained by integrating (\ref{e:HSmasterfano68ph1}) over the baryonic fugacities $b_1$ and $b_2$:
\bea
\gm (t, x_1, x_2, q; \cC^{(I)}_5) &=& \oint \limits_{|b_1| =1} {\frac{\ud b_1}{2 \pi i b_1}} \oint \limits_{|b_2| =1} {\frac{\ud b_2}{2 \pi i b_2}} g^{\firr{}} (t, x_1, x_2, q, b_1, b_2; \cC^{(I)}_5) \nn \\
&=& \frac{P(t,x_1, x_2, q; \cC^{(I)}_5)}{\left(1-\frac{t^{22} x^3_1 q^2}{x_2}\right)\left(1-\frac{t^{22} x_2 q^2}{x^3_1}\right)\left(1-\frac{t^{22} q^2}{x^3_1 x_2}\right)\left(1- t^{22} x^3_1 x_2 q^2\right)} \nn \\
&\times& \frac{1}{\left(1-\frac{t^{22} x_1}{q^2 x^3_2}\right)\left(1-\frac{t^{22} x^3_2}{x_1 q^2}\right)\left(1-\frac{t^{22}}{x^3_2 x_1 q^2}\right)\left(1-\frac{t^{22} x_1 x^3_2}{q^2}\right)}~.
\eea
The unrefined Hilbert series can be written as:
\bea
\gm(t,1,1,1;\cC^{(I)}_5) = \frac{1+ 21t^{22}+ 21t^{44}+t^{66}}{\left(1-t^{22}\right)^4}~.
\eea
The plethystic logarithm of this Hilbert series is given by:
\bea
\PL[\gm(t,x_1, x_2, q; \cC^{(I)}_5)] = \left([3;1]q^2 + [2;2] + \frac{1}{q^2}[1;3]\right)t^{22} - O(t^{44})
\eea
The generators of the mesonic moduli space are:
\bea
p_i p_j p_k q_l r^2_1 v_1 v_2, \qquad p_i p_j q_k q_l r_1 r_2 v_1 v_2, \qquad p_i q_j q_l q_k r^2_2 v_1 v_2~, 
\eea
where $i,j,k,l = 1,2$~.  Each generator carries R-charge equal to 2.  The lattice of generators is drawn in \fref{f:latc5}.

\comment{
The R-charges of generators are presented in Table \ref{t:Rgenfano68Ph1}.
\begin{table}[h]
 \begin{center}  
  \begin{tabular}{|c||c|}
  \hline
  \; Generators &$U(1)_R$\\
  \hline  \hline 
  $p_i p_j p_k q_l r^2_1 v_1 v_2$ & 2 \\
  \hline
  $p_i p_j q_k q_l r_1 r_2 v_1 v_2$ & 2 \\
  \hline
  $p_i q_j q_l q_k r^2_2 v_1 v_2$ & 2 \\
  \hline
  \end{tabular}
  \end{center}
\caption{R-charges of the generators of the mesonic moduli space for the $\cC^{(I)}_5$ Model.}
\label{t:Rgenfano68Ph1}
\end{table}}

\subsection{Phase II of The $\cC_5$ Theory}
The quiver diagram and tiling of this model are identical to those of Phase II of $Q^{1,1,1}/\BZ_2$ (\fref{f:phase2f0}).  However, for this model the CS levels are chosen to be $\vec{k} =(0,0,1,-1)$.

\paragraph{The toric diagram.} Two methods of computing the toric diagram for this model are discussed.
\begin{itemize}
\item {\bf The Kasteleyn matrix.} 
\comment{The integers $n_i$ are assigned to the edges of the tiling as shown in Figure \ref{}.
Therefore,
\bea
\begin{array}{rcl}
\text{Gauge group 1~:} \qquad k_1  =&  0 &= - n_1 - n_2 - n_3 - n_4 + n_9 + n_{10} + n_{11} + n_{12} ~, \nn \\
\text{Gauge group 2~:} \qquad k_2  =&  0 &=   n_5 + n_6 + n_7 + n_8  - n_9 - n_{10} - n_{11} - n_{12}~, \nn \\
\text{Gauge group 3~:} \qquad k_3  =&  1 &=   n_1 + n_2 - n_7 - n_8 ~, \nn \\
\text{Gauge group $3'$~:} \qquad k_{3'} =& -1 &=  n_3 + n_4 - n_5 - n_6 ~.
\label{e:kafano68ph2}
\end{array}
\eea}
The Chern-Simons levels can be parametrized in terms of integers as according to (\ref{e:csfano62phII}). For this model let us choose:
\bea
n^{1}_{31} = - n^{1}_{3'1} = 1,\quad n^i_{kl}=n^{ij}_{kl}=0 \; \text{otherwise}~.
\eea
Since the fundamental domain contains four pairs of black nodes and white nodes, the Kasteleyn matrix $K$ is a $4\times 4$ matrix\footnote{The weight assignment for the edges crossing the fundamental domain is different from Phase II of $Q^{1,1,1}/\BZ_2$. This choice will make the non-abelian mesonic symmetries more evident in the $G_K$ matrix}:
\be
K= \left(
\begin{array}{c|cccc}
& w_1 & w_2 & w_3 & w_4\\
\hline
b_1 & y z^{n^{2}_{23}} &\ \frac{1}{x} z^{n^{1}_{31}}      &\           0           &\ z^{n^{21}_{12}} \\
b_2 & x z^{n^{2}_{31}} &\ \frac{1}{y} z^{n^{1}_{23}}      &\ z^{n^{12}_{12}}  &\         0            \\
b_3 &         0          &\ z^{n^{22}_{12}}  &\ z^{n^{1}_{3'1}}   &\ z^{n^{1}_{23'}} \\
b_4 & z^{n^{11}_{12}} &\       0              &\ z^{n^2_{23'}}     &\ z^{n^{2}_{3'1}}
\end{array}
\right) ~. \label{e:kastfano68ph2}
\ee
The permanent of the Kasteleyn matrix is:
\bea
\perm~K &=&  
 x z^{(n^{2}_{31} + n^{2}_{23'} + n^{21}_{12} + n^{22}_{12})} + x^{-1} z^{(n^{1}_{31} + n^{1}_{23'} + n^{11}_{12} + n^{12}_{12})} +  y z^{(n^{2}_{3'1} + n^{2}_{23} + n^{12}_{12} + n^{22}_{12})}\nn \\
&+&  y^{-1} z^{(n^{1}_{3'1} + n^{1}_{23} + n^{11}_{12} + n^{21}_{12})} + z^{(n^{1}_{31} + n^{2}_{31} + n^{1}_{23'} + n^{2}_{23'})} +  z^{(n^{1}_{3'1} + n^{2}_{3'1} + n^{1}_{23} + n^{2}_{23})} \nn \\
&+&  z^{(n^{11}_{12} + n^{21}_{12} + n^{12}_{12} + n^{22}_{12})} +  z^{(n^{1}_{23'} + n^{2}_{23'} + n^{1}_{23} + n^{2}_{23})} +  z^{(n^{1}_{31} + n^{2}_{31} + n^{1}_{3'1} + n^{2}_{3'1})} \nn \\
&=&  x + x^{-1} z +  y + y^{-1} z^{-1} +  z +  z^{-1} + 3\nn \\
&&  \text{(for $n^{1}_{31} = - n^{1}_{3'1} = 1,\quad n^i_{kl}=n^{ij}_{kl}=0 \; ~ \text{otherwise}$)} ~.
\label{e:permKfano68ph2}
\eea
The coordinates of the toric diagram are contained in the columns of the following matrix:
\bea
G_K = \left(
\begin{array}{ccccccccc}
   1 & -1 &  0 &  0 & 0 &  0 & 0 & 0 & 0\\
   0 &  0 &  1 & -1 & 0 &  0 & 0 & 0 & 0\\
   0 &  1 &  0 & -1 & 1 & -1 & 0 & 0 & 0
\end{array}
\right)~.
\eea
Note that the first two rows of the $G_K$ matrix contain the weights of $SU(2)$; this imply that the non-abelian part of the mesonic symmetry is $SU(2)\times SU(2)$. The permanent of the Kasteleyn matrix (\ref{e:permKfano68ph2}) gives us the possibility to write down the perfect matchings in terms of the chiral fields of the model:
\bea 
&&   p_1 = \left\{X^2_{31}, X^2_{23'},X^{21}_{12}, X^{22}_{12}\right\}, 
\;\; p_2 = \left\{X^1_{31}, X^1_{23'},X^{11}_{12}, X^{12}_{12}\right\},\nn \\
&&   q_1 = \left\{X^2_{3'1}, X^2_{23},X^{12}_{12}, X^{22}_{12}\right\},
\;\; q_2 = \left\{X^1_{3'1}, X^1_{23},X^{11}_{12}, X^{21}_{12}\right\},\nn \\
&&   r_1 = \left\{X^1_{31}, X^2_{31},X^{1}_{23'}, X^2_{23'}\right\},
\;\; r_2 = \left\{X^1_{3'1}, X^2_{3'1},X^1_{23}, X^2_{23}\right\},\nn \\
&&   v_1 = \left\{X^{11}_{12}, X^{12}_{12},X^{21}_{12}, X^{22}_{12}\right\},
\;\; v_2 = \left\{X^1_{23'}, X^2_{23'},X^1_{23}, X^2_{23}\right\},\nn \\
&&   v_3 = \left\{X^1_{31}, X^2_{31},X^1_{3'1}, X^2_{3'1}\right\}\ . \qquad
\eea
From (\ref{e:permKfano68ph2}), the perfect matchings $p_1, p_2, q_1, q_2, r_1, r_2$ correspond to external points in the toric diagram, whereas $v_1, v_2$ and $v_3$ correspond to the internal point.
The chiral fields can be parametrized in terms of perfect matchings as follows:
\bea
&& X^2_{31} = p_1 r_1 v_3, \quad X^1_{31} = p_2 r_1 v_3, \quad  X^2_{3'1} = q_1 r_2 v_3 , \quad X^1_{3'1} = q_2 r_2 v_3,\nn \\
&& X^2_{23'} = p_1 r_1 v_2, \quad X^1_{23'} = p_2 r_1 v_2, \quad  X^2_{23} = q_1 r_2 v_2, \quad X^1_{23} = q_2 r_2 v_2 ,\nn \\
&& X^{22}_{12} = p_1 q_1 v_1, \quad X^{21}_{12} = p_1 q_2 v_1 \quad  X^{12}_{12} = p_2 q_1 v_1, \quad X^{11}_{12} = p_2 q_2 v_1~.
\eea
The information above can be summarized in the perfect matching matrix:
\beq
P=\left(\begin{array} {c|ccccccccc}
  \;& p_1 & p_2 & q_1 & q_2 & r_1 & r_2 & v_1 & v_2 & v_3\\
  \hline 
  X^{2}_{31} & 1&0&0&0&1&0&0&0&1\\
  X^{1}_{31} & 0&1&0&0&1&0&0&0&1\\
  X^{2}_{3'1}& 0&0&1&0&0&1&0&0&1\\
  X^{1}_{3'1}& 0&0&0&1&0&1&0&0&1\\
  X^{2}_{23'}& 1&0&0&0&1&0&0&1&0\\
  X^{1}_{23'}& 0&1&0&0&1&0&0&1&0\\
  X^{2}_{23} & 0&0&1&0&0&1&0&1&0\\
  X^{1}_{23} & 0&0&0&1&0&1&0&1&0\\
  X^{22}_{12}& 1&0&1&0&0&0&1&0&0\\
  X^{21}_{12}& 1&0&0&1&0&0&1&0&0\\
  X^{12}_{12}& 0&1&1&0&0&0&1&0&0\\
  X^{11}_{12}& 0&1&0&1&0&0&1&0&0\\
  \end{array}
\right).
\eeq
A basis of the null space of $P$ is given in the rows of
\be
Q_F =   \left(
\begin{array}{ccccccccc}
1 & 1 & 0 & 0 &-1 & 0 &-1 & 0 &  0\\
0 & 0 & 1 & 1 & 0 &-1 &-1 & 0 &  0\\
0 & 0 & 0 & 0 & 1 & 1 & 0 &-1 & -1
\end{array}
\right)~.  \label{e:qffano68ph2}
\ee
Hence, there are three relations among the perfect matchings:
\bea
p_1 + p_2 - r_1 - v_1 = 0,\\
q_1 + q_2 - r_2 - v_1 = 0,\\
r_1 + r_2 - v_2 - v_3 = 0.
\label{e:relpmfano68ph2}
\eea

\item {\bf The charge matrices.}
Since the number of gauge groups of this model is $G = 4$, there are $G-2 =2$ baryonic symmetries coming from the D-terms. The charges of the perfect matchings can be collected in the $Q_D$ matrix:
\be
Q_D =   \left(
\begin{array}{ccccccccc}
 0 &  0 & 0 &  0 & 1 & 1 & 0 & -2 & 0\\
 0 &  0 & 0 &  0 & 0 & 0 & 1 & -1 & 0
\end{array}
\right). \label{e:qdfano68ph2}
\ee
The total charge matrix that contains all the baryonic charges comes from the combination of the $Q_F$ matrix and the $Q_D$ matrix and, accordingly, can be written as:
\be
Q_t = { \Blue Q_F \choose \Green Q_D \Black } =   \left( 
\begin{array}{ccccccccc} \Blue
  1 &  1 & 0 & 0 & -1 & 0 &-1 &  0 & 0\\
  0 &  0 & 1 & 1 &  0 &-1 &-1 &  0 & 0\\
  0 &  0 & 0 & 0 &  1 & 1 & 0 & -1 &-1\\ \Green
  0 &  0 & 0 &  0 & 1 & 1 & 0 & -2 & 0\\
  0 &  0 & 0 &  0 & 0 & 0 & 1 & -1 & 0\Black
\end{array}
\right). 
\label{e:qtfano68ph2}
\ee
The $G_t$ matrix is the kernel of $Q_t$.  After eliminating the first trivial row the $G'_t$ matrix is obtained, which has the coordinates of the points of the toric diagram as its columns:
\bea
G'_t = \left(
\begin{array}{cccccccccc}
   1 & -1 &  0 &  0 & 0 &  0 & 0 & 0 & 0\\
   0 &  0 &  1 & -1 & 0 &  0 & 0 & 0 & 0\\
   0 &  1 &  0 & -1 & 1 & -1 & 0 & 0 & 0
\end{array}
\right) = G_K~. \label{e:gtfano68ph2}
\eea
The toric diagram constructed from (\ref{e:gtfano68ph2}) coincides with the one presented in \ref{f:tdtoricfano68ph1}.
\end{itemize}

\paragraph{The baryonic charges.} The toric diagram of this model has 6 external points and, accordingly, this theory has $6-4=2$ baryonic symmetries, which will be called $U(1)_{B_1}$ and $U(1)_{B_2}$. The charges of the perfect matchings under these two baryonic symmetries can be read off from the $Q_D$ matrix.

\paragraph{The global symmetry.} From the $Q_t$ matrix (\ref{e:qtfano68ph2}), it can be seen that the mesonic symmetry of this theory is the same as that of Phase I, \emph{i.e.} $SU(2)\times SU(2)\times U(1)\times U(1)_R$. The baryonic symmetry is also the same. The perfect matchings $p_1$ and $p_2$ transform as a doublet under the first $SU(2)$ and $q_1$ and $q_2$ as a doublet under the second $SU(2)$. Note that, apart from $v_3$, the perfect matchings of Phase I and Phase II of the theory are in one-to-one correspondence.  Computations show that the R-charges of perfect matchings of Phase I and Phase II are identical. The charge assignment is given in Table \ref{t:chargefano68ph1}.

\begin{table}[h]
 \begin{center}  
  \begin{tabular}{|c||c|}
  \hline
  \; Quiver fields &R-charge\\
  \hline  \hline 
  $X^i_{31}, X^i_{3'1}, X^i_{23}, X^i_{23'}$ &  $7/11$\\
  \hline
  $X^{ij}_{12}$ &  $8/11$\\
  \hline
  \end{tabular}
  \end{center}
\caption{R-charges of the quiver fields of $\cC_5$, Phase II.}
\label{t:Rgenfano68phII}
\end{table}

\paragraph{The Hilbert series.} The Hilbert series for the Master space is
{\small
\bea
g^{\firr{}} (t, x_1, x_2, q, b_1, b_2; \cC^{(II)}_5) &=& \prod_{j=1}^3\oint \limits_{|z_i| =1} {\frac{\ud z_i}{2 \pi i z_i}}  \frac{1}{\left(1- t^4 x_1 q z_1\right)\left(1- \frac{t^4 q z_1}{x_1}\right)\left(1- \frac{t^4 x_2 z_2}{q}\right)\left(1- \frac{t^4 z_2}{x_2 q}\right)}\nn \\
& \times & \frac{1}{\left(1- \frac{t^3 b_1 z_3}{z_1} \right)\left(1- \frac{t^3 b_1 z_3}{z_2}\right)\left(1- \frac{b_2}{z_1 z_2}\right)\left(1-\frac{1}{b^2_1 b_2 z_3}\right)\left(1- \frac{1}{z_3}\right) }\nn \\
&=& \frac{\cP(t,x_1, x_2, q;  \cC^{(II)}_5)}{\left(1 - t^7 x_1 q b_1\right)\left(1-\frac{t^7 q b_1}{x_1}\right)\left(1-\frac{t^7 x_1 q}{b_1 b_2}\right)\left(1-\frac{t^7 q}{x_1 b_1 b_2}\right)\left(1-\frac{t^7 x_2 b_1}{q}\right)}\nn \\
&\times& \frac{1}{\left(1-\frac{t^7 b_1}{x_2 q}\right)\left(1-\frac{t^7 x_2}{q b_1 b_2}\right)\left(1-\frac{t^7}{x_2 q b_1 b_2}\right)\left(1-t^8 x_1 x_2 b_2\right)\left(1-\frac{t^8 x_1 b_2}{x_2}\right)}\nn \\
&\times& \frac{1}{\left(1-\frac{t^8 x_2 b_2}{x_1}\right)\left(1-\frac{t^8 b_2}{x_1 x_2}\right)}~.
\label{e:HSmasterfano68ph2}
\eea}
where $\cP(t,x_1, x_2, q; \cC^{(II)}_5)$ is a polynomial which is not reported here.
The Hilbert series of the mesonic moduli space can be computed integrating (\ref{e:HSmasterfano68ph2}) over the two baryonic fugacities $b_1$ and $b_2$:
{\small
\bea
\gm (t, x_1, x_2, q; \cC^{(II)}_5) &=& \oint \limits_{|b_1| =1} {\frac{\ud b_1}{2 \pi i b_1}} \oint \limits_{|b_2| =1} {\frac{\ud b_2}{2 \pi i b_2}} g^{\firr{}} (t, x_1, x_2, q, b_1, b_2; \cC_1) \nn \\
&=& \frac{P(t,x_1, x_2, q; \cC^{(II)}_5)}{\left(1-\frac{t^{22} x^3_1 q^2}{x_2}\right)\left(1-\frac{t^{22} x_2 q^2}{x^3_1}\right)\left(1-\frac{t^{22} q^2}{x^3_1 x_2}\right)\left(1- t^{22} x^3_1 x_2 q^2\right)}\nn \\
&\times& \frac{1}{\left(1-\frac{t^{22} x_1}{q^2 x^3_2}\right)\left(1-\frac{t^{22} x^3_2}{x_1 q^2}\right)\left(1-\frac{t^{22}}{x^3_2 x_1 q^2}\right)\left(1-\frac{t^{22} x_1 x^3_2}{q^2}\right)}~,
\eea}
where $P(t,x_1, x_2, q; \cC^{(II)}_5)$ is a polynomial which is not reported here.
The unrefined mesonic Hilbert series of this equation is
\bea
\gm(t,1,1,1;\cC^{(II)}_5) &=& \frac{1+ 21t^{22}+ 21t^{44}+t^{66}}{\left(1-t^{22}\right)^4}~.
\eea
As was to be expected, this is identical to the mesonic Hilbert series of Phase I.
The plethystic logarithm is given by
\bea
\PL[\gm(t,x_1, x_2, q; \cC^{(II)}_5)] = \left([3;1]q^2 + [2;2] + \frac{1}{q^2}[1;3]\right)t^{22} - O(t^{22})~.
\eea
Therefore, the generators of the mesonic moduli space can be written in terms of perfect matchings as:
\bea
p_i p_j p_k q_l r^2_1 v_1 v_2 v_3, \qquad p_i p_j q_k q_l r_1 r_2 v_1 v_2 v_3, \qquad p_i q_j q_l q_k r^2_2 v_1 v_2 v_3~, 
\eea
with $i,j,k,l = 1,2$.
Each generator carries R-charge 2.

\comment{
\begin{table}[h]
 \begin{center}  
  \begin{tabular}{|c||c|}
  \hline
  \; $Generators$ &$U(1)_R$\\
  \hline  \hline 
  $p_i p_j p_k q_l r^2_1 v_1 v_2 v_3$ & 2 \\
  \hline
  $p_i p_j q_k q_l r_1 r_2 v_1 v_2 v_3$ & 2 \\
  \hline
  $p_i q_j q_l q_k r^2_2 v_1 v_2 v_3$ & 2 \\
  \hline
  \end{tabular}
  \end{center}
\caption{R-charges of the generators of the mesonic moduli space for the $\cC^{(II)}_5$ Model.}
\label{t:Rgenfano68Ph2}
\end{table}}

\section{$\cC_1$ (Toric Fano 105): $\BP(\mathcal{O}_{\BP^1 \times \BP^1} \oplus \mathcal{O}_{\BP^1 \times \BP^1}(1,1))$}
This theory has 4 gauge groups and 12 chiral fields, which are denoted by $X_{12}^{ij}$, $X_{23}^i$, $X_{23'}^i$, $X_{31}^i$ and $X^{i}_{3'1}$ (with $i,j=1,2$). The quiver diagram and tiling are given in Figure \ref{f:phase2f0}, with CS levels $\vec{k} = (2,0,-1,-1)$.  The superpotential of this model is shown in (\ref{e:spotfano62phII}).

\comment{
\bea
W &=& \epsilon_{ij}\epsilon_{kl} \tr(X^{ik}_{12}X^{l}_{23} X^{j}_{31}) - \epsilon_{ij}\epsilon_{kl} \tr(X^{ki}_{12}X^{l}_{23'}X^{j}_{3'1})~.
\eea
\begin{figure}[ht]
\begin{center}
  \hskip -5cm
  \includegraphics[totalheight=4cm]{quivph2f0.pdf}
    \vskip -3.5cm
  \hskip 8cm
  \includegraphics[totalheight=4cm]{tilph2f0.pdf}
 \caption{(i) Quiver diagram for the $\cC_1$ theory.\ (ii) Tiling for the $\cC_1$ theory.}
  \label{f:tqtoricfano105}
\end{center}
\end{figure}
}

\paragraph{The toric diagram.} The toric diagram for this model is constructed using two different methods.
\begin{itemize}
\item {\bf The Kasteleyn matrix.} The Chern-Simons levels for this model can be parametrized in terms of integers as shown in \ref{e:csfano62phII}.

\comment{
\bea
\begin{array}{rcl}
\text{Gauge group 1~:} \qquad k_1  =&  2 &=    n_1 + n_2 + n_3 + n_4 - n_5 - n_{6} - n_{9} - n_{10} ~, \nn \\
\text{Gauge group 2~:} \qquad k_2  =&  0 &=  - n_1 - n_2 - n_3 - n_4 +  n_7 + n_8 + n_{11} + n_{12}~, \nn \\
\text{Gauge group 3~:} \qquad k_3  =&  -1 &= - n_7 - n_8 + n_9 + n_{10} ~, \nn \\
\text{Gauge group $3'$:} \qquad k_{3'} =& -1 &=   n_5 + n_6 - n_{11} - n_{12} ~.
\label{e:kafano105}
\end{array}
\eea
}

Let us choose:
\bea
n^{12}_{12} = n^{21}_{12} = n^2_{23'} = -n^{22}_{12} = -n^{2}_{31} = 1,\quad n^i_{kl}=n^{ij}_{kl}=0 \; \text{otherwise}~.
\eea
The Kasteleyn matrix for this model can be written as:
\be
K= \left(
\begin{array}{c|cccc}
& w_1 & w_2 & w_3 & w_4\\
\hline
b_1 &  y z^{n^{2}_{23}}  &\  \frac{1}{x}  z^{n^{1}_{31}}      &\           0           &\  z^{n^{21}_{12}} \\
b_2 &  x z^{n^{2}_{31}}  &\   \frac{1}{y} z^{n^{1}_{23}}      &\  z^{n^{12}_{12}}      &\       0            \\
b_3 &         0          &\   z^{n^{22}_{12}}                &\   z^{n^{1}_{3'1}}      &\ z^{n^{1}_{23'} } \\
b_4 &    z^{n^{11}_{12}} &\       0                           &\   z^{n^2_{23'} }      &\  z^{n^{2}_{3'1}}
\end{array}
\right) ~. \label{e:kastfano105}
\ee
The permanent of the Kasteleyn matrix is
\bea
\perm~K &=&  y  z^{(n^{2}_{23} + n^{2}_{3'1} + n^{12}_{12} + n^{22}_{12})} +  \frac{1}{y} z^{(n^{1}_{23} + n^{1}_{3'1} + n^{21}_{12} + n^{11}_{12})} + x z^{(n^{2}_{23'} + n^{2}_{31} + n^{21}_{12} + n^{22}_{12})} \nn \\
&+&  \frac{1}{x} z^{(n^{1}_{23'} + n^{1}_{31} + n^{11}_{12} + n^{12}_{12})} + z^{(n^{1}_{31} + n^{2}_{31} + n^{1}_{3'1} + n^{2}_{3'1})} +  z^{(n^{1}_{23'} + n^{2}_{23'} + n^{1}_{23} + n^{2}_{23})}\nn \\
&+&  z^{(n^{11}_{12} + n^{21}_{12} + n^{12}_{12} + n^{22}_{12})} +  z^{(n^{1}_{23} + n^{2}_{23} + n^{1}_{3'1} + n^{2}_{3'1})} + z^{(n^{1}_{31} + n^{2}_{31} + n^{1}_{23'} + n^{2}_{23'})}\nn \\
&=&  y + y^{-1} z + x + x^{-1} z  + z^{-1} + 2z + 2 \nn \\
&& \text{(for $n^{12}_{12} = n^{21}_{12} = n^2_{23'} = -n^{22}_{12} = -n^{2}_{31} = 1,\quad n^i_{kl}=n^{ij}_{kl}=0 \; ~ \text{otherwise}$)} ~.\nn \\
\label{e:permKfano105}
\eea

The coordinates of the toric diagram are given by the powers of each monomial in \eref{e:permKfano105}:
\bea
G_K = \left(
\begin{array}{ccccccccc}
   1 & -1 &  0 &  0 & 0 &  0 & 0 & 0 & 0\\
   0 &  0 &  1 & -1 & 0 &  0 & 0 & 0 & 0\\
   0 &  1 &  0 &  1 &-1 &  1 & 1 & 0 & 0
\end{array}
\right).~
\eea
The first two rows of the $G_K$ matrix contain the weights of the $SU(2)$'s, signifying that the non-abelian part of the mesonic symmetry is $SU(2)\times SU(2)$. 

The permanent of the Kasteleyn matrix (\ref{e:permKfano105}) gives us the possibility to write down the perfect matchings in terms of the chiral fields of the model:
\bea 
&&   p_1  = \left\{X^{12}_{12},X^{22}_{12},X^{2}_{23},X^{2}_{3'1}\right\}, 
\;\; p_2  = \left\{X^{11}_{12},X^{21}_{12},X^{1}_{23},X^{1}_{3'1}\right\},\nn \\
&&   q_1  = \left\{X^{21}_{12},X^{22}_{12},X^{2}_{31},X^{2}_{23'}\right\},
\;\; q_2  = \left\{X^{11}_{12},X^{12}_{12},X^{1}_{31},X^{1}_{23'}\right\},\nn \\
&&   r_1  = \left\{X^{1}_{31},X^{2}_{31},X^{1}_{3'1},X^{2}_{3'1}\right\},
\;\; r_2  = \left\{X^{1}_{23},X^{2}_{23},X^{1}_{23'},X^{2}_{23'}\right\},\nn \\
&&   r'_2 = \left\{X^{11}_{12},X^{12}_{12},X^{21}_{12},X^{22}_{12}\right\},
\;\; v_1  = \left\{X^{1}_{23},X^{2}_{23},X^{1}_{3'1},X^{2}_{3'1}\right\},\nn \\
&&   v_2  = \left\{X^{1}_{31},X^{2}_{31},X^{1}_{23'},X^{2}_{23'}\right\}\ . \qquad
\eea
From (\ref{e:permKfano105}), it can be seen that the perfect matchings $p_1, p_2, q_1, q_2, r_1, r_2$ and $r'_2$ correspond to external points in the toric diagram, while $v_1$ and $v_2$ correspond to internal points. Also note that $r_2$ and $r'_2$ correspond to the same external point in the toric diagram.
The chiral fields can be parametrised in terms of perfect matchings:
\bea
 X^{22}_{12} &=& p_1 q_1 r'_2, \quad X^{12}_{12} = p_1 q_2 r'_2, \quad  X^{21}_{12} = p_2 q_1 r'_2 , \quad X^{11}_{12} = p_2 q_2 r'_2,\nn \\
  X^2_{3'1} &=& p_1 r_1 v_1, \quad X^1_{3'1} = p_2 r_1 v_1, \quad  X^2_{23} = p_1 r_2 v_1, \quad X^1_{23} = p_2 r_2 v_1 ,\nn \\
X^{2}_{31} &=& q_1 r_1 v_2, \quad X^{1}_{31} = q_2 r_1 v_2, \quad  X^{2}_{23'} = q_1 r_2 v_2, \quad X^{1}_{23'} = q_2 r_2 v_2~.
\eea
This information can be summarised in the perfect matching matrix:
\beq
P=\left(\begin{array} {c|ccccccccc}
  \;& p_1 & p_2 & q_1 & q_2 & r_1 & r_2 & r'_2 & v_1 & v_2\\
  \hline 
  X^{22}_{12} & 1&0&1&0&0&0&1&0&0\\
  X^{12}_{12} & 1&0&0&1&0&0&1&0&0\\
  X^{21}_{12} & 0&1&1&0&0&0&1&0&0\\
  X^{11}_{12} & 0&1&0&1&0&0&1&0&0\\
  X^{2}_{3'1} & 1&0&0&0&1&0&0&1&0\\
  X^{1}_{3'1} & 0&1&0&0&1&0&0&1&0\\
  X^{2}_{23}  & 1&0&0&0&0&1&0&1&0\\
  X^{1}_{23}  & 0&1&0&0&0&1&0&1&0\\
  X^{2}_{31}  & 0&0&1&0&1&0&0&0&1\\
  X^{1}_{31}  & 0&0&0&1&1&0&0&0&1\\
  X^{2}_{23'} & 0&0&1&0&0&1&0&0&1\\
  X^{1}_{23'} & 0&0&0&1&0&1&0&0&1\\
  \end{array}
\right).
\eeq
The three vectors that span the null space of $P$ can be written as rows of the following charge matrix:
\be
Q_F =   \left(
\begin{array}{ccccccccc}
1 & 1 & 0 & 0 & 0 & 0 &-1 &-1 &  0\\
0 & 0 & 1 & 1 & 0 & 0 &-1 & 0 & -1\\
0 & 0 & 0 & 0 & 1 & 1 & 0 &-1 & -1
\end{array}
\right)~.  \label{e:qffano105}
\ee
Hence, among the perfect matchings there are three relations, which are given by:
\bea
p_1 + p_2 - r'_2 - v_1 = 0,\\
q_1 + q_2 - r'_2 - v_2 = 0,\\
r_1 + r_2 - v_1 - v_2 = 0.
\label{e:relpmfano105}
\eea
\paragraph{ The charge matrices.}
Since this model has 4 gauge groups, the total number of baryonic symmetries coming from the D-terms is 2. The charges of the perfect matchings under these four symmetries are collected in the columns of the $Q_D$ matrix:
\be
Q_D =   \left(
\begin{array}{ccccccccc}
 0 &  0 & 0 &  0 & 0 & 1 & -1 & 0 & 0\\
 0 &  0 & 0 &  0 & 0 & 0 & 0 &  1 &-1
\end{array}
\right). 
\label{e:qdfano105}
\ee
Combining the $Q_F$ and $Q_D$ matrices, the total charge matrix can be written as:
\be
Q_t = { \Blue Q_F \choose \Green Q_D \Black } =   \left( 
\begin{array}{ccccccccc} \Blue
1 & 1 & 0 & 0 & 0 & 0 &-1 &-1 &  0\\
0 & 0 & 1 & 1 & 0 & 0 &-1 & 0 & -1\\
0 & 0 & 0 & 0 & 1 & 1 & 0 &-1 & -1\\ \Green
0 &  0 & 0 &  0 & 0 & 1 & -1 & 0 & 0\\
0 &  0 & 0 &  0 & 0 & 0 & 0 &  1 &-1 \Black
\end{array}
\right). 
\label{e:qtfano105}
\ee
The kernel of the $Q_t$ matrix, after the removal of the first row, contains the coordinates of the toric diagram in its columns, and can be written as:
\bea
G'_t = \left(
\begin{array}{ccccccccc}
   1 & -1 &  0 &  0 & 0 &  0 & 0 & 0 & 0\\
   0 &  0 &  1 & -1 & 0 &  0 & 0 & 0 & 0\\
   0 &  1 &  0 &  1 &-1 &  1 & 1 & 0 & 0
\end{array}
\right) = G_K~. \label{e:toricdiafano105}
\eea
The toric diagram constructed from (\ref{e:toricdiafano105}) is presented in Figure \ref{f:tdtoricfano105}.
\begin{figure}[ht]
\begin{center}
\includegraphics[totalheight=3.0cm]{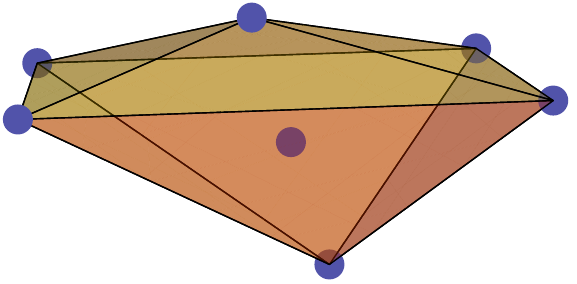}
 \caption{The toric diagram of $\cC_1$.}
  \label{f:tdtoricfano105}
\end{center}
\end{figure}
\end{itemize}

\paragraph{The baryonic charges.} The toric diagram of this model contains 6 external points. It follows that the number of baryonic symmetries of the gauge theory is $6-4=2$. These symmetries come from the D-terms and will be denoted as $U(1)_{B_1}$ and $U(1)_{B_2}$.

\paragraph{The global symmetry.} The total charge matrix contains two pairs of repeated columns, confirming that the mesonic symmetry of this model is $SU(2)^2\times U(1)^2$, where one of the abelian factors is precisely the R-symmetry. The perfect matchings $p_1$ and $p_2$ transform as a doublet under one of the non-abelian factors of the mesonic symmetry, while $q_1$ and $q_2$ transform as a doublet under the other non-abelian factor. Note that, since $v_1$ and $v_2$ correspond to the internal point in the toric diagram, they carry 0 R-charge.
Let us now proceed and minimise the volume of the Calabi-Yau. This will allow us to compute the R-charges of all the perfect matchings.
Assigning fugacities $s_1$ to the perfect matchings $p_1$ and $p_2$, $s_2$ to the perfect matchings $q_1$ and $q_2$, and fugacities $s_3,s_4,s_5$ to the perfect matchings $r_1,r_2$ and $r'_2$ respectively, the Hilbert series of the mesonic moduli space can be written as:
\bea
\gm (s_{\alpha}; \cC_1) &=& \left( \prod_{i=1}^2 \oint \limits_{|b_i|=1} \frac{\ud b_i}{2\pi i b_i}  \right)   \left(\prod_{j=1}^3 \oint \limits_{|z_j|=1} \frac{\ud z_j}{2\pi i z_j} \right) \frac{1}{\left(1-s_1 z_1\right)^2\left(1- s_2 z_2\right)^2\left(1- s_3 z_3\right)}\nn \\
&\times& \frac{1}{\left(1- s_4 z_3 b_1\right)\left(1-\frac{s_5}{z_1 z_2 b_1}\right)\left(1-\frac{b_2}{z_1 z_3}\right)\left(1-\frac{1}{z_2 z_3 b_2}\right)} \nn \\
&=& \frac{\cP(s_{\alpha}; \cC_1)}{\left(1-s_1 s_2 s^2_3\right)^3\left(1-s^3_1 s^3_2 s^2_4 s^2_5\right)^3},
\label{e:hsvolfano105}
\eea
where $\cP\left(s_{\alpha}; \cC_1\right)$ is polynomial that is not reported here. The Hilbert series depends only on two combinations of perfect matchings. In order to see this explicitly, let us define:
\bea
 t^2_1 = s_1 s_2 s^2_3, \qquad t^2_2 = s^3_1 s^3_2 s^2_4 s^2_5 
\eea
and rewrite (\ref{e:hsvolfano105}) in terms of these new variables:
{\small
\bea
\gm (t_1, t_2; \cC_1) &=& \frac{P(t_1, t_2; \cC_1)}{\left(1-t^2_1\right)^3\left(1-t^2_2\right)^3},
\label{e:hsvolfano105T}
\eea}.
where $P\left(t_1, t_2; \cC_1\right)$ is a polynomial which can be written as:
{\small
\bea
P(t_1, t_2;\cC_1) &=& 1 + t_1^2 + 9 t_1 t_2 - 11 t_1^3 t_2 + 4 t_1^5 t_2 + 13 t_2^2 - 26 t_1^2 t_2^2 + 9 t_1^4 t_2^2 +9 t_1 t_2^3 -\nn \\
&&26 t_1^3 t_2^3 +  13 t_1^5 t_2^3 + 4 t_2^4 - 11 t_1^2 t_2^4 + 9 t_1^4 t_2^4 + t_1^3 t_2^5 +  t_1^5 t_2^5~.
\eea}

Denoting by $R_i$ the R-charge relative to the fugacity $t_i$, the requirement that superpotential has R-charge 2 imposes that
\bea
R_1 + R_2 = 2.
\label{e:cyfano105}
\eea
Using this condition, the volume of $\cC_1$ can be expressed in terms of one variable only, and can be written as:
{\small
\bea
\lim_{\mu \rightarrow 0} \mu^4 \gm (e^{-\mu R_1}, e^{-\mu (2-R_1)}; \cC_1)= \frac{7 R^2_1 + 2R_1 + 4}{4 R^3_1 (R_1 -2)^3}~. \label{e:volfano105}
\eea}
This functions has a minimum at:
\bea
R_1 = \frac{1}{21}\left(1- \frac{83}{(3844 + 63\sqrt{3867})^{1/3}}+(3844 + 63\sqrt{3867})^{1/3} \right) \approx 0.791~.\nn \\
\eea

Let us now compute the R-charge of each particular perfect matching by computing the normalized volume of the corresponding divisor.  As an example, the Hilbert series of the divisor $D_1$ is given by:
\bea
g (D_1;  s_\alpha ;\cC_1) &=& \left( \prod_{i=1}^2 \oint \limits_{|b_i|=1} \frac{\ud b_i}{2\pi i b_i}  \right)   \left(\prod_{j=1}^3 \oint \limits_{|z_j|=1} \frac{\ud z_j}{2\pi i z_j} \right) \frac{(s_1 z_1)^{-1}}{\left(1-s_1 z_1\right)^2\left(1- s_2 z_2\right)^2\left(1- s_3 z_3\right)}\nn \\
&& \times \frac{1}{\left(1- s_4 z_3 b_1\right)\left(1-\frac{s_5}{z_1 z_2 b_1}\right)\left(1-\frac{b_2}{z_1 z_3}\right)\left(1-\frac{1}{z_2 z_3 b_2}\right)}~.
\eea
In terms of the fugacities $t_1$ and $t_2$ introduced above, the result of the integration can be written as:
\bea
g (D_1;  t_1, t_2 ;\cC_1) &=& \frac{1}{\left(1-t_1^2\right)^3 \left(1- t_ 2^2\right)^3} \times (2+12 t_1 t_2-16 t_1^3 t_2+6 t_1^5 t_2+14 t_2^2-30 t_1^2 t_2^2 +\nn \\
&&12 t_1^4 t_2^2+6 t_1 t_2^3-22 t_1^3 t_2^3+12 t_1^5 t_2^3+2 t_2^4-6 t_1^2 t_2^4+ t_1^4 t_2^4 +2 t_1^3 t_2^5)~, \nn \\
\eea
and the R-charge of the perfect matching $p_1$ is given by
\bea
\lim_{\mu\rightarrow0}\frac{1}{\mu} \left[ \frac{g(D_1; e^{- \mu R_1 }, e^{- \mu R_2 }; \cC_1) }{\gm(e^{-\mu R_1}, e^{- \mu R_2}; \cC_1)}- 1 \right] \approx 0.344~.
\eea 
The results for the other perfect matchings are presented in Table \ref{t:chargefano105}.
The other charges can be assigned with the conditions that the superpotential remains uncharged and that the charge vectors are linearly independent. In \tref{t:chargefano105}, our choice is presented.

\begin{table}[h!]
 \begin{center}  
  \begin{tabular}{|c||c|c|c|c|c|c|c|}
  \hline
  \;& $SU(2)_1$&$SU(2)_2$&$U(1)_q$&$U(1)_R$&$U(1)_{B_1}$&$U(1)_{B_2}$&fugacity\\
  \hline\hline  
   
  $p_1$&$  1$&$  0$&$ 0$&$0.344$&$ 0$&$ 0$ & $s_1 x_1$\\
  \hline
  
  $p_2$&$ -1$&$  0$&$ 0$&$0.344$&$ 0$&$ 0$ & $s_1 / x_1 $\\
  \hline  
  
  $q_1$&$  0$&$  1$&$ 0$&$0.344$&$ 0$&$ 0$ & $s_1 x_2 $\\
  \hline
  
  $q_2$&$  0$&$ -1$&$ 0$&$0.344$&$ 0$&$ 0$ & $s_1 / x_2$\\
  \hline
   
  $r_1$&$  0$&$  0$&$ 1$&$0.447$&$ 0$&$ 0$ & $s_2 q$\\
  \hline
   
  $r_2$&$  0$&$  0$&$-1$&$0.177$&$ 1$&$ 0$ &$s_3 b_1/ q$\\
  \hline
  
  $r'_2$&$  0$&$  0$&$ 0$&$0$&$   -1$&$ 0$ & $ 1/b_1$\\
  \hline
  
  $v_1$&$  0$&$  0$&$ 0$&$0$&$      0$&$ 1$ & $b_2$\\
  \hline
 
  $v_2$&$  0$&$  0$&$ 0$&$0$&$      0$&$-1$ & $ 1/b_2 $\\  
  
  \hline
   \end{tabular}
  \end{center}
\caption{Charges of the perfect matchings under the global symmetry of the $\cC_1$ model. Here $s_i$ is the fugacity of the R-charge, $x_1$ and $x_2$ are the weights of the $SU(2)$ symmetries, $q, b_1$ and $b_2$ are, respectively, the charges under the mesonic abelian symmetry $U(1)_q$ and of the two baryonic $U(1)_{B_1}$ and $U(1)_{B_2}$.}
\label{t:chargefano105}
\end{table}

\begin{table}[h]
 \begin{center}  
  \begin{tabular}{|c||c|}
  \hline
  \; Quiver fields &R-charge\\
  \hline  \hline 
  $ X^{i}_{23}, X^{i}_{23'} $ & 0.521 \\
  \hline
  $ X^{i}_{31}, X^{i}_{3'1} $ & 0.791\\
  \hline
  $ X^{ij}_{12} $ & 0.688\\
  \hline
  \end{tabular}
  \end{center}
\caption{R-charges of the quiver fields for the $\cC_1$ Model.}
\label{t:Rgenchfano105}
\end{table}

\begin{figure}[ht]
\begin{center}
\includegraphics[totalheight=4cm]{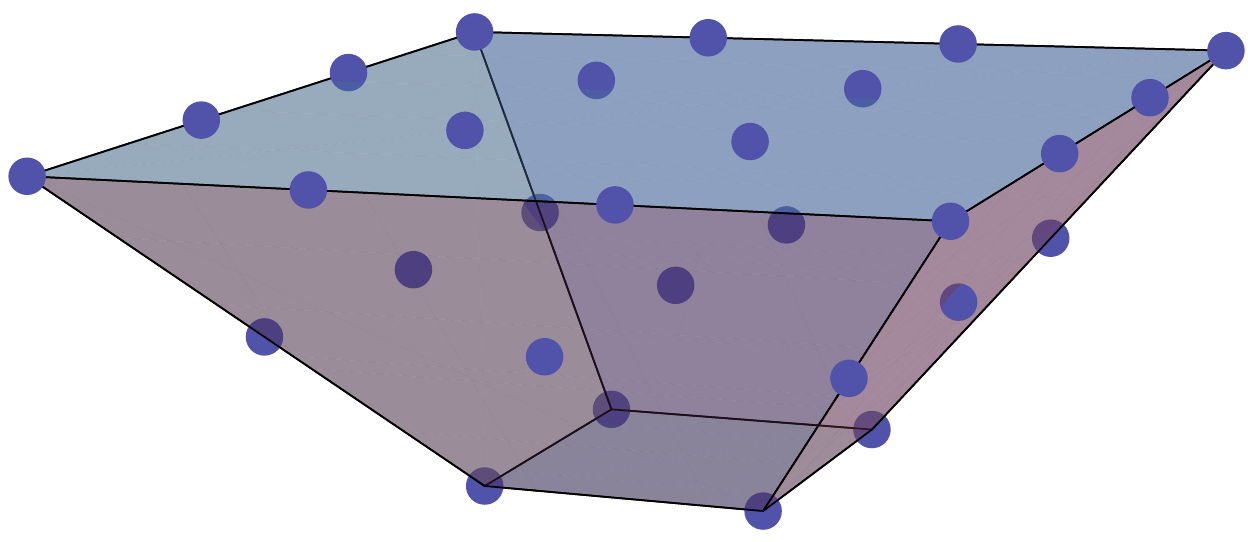}
\caption{The lattice of generators of the $\cC_1$ theory.}
  \label{f:latc1}
 \end{center}
\end{figure}

\paragraph{The Hilbert series.} Having assigned all the charges of the perfect matchings, it is now possible to compute the Hilbert series of the Master space and of the mesonic moduli space. In order to determine the former, one needs to integrate the Hilbert series of the space of perfect matchings, $\BC^9$, over the fugacities $z_1, z_2$ and $z_3$
{\small
\bea
g^{\firr{}} (s_\alpha, x_1, x_2, q, b_1,b_2; \cC_1) &=& \left( \prod_{j=1}^3\oint \limits_{|z_i| =1} {\frac{\ud z_i}{2 \pi i z_i}} \right)  \frac{1}{\left(1- s_1 x_1 z_1\right)\left(1- \frac{s_1 z_1}{x_1}\right)\left(1- s_1 x_2 z_2\right)\left(1- \frac{s_1 z_2}{x_2}\right)}\nn \\
& \times & \frac{1}{\left(1- s_2 q z_3\right)\left(1- \frac{s_3 z_3 b_1}{q}\right)\left(1- \frac{1}{z_1 z_2 b_1}\right)\left(1-\frac{b_2}{z_1 z_3}\right)\left(1- \frac{1}{b_2 z_2 z_3}\right) }\nn \\
&=& \frac{\cP(s_\alpha,x_1, x_2, q, b_1,b_2;  \cC_1)}{\left(1-\frac{s^2_1 x_1 x_2}{b_1}\right)\left(1-\frac{s^2_1 x_1}{x_2 b_1}\right)\left(1-\frac{s^2_1 x_2}{x_1 b_1}\right)\left(1-\frac{s^2_1}{x_1 x_2 b_1}\right)\left(1- s_1 s_2 x_1 q b_2\right)}\nn \\
&\times& \frac{1}{\left(1-\frac{s_1 s_2 q b_2}{x_1}\right)\left(1-\frac{s_1 s_2 x_2 q}{b_2}\right)\left(1-\frac{s_1 s_2 q}{x_2 b_2}\right)\left(1-\frac{s_1 s_3 x_1 b_1 b_2}{q}\right)}\nn \\
&\times& \frac{1}{\left(1-\frac{s_1 s_3 b_1 b_2}{x_1 q}\right)\left(1-\frac{s_1 s_3 x_2 b_1}{q b_2}\right)\left(1-\frac{s_1 s_3 b_1}{x_2 q b_2}\right)}~.
\label{e:HSmasterfano105}
\eea}
where $\cP(s_\alpha,x_1, x_2, q,b_1,b_2; \cC_1)$ is a polynomial that is not reported here.  The unrefined Hilbert series of the Master space can be written as:
\bea
g^{\firr{}} (s_\alpha,1,1,1,1,1; \cC_1) &=& \frac{\cP(s_\alpha,1,1,1,1,1; \cC_1)}{\left(1 - s_1^2\right)^4 \left(1 - s_1 s_2\right)^4 \left(1 - s_1 s_3\right)^4} ~,
\eea
where $\cP(s_\alpha,1,1,1,1,1; \cC_1)$ is a polynomial of degree 16.
Finally, the Hilbert series of the mesonic moduli space can be computed integrating (\ref{e:HSmasterfano105}) over the two baryonic fugacities $b_1$ and $b_2$.
{\small
\bea
\gm (s_\alpha, x_1, x_2, q; \cC_1) &=& \oint \limits_{|b_1| =1} {\frac{\ud b_1}{2 \pi i b_1}} \oint \limits_{|b_2| =1} {\frac{\ud b_2}{2 \pi i b_2}} g^{\firr{}} (s_\alpha, x_1, x_2, q, b_1, b_2; \cC_1) \nn \\
&=& \frac{P(s_\alpha,x_1, x_2, q; \cC_1)}{\left(1- s^2_1 s^2_2 x_1 x_2 q^2\right)\left(1-\frac{s^2_1 s^2_2 x_1 q^2}{x_2}\right)\left(1-\frac{s^2_1 s^2_2 x_2 q^2}{x_1}\right)\left(1-\frac{s^2_1 s^2_2 q^2}{x_1 x_2}\right)}\nn \\
&\times& \frac{1}{\left(1-\frac{s^6_1 s^2_3 x^3_1 x^3_2}{q^2}\right)\left(1-\frac{s^6_1 s^2_3 x^3_2}{x^3_1 q^2}\right)\left(1-\frac{s^6_1 s^2_3 x^3_1}{x^3_2 q^2}\right)\left(1-\frac{s^6_1 s^2_3}{x^3_1 x^3_2 q^2}\right)}~.\nn \\
\eea}
The plethystic logarithm can be written in terms of the fugacities $t_1$ and $t_2$ as
{\small
\bea
\PL[\gm(t_1,t_2,x_1, x_2, q; \cC_1)] = \frac{1}{q^2} [3;3] t^2_2 + [2;2] t_1 t_2 + q^2 [1;1] t^2_1 - O(t_1^3 ) O( t_2^3)
\eea}
From this function, it can be deduced that the generators of the mesonic moduli space can be written in terms of perfect matchings as:
\bea
p_i p_j p_k q_l q_m q_n r^2_2 r'^2_2 v_1 v_2, \qquad p_i p_j q_k q_l r_1 r_2 r'_2 v_1 v_2, \qquad p_i q_j r^2_1 v_1 v_2, 
\eea
with $i,j,k,l,m,n = 1,2$.
The R-charges of these generators are presented in Table \ref{t:Rgenfano105}, and the lattice of generators is drawn in \fref{f:latc1}.

\begin{table}[h]
 \begin{center}  
  \begin{tabular}{|c||c|}
  \hline
  \; Generators &R-charge\\
  \hline  \hline 
  $p_i p_j p_k q_l q_m q_n r^2_2 r'^2_2 v_1 v_2$ & 2.418 \\
  \hline
  $p_i p_j q_k q_l r_1 r_2 r'_2 v_1 v_2$ & 2 \\
  \hline
  $p_i q_j r^2_1 v_1 v_2$ & 1.582 \\
  \hline
  \end{tabular}
  \end{center}
\caption{R-charges of the generators of the mesonic moduli space for the $\cC_1$ theory.}
\label{t:Rgenfano105}
\end{table}

\section{$\cC_2$ (Toric Fano 136): $\BP(\mathcal{O}_{dP_1} \oplus \mathcal{O}_{dP_1} (l)),~l^2|_{dP_1}=1$}
\begin{figure}[ht]
\begin{center}
\hskip -0.5cm
\includegraphics[totalheight=3.7cm]{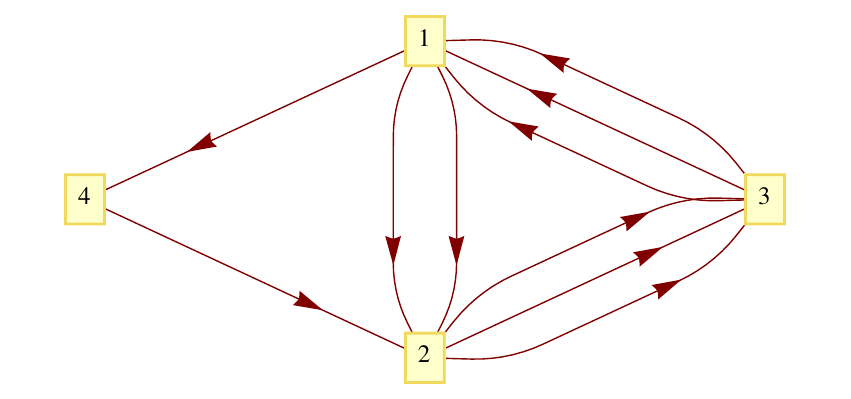}
\hskip 0.5cm
\includegraphics[totalheight=5cm]{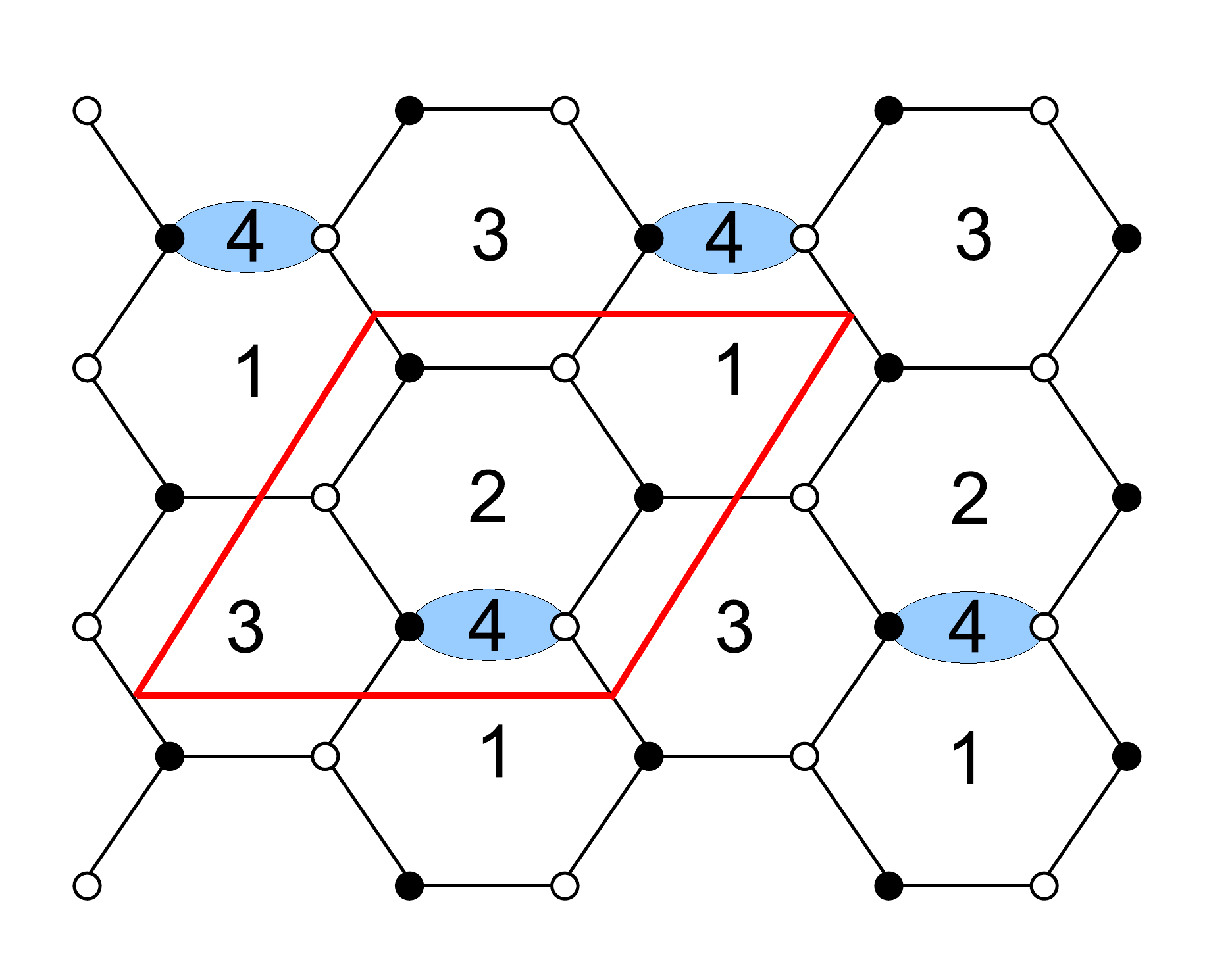}
 \caption{(i) Quiver diagram of the $\cC_2$ model.\ (ii) Tiling of the $\cC_2$ model.}
  \label{f:tqfano136}
\end{center}
\end{figure}

This model has 4 gauge groups and chiral fields $X^i_{23}$, $X^i_{31}$ (with $i=1,2,3$), $X^j_{12}$ (with $j = 1,2$), $X_{14}$ and $X_{42}$. The tiling and the quiver diagram are presented in Figure \ref{f:tqfano136}.  Note that the former can be obtained by  `double bonding' the tiling of $\cB_4$.  The superpotential of this model can be written as
\bea
W = \epsilon_{ij} \tr (X^{i}_{31}X^{j}_{12}X^3_{23})+\epsilon_{ij} \tr (X^{i}_{12}X^{j}_{23}X^3_{31})+\epsilon_{ij}\tr (X^{i}_{23}X^{j}_{31}X_{14}X_{42})~.
\label{e:spotfano136}
\eea
The CS levels are $\vec{k} = (-1,2,0,-1)$.

\comment{
\begin{figure}[ht]
\begin{center}
\vskip -1cm
   \includegraphics[totalheight=5cm]{fdc2.pdf}
 \caption{The fundamental domain of the tiling for $\cC_2$.}
  \label{f:fdtoricfano136}
\end{center}
\end{figure}
}

\paragraph{The Kasteleyn matrix.} The Chern-Simons levels can be parametrized in terms of integers $n^i_{jk}$ and $n_{jk}$ as follows:
\bea
\begin{array}{ll}
\text{Gauge group 1~:} \qquad k_1  &=   n_{14} + n^{1}_{12} + n^{2}_{12} - n^{1}_{31} - n^{2}_{31} - n^{3}_{31} ~, \nn \\
\text{Gauge group 2~:} \qquad k_2  &=   n^{1}_{23} + n^{2}_{23} + n^{3}_{23} - n^{1}_{12} - n^{2}_{12} - n_{42}  ~, \nn \\
\text{Gauge group 3~:} \qquad k_3  &=   n^{1}_{31} + n^{2}_{31} + n^{3}_{31} - n^{1}_{23} - n^{2}_{23} - n^{3}_{23}  ~, \nn \\
\text{Gauge group 4~:} \qquad k_4  &=   n_{42} - n_{14}  ~.
\label{e:kafano136}
\end{array}
\eea
Let us choose
\bea
n^2_{31} = n^3_{23} = - n_{42} = 1,\quad n^i_{jk}=n_{jk}=0 \; \text{otherwise}~.
\eea
The Kasteleyn matrix $K$ for this model can be computed. Since the fundamental domain contains six nodes in total, $K$ is a $3\times 3$ matrix:
\bea
K =   \left(
\begin{array}{c|ccc}
& b_1 & b_2 & b_3\\
\hline
w_1 &  z^{n_{42}} +  z^{n_{14}} &  z^{n^2_{23}} &  \frac{y}{x} z^{n^1_{31}} \\
w_2 &  x z^{n^2_{31}} & z^{n^1_{12}} & z^{n^3_{23}} \\
w_3 &  z^{n^1_{23}} &  \frac{1}{y} z^{n^3_{31}} &  z^{n^2_{12}} \end{array}
\right) ~. \label{e:kastfano136}
\eea
The permanent of this matrix is given by
\bea
\perm~K &=& 
 x z^{(n^{2}_{31} + n^{2}_{12} + n^{2}_{23})} +  x^{-1} y z^{(n^{1}_{31} + n^{1}_{12} + n^{1}_{23})} +   y^{-1} z^{(n^{3}_{31} + n_{42} + n^{3}_{23})}\nn \\
&+&  y^{-1} z^{(n^{3}_{31} + n_{14} + n^3_{23})} +  z^{(n^{1}_{12} + n^{2}_{12} + n_{42})}+  z^{(n^{1}_{31} + n^{2}_{31} + n^{3}_{31})}\nn \\
&+&  z^{(n^{1}_{23} + n^{2}_{23} + n^{3}_{23})} +  z^{(n^{1}_{12} + n^{2}_{12} + n_{14})} \nn \\
&=&  x z+ x^{-1} y +y^{-1} + y^{-1} z + z^{-1} +  2z + 1\nn\\ 
&& \text{(for $n^2_{31} = n^3_{23} = - n_{42} = 1,\quad n^i_{jk}=n_{jk}=0 \; ~ \text{otherwise}$)} ~.
\label{e:permKfano136}
\eea

\paragraph{The perfect matchings.} From (\ref{e:permKfano136}), the perfect matchings can be written in terms of the chiral fields as
\bea 
&&   p_1 = \left\{X^2_{31}, X^2_{12},X^2_{23}\right\}, 
\;\; p_2 = \left\{X^1_{31}, X^1_{12},X^1_{23}\right\},
\;\; q_1 = \left\{X^3_{31}, X_{42},X^3_{23}\right\},\nn \\
&&   q_2 = \left\{X^3_{31}, X_{14},X^3_{23}\right\},
\;\; r_1 = \left\{X^1_{12}, X^2_{12},X_{42}\right\},
\;\; r_2 = \left\{X^1_{31}, X^2_{31},X^3_{31}\right\},\nn \\
&&   r'_2 = \left\{X^1_{23}, X^2_{23},X^3_{23}\right\},
\;\; v_1 = \left\{X^1_{12}, X^2_{12},X_{14}\right\}\ . \qquad
\eea
The chiral fields can be parametrised in terms of perfect matchings:
\bea
X^2_{31} &=& p_1 r_2, \quad X^2_{12} = p_1 r_1 v_1, \quad X^2_{23} = p_1 r'_2, \quad X^1_{31} = p_2 r_2, \quad X^1_{12} = p_2 r_1 v_1,  \nn \\
X^1_{23} &=& p_2 r'_2, \quad  X^3_{31} = q_1 q_2 r_2, \quad X_{42} = q_1 r_1, \quad X_{14} = q_2 v_1, \quad X^3_{23} = q_1 q_2 r'_2~.
\eea
This information can be presented in the perfect matching matrix:
\beq
P=\left(\begin{array} {c|cccccccc}
  \;& p_1 & p_2 & q_1 & q_2 & r_1 & r_2 & r'_2 & v_1\\
  \hline 
  X^{2}_{31} & 1&0&0&0&0&1&0&0\\
  X^{2}_{12} & 1&0&0&0&1&0&0&1\\
  X^{2}_{23} & 1&0&0&0&0&0&1&0\\
  X^{1}_{31} & 0&1&0&0&0&1&0&0\\
  X^{1}_{12} & 0&1&0&0&1&0&0&1\\
  X^{1}_{23} & 0&1&0&0&0&0&1&0\\
  X^{3}_{31} & 0&0&1&1&0&1&0&0\\
  X_{42}     & 0&0&1&0&1&0&0&0\\
  X_{14}     & 0&0&0&1&0&0&0&1\\
  X^3_{23}   & 0&0&1&1&0&0&1&0\\
  \end{array}
\right).
\eeq
The null space of the $P$ matrix is spanned by two vectors that can be cast in the rows of the following matrix
\be
Q_F =   \left(
\begin{array}{cccccccc}
1 & 1 & 1 &  0 & -1 & -1 & -1 &  0 \\
0 & 0 & 1 & -1 & -1 &  0 &  0 &  1 
\end{array}
\right)~.  \label{e:qffano136}
\ee
Hence, among the perfect matchings there are two relations which are given by
\bea
p_1 + p_2 + q_1 - r_1 - r_2 - r'_2 &=& 0~,\nn \\
q_1 - q_2 - r_1 + v_1 &=& 0~.
\label{e:relpmfano136}
\eea

\paragraph{The toric diagram.} Two methods of computing the toric diagram for this model are presented
\begin{itemize}
\item {\bf The Kasteleyn matrix.} The powers of $x, y, z$ in each term of \eref{e:permKfano136} give the coordinates of each point in the toric diagram. These these points can be written as columns of the following matrix: 
\bea
 \left(
\begin{array}{cccccccc}
   1 & -1 &  0 &  0 & 0 & 0 & 0 & 0 \\
   0 &  1 & -1 & -1 & 0 & 0 & 0 & 0 \\
   1 &  0 &  0 &  1 &-1 & 1 & 1 & 0 
\end{array}
\right)~. \nn
\eea
By multiplying on the left by {\footnotesize $\left( \begin{array}{ccc} 1&0&0\\0&1&0\\-1&-1&1 \end{array} \right) \in GL(3, \BZ)$}, the following matrix is obtained:
\bea
G_K = \left(
\begin{array}{cccccccc}
   1 & -1 &  0 &  0 & 0 & 0 & 0 & 0 \\
   0 &  1 & -1 & -1 & 0 & 0 & 0 & 0 \\
   0 &  0 &  1 &  2 &-1 & 1 & 1 & 0 
\end{array}
\right)~.
\eea
The first row of this matrix contains the weights of the fundamental representation of $SU(2)$, which implies that the non-abelian part of the mesonic symmetry contains only one $SU(2)$ factor.

\item {\bf The charge matrices.}
Since the number of gauge groups of this model is $G = 4$, there are $G-2 =2$ baryonic symmetries coming from the D-terms. The charges of the perfect matchings under these symmetries can be collected in the rows of the $Q_D$ matrix:
\be
Q_D =   \left(
\begin{array}{cccccccc}
 0 &  0 & 0 &  0 &  1 &  1 & 0 & -2\\
 0 &  0 & 0 &  0 &  0 &  1 &-1 &  0 
 \end{array}
\right). \label{e:qdfano136}
\ee
It is possible to obtain the total charge matrix $Q_t$ by combining the $Q_F$ and the $Q_D$ matrix:
\be
Q_t = { \Blue Q_F \choose \Green Q_D \Black } =   \left( 
\begin{array}{cccccccc} \Blue
1 & 1 & 1 &  0 & -1 & -1 & -1 &  0 \\
0 & 0 & 1 & -1 & -1 &  0 &  0 &  1 \\ \Green
0 &  0 & 0 &  0 &  1 &  1 & 0 & -2\\
0 &  0 & 0 &  0 &  0 &  1 &-1 &  0 \Black
\end{array}
\right). 
\label{e:qtfano136}
\ee
The $G_t$ matrix is given by the kernel of the $Q_t$ matrix.  After eliminating the trivial row, the $G'_t$ matrix is obtained, whose columns give the coordinates of the toric diagram:
\bea
G'_t = \left(
\begin{array}{cccccccc}
   1 & -1 &  0 &  0 & 0 & 0 & 0 & 0 \\
   0 &  1 & -1 & -1 & 0 & 0 & 0 & 0 \\
   0 &  0 &  1 &  2 &-1 & 1 & 1 & 0 
\end{array}
\right)= G_K~. \label{e:gtfano136}
\eea
The toric diagram constructed from (\ref{e:gtfano136}) is presented in Figure \ref{f:tdtoricfano136}:
\begin{figure}[ht]
\begin{center}
  \includegraphics[totalheight=1.8cm]{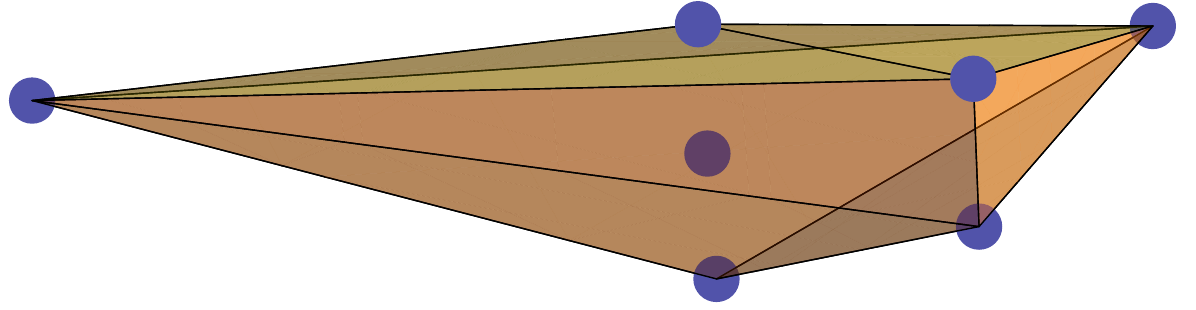}
 \caption{The toric diagram of $\cC_2$.}
  \label{f:tdtoricfano136}
\end{center}
\end{figure}

\end{itemize}

\paragraph{The baryonic charges.} The toric diagram of this model has 6 external points and, therefore, the total number of baryonic symmetries is $6-4 = 2$. These shall be referred to as $U(1)_{B_1}$ and $U(1)_{B_2}$. The charges of the perfect matchings under these baryonic symmetries can be read off from the rows of the $Q_D$ matrix (\ref{e:qdfano136}).

\paragraph{The global symmetry.} Since there is only one pair of repeated columns in the $Q_t$ matrix (\ref{e:qtfano136}), the mesonic symmetry contains only one non-abelian $SU(2)$ factor, under which the perfect matchings $p_1$ and $p_2$ transform as a doublet.  Since the total rank of the mesonic symmetry is 4, the mesonic symmetry of this model is $SU(2) \times U(1)^3$, where one of the three abelian symmetries can be identified with the R-symmetry. The perfect matching $v_1$ corresponds to the internal point of the toric diagram and, therefore, it carries zero R-charge. Note also that the perfect matchings $r_2$ and $r'_2$ correspond to the same point in the toric diagram.

Since the mesonic symmetry has 3 abelian factors, in order to determine the correct R-charge of each perfect matching in the IR, the volume minimisation of $\cC_2$ must be performed.  Let us assign the R-charge fugacity $s_1$ to the perfect matchings $p_1$ and $p_2$, and the R-charge fugacities $s_2, s_3, s_4, s_5, s_6$ respectively to the perfect matchings $q_1, q_2, r_1, r_2, r'_2$. Since the perfect matching $v_1$ corresponds to an internal point in the toric diagram its fugacity is set to unity.

The charges of the perfect matchings that need to be integrated over are given in (\ref{e:qtfano136}). The Hilbert series of the mesonic moduli space is given by:
\bea
\gm (s_{\alpha}; \cC_2) &=& \oint \limits_{|z_1| =1} {\frac{\ud z_1}{2 \pi i z_1}} \oint \limits_{|z_2| =1} {\frac{\ud z_2}{2 \pi i z_2}} \oint \limits_{|b_1| =1} {\frac{\ud b_1}{2 \pi i b_1}} \oint \limits_{|b_2| =1} {\frac{\ud b_2}{2 \pi i b_2}} \frac{1}{\left(1-s_1 z_1\right)^2\left(1-s_2 z_1 z_2\right)}\nn \\
&\times& \frac{1}{\left(1-\frac{s_3}{z_2}\right)\left(1-\frac{s_4 b_1}{z_1 z_2}\right)\left(1-\frac{s_5 b_1 b_2}{z_1}\right)\left(1- \frac{s_6}{z_1 b_2}\right)\left(1-\frac{z_2}{b^2_1}\right)}
\label{e:hsvolfano136}
\eea

Since there are three abelian factors in the mesonic symmetry, the integral depends only on three specific combinations of perfect matchings, namely: 
\bea
t_1 = s_1 s_2 s^2_4, \quad t_2 = s^2_2 s_3 s^2_4, \quad t_3 = s^3_1 s_4 s_5 s_6~.
\eea
The Hilbert series (\ref{e:hsvolfano136}) can be written in terms of $t$'s as:
{\small
\bea
\gm (t_1, t_2, t_3; \cC_2) &=& \frac{P(t_1, t_2, t_3; \cC_2)} {\left(1-t_1\right)^2\left(1-t_2\right)\left(1-t_3\right)^2\left(1-\frac{t_2 t_3^2}{t^2_1}\right)^2\left(1-\frac{t^5_2 t^2_3}{t^6_1}\right)}~,
\eea}
where:
{\scriptsize
\bea
P(t_1, t_2, t_3; \cC_2) &=& 1+2 t_3-3 t_1 t_3-6 t_2 t_3+3 t_1 t_2 t_3+2 t_2^2 t_3-2 t_2 t_3^3+2 t_2^2 t_3^3-4 t_2^2 t_3^2+5 t_2 t_3^2+\frac{3 t_2 t_3}{t_1}+\nn \\
&&\frac{2 t_2^2 t_3}{t_1^2}-\frac{4 t_2^2 t_3}{t_1}+\frac{t_2^3 t_3}{t_1^3}-\frac{2 t_2^3 t_3}{t_1^2}+\frac{t_2^3 t_3}{t_1}+\frac{3 t_2 t_3^2}{t_1^2}-\frac{10 t_2 t_3^2}{t_1}+\frac{4 t_2^2 t_3^2}{t_1^3}- \frac{12 t_2^2 t_3^2}{t_1^2}+\frac{12 t_2^2 t_3^2}{t_1}+\nn \\
&&\frac{3 t_2^3 t_3^2}{t_1^4}-\frac{8 t_2^3 t_3^2}{t_1^3}+\frac{7 t_2^3 t_3^2}{t_1^2}-\frac{2 t_2^3 t_3^2}{t_1}+\frac{2 t_2^4 t_3^2}{t_1^5}-\frac{4 t_2^4 t_3^2}{t_1^4}+\frac{2 t_2^4 t_3^2}{t_1^3}-\frac{2 t_2^5 t_3^2}{t_1^5}+\frac{t_2^5 t_3^2}{t_1^4}-\frac{6 t_2 t_3^3}{t_1^2}+\nn \\
&&\frac{10 t_2 t_3^3}{t_1}- \frac{7 t_2^2 t_3^3}{t_1^3}+\frac{16 t_2^2 t_3^3}{t_1^2}-\frac{11 t_2^2 t_3^3}{t_1}-\frac{4 t_2^3 t_3^3}{t_1^4}+\frac{9 t_2^3 t_3^3}{t_1^3}-\frac{6 t_2^3 t_3^3}{t_1^2}+\frac{t_2^3 t_3^3}{t_1}-\frac{t_2^4 t_3^3}{t_1^5}+\frac{2 t_2^4 t_3^3}{t_1^4}-\nn \\
&&\frac{t_2^4 t_3^3}{t_1^3}-\frac{t_2^5 t_3^3}{t_1^5}+\frac{2 t_2^5 t_3^3}{t_1^4}+\frac{2 t_2^2 t_3^4}{t_1^3}-\frac{t_2^2 t_3^4}{t_1^2}-\frac{t_2^3 t_3^4}{t_1^4}+\frac{2 t_2^3 t_3^4}{t_1^3}-\frac{t_2^3 t_3^4}{t_1^2}+\frac{t_2^4 t_3^4}{t_1^6}-\frac{6 t_2^4 t_3^4}{t_1^5}+\frac{9 t_2^4 t_3^4}{t_1^4}-\nn \\
&&\frac{4 t_2^4 t_3^4}{t_1^3}+\frac{2 t_2^5 t_3^4}{t_1^7}-\frac{11 t_2^5 t_3^4}{t_1^6}+\frac{16 t_2^5 t_3^4}{t_1^5}-\frac{7 t_2^5 t_3^4}{t_1^4}-\frac{2 t_2^6 t_3^4}{t_1^7}+\frac{10 t_2^6 t_3^4}{t_1^6}-\frac{6 t_2^6 t_3^4}{t_1^5}+\frac{t_2^2 t_3^5}{t_1^3}-\frac{2 t_2^2 t_3^5}{t_1^2}+\nn \\
&&\frac{2 t_2^3 t_3^5}{t_1^4}-\frac{4 t_2^3 t_3^5}{t_1^3}+\frac{2 t_2^3 t_3^5}{t_1^2}-\frac{2 t_2^4 t_3^5}{t_1^6}+\frac{7 t_2^4 t_3^5}{t_1^5}-\frac{8 t_2^4 t_3^5}{t_1^4}+\frac{3 t_2^4 t_3^5}{t_1^3}-\frac{4 t_2^5 t_3^5}{t_1^7}+\frac{12 t_2^5 t_3^5}{t_1^6}-\frac{12 t_2^5 t_3^5}{t_1^5}+\nn \\
&&\frac{4 t_2^5 t_3^5}{t_1^4}+\frac{5 t_2^6 t_3^5}{t_1^7}-\frac{10 t_2^6 t_3^5}{t_1^6}+\frac{3 t_2^6 t_3^5}{t_1^5}+\frac{t_2^4 t_3^6}{t_1^6}-\frac{2 t_2^4 t_3^6}{t_1^5}+\frac{t_2^4 t_3^6}{t_1^4}+\frac{2 t_2^5 t_3^6}{t_1^7}-\frac{4 t_2^5 t_3^6}{t_1^6}+\frac{2 t_2^5 t_3^6}{t_1^5}+\nn \\
&&\frac{3 t_2^6 t_3^6}{t_1^8}-\frac{6 t_2^6 t_3^6}{t_1^7}+\frac{3 t_2^6 t_3^6}{t_1^6}-\frac{3 t_2^7 t_3^6}{t_1^8}+\frac{2 t_2^7 t_3^6}{t_1^7}+\frac{t_2^7 t_3^7}{t_1^7}~.
\eea}
The superpotential must have R-charge equal to 2 and, since each monomial has fugacity $t_3 t_2 / t_1$,
\bea
R_3 + R_2 - R_1 = 2, 
\label{e:cycondfano136}
\eea
where $R_i$ is the R-charge associated with the fugacity $t_i$. From (\ref{e:cycondfano136}), the volume of $\cC_2$ is given by:
{\small
\bea
\lim_{\mu \rightarrow 0} \mu^4 \gm (e^{-\mu R_1}, e^{-\mu R_2}, e^{-\mu (2+ R_1 - R_2)}; \cC_2)= \frac{p(R_1, R_2; \cC_2)}{R^2_1 R_2\left(2+ R_1 - R_2\right)^2\left(R_2 -4\right)^2\left(4R_1 - 3R_2- 4\right)},\nn \\
\eea}
where:
{\small
\bea
p(R_1, R_2; \cC_2) &=& 128 + 128 R_1 + 32 R_1^2 - 32 R_2 + 192 R_1 R_2 + 64 R_1^2 R_2 + 16 R_1^3 R_2 - 56 R_2^2-\nn \\
&&120 R_1 R_2^2 - 26 R_1^2 R_2^2 + 26 R_2^3 + 16 R_1 R_2^3 - 3 R_2^4 ~.
\eea}
The values of R-charges which minimise the above volume are:
\bea
R_1 \approx 1.501, \qquad R_2 \approx 1.647, \qquad R_3 \approx 1.854~.
\eea

\paragraph{R-charges of perfect matchings.} The Hilbert series $g(D_\alpha,s_\alpha, \cC_2)$ of the divisor $D_\alpha$ can be obtained by inserting the inverse of fugacity for the corresponding perfect matching and then integrating over the baryonic fugacities.  This can be rewritten in terms of the variables $t_1, t_2, t_3$ and, accordingly, will be denoted as $g(D_\alpha,t_1,t_2,t_3, \cC_2)$.  The R-charge of the perfect matching $p_\alpha$ is then given by:
\bea
\lim_{\mu\rightarrow0}\frac{1}{\mu} \left[ \frac{g(D_1; e^{- \mu R_1 }, e^{- \mu R_2 }, e^{- \mu R_3 }; \cC_2) }{\gm(e^{- \mu R_1 }, e^{- \mu R_2 }, e^{- \mu R_3 }; \cC_2)}- 1 \right]. 
\eea
The results are presented in Table \ref{t:chargefano136}.
The other charges can be assigned with the conditions that the superpotential remains uncharged and that the charge vectors are linearly independent. Our choice is reported in Table \ref{t:chargefano136}.

\begin{table}[h!]
 \begin{center}  
  \begin{tabular}{|c||c|c|c|c|c|c|c|}
  \hline
  \;& $SU(2)_1$&$U(1)_1$&$U(1)_2$&$U(1)_R$&$U(1)_{B_1}$&$U(1)_{B_2}$&fugacity\\
  \hline\hline  
   
  $p_1$&$  1$&$  0$&$ 0$&$0.458$&$ 0$&$ 0$ & $s_1 x$\\
  \hline
  
  $p_2$&$ -1$&$  0$&$ 0$&$0.458$&$ 0$&$ 0$ & $s_1 /x$\\
  \hline  
  
  $q_1$&$  0$&$  1$&$ 0$&$0.291$&$ 0$&$ 0$ & $s_2 q_1$\\
  \hline
  
  $q_2$&$  0$&$ -1$&$ 0$&$0.314$&$ 0$&$ 0$ & $s_3/  q_1 $\\
  \hline
   
  $r_1$&$  0$&$  0$&$ 1$&$0.376$&$ 1$&$ 0$ & $s_4 q_2 b_1$\\
  \hline
      
  $r_2$&$  0$&$  0$&$-1$&$0.103$&$ 1$&$ 1$ & $s_5 b_1 b_2/ q_2 $\\
  \hline
  
  $r'_2$&$  0$&$  0$&$ 0$&$   0$&$ 0$&$-1$ & $1 / b_2  $\\
  \hline

  $v_1$&$  0$&$  0$&$ 0$&$    0$&$ -2$&$ 0$ & $1 / b^2_1 $\\
  \hline
 
     \end{tabular}
  \end{center}
\caption{Charges of the perfect matchings under the global symmetry of the $\cC_2$ model. Here $s_i$ are the fugacities of the R-charges, $x$ is the weight of the $SU(2)$ symmetry, $q_1, q_2, b_1$ and $b_2$ are, respectively, the charges under the mesonic abelian symmetries $U(1)_1$, $U(1)_2$ and under the two baryonic $U(1)_{B_1}$ and $U(1)_{B_2}$. The fugacity $s_6$ is set to 1 as it corresponds to a perfect matching with zero R-charge.}
\label{t:chargefano136}
\end{table}

\begin{table}[h]
 \begin{center}  
  \begin{tabular}{|c||c|}
  \hline
  \; Quiver fields &R-charge\\
  \hline  \hline 
  $X^{i}_{12}$ &  $0.835$\\
  \hline
  $X^{1}_{23}, X^{2}_{23}$ &  $0.458$\\
  \hline
  $X^{1}_{31}, X^{2}_{31}$ &  $0.561$\\
  \hline
  $X^{3}_{23}$ &  $0.604$\\
  \hline
  $X^{3}_{31}$ &  $0.707$\\
  \hline
  $X_{42}$ &  $0.667$\\
  \hline
  $X_{14}$ &  $0.314$\\
  \hline
  \end{tabular}
  \end{center}
\caption{R-charges of the quiver fields of $\cC_2$.}
\label{t:Rgenchfano136}
\end{table}

\begin{figure}[ht]
\begin{center}
 \includegraphics[totalheight=4cm]{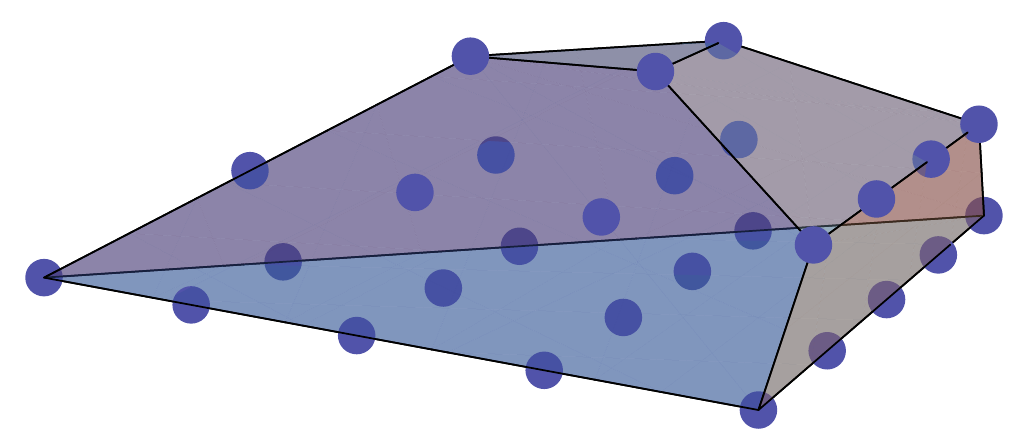}
\caption{The lattice of generators of the $\cC_2$ theory.}
  \label{f:latc2}
  \end{center}
\end{figure}

\paragraph{The Hilbert series.} The Hilbert series of the Master space is given by
{\small
\bea
g^{\firr{}}  (s_{\alpha}, x, q_1, q_2, b_1, b_2; \cC_2) &=& \oint \limits_{|z_1| =1} {\frac{\ud z_1}{2 \pi i z_1}} \oint \limits_{|z_2| =1} {\frac{\ud z_2}{2 \pi i z_2}}  \frac{1}{\left(1- s_1 x z_1\right)\left(1-\frac{s_1 z_1}{x}\right)\left(1-s_2 q_1 z_1 z_2\right)} \nn \\
&\times& \frac{1}{\left(1-\frac{s_3}{q_1 z_2}\right)\left(1-\frac{s_4 q_2 b_1}{z_1 z_2}\right)\left(1-\frac{s_5 b_1 b_2}{q_2 z_1}\right)\left(1- \frac{1}{z_1 b_2}\right)\left(1-\frac{z_2}{b^2_1}\right)}\nn \\
&=& \frac{\cP(s_{\alpha}, x, q_1, q_2 b_1, b_2; \cC_2)}{\left(1- \frac{s_1 x}{b_2}\right)\left(1-\frac{s_1}{x b_2}\right)\left(1- \frac{s_1 s_4 x q_2}{b_1}\right)\left(1-\frac{s_1 s_4 q_2}{x b_1}\right)\left(1-\frac{s_1 s_5 x b_1 b_2}{q_2}\right)}\nn \\
&\times& \frac{1}{\left(1-\frac{s_1 s_5 b_1 b_2}{x q_2}\right)\left(1-\frac{s_3}{q_1 b^2_1}\right)\left(1-\frac{s_2 s_3}{b_2}\right)\left(1- \frac{s_2 s_3 s_5 b_1 b_2}{q_2}\right)\left(1- s_2 s_4 q_1 q_2 b_1\right)},\nn \\
\label{e:HSmasterfano136}
\eea}
where $\cP(s_{\alpha}, x, q_1, q_2, b_1, b_2; \cC_2)$ is a polynomial that is not reported here. The integration over the two baryonic fugacities $b_1$ and $b_2$ gives the Hilbert series of the mesonic moduli space:
\bea
\gm (s_{\alpha}, x, q_1, q_2; \cC_2) &=& \oint \limits_{|b_1| =1} {\frac{\ud b_1}{2 \pi i b_1}} \oint \limits_{|b_2| =1} {\frac{\ud b_2}{2 \pi i b_2}} g^{\firr{}}  (s_{\alpha}, x, q_1, q_2, b_1, b_2; \cC_2) \nn \\
&=& \frac{P(s_{\alpha}, x, q_1, q_2; \cC_2)}{\left(1-\frac{s^4_1 s_3 s^2_5 x^4}{q_1 q^2_2}\right)\left(1-\frac{s^4_1 s_3 s^2_5}{x^4 q_1 q^2_2}\right)\left(1-s^3_1 s_4 s_5 x^3\right)\left(1-\frac{s^3_1 s_4 s_5}{x^3}\right)} \nn \\
&\times& \frac{1}{\left(1-s_1 s_2 s^2_4 x q_1 q^2_2\right)\left(1-\frac{s_1 s_2 s^2_4 q_1 q^2_2}{x}\right)\left(1-\frac{s^4_2 s^5_3 s^2_5}{q_1 q^2_2}\right)\left(1-s^2_2 s_3 s_4^2 q_1 q^2_2\right)}~, \qquad \quad
\label{e:HSmesonicfano136}
\eea
where $P(s_{\alpha}, x, q_1, q_2; \cC_2)$ is a polynomial that is not reported here.
The plethystic logarithm of (\ref{e:HSmesonicfano136}) can be written as:
\bea
\PL[\gm(t_{\alpha}, x, q_1, q_2; \cC_2)] &=& [4] \frac{t_2 t^2_3}{q_1 q^2_2 t^2_1} + [3]\left(t_3 + \frac{t^2_2 t^2_3}{q_1 q_2^2 t^3_1}\right) + [2]\frac{t_2 t_3}{t_1}\nn \\
&+& [2]\frac{t^3_2 t^2_3}{q_1 q^2_2 t^4_1} + [1] \left(q_1 q^2_2 t_1 + \frac{t^4_2 t^2_3}{q_1 q^2_2 t^5_1} + \frac{t^2_2 t_3}{t^2_1}\right)\nn \\
&+& q_1 q^2_2 t_2 + \frac{t^3_2 t_3}{t^3_1} + \frac{t^5_2 t^2_3}{q_1 q_2^2 t^6_1} - O(t_1) O(t_2) O(t_3)
\eea
The generators of the mesonic moduli space are
\bea
\begin{array}{llll}
p_i p_j p_k p_l q_2 r^2_2 r'^2_2 v_1, \quad & p_i p_j p_k r_1 r_2 r'_2 v_1, \quad & p_i p_j p_k q_1 q^2_2 r^2_2 r'^2_2 v_1, \quad &p_i p_j q_1 q_2 r_1 r_2 r'_2 v_1, \nn \\
p_i p_j q^2_1 q^3_2 r^2_2 r'^2_2 v_1, \quad & p_i q^2_1 q^2_2 r_1 r_2 r'_2 v_1, \quad & p_i q_1 r^2_1 v_1, \quad & p_i q^3_1 q^4_2 r^2_2 r'^2_2 v_1, \nn \\
q^2_1 q_2 r^2_1 v_1, \quad & q^3_1 q^3_2 r_1 r_2 r'_2 v_1,  \quad &q^4_1 q^5_2 r^2_2 r'^2_2 v_1~. &
\label{e:genefano136}
\end{array}
\eea
with $i, j, k,l=1,2$. The R-charges of the generators of the mesonic moduli space are presented in Table \ref{t:Rgenfano136}.
\begin{table}[h]
 \begin{center}  
  \begin{tabular}{|c||c|}
  \hline
  \; Generators &$U(1)_R$\\
  \hline  \hline 
  $p_i p_j p_k p_l q_2 r^2_2 r'^2_2 v_1$ & 2.353\\
  \hline
  $p_i p_j p_k r_1 r_2 r'_2 v_1$ & 1.854\\
  \hline
  $p_i p_j p_k q_1 q^2_2 r^2_2 r'^2_2 v_1$ & 2.499\\
  \hline
  $p_i p_j q_1 q_2 r_1 r_2 r'_2 v_1$ & 2\\
  \hline
  $p_i p_j q^2_1 q^3_2 r^2_2 r'^2_2 v_1$ & 2.644\\
  \hline
  $p_i q^2_1 q^2_2 r_1 r_2 r'_2 v_1$ & 2.146\\
  \hline
  $p_i q_1 r^2_1 v_1$ & 1.501\\
  \hline
  $p_i q^3_1 q^4_2 r^2_2 r'^2_2 v_1$ & 2.790\\
  \hline
  $q^2_1 q_2 r^2_1 v_1$ & 1.647\\
  \hline
  $q^3_1 q^3_2 r_1 r_2 r'_2 v_1$ & 2.292\\
  \hline
  $q^4_1 q^5_2 r^2_2 r'^2_2 v_1$ & 2.936\\
  \hline
  \end{tabular}
  \end{center}
\caption{R-charges of the generators of the mesonic moduli space for the $\cC_2$ Model.}
\label{t:Rgenfano136}
\end{table}
The lattice of generators is drawn \fref{f:latc2}.


\section{$\cD_1$ (Toric Fano 131): $\BP^1$-blowup of $\mathcal{B}_2$}
This theory has 4 gauge groups and chiral fields $X_{13}$, $X_{12}$, $X_{42}$, $X^i_{34}$, $X^j_{23}$ and $X^j_{41}$ (with $i=1,2,3$ and $j=1,2$).
 The tiling and the quiver diagram coincide with those presented in Figure \ref{f:tqfano123}, with CS levels $\vec{k} = (-1,-1,0,2)$.
The superpotential coincide with that presented in (\ref{e:sptoric123}).
\comment{
 \bea
W = \tr \left[ \epsilon_{ij} \left( X_{13}X^{i}_{34}X^{j}_{41} + X_{42}X^{i}_{23}X^{j}_{34} - X_{12}X^{i}_{23}X^{3}_{34}X^{j}_{41} \right) \right]~.
\label{e:sptoric131}
\eea
\begin{figure}[ht]
\begin{center}
 \centerline{  \epsfxsize = 5cm \epsfbox{QuiverdP1.pdf}\hskip 10mm \epsfxsize = 6.5cm \epsfbox{tilc4.pdf}}
 \caption{(i) Quiver diagram of the $\cD_1$  model.\ (ii) Tiling of the $\cD_1$  model.}
  \label{f:tqfano123}
\end{center}
\end{figure}}

\comment{
\begin{figure}[ht]
\begin{center}
   \includegraphics[totalheight=5cm]{fdd1.pdf}
 \caption{The fundamental domain of the tiling for the $\cD_1$ model.}
  \label{f:fdtoricfano131}
\end{center}
\end{figure}
}

\paragraph{The Kasteleyn matrix.} The Chern-Simons levels can be written in terms of the integers $n^i_{jk}$ and $n_{jk}$ as shown below

\bea
\begin{array}{ll}
\text{Gauge group 1~:} \qquad k_1  &=   n_{12} + n_{13} - n^{1}_{41} - n^{2}_{41}~, \nn \\
\text{Gauge group 2~:} \qquad k_2  &=   n^{1}_{23} + n^{2}_{23} - n_{12} - n_{42}~, \nn \\
\text{Gauge group 3~:} \qquad k_3  &=   n^{1}_{34} + n^{2}_{34} + n^{3}_{34} - n^{1}_{23} - n^{2}_{23} - n_{13}~, \nn \\
\text{Gauge group 4~:} \qquad k_4  &=   n_{42} + n^{1}_{41} + n^{2}_{41} - n^{1}_{34} - n^{2}_{34} - n^{3}_{34} ~.
\label{e:kafano131}
\end{array}
\eea

For this model, let us choose:
\bea
n^1_{34} = n_{13} = - n^1_{41} = - n_{12} =1,\quad n^i_{jk}=n_{jk}=0 \; \text{otherwise}~.
\eea
The fundamental domain contains three pairs of black and white nodes, and so the Kasteleyn matrix $K$ is a $3\times 3$ matrix\footnote{Note that, in order to make the non-abelian mesonic symmetry more apparent in the $G_K$ matrix, the weight assignment is different to \ref{e:kastfano123}}:
\bea
K =   \left(
\begin{array}{c|ccc}
& b_1 & b_2 & b_3\\
\hline
w_1 & z^{n_{13}} & z^{n^1_{34}} & \frac{y}{x} z^{n^2_{41}} \\
w_2 & x z^{n^1_{41}} & z^{n^2_{23}} & z^{n^3_{34}} + y z^{n_{12}} \\
w_3 & z^{n^2_{34}} & \frac{1}{y} z^{n_{42}} & z^{n^1_{23}} \end{array}
\right) ~.
\label{e:kastfano131}
\eea
The permanent of this matrix is given by
\bea
\perm~K &=& x z^{(n^1_{41} + n^1_{23} + n^1_{34} )} +  x^{-1} y z^{(n^2_{41} + n^2_{23} + n^2_{34})} +  y z^{(n^1_{34} + n^2_{34} + n_{12})}\nn \\
&+&  y^{-1} z^{(n^3_{34} + n_{42} + n_{13})} +  z^{(n^1_{41} + n^2_{41} + n_{42})} +  z^{(n^1_{23} + n^2_{23} + n_{13})}\nn \\
&+&  z^{(n^1_{34} + n^2_{34} + n^3_{34})} +  z^{(n_{12} + n_{42} + n_{13})}\nn \\
&=&  x + x^{-1} y + y  + y^{-1} z + z^{-1} + 2z + 1\nn\\ 
&&  \text{(for $n^1_{34} = n_{13} = - n^1_{41} = - n_{12} =1,\quad n^i_{jk}=n_{jk}=0 \; ~ \text{otherwise}$)} ~.\nn \\
\label{e:permKfano131}
\eea
The permanent of the Kasteleyn matrix can be used to write the perfect matchings in terms of the chiral fields of the model:
\bea 
&&   p_1 = \left\{X^1_{41}, X^1_{23},X^1_{34}\right\},
\;\; p_2 = \left\{X^2_{41}, X^2_{23},X^2_{34}\right\},
\;\; q_1 = \left\{X^1_{34}, X^2_{34},X_{12}\right\},\nn \\
&&   q_2 = \left\{X^3_{34}, X_{42},X_{13}\right\},
\;\; r_1 = \left\{X^1_{41}, X^2_{41},X_{42}\right\},
\;\; r_2 = \left\{X^1_{23}, X^2_{23},X_{13}\right\},\nn \\
&&   r'_2 = \left\{X^1_{34}, X^2_{34},X^3_{34}\right\},
\;\; v_1 = \left\{X_{12}, X_{42},X_{13}\right\}\ . \qquad
\eea
In turn, let us parametrize the chiral fields in terms of perfect matchings:
\bea \begin{array}{llll}
X^1_{41} = p_1 r_1,  \quad & X^2_{41} = p_2 r_1,  \quad & X^1_{23} = p_1 r_2, \quad & X^2_{23} = p_2 r_2,  \\
X^1_{34} = p_1 q_1 r'_2,  \quad &  X^2_{34} = p_2 q_1 r'_2,  \quad &  X^3_{34} = q_2 r'_2, \quad & X_{12} = q_1 v_1,\nn \\
X_{42} = q_2 r_1 v_1, \quad & X_{13} = q_2 r_2 v_1~.
\end{array} \eea
This information can be summarised in the perfect matching matrix:
\beq
P=\left(\begin{array} {c|cccccccc}
  \;& p_1 & p_2 & q_1 & q_2 & r_1 & r_2 & r'_2 & v_1\\
  \hline 
  X^{1}_{41} & 1&0&0&0&1&0&0&0\\
  X^{2}_{41} & 0&1&0&0&1&0&0&0\\
  X^{1}_{23} & 1&0&0&0&0&1&0&0\\
  X^{2}_{23} & 0&1&0&0&0&1&0&0\\
  X^{1}_{34} & 1&0&1&0&0&0&1&0\\
  X^{2}_{34} & 0&1&1&0&0&0&1&0\\
  X^{3}_{34} & 0&0&0&1&0&0&1&0\\
  X_{12}     & 0&0&1&0&0&0&0&1\\
  X_{42}     & 0&0&0&1&1&0&0&1\\
  X_{13}     & 0&0&0&1&0&1&0&1  
  \end{array}
\right).
\eeq
The null space of the $P$ matrix is spanned by two vectors that can be cast in the rows of the following matrix:
\be
Q_F =   \left(
\begin{array}{cccccccc}
1 & 1 & -1 & 0 & -1 & -1 & 0 & 1 \\
0 & 0 &  1 & 1 &  0 &  0 &-1 &-1 
\end{array}
\right)~.  \label{e:qffano131}
\ee
Hence, among the perfect matchings there are two relations, which are given by:
\bea
p_1 + p_2 - q_1 - r_1 - r_2 + v_1 &=& 0~, \nn \\
q_1 + q_2 - r'_2 - v_1 &=& 0~.
\label{e:relpmfano131}
\eea

\paragraph{The toric diagram.} Two methods of computing the toric diagram for this model are presented
\begin{itemize}
\item {\bf The Kasteleyn matrix.} The coordinates of the toric diagram are collected in the columns of the following matrix:
\bea
G_K = \left(
\begin{array}{cccccccc}
  1 &-1 & 0 &  0 & 0 &  0 & 0 & 0\\
  0 & 1 & 1 & -1 & 0 &  0 & 0 & 0\\
  0 & 0 & 0 &  1 &-1 &  1 & 1 & 0
\end{array}
\right)~.
\eea
The first row of this matrix contains weights of the fundamental representation of $SU(2)$.  Therefore, the mesonic symmetry contains only one non-abelian factor. Given that the total rank of the mesonic symmetry is 4, this can be identified with $SU(2)\times U(1)^3$, where one of the abelian factor corresponds to the R-symmetry.

\item {\bf The charge matrices.}
Since the number of gauge groups of this model is $G = 4$, there are $G-2 =2$ baryonic symmetries coming from the D-terms. The charges of the perfect matchings can be collected in the rows of the $Q_D$ matrix:
\be
Q_D =   \left(
\begin{array}{cccccccc}
 0 & 0 &  0 &  0 &  1 & 1 & 0 & -2\\
 0 & 0 &  0 &  0 &  0 & 1 &-1 &  0
\end{array}
\right). \label{e:qdfano131}
\ee
The total charge matrix $Q_t$, which is  the combination of the $Q_F$ and $Q_D$ matrices, can be written as:
\be
Q_t = { \Blue Q_F \choose \Green Q_D \Black } =   \left( 
\begin{array}{cccccccc} \Blue
1 & 1 & -1 & 0 & -1 & -1 & 0 & 1 \\
0 & 0 &  1 & 1 &  0 &  0 &-1 &-1 \\ \Green
 0 & 0 &  0 &  0 &  1 & 1 & 0 & -2\\
 0 & 0 &  0 &  0 &  0 & 1 &-1 &  0 \Black
\end{array}
\right). 
\label{e:qtfano131}
\ee
The kernel of this matrix gives the $G_t$ matrix. After eliminating the first row, the $G'_t$ matrix is obtained. The coordinates of the toric diagram are given in the column of:
\bea
G'_t = \left(
\begin{array}{cccccccc}
  1 &-1 & 0 &  0 & 0 &  0 & 0 & 0\\
  0 & 1 & 1 & -1 & 0 &  0 & 0 & 0\\
  0 & 0 & 0 &  1 &-1 &  1 & 1 & 0
\end{array}
\right) = G_K~. 
\label{e:gtfano131}
\eea
The toric diagram constructed from (\ref{e:gtfano131}) is presented in Figure \ref{f:tdtoricfano131}.
\begin{figure}[ht]
\begin{center}
  \includegraphics[totalheight=3.0cm]{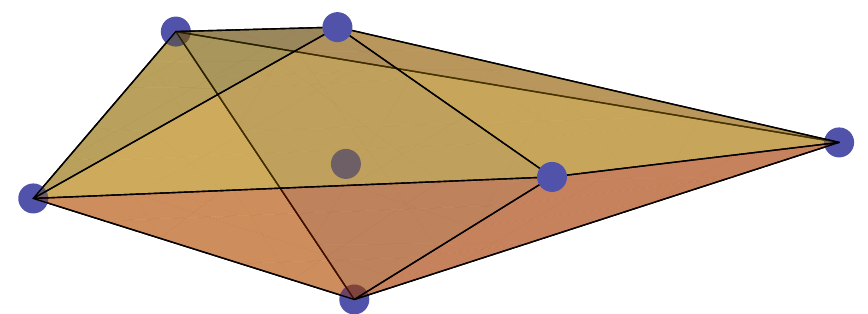}
 \caption{The toric diagram of $\cD_1$.}
  \label{f:tdtoricfano131}
\end{center}
\end{figure}
\end{itemize}

\paragraph{The baryonic charges.} The toric diagram of this model is characterized by 6 external points. Thus, there are 2 baryonic symmetries which shall be denoted as $U(1)_{B_1}$ and $U(1)_{B_2}$. The charges of the perfect matchings under these two symmetries are collected in the rows of the $Q_D$ matrix presented in (\ref{e:qdfano131}).

{\small \paragraph{The global symmetry.}  The total charge matrix in (\ref{e:qtfano131}) shows that there is one pair of repeated columns, thus confirming that the mesonic symmetry of this model is $SU(2)\times U(1)^3$. One of the $U(1)$'s can be identified with the R-symmetry. The perfect matchings $p_1$ and $p_2$ transform as a doublet under the non-abelian symmetry. Note that the perfect matchings $r_2$ and $r'_2$ correspond to the same point in the toric diagram. The perfect matching $v_1$ corresponds to the internal point in the toric diagram and has carries zero R-charge.
Let $s_1$ be the fugacity for the R-charge of the perfect matchings $p_1$ and $p_2$, and $s_2,s_3,s_4,s_5,s_6$ be those for $q_1,q_2,r_1,r_2,r'_2$ respectively.  Assigning the same fugacity to perfect matchings $p_1$ and $p_2$ is justified by the fact that the non-abelian symmetries do not play any role in the volume minimisation. 
The Hilbert series of the mesonic moduli space is given by:
\bea
\gm (t_\alpha; \cD_1) &=& \oint \limits_{|z_1|=1}{\frac{\ud z_1}{2\pi i z_1}}\oint \limits_{|z_2|=1}{\frac{\ud z_2}{2\pi i z_2}}\oint \limits_{|b_1|=1}{\frac{\ud b_1}{2\pi i b_1}}\oint \limits_{|b_2|=1}{\frac{\ud b_2}{2\pi i b_2}}
\frac{1}{\left(1-s_1 z_1\right)^2\left(1-\frac{s_2 z_2}{z_1}\right)}\nn \\
&\times& \frac{1}{\left(1- s_3 z_2\right)\left(1- \frac{s_4 b_1}{z_1}\right)\left(1- \frac{s_5 b_1 b_2}{z_1}\right)\left(1-\frac{s_6}{z_2 b_2}\right)\left(1-\frac{z_1}{z_2 b^2_1}\right)}~.
\label{e:hsvolfano131}
\eea
Since there are three $U(1)$ factors in the mesonic symmetry, the mesonic Hilbert series depends on only three combinations of $s_\alpha$s. Setting
 \bea
 t^2_1 = s_1 s_3 s^2_4, \qquad t^2_2 = s^2_1 s_2 s^2_4, \qquad t^2_3 = s_1 s^3_3 s^2_5 s^2_6~,
 \eea
gives us
\bea
\gm (t_1, t_2,t_3; \cD_1) &=& \frac{P\left(t_1,t_2,t_3\right)}{\left(1-t^2_1\right)^2\left(1-t^2_2\right)^2\left(1-t^2_3\right)^2\left(1-\frac{t^6_2 t^2_3}{t^6_1}\right)^2}~,
\eea
where }
{\scriptsize
\bea
P\left(t_1,t_2,t_3\right) &=& 1+t_2^2-2 t_1^2 t_2^2+2 t_1 t_3-t_1^3 t_3-6 t_1 t_2^2 t_3+2 t_1^3 t_2^2 t_3+ 6 t_1 t_2^4 t_3- t_1^3 t_2^4 t_3+ 6 t_1^2 t_2^2 t_3^2+6 t_2^4 t_3^2- 2 t_1 t_2^6 t_3- \nn \\
&&t_1^2 t_3^2 - 8 t_2^2 t_3^2+3 t_1 t_2^2 t_3^3-3 t_1 t_2^4 t_3^3- t_1^2 t_2^4 t_3^2-2 t_2^6 t_3^2-t_1 t_3^3 +t_1 t_2^6 t_3^3 + 4 t_2^2 t_3^4 -2 t_1^2 t_2^2 t_3^4-3 t_2^4 t_3^4 +t_2^6 t_3^4 +\nn \\
&&\frac{3 t_2^2 t_3}{t_1}+ \frac{4 t_2^4 t_3}{t_1^3} -  \frac{9 t_2^4 t_3}{t_1}- \frac{2 t_2^6 t_3}{t_1^3}+\frac{4 t_2^6 t_3}{t_1}+ \frac{3 t_2^2 t_3^2}{t_1^2}+\frac{4 t_2^4 t_3^2}{t_1^4}- \frac{9 t_2^4 t_3^2}{t_1^2}+\frac{3 t_2^6 t_3^2}{t_1^6}-  \frac{12 t_2^6 t_3^2}{t_1^4}+\frac{9 t_2^6 t_3^2}{t_1^2}-\nn \\
&&\frac{5 t_2^8 t_3^2}{t_1^6}+\frac{10 t_2^8 t_3^2}{t_1^4}-  \frac{3 t_2^8 t_3^2}{t_1^2}- \frac{2 t_2^2 t_3^3}{t_1}- \frac{3 t_2^4 t_3^3}{t_1^3}+\frac{6 t_2^4 t_3^3}{t_1}+\frac{2 t_2^6 t_3^3}{t_1^5}- \frac{9 t_2^6 t_3^3}{t_1^3}+\frac{4 t_2^6 t_3^3}{t_1}+  \frac{t_2^8 t_3^3}{t_1^7}- \frac{6 t_2^8 t_3^3}{t_1^5}+\nn \\
&&\frac{11 t_2^8 t_3^3}{t_1^3}-\frac{4 t_2^8 t_3^3}{t_1}-\frac{3 t_2^{10} t_3^3}{t_1^7}+\frac{6 t_2^{10} t_3^3}{t_1^5}-\frac{3 t_2^{10} t_3^3}{t_1^3}-\frac{2 t_2^{12} t_3^3}{t_1^9}+ \frac{4 t_2^{12} t_3^3}{t_1^7}-\frac{2 t_2^{12} t_3^3}{t_1^5}-\frac{2 t_2^2 t_3^4}{t_1^2}-\frac{3 t_2^4 t_3^4}{t_1^4}+\frac{6 t_2^4 t_3^4}{t_1^2}-\nn \\
&& \frac{4 t_2^6 t_3^4}{t_1^6}+\frac{11 t_2^6 t_3^4}{t_1^4}-  \frac{9 t_2^8 t_3^4}{t_1^4}+\frac{2 t_2^8 t_3^4}{t_1^2}-\frac{3 t_2^{10} t_3^4}{t_1^8}+ \frac{6 t_2^{10} t_3^4}{t_1^6}-\frac{3 t_2^{10} t_3^4}{t_1^4}+\frac{3 t_2^{12} t_3^4}{t_1^8}-  \frac{2 t_2^{12} t_3^4}{t_1^6}-\frac{t_2^{14} t_3^4}{t_1^8}-\frac{3 t_2^6 t_3^5}{t_1^5}+\nn \\
&&\frac{10 t_2^6 t_3^5}{t_1^3}-\frac{5 t_2^6 t_3^5}{t_1}-\frac{2 t_2^8 t_3^5}{t_1^7}+\frac{9 t_2^8 t_3^5}{t_1^5}-\frac{12 t_2^8 t_3^5}{t_1^3}+\frac{3 t_2^8 t_3^5}{t_1}-  \frac{t_2^{10} t_3^5}{t_1^9}+\frac{6 t_2^{10} t_3^5}{t_1^7}-\frac{9 t_2^{10} t_3^5}{t_1^5}+\frac{4 t_2^{10} t_3^5}{t_1^3}+\frac{6 t_2^{12} t_3^5}{t_1^9}-\nn \\
&&\frac{8 t_2^{12} t_3^5}{t_1^7}+\frac{3 t_2^{12} t_3^5}{t_1^5}-\frac{t_2^{14} t_3^5}{t_1^9}-\frac{2 t_2^8 t_3^6}{t_1^8}+  \frac{4 t_2^8 t_3^6}{t_1^6}-\frac{2 t_2^8 t_3^6}{t_1^4}-\frac{t_2^{10} t_3^6}{t_1^{10}}+\frac{6 t_2^{10} t_3^6}{t_1^8}-\frac{9 t_2^{10} t_3^6}{t_1^6}+\frac{4 t_2^{10} t_3^6}{t_1^4}+\frac{2 t_2^{12} t_3^6}{t_1^{10}}-\nn \\
&&\frac{6 t_2^{12} t_3^6}{t_1^8}+\frac{3 t_2^{12} t_3^6}{t_1^6}-\frac{t_2^{14} t_3^6}{t_1^{10}}+\frac{2 t_2^{14} t_3^6}{t_1^8}-\frac{2 t_2^{12} t_3^7}{t_1^9}+\frac{t_2^{12} t_3^7}{t_1^7}+\frac{t_2^{14} t_3^7}{t_1^7}
-6\frac{t_2^6 t_3^4}{t_1^2} + \frac{t_2^8 t_3^4}{t_1^8} + 4\frac{t_2^8 t_3^4}{t_1^6}
~.
\eea}
Since the superpotential scales like $t^2_2 t_3 / t_1$, if $R_i$ is defined to be the R-charge corresponding to $t_i$, it follows that:
\bea
2R_2 + R_3 - R_1=2~.
\eea
The volume of $\cD_1$ is given by
{\small
\bea
\lim_{\mu \rightarrow 0} \;\mu^4 \gm (e^{- \mu R_1}, e^{- \mu R_2}, e^{- \mu (2 + R_1 - 2 R_2)}; \cD_1) = \frac{p(R_1, R_2; \cD_1)}{8 R^2_1 R^2_2 \left(R_1 - 2R_2 + 2\right)^2\left(2R_1 - R_2 - 2\right)^2},\nn \\
\label{e:volfano131}
\eea}
where:
{\small
\bea 
p(R_1, R_2; \cD_1) &=&  16 R_1 - 12 R_1^3 - 4 R_1^4 + 8 R_2 - 8 R_1 R_2 + 34 R_1^2 R_2+ 20 R_1^3 R_2 +  4 R_1 R_2^2 -\nn \\
&&  31 R_1^2 R_2^2 - 6 R_2^3 + 6 R_1 R_2^3 - 2 R_2^4~.
\eea}
This function has a minimum at
\bea
R_1 \approx 0.796, \qquad R_2 \approx 0.900~.
\eea

The R-charge of the external perfect matching corresponding to the divisor $D_\alpha$ is given by
\bea
\lim_{\mu\rightarrow0}\frac{1}{\mu} \left[ \frac{g(D_\alpha; e^{- \mu R_1}, e^{- \mu R_2 }, e^{- \mu R_3 }; \cD_1) }{\gm(e^{-\mu R_1}, e^{- \mu R_2 }, e^{- \mu R_3 };\cD_1)}- 1 \right]~,
\eea
where $g(D_\alpha; e^{- \mu R_1}, e^{- \mu R_2 }, e^{- \mu R_3 }; \cD_1)$ is the Molien-Weyl integral with the insertion of the inverse of the weight corresponding to the divisor $D_\alpha$. The results are shown in Table \ref{t:chargefano131}.

The assignment of charges under the remaining abelian symmetries can be done by requiring that the superpotential is not charged under them and that the charge vectors are linearly independent. The assignments are shown in Table \ref{t:chargefano131}.
\begin{table}[h!]
 \begin{center}  
  \begin{tabular}{|c||c|c|c|c|c|c|c|}
  \hline
  \;& $SU(2)_1$&$U(1)_1$&$U(1)_2$&$U(1)_R$&$U(1)_{B_1}$&$U(1)_{B_2}$&fugacity\\
  \hline\hline  
   
  $p_1$&$  1$&$  0$&$ 0$&$0.354$&$ 0$&$ 0$ & $s_1 x$\\
  \hline
  
  $p_2$&$ -1$&$  0$&$ 0$&$0.354$&$ 0$&$ 0$ & $s_1 / x$\\
  \hline  
  
  $q_1$&$  0$&$  1$&$ 0$&$0.255$&$ 0$&$ 0$ & $s_2 q_1$\\
  \hline
  
  $q_2$&$  0$&$ -1$&$ 0$&$0.401$&$ 0$&$ 0$ & $s_3 / q_1$\\
  \hline
   
  $r_1$&$  0$&$  0$&$ 1$&$0.419$&$ 1$&$ 0$ & $s_4 q_2 b_1$\\
  \hline
      
  $r_2$&$  0$&$  0$&$-1$&$0.217$&$ 1$&$ 1$ & $s_5 b_1 b_2 / q_2$\\
  \hline
  
  $r'_2$&$ 0$&$  0$&$ 0$&$    0$&$ 0$&$-1$ & $ 1/ b_2$\\
  \hline

  $v_1$&$  0$&$  0$&$ 0$&$    0$&$-2$&$ 0$ & $1/ b^2_1$\\
  \hline
 
     \end{tabular}
  \end{center}
\caption{Charges of the perfect matchings under the global symmetry of the $\cD_1$ theory. Here $s_i$ are the fugacities of the R-charges, $x$ is the weight of the $SU(2)$ symmetry, $q_1, q_2, b_1$ and $b_2$ are respectively the charges under the mesonic abelian symmetries $U(1)_1, U(1)_2$, and under the two baryonic $U(1)_{B_1}$ and $U(1)_{B_2}$. The perfect matching $r'_2$ is found to have zero R-charge and, for this reason, its R-charge fugacity $(s_6)$ is set to 1.}
\label{t:chargefano131}
\end{table}

\begin{table}[h]
 \begin{center}  
  \begin{tabular}{|c||c|}
  \hline
  \; Quiver fields & R-charge\\
  \hline  \hline 
   $X^{1}_{23}, X^{2}_{23} $ & 0.571 \\
  \hline
   $X^{1}_{41}, X^{2}_{41}$ & 0.773\\
  \hline
   $X^{1}_{34}, X^{2}_{34}$ & 0.609\\
  \hline
   $X^{3}_{34}$ & 0.401\\
  \hline
   $X_{12}$ & $ $ 0.255\\
  \hline
   $X_{42}$ & $ $ 0.819\\
  \hline
   $X_{13}$ & $ $ 0.618\\
  \hline
  \end{tabular}
  \end{center}
\caption{R-charges of the quiver fields for the $\cD_1$ Model.}
\label{t:Rquivfano131}
\end{table}

\paragraph{The Hilbert series.} Given the charge assignment shown in Table \ref{t:chargefano131}, it is possible to compute the Hilbert series of the Master space and of the mesonic moduli space. The former one can be determined by integrating the Hilbert series of the space of perfect matchings over the fugacities $z_1$ and $z_2$
\bea
g^{\firr{}}  (s_{\alpha},x,q_1,q_2, b_1, b_2; \cD_1) &=& \oint \limits_{|z_1| =1} {\frac{\ud z_1}{2 \pi i z_1}} \oint \limits_{|z_2| =1} {\frac{\ud z_2}{2 \pi i z_2}} \frac{1}{\left(1-s_1 x z_1\right)\left(1-\frac{s_1 z_1}{x}\right)\left(1-\frac{s_2 q_1 z_2}{z_1}\right)}\nn\\
&\times& \frac{1}{\left(1-\frac{s_3 z_2}{q_1}\right)\left(1-\frac{s_4 q_2 b_1}{z_1}\right)\left(1-\frac{s_5 b_1 b_2}{q_2 z_1}\right)\left(1- \frac{1}{b_2 z_2}\right)\left(1-\frac{z_1}{b^2_1 z_2}\right)} \nn \\
&=& \frac{\cP\left(s_{\alpha},x,q_1,q_2, b_1, b_2; \cD_1\right)}{\left(1- \frac{s_1 s_2 x q_1}{b_2} \right)\left(1-\frac{s_1 s_2 q_1}{x b_2}\right)\left(1- s_1 s_4 x q_2 b_1\right)\left(1-\frac{s_1 s_4 q_2 b_1}{x}\right)}\nn \\
&\times& \frac{1}{\left(1-\frac{s_1 s_5 x b_1 b_2}{q_2}\right)\left(1-\frac{s_1 s_5 b_1 b_2}{x q_2}\right)\left(1-\frac{s_2 q_1}{b^2_1}\right)\left(1-\frac{s_3}{q_1 b_2}\right)}\nn \\
&\times& \frac{1}{\left(1-\frac{s_3 s_4 q_2}{q_1 b_1}\right)\left(1-\frac{s_3 s_5 b_2}{q_1 q_2 b_1}\right)},
\label{e:HSmasterfano131}
\eea
where $\cP\left(s_{\alpha},x,q_1,q_2, b_1, b_2; \cD_1\right)$ is a polynomial that is not reported here.
The Hilbert series of the mesonic moduli space can be obtained by integrating (\ref{e:HSmasterfano131}) over the two baryonic fugacities $b_1$ and $b_2$:
{\small
\bea
\gm (s_{\alpha}, x, q_1, q_2; \cD_1) &=& \oint \limits_{|b_1| =1} {\frac{\ud b_1}{2 \pi i b_1}} \oint \limits_{|b_2| =1} {\frac{\ud b_2}{2 \pi i b_2}}g^{\firr{}}  (s_{\alpha},x,q_1,q_2, b_1, b_2; \cD_1)\nn \\
&=& \frac{P\left(s_{\alpha},x, q_1, q_2; \cD_1\right)}{\left(1- \frac{s^4_1 s^3_2 s^2_5 q^3_1 x^4 }{q^2_2}\right)\left(1- \frac{s^4_1 s^3_2 s^2_5 q^3_1}{x^4 q^2_2}\right)\left(1- s^2_1 s_2 s^2_4 x^2 q_1 q^2_2\right)\left(1- \frac{s^2_1 s_2 s^2_4 q_1 q^2_2}{x^2}\right)}\nn \\
&\times& \frac{1}{\left(1- \frac{s_1 s^3_3 s^2_5 x}{q^3_1 q^2_2}\right)\left(1- \frac{s_1 s^3_3 s^2_5}{x q^3_1 q^2_2}\right)\left(1- \frac{s_1 s_3 s^2_4 x q^2_2}{q_1}\right)\left(1- \frac{s_1 s_3 s^2_4 q^2_2}{x q_1}\right)},
\label{e:HSmesfano131}
\eea}
where $P\left(s_{\alpha},x, q_1, q_2; \cD_1\right)$ is a polynomial that is not reported here.
The plethystic logarithm of the mesonic Hilbert series is given by
{\footnotesize
\bea
\PL [\gm (t_{\alpha}, x, q_1, q_2; \cD_1)] &=& [4] \frac{q^3_1 t^6_2 t^2_3}{q^2_2 t^6_1} + [3]\left(\frac{q^2_1 t^4_2 t_3}{t^3_1} + \frac{q_1 t^4_2 t^2_3}{q^2_2 t^4_1}\right) + [2]\left(q_1 q^2_2 t^2_2 + \frac{t^2_2 t_3}{t_1} + \frac{t^2_2 t^2_3}{q_1 q^2_2 t^2_1}\right) + \nn \\
&&  [1]\left(\frac{q^2_2 t^2_1}{q_1} + \frac{t_1 t_3}{q^2_1} + \frac{t^2_3}{q^3_1 q^2_2}\right) - O(t^2_1)O(t^2_2)O(t_3)
\eea}
Therefore, the generators of the mesonic moduli space are
\bea
\begin{array}{lll}
p_i p_j p_k p_l q^3_1 r^2_2 r'^2_2 v_1,\qquad & p_i p_j p_k q^2_1 r_1 r_2 r'_2 v_1,\qquad & p_i p_j p_k q_1^2 q_2 r^2_2 r'^2_2 v_1~, \nn \\
p_i p_j q_1 r^2_1 v_1,\qquad & p_i p_j q_1 q_2 r_1 r_2 r'_2 v_1,\qquad & p_i p_j q_1 q^2_2 r^2_2 r'^2_2 v_1~,\nn \\
p_i q_2 r^2_1 v_1,\qquad & p_i q^2_2 r_1 r_2 r'_2 v_1,\qquad & p_i q^3_2 r^2_2 r'^2_2 v_1~.\nn \\
\end{array}
\eea
with $i,j,k,l=1,2$. The R-charges of the generators of the mesonic moduli space are presented in \tref{t:Rgenfano131}. 

\begin{table}[h]
 \begin{center}  
  \begin{tabular}{|c||c|}
  \hline
  \; Generators & R-charge\\
  \hline  \hline 
   $p_i p_j p_k p_l q^3_1 r^2_2 r'^2_2 v_1 $ & 2.616 \\
  \hline
   $p_i p_j p_k q^2_1 r_1 r_2 r'_2 v_1$ & 2.208\\
  \hline
   $p_i p_j p_k q_1^2 q_2 r^2_2 r'^2_2 v_1$ & 2.408\\
  \hline
   $p_i p_j q_1 r^2_1 v_1$ & 1.800\\
  \hline
   $p_i p_j q_1 q_2 r_1 r_2 r'_2 v_1$ &  2\\
  \hline
   $p_i p_j q_1 q^2_2 r^2_2 r'^2_2 v_1$ & 2.200\\
  \hline
   $p_i q_2 r^2_1 v_1$ & 1.592\\
  \hline
    $p_i q^2_2 r_1 r_2 r'_2 v_1$ & 1.792\\
  \hline
   $p_i q^3_2 r^2_2 r'^2_2 v_1$ & 1.992\\
  \hline
  \end{tabular}
  \end{center}
\caption{R-charges of the generators for the $\cD_1$ Model.}
\label{t:Rgenfano131}
\end{table}

\begin{figure}[ht]
\begin{center}
\includegraphics[totalheight=4cm]{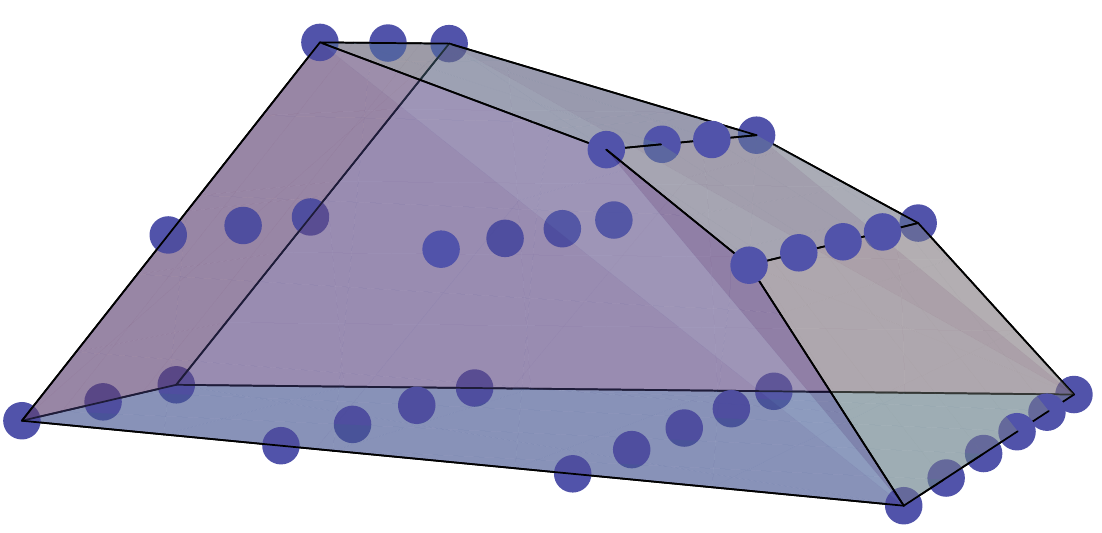}
\caption{The lattice of generators of the $\cD_1$ theory.}
  \label{f:latd1}
  \end{center}
\end{figure}


\section{$\cD_2$ (Toric Fano 139): $\BP^1$-blowup of $\cB_4$}
This theory has 4 gauge groups and chiral fields $X^i_{23}$, $X^i_{31}$ (with $i=1,2,3$), $X^j_{12}$ (with $j = 1,2$), $X_{14}$ and $X_{42}$. The tiling and the quiver diagram are presented in Figure \ref{f:tqfano136}.
Note that they are the identical to those of the $\cC_2$ theory (\emph{i.e.} the `double bonding' of the $M^{1,1,1}$ tiling).
However, the CS levels of this theory are $\vec{k} = (-1,1,1,-1)$.  The superpotential is given by \eref{e:spotfano136}.

\paragraph{The Kasteleyn matrix.} The Chern-Simons levels can be parametrized in terms of integers as according to (\ref{e:kafano139}).

\bea
\begin{array}{ll}
\text{Gauge group 1~:} \qquad k_1  &=   n_{14} + n^{1}_{12} + n^{2}_{12} - n^{1}_{31} - n^{2}_{31} - n^{3}_{31} ~, \nn \\
\text{Gauge group 2~:} \qquad k_2  &=   n^{1}_{23} + n^{2}_{23} + n^{3}_{23} - n^{1}_{12} - n^{2}_{12} - n_{42}  ~, \nn \\
\text{Gauge group 3~:} \qquad k_3  &=   n^{1}_{31} + n^{2}_{31} + n^{3}_{31} - n^{1}_{23} - n^{2}_{23} - n^{3}_{23}  ~, \nn \\
\text{Gauge group 4~:} \qquad k_4  &=   n_{42} - n_{14}  ~.
\label{e:kafano139}
\end{array}
\eea

For this model let us choose:
\bea
n^3_{31} = - n_{42} = 1,\quad n^i_{jk}=n_{jk}=0 \; \text{otherwise}~.
\eea
The Kasteleyn matrix $K$ for this model can be calculated. The fundamental domain contains six nodes in total, hence $K$ is a $3\times 3$ matrix:
\bea
K =   \left(
\begin{array}{c|ccc}
& b_1 & b_2 & b_3\\
\hline
w_1 & z^{n_{42}} + z^{n_{14}} &  z^{n^2_{23}} & \frac{y}{x} z^{n^1_{31}} \\
w_2 & x z^{n^2_{31}} & z^{n^1_{12}} & z^{n^3_{23}} \\
w_3 & z^{n^1_{23}} & \frac{1}{y} z^{n^3_{31}} & z^{n^2_{12}} \end{array}
\right) ~. \label{e:kastfano139}
\eea
The permanent of this matrix is given by:
\bea
\perm~K &=& 
 x z^{(n^{2}_{12} + n^{2}_{23} + n^{2}_{31})} +  x^{-1} y z^{(n^{1}_{12} + n^{1}_{23} + n^{1}_{31})} +  y^{-1} z^{(n^{3}_{23} + n^{3}_{31} + n_{42})}\nn \\
&+&  y^{-1} z^{(n^3_{23} + n^{3}_{31} + n_{14})} +  z^{(n^{1}_{31} + n^{2}_{31} + n^{3}_{31})}+  z^{(n^{1}_{12} + n^{2}_{12} + n_{42})}\nn \\
&+&  z^{(n^{1}_{12} + n^{2}_{12} + n_{14})} +  z^{(n^{1}_{23} + n^{2}_{23} + n^{3}_{23})} \nn \\
&=& x + x^{-1} y + y^{-1} +  y^{-1} z + z + z^{-1} + 2\nn\\ 
&&  \text{(for $n^3_{31} = - n_{42} = 1,\quad n^i_{jk}=n_{jk}=0 \; ~ \text{otherwise}$)} ~.
\label{e:permKfano139}
\eea
The perfect matchings can be written in terms of the chiral fields as:
\bea 
&&   p_1 = \left\{X^2_{31}, X^2_{12},X^2_{23}\right\}, 
\;\; p_2 = \left\{X^1_{31}, X^1_{12},X^1_{23}\right\},
\;\; q_1 = \left\{X^3_{31}, X_{42},X^3_{23}\right\},\nn \\
&&   q_2 = \left\{X^3_{31}, X_{14},X^3_{23}\right\},
\;\; r_1 = \left\{X^1_{31}, X^2_{31},X^3_{31}\right\},
\;\; r_2 = \left\{X^1_{12}, X^2_{12},X_{42}\right\},\nn \\
&&   v_1 = \left\{X^1_{12}, X^2_{12},X_{14}\right\},
\;\; v_2 = \left\{X^1_{23}, X^2_{23},X^3_{23}\right\}\ .
\eea
The perfect matchings $v_1$ and $v_2$ correspond to the internal point in the toric diagram, whereas all the others correspond to external points.
The chiral fields can be written in terms of perfect matchings:
\bea
\begin{array}{llll}
X^2_{31} = p_1 r_1, \quad & X^2_{12} = p_1 r_2 v_1, \quad &X^2_{23} = p_1 v_2, \quad & X^1_{31} = p_2 r_1,\nn \\
X^1_{12} = p_2 r_2 v_1, \quad &  X^1_{23} = p_2 v_2, \quad &  X^3_{31} = q_1 q_2 r_1, \quad & X_{42} = q_1 r_2 ,\nn \\
X_{14} = q_2 v_1, \quad & X^3_{23} = q_1 q_2 v_2~.
\end{array}
\eea
This information can be collected in the perfect matching matrix:
\beq
P=\left(\begin{array} {c|cccccccc}
  \;& p_1 & p_2 & q_1 & q_2 & r_1 & r_2 & v_1 & v_2\\
  \hline 
  X^{2}_{31} & 1&0&0&0&1&0&0&0\\
  X^{2}_{12} & 1&0&0&0&0&1&1&0\\
  X^{2}_{23} & 1&0&0&0&0&0&0&1\\
  X^{1}_{31} & 0&1&0&0&1&0&0&0\\
  X^{1}_{12} & 0&1&0&0&0&1&1&0\\
  X^{1}_{23} & 0&1&0&0&0&0&0&1\\
  X^{3}_{31} & 0&0&1&1&1&0&0&0\\
  X_{42}     & 0&0&1&0&0&1&0&0\\
  X_{14}     & 0&0&0&1&0&0&1&0\\
  X^3_{23}   & 0&0&1&1&0&0&0&1\\
  \end{array}
\right).
\eeq
The kernel of the $P$ matrix is given by:
\be
Q_F =   \left(
\begin{array}{cccccccc}
1 & 1 & 0 &  1 & -1 &  0 & -1 & -1 \\
0 & 0 & 1 & -1 &  0 & -1 &  1 &  0 
\end{array}
\right)~.  \label{e:qffano139}
\ee
Hence, among the perfect matchings there are two relations:
\bea
p_1 + p_2 + q_2 - r_1 - v_1 - v_2 &=& 0~, \nn \\
q_1 - q_2 - r_2 + v_1 &=&  0~.
\label{e:relpmfano139}
\eea

\paragraph{The toric diagram.} Two methods of computing the toric diagram for this model are presented.
\begin{itemize}
\item {\bf The Kasteleyn matrix.} The coordinates of the toric diagram are collected in the columns of the following matrix:
\bea
G_K = \left(
\begin{array}{cccccccc}
   1 & -1 &  0 &  0 & 0 & 0 & 0 & 0 \\
   0 &  1 & -1 & -1 & 0 & 0 & 0 & 0 \\
   0 &  0 &  0 &  1 & 1 &-1 & 0 & 0 
\end{array}
\right)~.
\eea
The first row of this matrix contains the weights of the fundamental representation of $SU(2)$, which implies that the non-abelian part of the mesonic symmetry contains only one $SU(2)$ factor.

\item {\bf The charge matrices.}
Since the number of gauge groups of this model is $G = 4$, there are $G-2 =2$ baryonic symmetries coming from the D-terms. The charges of the perfect matchings under this baryonic symmetry can be collected in the rows of the $Q_D$ matrix:
\be
Q_D =   \left(
\begin{array}{cccccccc}
 0 &  0 & 0 &  0 &  1 &  1 & 0 & -2\\
 0 &  0 & 0 &  0 &  0 &  0 & 1 & -1 
 \end{array}
\right)~. \label{e:qdfano139}
\ee
The combination of the $Q_F$ and the $Q_D$ matrix gives the $Q_t$ matrix, which contains the charges that must be integrated over in order to compute the Hilbert series of the mesonic moduli space:
\be
Q_t = { \Blue Q_F \choose \Green Q_D \Black } =   \left( 
\begin{array}{cccccccc} \Blue
 1 & 1 & 0 &  1 & -1 &  0 & -1 & -1 \\
 0 & 0 & 1 & -1 &  0 & -1 &  1 &  0\\ \Green
 0 &  0 & 0 &  0 &  1 &  1 & 0 & -2\\
 0 &  0 & 0 &  0 &  0 &  0 & 1 & -1 \Black
\end{array}
\right) ~.
\label{e:qtfano139}
\ee
The null space of this matrix gives the $G_t$ matrix. By eliminating the first row the $G'_t$ matrix is obtained whose rows give the coordinates of the toric diagram:
\bea
G'_t = \left(
\begin{array}{cccccccc}
   1 & -1 &  0 &  0 & 0 & 0 & 0 & 0 \\
   0 &  1 & -1 & -1 & 0 & 0 & 0 & 0 \\
   0 &  0 &  0 &  1 & 1 &-1 & 0 & 0 
\end{array}
\right)= G_K~. \label{e:gtfano139}
\eea
The toric diagram constructed from (\ref{e:gtfano139}) is presented in Figure \ref{f:tdtoricfano139}:
\begin{figure}[ht]
\begin{center}
  \includegraphics[totalheight=3.0cm]{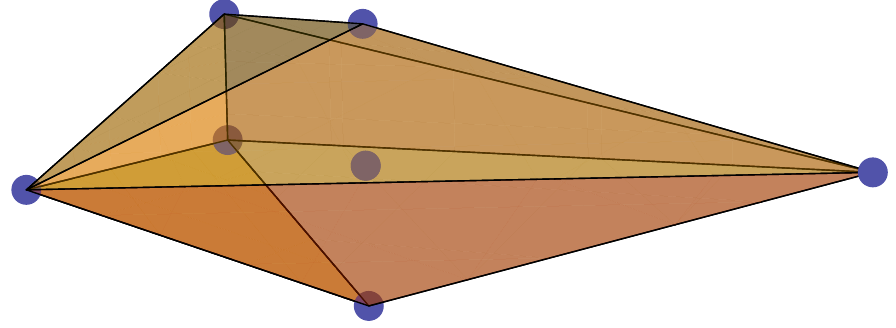}
 \caption{The toric diagram of $\cD_2$.}
  \label{f:tdtoricfano139}
\end{center}
\end{figure}
\end{itemize}

\paragraph{The baryonic charges.} The toric diagram of this model is characterized by 6 external points and, therefore, the total number of baryonic symmetries is 2. These symmetries shall be referred to as $U(1)_{B_1}$ and $U(1)_{B_2}$. The charges of the perfect matchings under these baryonic symmetries can be read off directly from the rows of the $Q_D$ matrix written in (\ref{e:qdfano139})

\paragraph{The global symmetry.} Since there is one pair of repeated columns in the $Q_t$ matrix (\ref{e:qtfano139}) the mesonic symmetry contains only one $SU(2)$ factor, under which the perfect matchings $p_1$ and $p_2$ transform as a doublet.  Since the mesonic symmetry has total rank 4, it can be identified with $SU(2) \times U(1)^3$, where one of the three abelian symmetries corresponds to the $R$-symmetry.  The perfect matchings $v_1, v_2$ correspond to the internal point of the toric diagram and both carry zero R-charge.

Since the mesonic symmetry has 3 abelian factors, in order to determine the correct R-charge of each perfect matching in the IR, the volume of $\cD_2$ must be minimised.  Let us assign the R-charge fugacity $s_1$ to the perfect matchings $p_1$ and $p_2$, and fugacities $s_2,s_3,s_4,s_5$ respectively to the perfect matchings $q_1, q_2, r_1$ and $r_2$. The charges of the perfect matchings that need to be integrated over are given in (\ref{e:qtfano139}). The Hilbert series of the mesonic moduli space is given by:
\bea
\gm (t_{\alpha}; \cD_2) &=& \oint \limits_{|z_1| =1} {\frac{\ud z_1}{2 \pi i z_1}} \oint \limits_{|z_2| =1} {\frac{\ud z_2}{2 \pi i z_2}} \oint \limits_{|b_1| =1} {\frac{\ud b_1}{2 \pi i b_1}} \oint \limits_{|b_2| =1} {\frac{\ud b_2}{2 \pi i b_2}} \frac{1}{\left(1-s_1 z_1\right)^2\left(1-s_2 z_2\right)}\nn \\
&&\times \frac{1}{\left(1-\frac{s_3 z_1}{z_2}\right)\left(1-\frac{s_4 b_1}{z_1}\right)\left(1-\frac{s_5 b_1}{z_2}\right)\left(1- \frac{ b_2 z_2}{z_1}\right)\left(1-\frac{1}{b^2_1 b_2 z_1}\right)}~.
\label{e:hsvolfano139}
\eea
Since there are 3 factors of $U(1)$ in the mesonic symmetry, this integral depends only on three combinations of $s_\alpha$'s. Defining:
\bea
t_1 = s^3_1 s_4 s_5, \quad t_2 = s^2_1 s_2 s^2_5, \quad t_3 = s^3_1 s_3 s^2_4~,
\eea
gives us
{\footnotesize
\bea
\gm (t_1, t_2, t_3; \cD_2) = \frac{P(t_1, t_2, t_3; \cD_2)} {\left(1-t_1\right)^2\left(1-t_2\right)^2\left(1-t_3\right)^2\left(1-\frac{t_2^3 t_3^2}{t^4_1}\right)\left(1-\frac{t^3_2 t^4_3}{t^6_1}\right)}~,
\eea}
where:
{\scriptsize
\bea
P(t_1, t_2, t_3; \cD_2) &=& 
1-2 {t_1}+{t_1}^2-2 {t_2}+4 {t_1} {t_2}-2 {t_1}^2 {t_2}+{t_2}^2-2 {t_1} {t_2}^2+{t_1}^2 {t_2}^2-2
{t_3}+4 {t_1} {t_3}-2 {t_1}^2 {t_3}+4 {t_2} {t_3}-8 {t_1} {t_2} {t_3}+ \nn \\
& & 4 {t_1}^2 {t_2} {t_3}-2
{t_2}^2 {t_3}+4 {t_1} {t_2}^2 {t_3}-2 {t_1}^2 {t_2}^2 {t_3}+{t_3}^2-2 {t_1} {t_3}^2+{t_1}^2 {t_3}^2-2
{t_2} {t_3}^2+4 {t_1} {t_2} {t_3}^2-2 {t_1}^2 {t_2} {t_3}^2+
\nn \\ & &
{t_2}^2 {t_3}^2-2 {t_1} {t_2}^2 {t_3}^2+{t_1}^2
{t_2}^2 {t_3}^2-\frac{{t_2}^3 {t_3}^2}{{t_1}^4}+\frac{2 {t_2}^3 {t_3}^2}{{t_1}^3}-\frac{{t_2}^3 {t_3}^2}{{t_1}^2}+
\frac{2
{t_2}^4 {t_3}^2}{{t_1}^4}-\frac{4 {t_2}^4 {t_3}^2}{{t_1}^3}+\frac{2 {t_2}^4 {t_3}^2}{{t_1}^2}-\frac{{t_2}^5
{t_3}^2}{{t_1}^4}+ \nn \\ & &
\frac{2 {t_2}^5 {t_3}^2}{{t_1}^3}-\frac{{t_2}^5 {t_3}^2}{{t_1}^2}+\frac{2 {t_2}^3 {t_3}^3}{{t_1}^4}-\frac{4
{t_2}^3 {t_3}^3}{{t_1}^3}+\frac{2 {t_2}^3 {t_3}^3}{{t_1}^2}-\frac{4 {t_2}^4 {t_3}^3}{{t_1}^4}+\frac{8 {t_2}^4
{t_3}^3}{{t_1}^3}-\frac{4 {t_2}^4 {t_3}^3}{{t_1}^2}
+\frac{2 {t_2}^5 {t_3}^3}{{t_1}^4}-\frac{4 {t_2}^5 {t_3}^3}{{t_1}^3}
\nn \\ & &
+\frac{2
{t_2}^5 {t_3}^3}{{t_1}^2}-\frac{{t_2}^3 {t_3}^4}{{t_1}^6}+\frac{2 {t_2}^3 {t_3}^4}{{t_1}^5}-\frac{2 {t_2}^3
{t_3}^4}{{t_1}^4}+\frac{2 {t_2}^3 {t_3}^4}{{t_1}^3}-\frac{{t_2}^3 {t_3}^4}{{t_1}^2}+\frac{2 {t_2}^4 {t_3}^4}{{t_1}^6}-\frac{4
{t_2}^4 {t_3}^4}{{t_1}^5}+\frac{4 {t_2}^4 {t_3}^4}{{t_1}^4}-\frac{4 {t_2}^4 {t_3}^4}{{t_1}^3}
\nn \\ & &
+ \frac{2 {t_2}^4
{t_3}^4}{{t_1}^2}-\frac{{t_2}^5 {t_3}^4}{{t_1}^6}+\frac{2 {t_2}^5 {t_3}^4}{{t_1}^5}-\frac{2 {t_2}^5 {t_3}^4}{{t_1}^4}+\frac{2
{t_2}^5 {t_3}^4}{{t_1}^3}-\frac{{t_2}^5 {t_3}^4}{{t_1}^2}+\frac{2 {t_2}^3 {t_3}^5}{{t_1}^6}-\frac{4 {t_2}^3
{t_3}^5}{{t_1}^5}+\frac{2 {t_2}^3 {t_3}^5}{{t_1}^4}-\frac{4 {t_2}^4 {t_3}^5}{{t_1}^6}
\nn \\ & &
+\frac{8 {t_2}^4 {t_3}^5}{{t_1}^5}-\frac{4
{t_2}^4 {t_3}^5}{{t_1}^4}+\frac{2 {t_2}^5 {t_3}^5}{{t_1}^6}-\frac{4 {t_2}^5 {t_3}^5}{{t_1}^5}+\frac{2 {t_2}^5
{t_3}^5}{{t_1}^4}-\frac{{t_2}^3 {t_3}^6}{{t_1}^6}+\frac{2 {t_2}^3 {t_3}^6}{{t_1}^5}-\frac{{t_2}^3 {t_3}^6}{{t_1}^4}+\frac{2
{t_2}^4 {t_3}^6}{{t_1}^6}-\frac{4 {t_2}^4 {t_3}^6}{{t_1}^5}
\nn \\ & &
+\frac{2 {t_2}^4 {t_3}^6}{{t_1}^4}-\frac{{t_2}^5
{t_3}^6}{{t_1}^6}+\frac{2 {t_2}^5 {t_3}^6}{{t_1}^5}-\frac{{t_2}^5 {t_3}^6}{{t_1}^4}+\frac{{t_2}^6 {t_3}^6}{{t_1}^{10}}-\frac{2
{t_2}^6 {t_3}^6}{{t_1}^9}+\frac{{t_2}^6 {t_3}^6}{{t_1}^8}-\frac{2 {t_2}^7 {t_3}^6}{{t_1}^{10}}+\frac{4 {t_2}^7
{t_3}^6}{{t_1}^9}-\frac{2 {t_2}^7 {t_3}^6}{{t_1}^8}
\nn \\ & &
+\frac{{t_2}^8 {t_3}^6}{{t_1}^{10}}-\frac{2 {t_2}^8 {t_3}^6}{{t_1}^9}+\frac{{t_2}^8
{t_3}^6}{{t_1}^8}-\frac{2 {t_2}^6 {t_3}^7}{{t_1}^{10}}+\frac{4 {t_2}^6 {t_3}^7}{{t_1}^9}-\frac{2 {t_2}^6 {t_3}^7}{{t_1}^8}+\frac{4
{t_2}^7 {t_3}^7}{{t_1}^{10}}-\frac{8 {t_2}^7 {t_3}^7}{{t_1}^9}+\frac{4 {t_2}^7 {t_3}^7}{{t_1}^8}-\frac{2 {t_2}^8
{t_3}^7}{{t_1}^{10}}
\nn \\ & &
+\frac{4 {t_2}^8 {t_3}^7}{{t_1}^9}-\frac{2 {t_2}^8 {t_3}^7}{{t_1}^8}+\frac{{t_2}^6 {t_3}^8}{{t_1}^{10}}-\frac{2
{t_2}^6 {t_3}^8}{{t_1}^9}+\frac{{t_2}^6 {t_3}^8}{{t_1}^8}-\frac{2 {t_2}^7 {t_3}^8}{{t_1}^{10}}+\frac{4 {t_2}^7
{t_3}^8}{{t_1}^9}-\frac{2 {t_2}^7 {t_3}^8}{{t_1}^8}+\frac{{t_2}^8 {t_3}^8}{{t_1}^{10}}-\frac{2 {t_2}^8 {t_3}^8}{{t_1}^9}
\nn \\ & &+\frac{{t_2}^8
{t_3}^8}{{t_1}^8}
\eea}
Since the R-charge fugacity of the superpotential is $t_3 t_2/ t_1$, the requirement that this has R-charge 2 implies that: 
\bea
R_3 + R_2 - R_1 = 2~, 
\label{e:cycondfano139}
\eea
where $R_i$ is the R-charge corresponding to the fugacity $t_i$. The volume of $\cD_2$ is given by:
{\footnotesize
\bea
\lim_{\mu \rightarrow 0} \mu^4 \gm (e^{-\mu R_1}, e^{-\mu (2+ R_1 - R_3)}, e^{-\mu R_3}; \cD_2)= \frac{p\left(R_1, R_3; \cD_2 \right)}{R^2_1 R^2_3\left(2+ R_1 - R_3\right)^2\left(6 - 3R_1 + R_3\right)\left(6 - R_1 - R_3\right)}~,\nn \\
\eea}
where:
\bea
p \left( R_1, R_3; \cD_2 \right) &=& 216 R_1 + 180 R_1^2 + 18 R_1^3 - 9 R_1^4 + 216 R_3 - 72 R_1 R_3 -  90 R_1^2 R_3 - \nn \\
&&108 R_3^2 +   78 R_1 R_3^2 + 6 R_1^2 R_3^2 - 6 R_3^3 -  8 R_1 R_3^3 + 3 R_3^4~.
\label{e:numvolfano139}
\eea
This function has a minimum at:
\bea
R_1 \approx 1.844, \qquad R_2 \approx 1.790, \qquad R_3 \approx 2.054~.
\eea
The R-charge of the external perfect matching corresponding to the divisor $D_\alpha$ is given by:
\bea
\lim_{\mu\rightarrow0}\frac{1}{\mu} \left[ \frac{g(D_\alpha; e^{- \mu R_1}, e^{- \mu R_2 }, e^{- \mu R_3 }; \cD_2) }{\gm(e^{-\mu R_1}, e^{- \mu R_2 }, e^{- \mu R_3 };\cD_2)}- 1 \right]~,
\eea
where $g(D_\alpha; e^{- \mu R_1}, e^{- \mu R_2 }, e^{- \mu R_3 }; \cD_2)$ is the Molien-Weyl integral with the insertion of the inverse of the weight corresponding to the divisor $D_\alpha$. The results are shown in Table \ref{t:chargefano139}. The assignment of charges under the remaining abelian symmetries can be done by requiring that the superpotential is not charged under them and that the charge vectors are linearly independent. The assignments are shown in Table \ref{t:chargefano139}.
\begin{table}[h!]
 \begin{center}  
  \begin{tabular}{|c||c|c|c|c|c|c|c|}
  \hline
  \;& $SU(2)_1$&$U(1)_1$&$U(1)_2$&$U(1)_R$&$U(1)_{B_1}$&$U(1)_{B_2}$&fugacity\\
  \hline\hline  
   
  $p_1$&$  1$&$  0$&$ 0$&$0.441$&$ 0$&$ 0$ & $s_1 x$\\
  \hline
  
  $p_2$&$ -1$&$  0$&$ 0$&$0.441$&$ 0$&$ 0$ & $s_1 /x$\\
  \hline  
  
  $q_1$&$  0$&$  1$&$ 0$&$0.295$&$ 0$&$ 0$ & $s_2 q_1$\\
  \hline
  
  $q_2$&$  0$&$ 0$&$ 1$&$0.301$&$ 0$&$ 0$ & $s_3 q_2 $\\
  \hline
   
  $r_1$&$  0$&$  -1$&$0$&$0.215$&$ 1$&$ 0$ & $s_4  b_1/q_1$\\
  \hline
      
  $r_2$&$  0$&$  0$&$-1$&$0.306$&$ 1$&$ 0$ & $s_5 b_1/ q_2 $\\
  \hline
  
  $v_1$&$  0$&$  0$&$ 0$&$    0$&$ 0$&$ 1$ & $b_2  $\\
  \hline

  $v_2$&$  0$&$  0$&$ 0$&$    0$&$-2$&$-1$ & $1/(b^2_1 b_2) $\\
  \hline
 \end{tabular}
  \end{center}
\caption{Charges of the perfect matchings under the global symmetry of the $\cD_2$ model. Here $s_i$ are the fugacities of the R-charges, $x$ is the weight of the $SU(2)$ symmetry, $q_1, q_2, b_1$ and $b_2$ are, respectively, the charges under the mesonic abelian symmetries $U(1)_1$, $U(1)_2$ and under the two baryonic $U(1)_{B_1}$ and $U(1)_{B_2}$.}
\label{t:chargefano139}
\end{table}

\begin{table}[h]
 \begin{center}  
  \begin{tabular}{|c||c|}
  \hline
  \; Quiver fields & R-charge\\
  \hline  \hline 
  $X^1_{12}, X^2_{12}$ &  0.747\\
  \hline
  $X^1_{23}, X^2_{23}$ &  0.441\\
  \hline
  $X^3_{23}$ &  0.597\\
  \hline
  $X^1_{31}, X^2_{31}$ &  0.656\\
  \hline
  $X^3_{31}$ &  0.812\\
  \hline
  $X_{42}$ &  0.602\\
  \hline
  $X_{14}$ &  0.301\\
  \hline
  \end{tabular}
  \end{center}
\caption{R-charges of the quiver fields for the $\cD_2$ model.}
\label{t:Rgenchfano139}
\end{table}

\begin{figure}[ht]
\begin{center}
\includegraphics[totalheight=8cm]{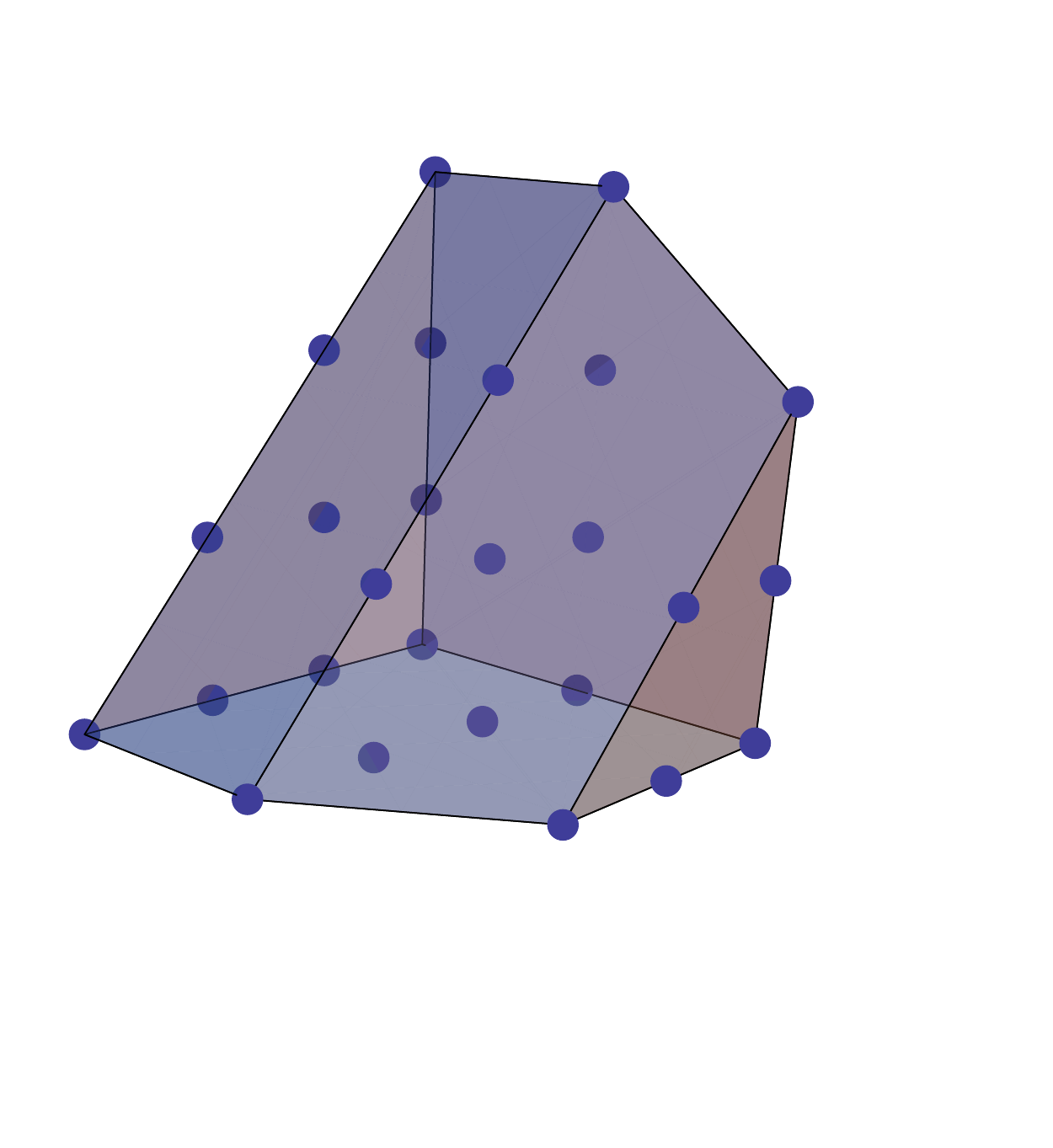}
\vskip -2cm
\caption{The lattice of generators of the $\cD_2$ theory.}
  \label{f:lat139}
  \end{center}
\end{figure}

\paragraph{The Hilbert series.} The Hilbert series of the Master space can be obtained by integrating the space of perfect matchings Hilbert series over the fugacities $z_1$ and $z_2$:
{\small
\bea
g^{\firr{}}  (s_{\alpha}, x, q_1, q_2, b_1, b_2; \cD_2) &=& \oint \limits_{|z_1| =1} {\frac{\ud z_1}{2 \pi i z_1}} \oint \limits_{|z_2| =1} {\frac{\ud z_2}{2 \pi i z_2}}  
\frac{1}{ \left(1-\frac{1}{b_1^2 b_2 z_1}\right) \left(1-\frac{b_2 z_2}{z_1}\right) \left(1-\frac{q_2 s_3 z_1}{z_2}\right) } \nn \\
& \times & \frac{1}{ \left(1-q_1 s_2
   z_2\right) \left(1-\frac{s_1 z_1}{x}\right) \left(1-s_1 x z_1\right) \left(1-\frac{b_1 s_4}{q_1 z_1}\right) \left(1-\frac{b_1 s_5}{q_2
   z_2}\right)} \nn \\
&=& \frac{\cP(s_{\alpha}, x, q_1, q_2, b_1, b_2; \cD_2)}{
\left(1-b_2 q_2 s_3\right)
 \left(1 -\frac{q_1 q_2 s_2 s_3}{b_1^2 b_2}\right)
 \left(1-b_1 q_2 s_2 s_3 s_4\right)
 \left(1-\frac{b_1 q_1 s_2 s_5}{q_2}\right)
   \left(1-\frac{s_1}{b_1^2 b_2 x}\right)} \nn \\ & \times &\frac{1}{
 \left(1-\frac{s_1 x}{b_1^2 b_2}\right)
 \left(1- \frac{b_1 s_1 s_4}{q_1 x}\right)
 \left(1-\frac{b_1 s_1 s_4 x}{q_1}\right)
 \left(1-\frac{b_1 b_2 s_1 s_5}{q_2 x}\right)
 \left(1-\frac{b_1 b_2 s_1 s_5 x}{q_2}\right)} \qquad \qquad
\label{e:HSmasterfano139}
\eea}
where $\cP(s_{\alpha}, x, q_1, q_2, b_1, b_2; \cD_2)$ is a polynomial that is not reported here. Integrating over the two baryonic fugacities $b_1$ and $b_2$ gives the Hilbert series of the mesonic moduli space:
\bea
\gm (s_{\alpha}, x, q_1, q_2; \cD_2) &=& \oint \limits_{|b_1| =1} {\frac{\ud b_1}{2 \pi i b_1}} \oint \limits_{|b_2| =1} {\frac{\ud b_2}{2 \pi i b_2}} g^{\firr{}}  (s_{\alpha}, x, q_1, q_2, b_1, b_2; \cD_2) \nn \\
&=& \frac{P(s_{\alpha}, x, q_1, q_2; \cD_2)}{
\left(1- q_1 q_2^4 s_2^3 s_3^4 s_4^2\right) 
\left(1-q_1^3 s_2^3 s_3^2 s_5^2\right)
 \left(1-\frac{q_2 s_1^3 s_3 s_4^2}{ q_1^2 x^3}\right)
 \left(1- \frac{q_2 s_1^3 s_3 s_4^2 x^3}{q_1^2}\right) } \nn \\
&\times& \frac{1}{
\left(1-\frac{s_1^3 s_4 s_5}{q_1 q_2 x^3}\right)
 \left(1-\frac{s_1^3 s_4 s_5 x^3}{q_1 q_2}\right)
 \left(1-\frac{q_1 s_1^2 s_2 s_5^2}{q_2^2 x^2}\right) 
\left(1-\frac{q_1 s_1^2 s_2 s_5^2 x^2}{q_2^2}\right)} \quad \quad \quad \qquad~
\label{e:HSmesonicfano139}
\eea
where $P(s_{\alpha}, x, q_1, q_2; \cD_2)$ is a polynomial that is not reported here.
The plethystic logarithm of (\ref{e:HSmesonicfano139}) is
\bea
\PL[\gm(t_1,t_2,t_3, x, q_1, q_2; \cD_2)] &=&
 [3] \left( \frac{t_1}{q_1 q_2}+\frac{q_2 t_3}{q_1^2}  \right) +
 [2]\left( \frac{q_2^2 t_2 t_3^2}{q_1 t_1^2}+\frac{q_1 t_2}{q_2^2}+\frac{t_2 t_3}{t_1}  \right)
\nn \\ &+&
 [1]\left(
\frac{q_2^3 t_2^2 t_3^3}{t_1^4}+\frac{q_1 q_2 t_2^2 t_3^2}{t_1^3}+\frac{q_1^2 t_2^2 t_3}{q_2 t_1^2}
\right) 
+ \frac{q_1 q_2^4 t_2^3 t_3^4}{t_1^6} \nn \\
&+& \frac{q_1^2 q_2^2 t_2^3 t_3^3}{t_1^5}+\frac{q_1^3 t_2^3 t_3^2}{t_1^4} - O(t^2_1)O(t^2_2)O(t^2_3)
\eea
The generators of the mesonic moduli space are
\bea
\begin{array}{llll}
p_i p_j p_k q_2 r^2_1 v_1 v_2, \quad & p_i p_j p_k r_1 r_2 v_1 v_2, \quad& p_i p_j q_1 q^2_2 r^2_1 v_1 v_2, \quad & p_i p_j q_1 q_2 r_1 r_2 v_1 v_2, \nn \\
p_i p_j q_1 r^2_2 v_1 v_2, \quad & p_i q^2_1 q^3_2 r^2_1 v_1 v_2, \quad & p_i q^2_1 q^2_2 r_1 r_2 v_1 v_2, \quad & p_i q^2_1 q_2 r^2_2 v_1 v_2, \nn \\
q^3_1 q^4_2 r^2_1 v_1 v_2, \quad & q^3_1 q^3_2 r_1 r_2 v_1 v_2, \quad &q^3_1 q^2_2 r^2_2 v_1 v_2~. &
\label{e:genefano139}
\end{array}
\eea
with $i, j, k=1,2$.
In Table \ref{t:Rgenfano139}, the R-charges of the generators of the mesonic moduli space are presented. The lattice of generators is give in \fref{f:lat139}.
\begin{table}[h]
 \begin{center}  
  \begin{tabular}{|c||c|}
  \hline
  \; Generators &$U(1)_R$\\
  \hline  \hline 
  $p_i p_j p_k q_2 r^2_1 v_1 v_2$ & 2.055 \\
  \hline
  $p_i p_j p_k r_1 r_2 v_1 v_2$ & 1.844 \\
  \hline
  $p_i p_j q_1 q^2_2 r^2_1 v_1 v_2$ & 2.211 \\
  \hline
  $p_i p_j q_1 q_2 r_1 r_2 v_1 v_2$ & 2 \\
  \hline
  $p_i p_j q_1 r^2_2 v_1 v_2$ & 1.789 \\
  \hline
  $p_i q^2_1 q^3_2 r^2_1 v_1 v_2$ & 2.367 \\
  \hline
  $p_i q^2_1 q^2_2 r_1 r_2 v_1 v_2$ & 2.156 \\
  \hline
  $p_i q^2_1 q_2 r^2_2 v_1 v_2$ & 1.945 \\
  \hline
  $q^3_1 q^4_2 r^2_1 v_1 v_2$ & 2.523 \\
  \hline
  $q^3_1 q^3_2 r_1 r_2 v_1 v_2$ & 2.312 \\
  \hline
  $q^3_1 q^2_2 r^2_2 v_1 v_2$ & 2.101 \\
  \hline

  \end{tabular}
  \end{center}
\caption{R-charges of the generators of the mesonic moduli space for the $\cD_2$ model.}
\label{t:Rgenfano139}
\end{table}

\section{$\cE_1 $(Toric Fano 218): $dP_2$ bundle over $\BP^1$}
\begin{figure}[ht]
\begin{center}
\hskip -1cm
\includegraphics[totalheight=4.5cm]{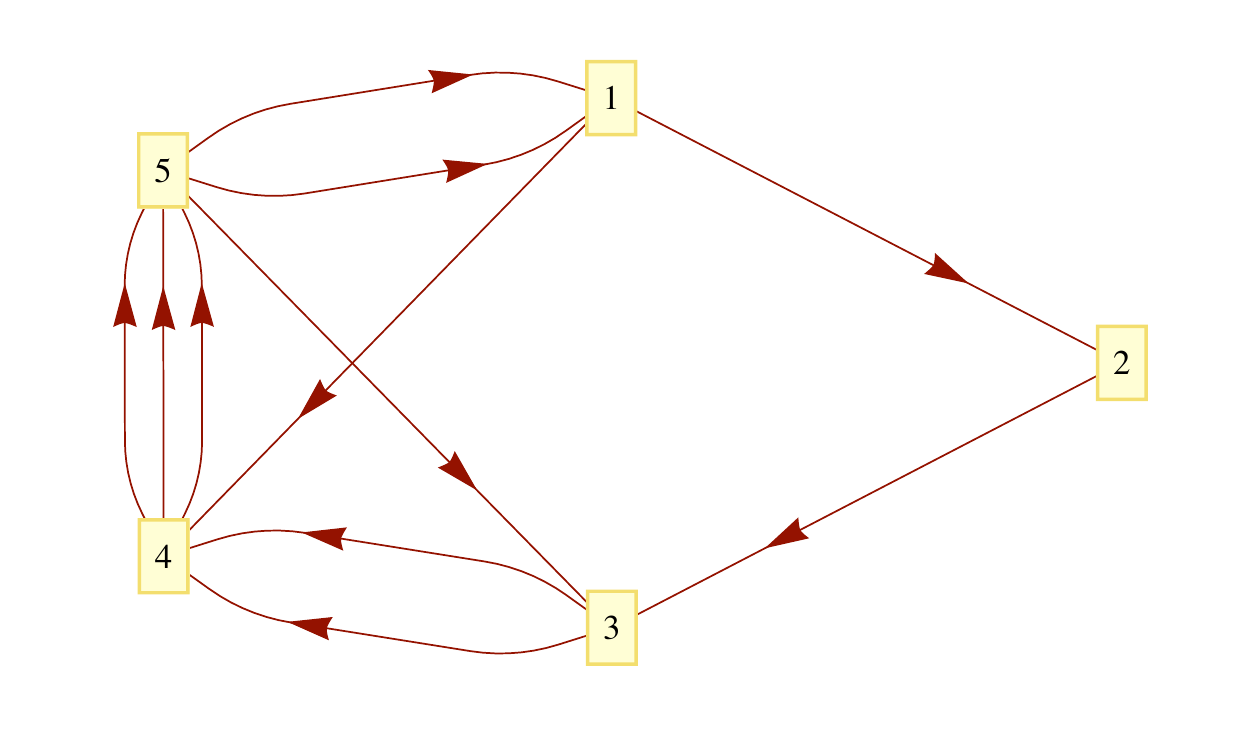}
\hskip 1cm
\includegraphics[totalheight=5.5cm]{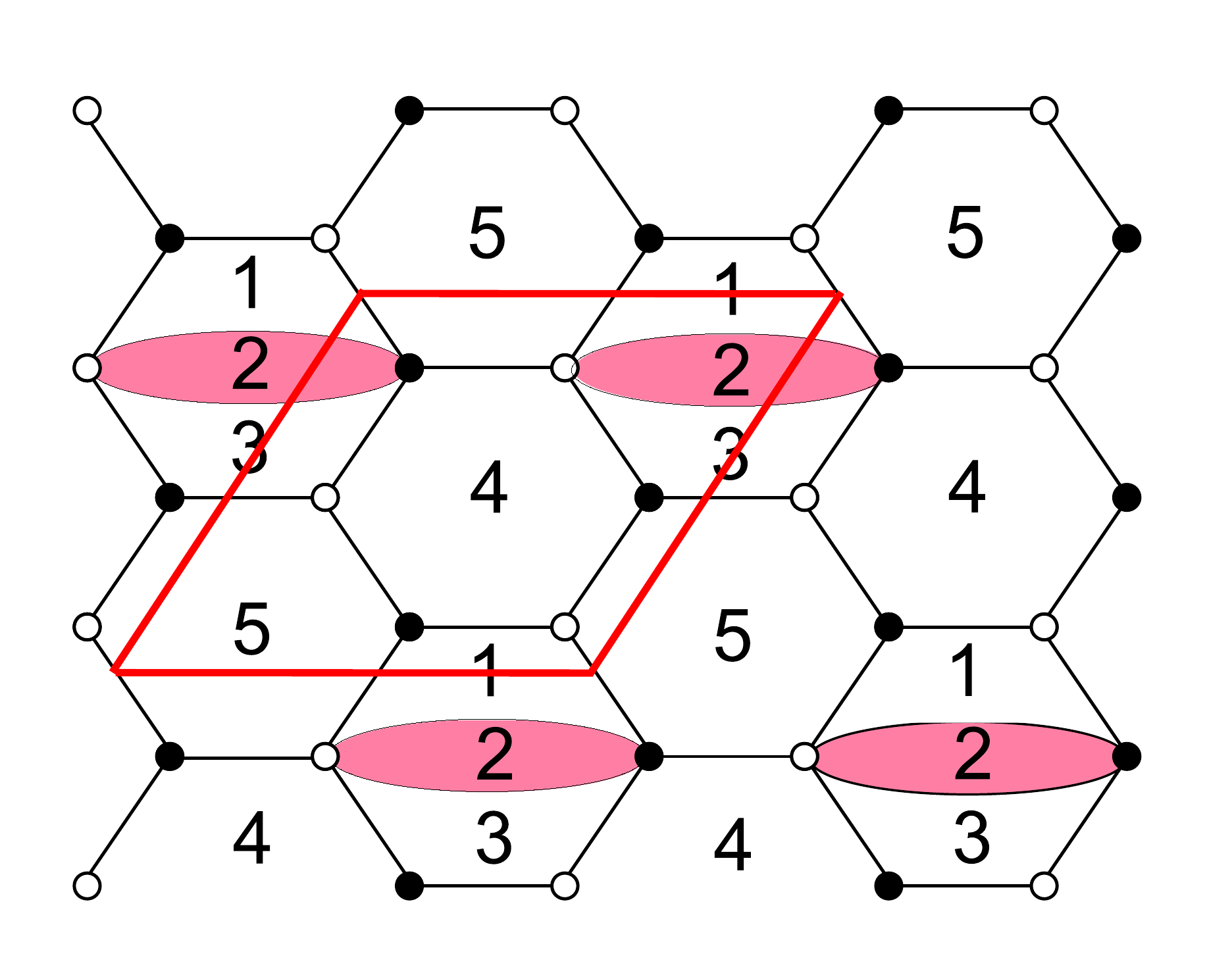}
 \caption{(i) Quiver diagram of the $\cE_1$  model. (ii) Tiling of the $\cE_1$ model.}
  \label{f:tqfano218}
\end{center}
\end{figure}

This theory has 5 gauge groups and chiral superfields $X^i_{45}$ (with $i=1,2,3$), $X^j_{51}$, $X^j_{34}$ (with $j=1,2$), $X_{14}$, $X_{12}$, $X_{53}$ and $X_{23}$.
The tiling and quiver of this theory are shown in Figure \ref{f:tqfano218}.  The superpotential can be read off from the tiling:
\bea
W = \tr \left[ \epsilon_{ij} \left( X^i_{51}X_{12}X_{23}X^j_{34} X^3_{45} +  X_{53}X^i_{34}X^j_{45} + X_{14} X^i_{45}X^j_{51} \right) \right]~.
\label{e:spfano218}
\eea
Let us choose the CS levels to be $\vec{k}=(1,-1,0,-1,1)$

\comment{
\begin{figure}[ht]
\begin{center}
   \includegraphics[totalheight=6cm]{fde1.pdf}
 \caption{The fundamental domain of the tiling for the $\cE_1$ model.}
  \label{f:fdtoricfano218}
\end{center}
\end{figure}
}

\paragraph{The Kasteleyn matrix.} The CS levels can be parametrized in terms of integers $n^i_{jk}$ and $n_{jk}$ as follows:
\bea
\begin{array}{ll}
\text{Gauge group 1~:} \qquad k_1 &=   n_{12} + n_{14} - n^{1}_{51} - n^{2}_{51} ~,  \nn \\
\text{Gauge group 2~:} \qquad k_2 &=   n_{23} - n_{12} ~,  \nn \\
\text{Gauge group 3~:} \qquad k_3 &=   n^{1}_{34} + n^{2}_{34} - n_{23} - n_{53} ~,  \nn \\
\text{Gauge group 4~:} \qquad k_4 &=   n^{1}_{45} + n^{2}_{45} + n^{3}_{45} - n^{1}_{34} - n^{2}_{34} - n_{14} ~,  \nn \\
\text{Gauge group 5~:} \qquad k_5 &=   n_{53} + n^{1}_{51} + n^{2}_{51} - n^{1}_{45} - n^{2}_{45} - n^{3}_{45} ~.
\label{e:kafano218}
\end{array}
\eea  
Let us choose
\bea
n_{12}=-n^3_{45}= 1,\qquad n^i_{jk}=n_{jk}=0 \;\text{otherwise}~.
\eea
The fundamental domain contains three pairs of black and white nodes and, therefore, the Kasteleyn matrix is a $3\times 3$ matrix:
\bea
K =   \left(
\begin{array}{c|ccc}
& b_1 & b_2 & b_3\\
\hline
w_1 & z^{n_{14}} & z^{n^1_{45}} & \frac{y}{x} z^{n^2_{51}} \\
w_2 & x z^{n^1_{51}} & z^{n^2_{34}} & z^{n^3_{45}} + y z^{n_{12}} + y z^{n_{23}} \\
w_3 & z^{n^2_{45}} & \frac{1}{y} z^{n_{53}} & z^{n^1_{34}} \end{array}
\right) ~.
\label{e:kastfano218}
\eea
The permanent of the Kasteleyn matrix is given by:
\bea
\mathrm{perm}(K) &=&  x z^{(n^1_{34} + n^1_{45} + n^1_{51})} +  x^{-1} y z^{(n^2_{34} + n^2_{45} + n^2_{51})} +  y z^{(n^1_{45} + n^2_{45} + n_{23})}\nn \\
&+&   y z^{(n^1_{45} + n^2_{45} + n_{12})} +  y^{-1} z^{(n^3_{45} + n_{53} + n_{14})} + z^{(n^1_{45} + n^2_{45} + n^3_{45})}\nn \\
&+& z^{(n_{53} + n_{14} + n_{12})} + z^{(n_{53} + n_{14} + n_{23})} + z^{(n^1_{51} + n^2_{51} + n_{53})} + z^{(n^1_{34} + n^2_{34} + n_{14})}\nn \\
&=&  x+ x^{-1} y + y + y z + y^{-1} z^{-1} + z^{-1} + z + 3\nn \\
&& \; \text{(for $n_{12}=-n^3_{45}= 1,\qquad n^i_{jk}=n_{jk}=0 \; \text{otherwise}$)}.
\label{e:charpolyfano218}
\eea
The perfect matchings can be written in terms of the chiral fields as:
\bea
\begin{array}{lll} 
p_1 = \left\{X^1_{51}, X^1_{34}, X^1_{45}\right\}, \;\; & p_2 = \left\{X^2_{51}, X^2_{34}, X^2_{45}\right\}, \;\; & u_1 = \left\{X^1_{45}, X^2_{45}, X_{23}\right\}, \nn \\  
q_1 = \left\{X^1_{45}, X^2_{45}, X_{12}\right\}, \;\; & r_1 = \left\{X^3_{45},X_{53}, X_{14}\right\}, \;\; & q_2 = \left\{X^1_{45}, X^2_{45}, X^3_{45}\right\}, \nn \\  
r_2 = \left\{X_{53}, X_{14}, X_{12}\right\}, \;\; & v_1 = \left\{X_{53}, X_{14}, X_{23}\right\}, \;\; & v_2 = \left\{X^1_{51}, X^2_{51}, X_{53}\right\}, \nn \\  
v_3 = \left\{X^1_{34}, X^2_{34}, X_{14}\right\}\ .
\end{array}
\eea
In turn, the chiral fields can be written as products of perfect matchings as
\bea
\begin{array}{lll} 
X^1_{51} = p_1 v_2 , \quad & X^2_{51} = p_2 v_2 , \quad & X^1_{34} = p_1 v_3~, \nn \\
X^2_{34} = p_2 v_3 , \quad & X^1_{45} = p_1 u_1 q_1 q_2  , \quad & X^2_{45} = p_2 u_1 q_1 q_2~, \nn \\
X_{53} = r_1 r_2 v_1 v_2, \quad & X_{14} = r_1 r_2 v_1 v_3 ,   \quad & X_{23} = u_1 v_1~, \nn \\
X_{12} = q_1 r_2,   \quad &  X^3_{45} = r_1 q_2~.
\end{array}
\eea
These pieces of information can be collected in the perfect matching $P$ matrix:
\beq
P=\left(\begin{array} {c|cccccccccc}
  \;& p_1 & p_2 & u_1 & q_1 & r_1 & q_2 & r_2 & v_1 & v_2 & v_3\\
  \hline 
  X^{1}_{51}& 1&0&0&0&0&0&0&0&1&0\\
  X^{2}_{51}& 0&1&0&0&0&0&0&0&1&0\\
  X^{1}_{34}& 1&0&0&0&0&0&0&0&0&1\\
  X^{2}_{34}& 0&1&0&0&0&0&0&0&0&1\\
  X^{1}_{45}& 1&0&1&1&0&1&0&0&0&0\\
  X^{2}_{45}& 0&1&1&1&0&1&0&0&0&0\\
  X_{53}    & 0&0&0&0&1&0&1&1&1&0\\
  X_{14}    & 0&0&0&0&1&0&1&1&0&1\\
  X_{23}    & 0&0&1&0&0&0&0&1&0&0\\
  X_{12}    & 0&0&0&1&0&0&1&0&0&0\\
  X^{3}_{45}& 0&0&0&0&1&1&0&0&0&0
  \end{array}
\right)~.
\label{e:pmatrifano218}
\eeq

\paragraph{The toric diagram.} The toric diagram for this model is constructed using two different methods.
\begin{itemize}
\item {\bf The Kasteleyn matrix.} 
The powers of $x,y$ and $z$ in each term of \eref{e:charpolyfano218} give the coordinates of the toric diagram. They are collected in the columns of the following matrix:
\bea
G_K = \left(
\begin{array}{cccccccccc}
  1 & -1 & 0 & 0 &  0 &  0 & 0 & 0 & 0 & 0 \\
  0 &  1 & 1 & 1 & -1 &  0 & 0 & 0 & 0 & 0 \\
  0 &  0 & 0 & 1 & -1 & -1 & 1 & 0 & 0 & 0 
\end{array}
\right)~.
\label{e:Gkfano218}
\eea
Since the first row contains the weights of the fundamental representation of $SU(2)$, the mesonic symmetry contains only one $SU(2)$ factor as its non-abelian symmetry.
The kernel of the $P$ matrix, called $Q_F$, encodes in its rows the charges of the perfect matchings under the F-terms:
\bea
Q_F =   \left(
\begin{array}{cccccccccc}
1 & 1 & 0 &  0 & 1 & -1 &  0 &  0 & -1 & -1\\
0 & 0 & 1 & -1 & 0 &  0 &  1 & -1 &  0 &  0\\
0 & 0 & 0 &  1 & 1 & -1 & -1 &  0 &  0 &  0
\end{array}
\right)~.  
\label{e:qffano218}
\eea
Thus, among the perfect matchings there are 3 relations:
\bea
 p_1 + p_2 + r_1 - q_2 - v_2 - v_3 &=& 0~,\nn \\
 u_1 - q_1 + r_2 - v_1 &=& 0~, \nn \\
 q_1 + r_1 - q_2 - r_2 &=& 0~.
\label{e:relpmfano218}
\eea

\item {\bf The charge matrices.}
Since this model has 5 gauge groups, there are $5-2 = 3$ baryonic symmetries coming from the D-terms. The charges of the perfect matchings under these three symmetries can be written in the rows of the $Q_D$ matrix:
\be
Q_D =   \left(
\begin{array}{cccccccccc}
0 & 0 & 0 & 0 & 0 & 1 & 1 & 0 & 0 & -2\\
0 & 0 & 0 & 0 & 0 & 0 & 0 & 1 & 0 & -1\\
0 & 0 & 0 & 0 & 0 & 0 & 0 & 0 & 1 & -1
\end{array}
\right). 
\label{e:qdfano218}
\ee
The total charge matrix $Q_t$ which is  the combination of $Q_F$ and $Q_D$ can be written as:
\be
Q_t = { \Blue Q_F \choose \Green Q_D \Black } =   \left( 
\begin{array}{cccccccccc} \Blue
1 & 1 & 0 &  0 & 1 & -1 &  0 &  0 & -1 & -1\\
0 & 0 & 1 & -1 & 0 &  0 &  1 & -1 &  0 &  0\\
0 & 0 & 0 &  1 & 1 & -1 & -1 &  0 &  0 &  0\\ \Green
0 & 0 & 0 & 0 & 0 & 1 & 1 & 0 & 0 & -2\\
0 & 0 & 0 & 0 & 0 & 0 & 0 & 1 & 0 & -1\\
0 & 0 & 0 & 0 & 0 & 0 & 0 & 0 & 1 & -1\Black
\end{array}
\right) ~.
\label{e:qtfano218}
\ee
The kernel of this matrix gives the $G_t$ matrix. After eliminating the first row, the $G'_t$ matrix is obtained. The coordinates of the toric diagram are given in the columns of:
\bea
G'_t = \left(
\begin{array}{cccccccccc}
  1 & -1 & 0 & 0 &  0 &  0 & 0 & 0 & 0 & 0 \\
  0 &  1 & 1 & 1 & -1 &  0 & 0 & 0 & 0 & 0 \\
  0 &  0 & 0 & 1 & -1 & -1 & 1 & 0 & 0 & 0 
\end{array}
\right) = G_K~. \label{e:toricdiafano218}
\eea
The toric diagram constructed from (\ref{e:toricdiafano218}) is presented in Figure \ref{f:tdtoricfano218}.
Note that the toric diagrams constructed from the $G_K$ and from the $G'_t$ matrix are the same.
\begin{figure}[ht]
\begin{center}
  \includegraphics[totalheight=2.2cm]{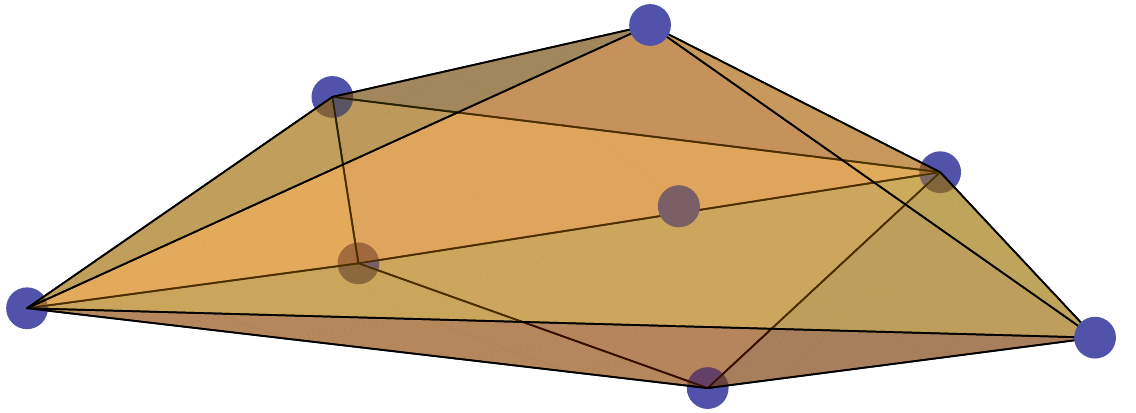}
 \caption{The toric diagram of $\cE_1$.}
  \label{f:tdtoricfano218}
\end{center}
\vskip -0.7cm
\end{figure}
\end{itemize}
\paragraph{The baryonic charges.} Since the toric diagram has 7 external points, the total number of baryonic symmetries is $7-4=3$. These symmetries, which will be denoted as $U(1)_{B_{1}}$,$U(1)_{B_{2}}$ and $U(1)_{B_{3}}$, come from the D-terms and the perfect matchings are charged under them as according to the rows of (\ref{e:qdfano218}).

\paragraph{The global symmetry.} The $Q_t$ has only one pair of repeated columns, thus confirming that the mesonic symmetry contains only one $SU(2)$ as non-abelian factor. Since the mesonic symmetry has total rank 4, it can be identified with $SU(2)\times U(1)^3$, where one of the abelian factors corresponds to the R-symmetry. The perfect matchings $p_1$ and $p_2$ transform as a doublet under the non-abelian factor.

Since there are three $U(1)$ factors in the mesonic symmetry, a volume minimisation problem needs to be solved in order to assign the R-charges to the perfect matchings.
Let us assign the R-charge fugacity $s_1$ to both $p_1$ and $p_2$ (note that the non-abelian symmetry does not play any role in the volume minimisation, so $p_1$ and $p_2$ carry the same R-charges), and the fugacities $s_2,s_3, s_4, s_5$ and $s_6$ to the perfect matchings $u_1, q_1, r_1, q_2$ and $r_2$ respectively.  Note that the three perfect matchings $v_1, v_2, v_3$ correspond to the internal point of the toric diagram and have found to have vanishing R-charge. Accordingly, their fugacities are be set to unity.
Therefore, the Hilbert series of the mesonic moduli space is:
\bea
\gm (s_{\alpha}; \cE_1) &=& \prod_{i=1}^3 \oint \limits_{|b_i|=1} \frac{\ud b_i}{2\pi i b_i}\prod_{j=1}^3 \oint \limits_{|z_j|=1} \frac{\ud z_j}{2\pi i z_j} \frac{1}{\left(1- s_1 z_1\right)^2\left(1- s_2 z_2\right)\left(1-\frac{s_3 z_3}{z_2}\right)\left(1- s_4 z_1 z_3\right)}\nn \\
&\times& \frac{1}{\left(1-\frac{s_5 b_1}{z_1 z_3}\right)\left(1- \frac{s_6 b_1 z_2}{z_3}\right)\left(1- \frac{b_2}{z_2}\right)\left(1-\frac{b_3}{z_1}\right)\left(1-\frac{1}{b^2_1 b_2 b_3 z_1}\right)}~.
\label{e:hsvolfano218}
\eea
Since there are 3 factors of $U(1)$ in the mesonic symmetry, the integral (\ref{e:hsvolfano218}) depends only on three combinations of $s_\alpha$'s. Defining: 
\bea
t_1 = s_1 s_3 s_4 s_6^2, \quad t_2 = s_1 s_4^2 s_5 s_6, \quad t_3 = s_1^2 s_2 s_4^2 s_5^2~,
\eea
the Hilbert series of the mesonic moduli space can then be written as:
{\small
\bea
\gm (t_1, t_2, t_3; \cE_1) = \frac{P\left(t_1, t_2, t_3; \cE_1 \right)}{\left(1- t_1\right)^2\left(1-t_2\right)^2\left(1-t_3\right)^2\left(1-\frac{t^2_1 t_3}{t^2_2}\right)^2\left(1-\frac{t_1^2 t^3_3}{t^4_2}\right)^2}~,
\label{e:hsvolTfano218}
\eea}
where:
{\scriptsize 
\bea
P\left(t_1, t_2, t_3; \cE_1 \right) &=& 1-t_1 t_2+t_3-8 t_1 t_3+7 t_1^2 t_3-2 t_1^3 t_3-  2 t_2 t_3+6 t_1 t_2 t_3-2 t_1^2 t_2 t_3+6 t_1 t_3^2-9 t_1^2 t_3^2+4 t_1^3 t_3^2- t_1 t_2 t_3^2-\nn \\
&&2 t_1 t_3^3+4 t_1^2 t_3^3-2 t_1^3 t_3^3+\frac{t_1^2 t_3}{t_2^2}-\frac{2 t_1^3 t_3}{t_2^2}+\frac{3 t_1 t_3}{t_2}-\frac{8 t_1^2 t_3}{t_2}+\frac{6 t_1^3 t_3}{t_2}+\frac{4 t_1^2 t_3^2}{t_2^3}-\frac{9 t_1^3 t_3^2}{t_2^3}+\frac{6 t_1^4 t_3^2}{t_2^3}- \frac{t_1^5 t_3^2}{t_2^3}+\nn \\
&&\frac{4 t_1 t_3^2}{t_2^2}-\frac{19 t_1^2 t_3^2}{t_2^2}+\frac{26 t_1^3 t_3^2}{t_2^2}-\frac{9 t_1^4 t_3^2}{t_2^2}-\frac{9 t_1 t_3^2}{t_2}+\frac{26 t_1^2 t_3^2}{t_2}-\frac{23 t_1^3 t_3^2}{t_2}+ \frac{4 t_1^4 t_3^2}{t_2}-\frac{2 t_1^4 t_3^3}{t_2^5}+\frac{4 t_1^5 t_3^3}{t_2^5}-\frac{2 t_1^6 t_3^3}{t_2^5}+\nn \\
&&\frac{3 t_1^2 t_3^3}{t_2^4}-  \frac{12 t_1^3 t_3^3}{t_2^4}+\frac{13 t_1^4 t_3^3}{t_2^4}-\frac{10 t_1^5 t_3^3}{t_2^4}+\frac{4 t_1^6 t_3^3}{t_2^4}-\frac{12 t_1^2 t_3^3}{t_2^3}+\frac{29 t_1^3 t_3^3}{t_2^3}-\frac{20 t_1^4 t_3^3}{t_2^3}+\frac{7 t_1^5 t_3^3}{t_2^3}- \frac{2 t_1^6 t_3^3}{t_2^3}-\frac{2 t_1 t_3^3}{t_2^2}+\nn \\
&&\frac{13 t_1^2 t_3^3}{t_2^2}-\frac{20 t_1^3 t_3^3}{t_2^2}+\frac{8 t_1^4 t_3^3}{t_2^2}+\frac{4 t_1 t_3^3}{t_2}- \frac{10 t_1^2 t_3^3}{t_2}+\frac{7 t_1^3 t_3^3}{t_2}-  \frac{5 t_1^4 t_3^4}{t_2^6}+\frac{10 t_1^5 t_3^4}{t_2^6}-\frac{3 t_1^6 t_3^4}{t_2^6}+\frac{t_1^3 t_3^4}{t_2^5}+\frac{8 t_1^4 t_3^4}{t_2^5}-\nn \\
&&\frac{17 t_1^5 t_3^4}{t_2^5}+\frac{6 t_1^6 t_3^4}{t_2^5}-\frac{5 t_1^2 t_3^4}{t_2^4}+\frac{8 t_1^3 t_3^4}{t_2^4}+  \frac{8 t_1^4 t_3^4}{t_2^4}-\frac{4 t_1^5 t_3^4}{t_2^4}-\frac{3 t_1^6 t_3^4}{t_2^4}+\frac{10 t_1^2 t_3^4}{t_2^3}-\frac{17 t_1^3 t_3^4}{t_2^3}-\frac{4 t_1^4 t_3^4}{t_2^3}+\frac{7 t_1^5 t_3^4}{t_2^3}-\nn \\
&&\frac{3 t_1^2 t_3^4}{t_2^2}+\frac{6 t_1^3 t_3^4}{t_2^2}-\frac{3 t_1^4 t_3^4}{t_2^2}-\frac{3 t_1^5 t_3^5}{t_2^7}+\frac{6 t_1^6 t_3^5}{t_2^7}-\frac{3 t_1^7 t_3^5}{t_2^7}+\frac{7 t_1^4 t_3^5}{t_2^6}- \frac{4 t_1^5 t_3^5}{t_2^6}-\frac{17 t_1^6 t_3^5}{t_2^6}+\frac{10 t_1^7 t_3^5}{t_2^6}-\frac{3 t_1^3 t_3^5}{t_2^5}-\nn \\
&& \frac{4 t_1^4 t_3^5}{t_2^5}+\frac{8 t_1^5 t_3^5}{t_2^5}+\frac{8 t_1^6 t_3^5}{t_2^5}-\frac{5 t_1^7 t_3^5}{t_2^5}+\frac{6 t_1^3 t_3^5}{t_2^4}-\frac{17 t_1^4 t_3^5}{t_2^4}+\frac{8 t_1^5 t_3^5}{t_2^4}+\frac{t_1^6 t_3^5}{t_2^4}- \frac{3 t_1^3 t_3^5}{t_2^3}+ \frac{10 t_1^4 t_3^5}{t_2^3}-\frac{5 t_1^5 t_3^5}{t_2^3}-\nn \\
&&\frac{2 t_1^6 t_3^6}{t_2^9}+\frac{4 t_1^7 t_3^6}{t_2^9}-\frac{2 t_1^8 t_3^6}{t_2^9}+\frac{7 t_1^6 t_3^6}{t_2^8}-\frac{10 t_1^7 t_3^6}{t_2^8}+\frac{4 t_1^8 t_3^6}{t_2^8}+\frac{8 t_1^5 t_3^6}{t_2^7}- \frac{20 t_1^6 t_3^6}{t_2^7}+\frac{13 t_1^7 t_3^6}{t_2^7}-\frac{2 t_1^8 t_3^6}{t_2^7}-\frac{2 t_1^3 t_3^6}{t_2^6}+\nn \\
&&\frac{7 t_1^4 t_3^6}{t_2^6}-\frac{20 t_1^5 t_3^6}{t_2^6}+\frac{29 t_1^6 t_3^6}{t_2^6}-\frac{12 t_1^7 t_3^6}{t_2^6}+\frac{4 t_1^3 t_3^6}{t_2^5}-\frac{10 t_1^4 t_3^6}{t_2^5}+ \frac{13 t_1^5 t_3^6}{t_2^5}-\frac{12 t_1^6 t_3^6}{t_2^5}+\frac{3 t_1^7 t_3^6}{t_2^5}-\frac{2 t_1^3 t_3^6}{t_2^4}+\frac{4 t_1^4 t_3^6}{t_2^4}-\nn \\
&&\frac{2 t_1^5 t_3^6}{t_2^4}-\frac{t_1^8 t_3^7}{t_2^{10}}+\frac{4 t_1^6 t_3^7}{t_2^9}-\frac{9 t_1^7 t_3^7}{t_2^9}+\frac{6 t_1^8 t_3^7}{t_2^9}+\frac{4 t_1^5 t_3^7}{t_2^8}- \frac{23 t_1^6 t_3^7}{t_2^8}+\frac{26 t_1^7 t_3^7}{t_2^8}-\frac{9 t_1^8 t_3^7}{t_2^8}-\frac{9 t_1^5 t_3^7}{t_2^7}+\frac{26 t_1^6 t_3^7}{t_2^7}-\nn \\
&& \frac{19 t_1^7 t_3^7}{t_2^7}+\frac{4 t_1^8 t_3^7}{t_2^7}-\frac{t_1^4 t_3^7}{t_2^6}+\frac{6 t_1^5 t_3^7}{t_2^6}-\frac{9 t_1^6 t_3^7}{t_2^6}+ \frac{4 t_1^7 t_3^7}{t_2^6}- \frac{2 t_1^7 t_3^8}{t_2^{10}}+\frac{6 t_1^8 t_3^8}{t_2^{10}}-\frac{2 t_1^9 t_3^8}{t_2^{10}}-\frac{2 t_1^6 t_3^8}{t_2^9}+\frac{7 t_1^7 t_3^8}{t_2^9}-\nn \\
&& \frac{8 t_1^8 t_3^8}{t_2^9}+\frac{t_1^9 t_3^8}{t_2^9}+\frac{6 t_1^6 t_3^8}{t_2^8}-\frac{8 t_1^7 t_3^8}{t_2^8}+\frac{3 t_1^8 t_3^8}{t_2^8}- \frac{2 t_1^6 t_3^8}{t_2^7}+\frac{t_1^7 t_3^8}{t_2^7}-\frac{t_1^8 t_3^9}{t_2^{10}}+\frac{t_1^9 t_3^9}{t_2^9}~.
\eea}
Let $R_i$ be the R-charge corresponding to the fugacity $t_i$.  Since the superpotential carries R-charge 2, it follows that:
\bea
R_1 + R_3 - R_2 = 2~.
\label{e:cyfano218}
\eea
Thus, the volume of $\cE_1$ is given by:
{\small
\bea \lim_{\mu \rightarrow 0} \mu^4 \gm (e^{-\mu (2 + R_2 - R_3)}, e^{-\mu R_2}, e^{-\mu R_3})= \frac{p\left(R_2, R_3; \cE_1\right)}{R_2^2 R_3^2 (2 + R_2 - R_3)^2 (4 - 2 R_2 + R_3)^2 (4 - R_3)^2}, \nn \\
\label{e:volfano218}
\eea}
where:
{\small
\bea
p\left(R_2, R_3; \cE_1\right) &=& 1024 R_2 + 512 R_2^2 - 256 R_2^3 - 128 R_2^4 + 512 R_3 - 512 R_2 R_3 + 384 R_2^2 R_3 +\nn \\
&& 512 R_2^3 R_3 +   64 R_2^4 R_3 -  256 R_3^2 + 256 R_2 R_3^2 - 512 R_2^2 R_3^2 - 192 R_2^3 R_3^2 -\nn \\
&& 16 R_2^4 R_3^2 - 64 R_3^3 - 32 R_2 R_3^3 +160 R_2^2 R_3^3 + 32 R_2^3 R_3^3 + 32 R_3^4 -20 R_2 R_3^4 -\nn \\
&&  20 R_2^2 R_3^4 + 2 R_3^5 + 4 R_2 R_3^5 - R_3^6 ~,
\eea}
This function has a minimum at:
\bea
R_3 = 2, \quad R_1=R_2 \approx 1.620~.
\eea
The R-charge of the perfect matching corresponding to the divisor $D_\alpha$ is given by: 
\bea
\lim_{\mu\rightarrow0}\frac{1}{\mu} \left[ \frac{g(D_\alpha; e^{- \mu R_1}, e^{- \mu R_2 }, e^{- \mu R_3 }; \cE_1) }{\gm(e^{-\mu R_1}, e^{- \mu R_2 }, e^{- \mu R_3 };\cE_1)}- 1 \right]~,
\eea
where $g(D_\alpha; e^{- \mu R_1}, e^{- \mu R_2 }, e^{- \mu R_3 }; \cE_1)$ is the Molien-Weyl integral with the insertion of the inverse of the weight corresponding to the divisor $D_\alpha$. The results are shown in Table \ref{t:chargefano218}.  The assignment of charges under the remaining abelian symmetries can be done by requiring that the superpotential is not charged under them and that the charge vectors are linearly independent. The assignments are shown in Table \ref{t:chargefano218}.
\begin{table}[h!]
 \begin{center}  
  \begin{tabular}{|c||c|c|c|c|c|c|c|c|}
  \hline
  \;& $SU(2)$&$U(1)_1$&$U(1)_2$&$U(1)_R$&$U(1)_{B_1}$&$U(1)_{B_2}$&$U(1)_{B_3}$&fugacity\\
  \hline\hline  
   
  $p_1$&$  1$&$  0$&$ 0$&$0.347$&$ 0$&$ 0$ &$ 0$ & $s_1 x $\\
  \hline
  
  $p_2$&$ -1$&$  0$&$ 0$&$0.347$&$ 0$&$ 0$ &$ 0$ & $s_1 / x $\\
  \hline  
  
  $q_1$&$  0$&$  1$&$ 0$&$0.201$&$ 0$&$ 0$ &$ 0$ & $s_2 q_1 $\\
  \hline
  
  $q_2$&$  0$&$ -1$&$ 0$&$0.201$&$ 1$&$ 0$ &$ 0$ & $s_2 b_1 / q_1 $\\
  \hline
   
  $r_1$&$  0$&$  0$&$ 1$&$0.357$&$ 0$&$ 0$ &$ 0$ & $s_3 q_2 $\\
  \hline
  
  $r_2$&$  0$&$  0$&$ 1$&$0.357$&$ 1$&$ 0$ &$ 0$ & $s_3 b_1 / q_2$\\
  \hline
  
  $u_1$&$  0$&$  0$&$-2$&$0.189$&$ 0$&$ 0$ &$ 0$ & $s_4 / q^2_2$\\
  \hline

  $v_1$&$  0$&$  0$&$ 0$&$    0$&$ 0$&$ 1$ &$ 0$ & $ b_2 $\\
  \hline
  
  $v_2$&$  0$&$  0$&$ 0$&$    0$&$ 0$&$ 0$ &$ 1$ & $ b_3 $\\
  \hline
 
  $v_3$&$  0$&$  0$&$ 0$&$    0$&$-2$&$-1$ &$-1$ & $1 / (b_1^2 b_2 b_3) $\\
  \hline
   \end{tabular}
  \end{center}
\caption{Charges of the perfect matchings under the global symmetry of the $\cE_1$ model. Here $s_\alpha$ is the fugacity of the R-charge, $x$ is the weight of the $SU(2)$ symmetry, $q_1, q_2, b_1, b_2$ and $b_3$ are, respectively, the charges under the mesonic abelian symmetries $U(1)_1, U(1)_2$ and of the three baryonic $U(1)_{B_1},U(1)_{B_2}$ and $U(1)_{B_3}$. 
}
\label{t:chargefano218}
\end{table}

\begin{table}[h]
 \begin{center}  
  \begin{tabular}{|c||c|}
  \hline
  \; Quiver fields & R-charge\\
  \hline  \hline 
  $ X^{i}_{51}, X^{i}_{34}  $ &  0.347\\
  \hline
  $ X^{1}_{45}, X^{2}_{45}, $ &  0.938\\
  \hline
  $  X^{3}_{45} , X_{12} $ &  0.558\\
  \hline
  $ X_{53},  X_{14}  $ &  0.715\\
  \hline
  $ X_{23} $ &  0.189\\
  \hline
  \end{tabular}
  \end{center}
\caption{R-charges of the quiver fields for the $ \cE_1$ model.}
\label{t:Rquivfano218}
\end{table}

\paragraph{The Hilbert series.}   The Hilbert series of the Master space can be obtained by integrating over the $z_i$ fugacities:
{\small
\bea
g^{\firr{}}  (s_{\alpha}, x, q_i, b_j; \cE_1) &=& \prod_{i=1}^3 \oint \limits_{|z_i|=1} \frac{\ud z_i}{2\pi i z_i} \frac{1}{\left(1 - s_1 x z_1\right)\left(1-\frac{s_1 z_1}{x}\right)\left(1- \frac{s_2 q_1 z_3}{z_2}\right)\left(1-\frac{s_2 b_1}{q_1 z_1 z_3}\right)}\nn \\
&\times& \frac{1}{\left(1-s_3 q_2 z_1 z_3\right)\left(1-\frac{s_3 q_2 b_1 z_2}{z_3}\right)\left(1-\frac{s_4 z_2}{q^2_2}\right)\left(1- \frac{b_2}{z_2}\right)\left(1-\frac{b_3}{z_1}\right)\left(1-\frac{1}{b^2_1 b_2 b_3 z_1}\right)}\nn \\
&=& \frac{\cP\left(s_{\alpha}, x, q_i, b_j; \cE_1\right)}{\left(1-s_2 s_3 q_1 q_2 b_1\right)\left(1-\frac{s_2 s_3 q_2 b_1}{q_1}\right)\left(1- s^2_3 q^2_2 b_1 b_2 b_3\right)\left(1-\frac{s^2_3 q^2_2}{b_1 b_3}\right) \left(1- \frac{s_1 s^2_2 s_4 x b_1}{q^2_2}\right)}\nn \\
&\times& \frac{1}{\left(1-\frac{s_1 s^2_2 s_4 b_1}{x q^2_2}\right)\left(1-\frac{s_1 x}{b^2_1 b_2 b_3}\right)\left(1-\frac{s_1}{x b^2_1 b_2 b_3}\right){\left(1- s_1 x b_3\right)\left(1-\frac{s_1 b_3}{x}\right)\left(1- \frac{s_4 b_2}{q^2_2}\right)}}~. \nn \\
\label{e:HSmasterfano218}
\eea}
where $\cP\left(s_{\alpha}, x, q_1, q_2, b_j; \cE_1\right)$ is polynomial which is not reported here. 

The Hilbert series of the mesonic moduli space can be obtained by integrating over the three baryonic fugacities
{\small
\bea
\gm (s_{\alpha}, x, q_1, q_2; \cE_1) &=& \left( \prod^3_{i=i} \oint\limits_{|b_i|=1} \frac{\ud b_i}{2 \pi i b_i} \right) g^{\firr{}}  (s_{\alpha}, x, q_i, b_j; \cE_1)\nn\\
&=& \frac{P \left(s_{\alpha}, x, q_1, q_2; \cE_1 \right)}{\left(1- \frac{s_1 s_2 s^3_3 x q_1}{q^3_2}\right)\left(1-\frac{s_1 s_2 s^3_3 q_1}{x q^3_2}\right)\left(1-\frac{s_1 s_2 s^3_3 x q^3_2}{q_1}\right)\left(1-\frac{s_1 s_2 s^3_3 q^3_2}{x q_1}\right)}\nn \\
&\times& \frac{1}{\left(1- s^2_1 s^2_2 s^2_3 s_4 x^2 q^2_1\right)\left(1-\frac{s^2_1 s^2_2 s^2_3 s_4 q^2_1}{x^2}\right)\left(1-\frac{s^2_1 s^2_2 s^2_3 s_4 x^2}{q^2_1}\right)\left(1-\frac{s^2_1 s^2_2 s^2_3 s_4}{x^2 q^2_1}\right)}\nn \\
&\times& \frac{1}{\left(1- \frac{s^4_1 s^4_2 s^3_4 x^4}{q^6_2}\right)\left(1-\frac{s^4_1 s^4_2 s^2_4}{x^4 q^6_2}\right)}~,\nn \\
\label{e:HSmesonicfano218}
\eea}
where $P \left(s_{\alpha}, x, q_1, q_2; \cE_1 \right)$ is polynomial which is not reported here. 

The plethystic logarithm of this function can be written as:
{\small
\bea
\PL[\gm (t_{\alpha}, x, q_1, q_2; \cE_1)] &=& [4] \frac{t^3_3}{q^6_2 t^2_1} + [3]\left(q_1+\frac{1}{q_1}\right) \frac{t^2_3}{q^3_2 t_1} + [2] \left(q^2_1 + 1 + \frac{1}{q^2_1}\right) t_3 \nn \\
&&+ [1] \left(q_1+\frac{1}{q_1}\right) q^3_2 t_1 - O(t_1^2)O(t^4_3) \nn \\
\label{e:plfano218}
\eea}
From the plethystic logarithm it's clear that in the mesonic moduli space the abelian symmetry $U(1)_1$ is enhanced to $SU(2)$.

\begin{figure}[ht]
\begin{center}
\includegraphics[totalheight=8cm]{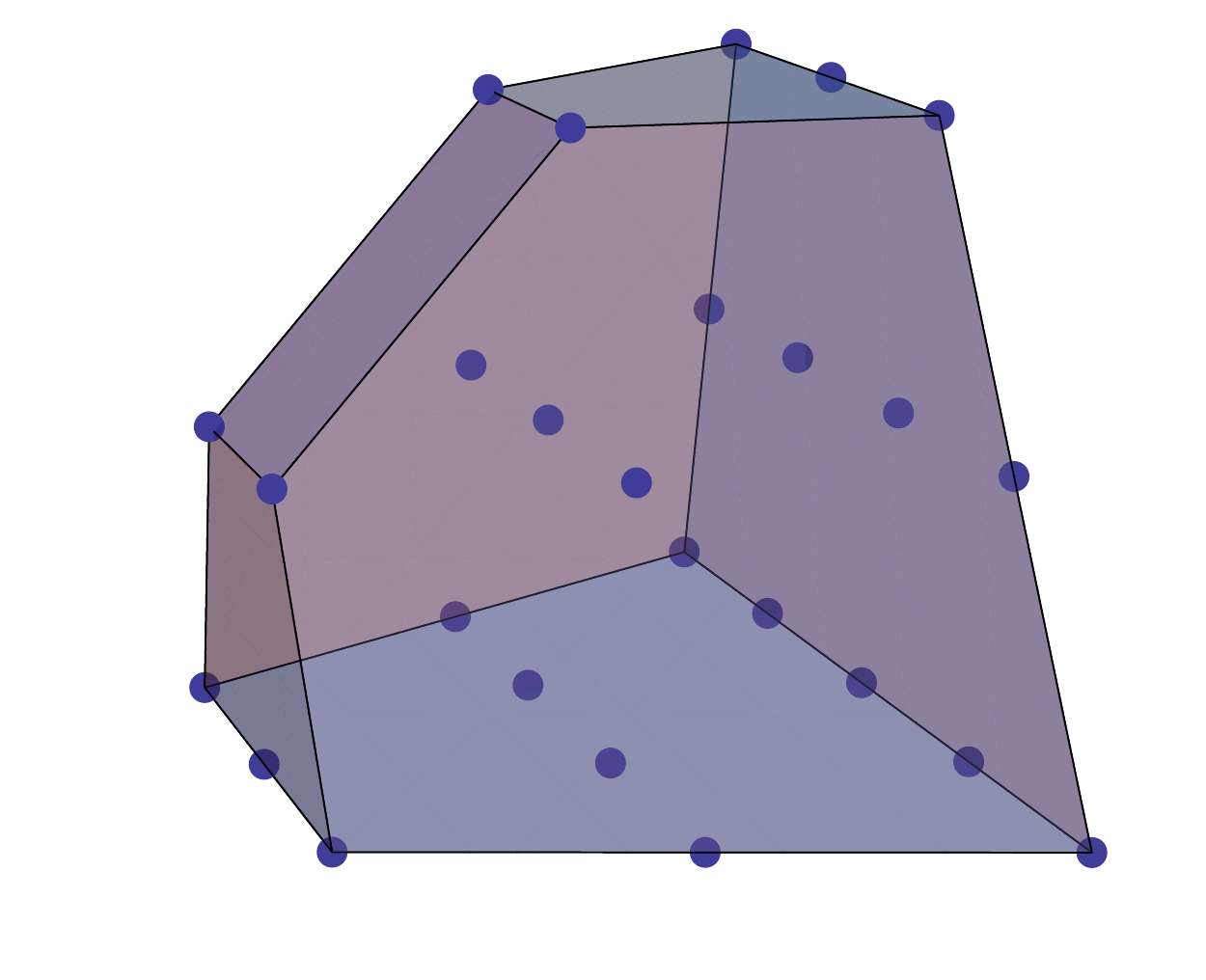}
\caption{The lattice of generators of the $\cE_1$ theory.}
  \label{f:late1}
  \end{center}
\end{figure}

\paragraph{The generators.}
The generators of the mesonic moduli space are
\bea
&& p_i p_j p_k p_l q^2_1 q^2_2 u^3_1 v_1 v_2 v_3, \qquad p_i p_j p_k q^2_m q_n r_n u^2_1 v_1 v_2 v_3, \qquad p_i p_j q^2_m r^2_n u_1 v_1 v_2 v_3, \nn \\
&& p_i p_j q_1 q_2 r_1 r_2 u_1 v_1 v_2 v_3, \qquad   p_i q_m r^2_n r_m v_1 v_2 v_3,
\label{e:genfano218}
\eea
with $i,j,k,l,m,n=1,2$ and $m \neq n$.
The R-charges of the generators of the mesonic moduli space listed in Table \ref{t:Rgenfano218}. The lattice of generators is given in \fref{f:late1}.
\begin{table}[h]
 \begin{center}  
  \begin{tabular}{|c||c|}
  \hline
  \; Generators & R-charge\\
  \hline  \hline 
  $p_i p_j p_k p_l q^2_1 q^2_2 u^3_1 v_1 v_2 v_3$ & 2.760 \\
  \hline
  $p_i p_j p_k q^2_m q_n r_n u^2_1 v_1 v_2 v_3$ & 2.380 \\
  \hline
  $p_i p_j q^2_m r^2_n u_1 v_1 v_2 v_3$ & 2 \\
  \hline
  $p_i p_j q_1 q_2 r_1 r_2 u_1 v_1 v_2 v_3$ & 2 \\
  \hline
  $p_i q_m r^2_n r_m v_1 v_2 v_3$ & 1.620 \\
  \hline
  \end{tabular}
  \end{center}
\caption{R-charges of the generators of the mesonic moduli space for the $ \cE_1$ Model.}
\label{t:Rgenfano218}
\end{table}

\section{$\cE_2$ (Toric Fano 275):  $dP_2$ bundle over $\mathbb{P}^1$}
\begin{figure}[ht]
\begin{center}
\vskip 0cm
\hskip -1cm
\includegraphics[totalheight=4.5cm]{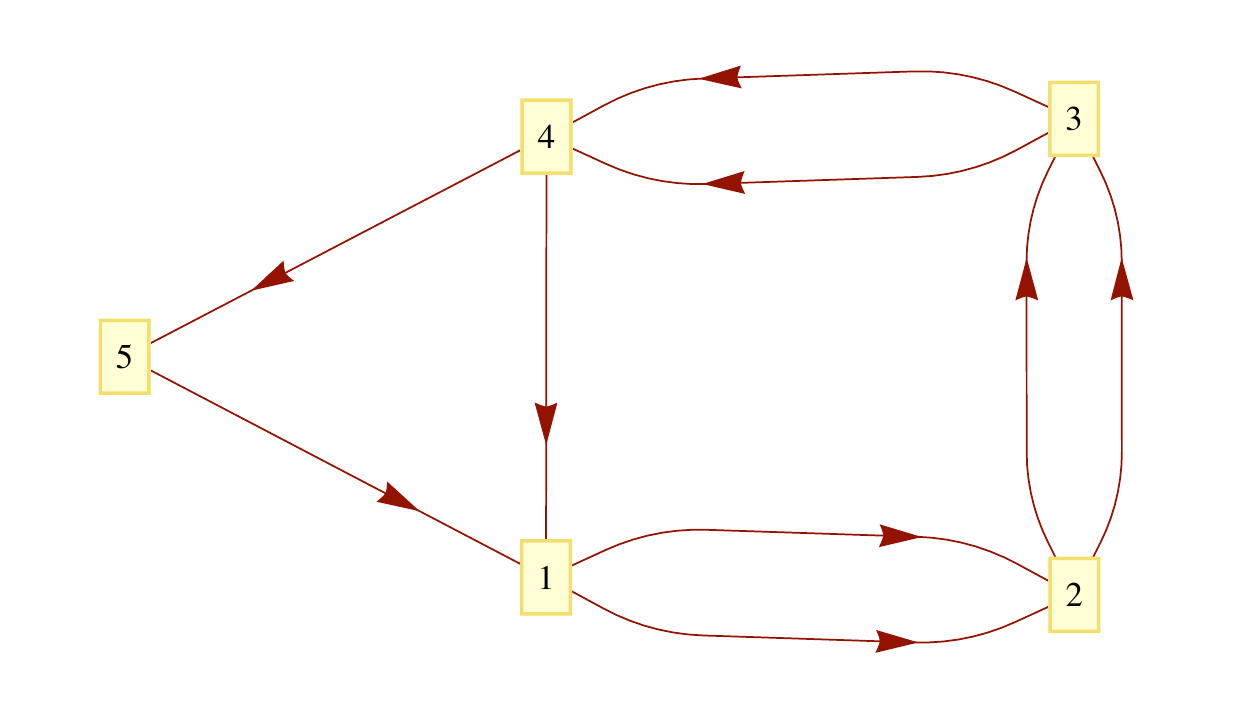}
\hskip 1cm
\includegraphics[totalheight=5.5cm]{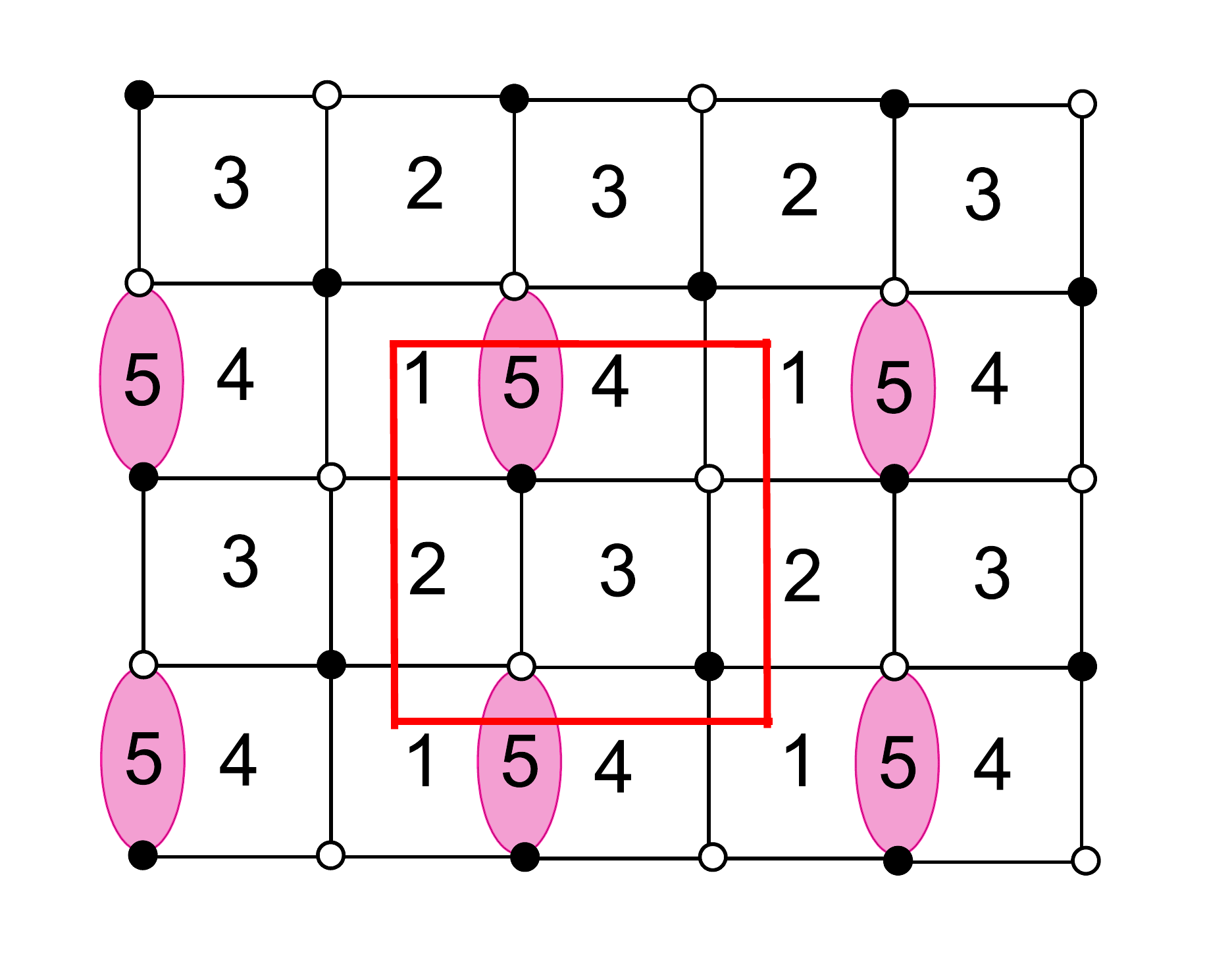}
  \caption{(i) Quiver of the $\cE_2$ model. \quad (ii) Tiling of the $\cE_2$ model.}
  \label{t:fano275tileandquiver}
  \end{center}
\end{figure}

This model has 5 gauge groups and bi-fundamental fields $X_{34}^i$, $X_{12}^i$, $X_{23}^i$, $X_{41}$, $X_{51}$, $X_{45}$ (with $i=1,2$). The quiver diagram and tiling are drawn in Figure \ref{t:fano275tileandquiver}. 
Note that the tiling of this model is a `double bonding' of the tiling of Phase I of the $\BF_0$ theory. 

The superpotential is given by
\bea
W = \tr \left[ \epsilon_{ij} ( X_{45} X_{51} X^i_{12}X^1_{23} X^j_{34} -  X_{41} X^i_{12} X^2_{23} X^j_{34}) \right]~.
\label{e:spotfano275}
\eea
The CS levels are chosen to be $\vec{k} = (1,0,-1,-1,1)$.

\comment{
\begin{figure}[ht]
\begin{center}
\includegraphics[totalheight=6.5cm]{fano266tilingfd.pdf}
  \caption{The fundamental domain of the tiling for the $\cE_2$ model.}
  \label{t:fano275funddom}
  \end{center}
 \end{figure} 
}

\paragraph{The Kasteleyn matrix.}   The Chern-Simons levels can be parametrized as follows:
\bea
\begin{array}{ll}
\text{Gauge group 1:} \qquad  k_1 &=   n^{1}_{12} + n^{2}_{12} - n_{41} - n_{51} ~, \nn \\
\text{Gauge group 2:} \qquad  k_2 &=   n^{1}_{23} + n^{2}_{23} - n^{1}_{12} - n^{2}_{12} ~, \nn \\
\text{Gauge group 3:} \qquad  k_3 &=   n^{1}_{34} + n^{2}_{34} - n^{1}_{23} - n^{2}_{23} ~,   \nn \\
\text{Gauge group 4:} \qquad  k_4 &=   n_{41} + n_{45} - n^{1}_{34} - n^{2}_{34} ~, \nn \\
\text{Gauge group 5:} \qquad  k_5 &=   n_{51} - n_{45} ~.
\label{e:kafano275}
\end{array}
\eea
Let us choose 
\bea
n^{1}_{12} = n^{2}_{23} = -n_{45} = 1,~ n^i_{jk}=n_{jk}=0 \; \text{otherwise}~.
\eea
The fundamental domain contains two pairs of black and white nodes and, therefore, the Kasteleyn matrix is a $2 \times 2$ matrix: 
\bea
K= \left(
\begin{array}{c|cc}
& w_1 & w_2 \\
\hline
b_1 & z^{n^{1}_{34}} + x z^{n^{2}_{12}} &\ z^{n^{1}_{23}} + \frac{1}{y} z^{n_{45}} + \frac{1}{y} z^{n_{51}}   \\
b_2 & z^{n^{2}_{23}} + y z^{n_{41}} &\ z^{n^{2}_{34}} + \frac{1}{x} z^{n^{1}_{12}}
\end{array}
\right)~.
\label{e:kastfano275}
\eea
The permanent of this matrix is given by:
\bea
\perm(K) &=&  x z^{(n^2_{12} + n^2_{34})} +  x^{-1}z^{(n^1_{12} + n^1_{34})} +  y z^{(n^1_{23} + n_{41})} +  y^{-1} z^{(n^2_{23} + n_{45})}\nn \\
&+& y^{-1} z^{(n^2_{23} + n_{51})} +  z^{(n_{41} + n_{45})} + z^{(n^1_{23} + n^2_{23})} + z^{(n^1_{12} + n^2_{12})}\nn \\
&+&z^{(n^1_{34} + n^2_{34})} + z^{(n_{41} + n_{51})}\nn \\
&=& x + x^{-1} z +y + y^{-1} + y^{-1} z +z^{-1} + 2z + 2\nn \\
&& \text{(for $n^{1}_{12} = n^{2}_{23} = -n_{45} = 1,~ n^i_{jk}=n_{jk}=0\; \text{otherwise}$)} ~.
\label{e:charpolyfano275}
\eea
The perfect matchings can be written in terms of the chiral fields:
\bea 
\begin{array}{llll}
p_1 = \left\{X^2_{12}, X^2_{34}\right\}, \;\; & p_2 = \left\{X^1_{12}, X^1_{34}\right\}, \;\; & r_1 = \left\{X_{41},X^1_{23}\right\}, \;\; & r_2 = \left\{X_{45}, X^2_{23}\right\}, \nn \\
r_3 = \left\{X^2_{23},X_{51}\right\}, \;\; & r_4 = \left\{X_{41}, X_{45}\right\},  \;\; & r_5 = \left\{X^1_{23}, X^2_{23}\right\}, \;\; & r'_5 = \left\{X^1_{12},X^2_{12}\right\},\nn \\   
v_1 = \left\{X^1_{34}, X^2_{34}\right\}, \;\; &  v_2 = \left\{X_{41},X_{51}\right\}~. & & 
\end{array}
\eea
Note that the perfect matchings $v_1$ and $v_2$ correspond to the internal point at the origin, and the perfect matchings $r_5$ and $r'_5$ correspond to a point at one of the corners of the toric diagram.
The chiral fields can be written as products of perfect matchings as
\bea
\begin{array}{lllll}
X^2_{34} = p_1 v_1, \quad & X^1_{34} = p_2 v_1, \quad & X^2_{12} = p_1 r'_5, \quad & X^1_{12} = p_2 r'_5, \quad & X^1_{23} = r_1 r_5,  \nn \\
X^2_{23} = r_2 r_3 r_5, \quad &  X_{41} = r_1 r_4 v_2, \quad  & X_{45} = r_2 r_4,\quad & X_{51} = r_3 v_2~. &
\end{array}
\eea

\paragraph{The perfect matching.}  From \eref{e:charpolyfano275}, the relationship between the quiver fields and the perfect matchings is summarised in the $P$ matrix:
\beq
P=\left(\begin{array} {c|cccccccccc}
&p_1 & p_2 & r_1 & r_2 & r_3 & r_4 & r_5 & r'_5 & v_1 & v_2\\
\hline
X^2_{34}  &1&0&0&0&0&0&0&0&1&0\\
X^1_{34}  &0&1&0&0&0&0&0&0&1&0\\
X^2_{12}  &1&0&0&0&0&0&0&1&0&0\\
X^1_{12}  &0&1&0&0&0&0&0&1&0&0\\
X^1_{23}  &0&0&1&0&0&0&1&0&0&0\\
X^2_{23}  &0&0&0&1&1&0&1&0&0&0\\
X_{41}    &0&0&1&0&0&1&0&0&0&1\\
X_{45}    &0&0&0&1&0&1&0&0&0&0\\
X_{51}    &0&0&0&0&1&0&0&0&0&1
\end{array}\right)~.
\label{e:pfano275}
\eeq
Basis vectors of the null space of $P$ are given in the rows of the charge matrix: 
\bea
Q_F= \left(
\begin{array}{cccccccccc}
 1 & 1 & 0 & 0 & 0 & 0 & 0 &-1 &-1 & 0 \\
 0 & 0 & 1 & 1 & 0 &-1 &-1 & 0 & 0 & 0 \\
 0 & 0 & 0 & 1 &-1 &-1 & 0 & 0 & 0 & 1
\end{array}
\right)~. \label{e:qffano275}
\eea
Therefore, the relations between the perfect matchings are given by
\bea \label{relperm275}
p_1 + p_2 - r'_5 - v_1 &=& 0~, \nn \\
r_1 + r_2 - r_4 - r_5 &=& 0~, \nn \\
r_2 - r_3 - r_4 + v_2 &=& 0~.
\eea

\paragraph{The toric diagram.} Two methods of constructing the toric diagram are presented.
\begin{itemize}
 \item {\bf The charge matrices.}  Since the number of gauge groups is $G = 5$, there are $G - 2 = 3$ baryonic symmetries coming from the D-terms.  The charges of the perfect matchings are collected in the $Q_D$ matrix: 
\bea
Q_D = \left( \begin{array}{cccccccccc}
 0 & 0 & 0 & 0 & 0 & 1 & 1 & 0 &-2 & 0 \\
 0 & 0 & 0 & 0 & 0 & 0 & 1 &-1 & 0 & 0 \\
 0 & 0 & 0 & 0 & 0 & 0 & 0 & 0 & 1 &-1
 \end{array} \right) ~.
 \label{e:qdfano275}
\eea
From \eref{e:qffano275} and \eref{e:qdfano275}, the total charge matrix is given by
\bea
Q_t =  { \Blue Q_F \choose \Green Q_D \Black } = 
\left( \begin{array}{cccccccccc} \Blue
 1 & 1 & 0 & 0 & 0 & 0 & 0 &-1 &-1 & 0 \\
 0 & 0 & 1 & 1 & 0 &-1 &-1 & 0 & 0 & 0 \\
 0 & 0 & 0 & 1 &-1 &-1 & 0 & 0 & 0 & 1 \\ \Green
 0 & 0 & 0 & 0 & 0 & 1 & 1 & 0 &-2 & 0 \\
 0 & 0 & 0 & 0 & 0 & 0 & 1 &-1 & 0 & 0 \\
 0 & 0 & 0 & 0 & 0 & 0 & 0 & 0 & 1 &-1 \Black
 \end{array} \right) ~.
 \label{e:qfanddfano275}
\eea
The kernel of this matrix is the $G_t$ matrix which, after removing the first row, contains the coordinates of the points in the toric diagram in its columns: 
\bea
G'_t = \left(
\begin{array}{cccccccccc}
 1 &-1 & 0 & 0 & 0 & 0 & 0 & 0 & 0 & 0 \\
 0 & 0 & 1 &-1 &-1 & 0 & 0 & 0 & 0 & 0 \\
 0 & 1 & 0 & 0 & 1 &-1 & 1 & 1 & 0 & 0
\end{array}
\right)~.
\eea
The toric diagram is drawn in \fref{f:fano275toric}. 

\begin{figure}[ht]
\begin{center}
\includegraphics[totalheight=3.5cm]{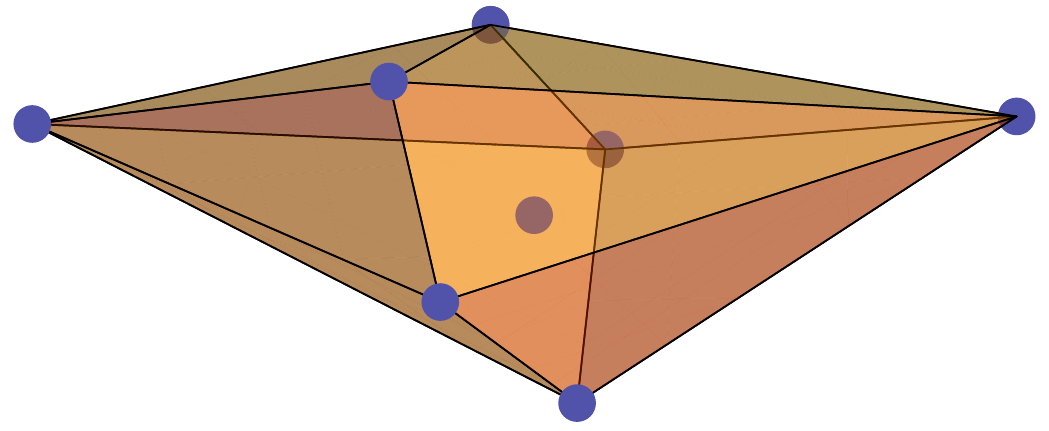}
  \caption{The toric diagram of the toric $\cE_2$.}
 \label{f:fano275toric}
 \end{center}
 \end{figure} 

\comment{
Observe that a projection of this 3 dimensional toric diagram down to 2 dimensions gives the toric diagram of the $\BF_0$ theory with an external multiplicity (\fref{f:fano275toric}). 
\begin{figure}[ht]
 \centerline{\epsfxsize =6cm \epsfbox{torf0.pdf}}
 \vskip -1cm
  \caption{The toric diagram of the $\BF_0$ theory.}
 \label{f:fano275toric}
 \end{figure} }

\item {\bf The Kasteleyn matrix.} The powers of $x, y, z$ in each term of \eref{e:charpolyfano275} give the coordinates of each point in the toric diagram. These points can be collected in the columns of the following $G_K$ matrix: 
\bea
G_K = \left(
\begin{array}{cccccccccc}
 1 &-1 & 0 & 0 & 0 & 0 & 0 & 0 & 0 & 0 \\
 0 & 0 & 1 &-1 &-1 & 0 & 0 & 0 & 0 & 0 \\
 0 & 1 & 0 & 0 & 1 &-1 & 1 & 1 & 0 & 0
\end{array}
\right) = G'_t ~.
\eea
The first row of the Kasteleyn matrix contains the weights of the fundamental representation of $SU(2)$. Hence, the mesonic symmetry has $SU(2)$ as the only non-abelian factor. 
\end{itemize}

\paragraph{The baryonic charges.} Since the toric diagram has 6 external points, this model has precisely $7-4 = 3$ baryonic charges which will be denoted by $U(1)_{B_1}, U (1)_{B_2}, U (1)_{B_3}$. The baryonic charges of the perfect matchings are given by the rows of the $Q_D$ matrix.

\paragraph{The global symmetry.} As mentioned above, the global symmetry of this model contains one $SU(2)$ as the only non-abelian factor. This is confirmed by the fact that $Q_t$ has one pair of repeated columns. Since the mesonic symmetry has total rank 4, it can be identified with $SU(2) \times U(1)^3$.  
The perfect matchings $p_1$ and $p_2$ transform as a doublet under the $SU(2)$.   From \eref{e:charpolyfano275} it can be seen that the perfect matchings $v_1, v_2$ correspond to the internal point of the toric diagrams and accordingly they carry zero R-charges. Note that the perfect matchings $r_5$ and $r'_5$ correspond to the same point in the toric diagram, but they carry different R-charges.

Since there are three $U(1)$ factors in the mesonic symmetry, a volume minimisation problem needs to be solved in order to assign the R-charges to the perfect matchings. Let us assign the R-charge fugacity $s_1$ to the perfect matchings $p_1$ and $p_2$ (note that the non-abelian symmetry does not play any role in the volume minimisation, so $p_1$ and $p_2$ carry the same R-charges), and the R-charge fugacities $s_2,s_3,s_4,s_5,s_6,s_7$ to the perfect matchings $r_1,r_2,r_3,r_4,r_5,r'_5$ respectively.  The Hilbert series of the mesonic moduli space is given by
{\small
\bea
\gm (t_{\alpha}; \cE_2) &=& \left( \prod^{3}_{i=1} \oint \limits_{|z_i|=1} {\frac{\ud z_i}{2 \pi i z_i}} \prod^{3}_{j=1} \oint \limits_{|b_j|=1} {\frac{\ud b_j}{2 \pi i b_j}} \right) \frac{1}{\left(1- s_1 z_1\right)^2\left(1- s_2 z_2\right)\left(1-s_3 z_2 z_3 \right)\left(1- \frac{s_4}{z_3}\right)}\nn \\
&&\times \frac{1}{\left(1- \frac{s_5 b_1}{z_2 z_3}\right)\left(1-\frac{s_6 b_1 b_2}{z_2}\right)\left(1-\frac{s_7}{b_2 z_1}\right)\left(1-\frac{b_3}{z_1 b^2_1}\right)\left(1-\frac{z_3}{b_3}\right)}~.
\label{e:HSmesvolfano275}
\eea}
Since there are three factors of $U(1)$ in the mesonic symmetry, the integral (\ref{e:HSmesvolfano275}) depends only on three combinations of perfect matchings. By defining:
\bea
t_1 = s_1 s_2 s_3 s^2_5, \quad t_2 = s_1 s^2_3 s_4 s^2_5, \quad t_3 = s^2_1 s^2_2 s_5 s_6 s_7~,
\eea
we have that the integral \eref{e:HSmesvolfano275} can be written as:
{\small
\bea
\gm (t_1, t_2, t_3; \cE_2) &=& \frac{P\left(t_1, t_2, t_3; \cE_2 \right)}{\left(1-t_1\right)^2\left(1-t_2\right)^2\left(1-t_3\right)^2\left(1-\frac{t_2 t^2_3}{t^2_1}\right)^2\left(1-\frac{t^3_2 t^2_3}{t^4_1}\right)^2}~, \qquad 
\eea}
where:
{\scriptsize
\bea
P \left(t_1, t_2, t_3; \cE_2\right) &=& 1 - t_1 t_2 + t_3 - 2 t_1 t_3 - 8 t_2 t_3  + 6 t_1 t_2 t_3 + 7 t_2^2 t_3 + 4 t_2^2 t_3^3  - 2 t_1 t_2^2 t_3 - 2 t_2^3 t_3 + 6 t_2 t_3^2- t_1 t_2 t_3^2- 2 t_2 t_3^3 -\nn \\
&& 9 t_2^2 t_3^2- 2 t_2^3 t_3^3+ 4 t_2^3 t_3^2 - \frac{2 t_2^3 t_3}{t_1^2} + \frac{4 t_2^3 t_3}{t_1} + \frac{2 t_2 t_3^2}{t_1^2} - \frac{9 t_2 t_3^2}{t_1}  + \frac{4 t_2^2 t_3^2}{t_1^3} - \frac{17 t_2^2 t_3^2}{t_1^2} + \frac{24 t_2^2 t_3^2}{t_1} +\frac{2 t_2^3 t_3^2}{t_1^4} - \nn \\
&& \frac{11 t_2^3 t_3^2}{t_1^3} + \frac{14 t_2^3 t_3^2}{t_1^2} - \frac{11 t_2^3 t_3^2}{t_1}+ \frac{3 t_2 t_3}{t_1} - \frac{3 t_2^4 t_3^2}{t_1^4} + \frac{8 t_2^4 t_3^2}{t_1^3} - \frac{3 t_2^4 t_3^2}{t_1^2}  - \frac{4 t_2 t_3^3}{t_1^2} + \frac{8 t_2 t_3^3}{t_1} - \frac{8 t_2^2 t_3^3}{t_1^3} +\frac{24 t_2^2 t_3^3}{t_1^2} - \nn \\
&& \frac{22 t_2^2 t_3^3}{t_1}  - \frac{10 t_2^3 t_3^3}{t_1^4}+ \frac{3 t_2^2 t_3}{t_1^2} - \frac{8 t_2^2 t_3}{t_1} + \frac{25 t_2^3 t_3^3}{t_1^3} - \frac{20 t_2^3 t_3^3}{t_1^2} + \frac{9 t_2^3 t_3^3}{t_1} + \frac{9 t_2^4 t_3^3}{t_1^4} - \frac{16 t_2^4 t_3^3}{t_1^3} + \frac{5 t_2^4 t_3^3}{t_1^2} -\frac{t_2^5 t_3^3}{t_1^5} +\nn \\
&&  \frac{2 t_2^5 t_3^3}{t_1^4} - \frac{t_2^5 t_3^3}{t_1^3} - \frac{t_2^6 t_3^3}{t_1^6} + \frac{2 t_2^6 t_3^3}{t_1^5} - \frac{t_2^6 t_3^3}{t_1^4} + \frac{4 t_2^2 t_3^4}{t_1^3} - \frac{5 t_2^2 t_3^4}{t_1^2} + \frac{2 t_2^2 t_3^4}{t_1} - \frac{t_2^3 t_3^4}{t_1^5} + \frac{6 t_2^3 t_3^4}{t_1^4} - \frac{10 t_2^3 t_3^4}{t_1^3} +\frac{4 t_2^3 t_3^4}{t_1^2} - \nn \\
&& \frac{5 t_2^4 t_3^4}{t_1^6} + \frac{8 t_2^4 t_3^4}{t_1^5} + \frac{10 t_2^4 t_3^4}{t_1^4} - \frac{8 t_2^4 t_3^4}{t_1^3} - \frac{t_2^4 t_3^4}{t_1^2} - \frac{t_2^5 t_3^4}{t_1^7} + \frac{8 t_2^5 t_3^4}{t_1^6} - \frac{7 t_2^5 t_3^4}{t_1^5} - \frac{14 t_2^5 t_3^4}{t_1^4} + \frac{10 t_2^5 t_3^4}{t_1^3} +\frac{2 t_2^6 t_3^4}{t_1^7} + \frac{t_2^6 t_3^4}{t_1^6} -\nn \\
&&  \frac{4 t_2^6 t_3^4}{t_1^5} + \frac{2 t_2^6 t_3^4}{t_1^4} - \frac{t_2^7 t_3^4}{t_1^7} - \frac{t_2^2 t_3^5}{t_1^2} + \frac{2 t_2^3 t_3^5}{t_1^5} - \frac{4 t_2^3 t_3^5}{t_1^4} + \frac{t_2^3 t_3^5}{t_1^3} + \frac{2 t_2^3 t_3^5}{t_1^2} + \frac{10 t_2^4 t_3^5}{t_1^6} - \frac{14 t_2^4 t_3^5}{t_1^5} -\frac{7 t_2^4 t_3^5}{t_1^4} + \frac{8 t_2^4 t_3^5}{t_1^3} -\nn \\
&&  \frac{t_2^4 t_3^5}{t_1^2} - \frac{t_2^5 t_3^5}{t_1^7} - \frac{8 t_2^5 t_3^5}{t_1^6} + \frac{10 t_2^5 t_3^5}{t_1^5} + \frac{8 t_2^5 t_3^5}{t_1^4} - \frac{5 t_2^5 t_3^5}{t_1^3} + \frac{4 t_2^6 t_3^5}{t_1^7} - \frac{10 t_2^6 t_3^5}{t_1^6} + \frac{6 t_2^6 t_3^5}{t_1^5} - \frac{t_2^6 t_3^5}{t_1^4} +\frac{2 t_2^7 t_3^5}{t_1^8} - \frac{5 t_2^7 t_3^5}{t_1^7} + \nn \\
&& \frac{4 t_2^7 t_3^5}{t_1^6} - \frac{t_2^3 t_3^6}{t_1^5} + \frac{2 t_2^3 t_3^6}{t_1^4} - \frac{t_2^3 t_3^6}{t_1^3} - \frac{t_2^4 t_3^6}{t_1^6} + \frac{2 t_2^4 t_3^6}{t_1^5} - \frac{t_2^4 t_3^6}{t_1^4} + \frac{5 t_2^5 t_3^6}{t_1^7} - \frac{16 t_2^5 t_3^6}{t_1^6} + \frac{9 t_2^5 t_3^6}{t_1^5} -\frac{2 t_2^6 t_3^6}{t_1^9} + \frac{9 t_2^6 t_3^6}{t_1^8} - \nn \\
&& \frac{20 t_2^6 t_3^6}{t_1^7} + \frac{25 t_2^6 t_3^6}{t_1^6} - \frac{10 t_2^6 t_3^6}{t_1^5} + \frac{4 t_2^7 t_3^6}{t_1^9} - \frac{22 t_2^7 t_3^6}{t_1^8} + \frac{24 t_2^7 t_3^6}{t_1^7} - \frac{8 t_2^7 t_3^6}{t_1^6} - \frac{2 t_2^8 t_3^6}{t_1^9} + \frac{8 t_2^8 t_3^6}{t_1^8} -\frac{ 4 t_2^8 t_3^6}{t_1^7} - \frac{3 t_2^5 t_3^7}{t_1^7} +\nn \\
&&  \frac{8 t_2^5 t_3^7}{t_1^6} - \frac{3 t_2^5 t_3^7}{t_1^5} + \frac{4 t_2^6 t_3^7}{t_1^9} - \frac{11 t_2^6 t_3^7}{t_1^8} + \frac{14 t_2^6 t_3^7}{t_1^7} - \frac{11 t_2^6 t_3^7}{t_1^6} + \frac{2 t_2^6 t_3^7}{t_1^5} - \frac{9 t_2^7 t_3^7}{t_1^9} + \frac{24 t_2^7 t_3^7}{t_1^8} - \frac{17 t_2^7 t_3^7}{t_1^7} + \frac{4 t_2^7 t_3^7}{t_1^6} -\nn \\
&& \frac{t_2^8 t_3^7}{t_1^{10}} + \frac{6 t_2^8 t_3^7}{t_1^9} - \frac{9 t_2^8 t_3^7}{t_1^8} + \frac{2 t_2^8 t_3^7}{t_1^7} - \frac{2 t_2^6 t_3^8}{t_1^9} + \frac{4 t_2^6 t_3^8}{t_1^8} - \frac{2 t_2^6 t_3^8}{t_1^7} - \frac{2 t_2^7 t_3^8}{t_1^{10}} + \frac{ 7 t_2^7 t_3^8}{t_1^9} - \frac{8 t_2^7 t_3^8}{t_1^8} +\frac{3 t_2^7 t_3^8}{t_1^7} + \frac{6 t_2^8 t_3^8}{t_1^{10}} -\nn \\
&&  \frac{8 t_2^8 t_3^8}{t_1^9} + \frac{3 t_2^8 t_3^8}{t_1^8} - \frac{2 t_2^9 t_3^8}{t_1^{10}} + \frac{t_2^9 t_3^8}{t_1^9} - \frac{t_2^8 t_3^9}{t_1^{10}} + \frac{t_2^9 t_3^9}{t_1^9}.
\eea}

Let $R_i$ be the charges corresponding to the fugacity $t_i$. Since the superpotential, which has fugacity $t_2 t_3 / t_1$, has to have R-charge equal to 2, the following condition must be imposed:
\bea
R_2 + R_3 - R_1 = 2~.
\eea
The volume of $\cE_2$ can be written as
{\footnotesize
\bea
\lim_{\mu \rightarrow 0} \mu^4 \gm (e^{-\mu R_1}, e^{-\mu R_2}, e^{-\mu (2 + R_1 - R_2)};\cE_2)= \frac{p\left(R_1, R_2; \cE_2\right)}{R^2_1 R^2_2 \left(2+R_1-R_2\right)^2\left(4-R_2\right)^2\left(4-2R_1+R_2\right)^2}~,\qquad \quad
\eea}
where
{\footnotesize
\bea
p\left(R_1, R_2; \cE_2\right) &=& 512 R_1 + 256 R_1^2 - 128 R_1^3 - 64 R_1^4 + 512 R_2 + 256 R_1^2 R_2 +  256 R_1^3 R_2+ 32 R_1^4 R_2 - \nn \\
&& 256 R_2^2 +320 R_1 R_2^2 - 320 R_1^2 R_2^2 -  112 R_1^3 R_2^2 - 16 R_1^4 R_2^2 - 64 R_2^3- 96 R_1 R_2^3 + \nn \\
&& 128 R_1^2 R_2^3 +32 R_1^3 R_2^3 + 32 R_2^4 - 22 R_1 R_2^4 - 23 R_1^2 R_2^4 +  2 R_2^5 + 6 R_1 R_2^5 - R_2^6~.\nn \\
\eea}
This function has a minimum at
\bea
R_1 \approx 1.614 , \qquad R_2 \approx 1.789, \qquad  R_3 \approx 1.825~. 
\eea
The R-charge of the perfect matching corresponding to the divisor $D_\alpha$ is given by
\bea
\lim_{\mu\rightarrow0}\frac{1}{\mu} \left[ \frac{g(D_\alpha; e^{-\mu R_1}, e^{-\mu R_2}, e^{-\mu R_3}; \cE_2) }{\gm(e^{-\mu R_1}, e^{-\mu R_2}, e^{-\mu R_3};\cE_2)}- 1 \right]~,
\eea
where $g(D_\alpha; e^{-\mu R_1}, e^{-\mu R_2}, e^{-\mu R_3}; \cE_2)$ is the Molien-Weyl integral with the insertion of the inverse of the weight corresponding to the divisor $D_\alpha$. The results are shown in Table \ref{t:chargefano275}.  The assignment of charges under the remaining abelian symmetries can be done by requiring that the superpotential is not charged under them and that the charge vectors are linearly independent. The assignments are shown in Table \ref{t:chargefano275}.

\begin{table}[h!]
 \begin{center}  
  \begin{tabular}{|c||c|c|c|c|c|c|c|c|}
  \hline
  \;& $SU(2)_1$&$U(1)_1$&$U(1)_2$&$U(1)_R$&$U(1)_{B_1}$&$U(1)_{B_2}$&$U(1)_{B_3}$&fugacity\\
  \hline\hline  
   
  $p_1$&$  1$&$  0$&$ 0$&$0.347$&$ 0$&$ 0$&$ 0$ & $s_1 x$\\
  \hline
  
  $p_2$&$ -1$&$  0$&$ 0$&$0.347$&$ 0$&$ 0$&$ 0$ & $s_1 / x$\\
  \hline  
  
  $r_1$&$  0$&$  1$&$ 0$&$0.304$&$ 0$&$ 0$&$ 0$ & $s_2 q_1$\\
  \hline
  
  $r_2$&$  0$&$ -1$&$ 0$&$0.245$&$ 0$&$ 0$&$ 0$ & $s_3 /q_1$\\
  \hline
   
  $r_3$&$  0$&$  0$&$ 1$&$0.234$&$ 0$&$ 0$&$ 0$ & $s_4 q_2$\\
  \hline
      
  $r_4$&$  0$&$  0$&$-1$&$0.359$&$ 1$&$ 0$&$ 0$ & $s_5 b_1/ q_2 $\\
  \hline
  
  $r_5$&$  0$&$  0$&$ 0$&$0.164$&$ 1$&$ 1$&$ 0$ & $s_6 b_1 b_2$\\
  \hline

  $r'_5$&$ 0$&$  0$&$ 0$&$    0$&$ 0$&$-1$&$ 0$ & $ 1/b_2 $\\
  \hline

  $v_1$&$  0$&$  0$&$ 0$&$    0$&$-2$&$ 0$&$ 1$ & $ b_3/ b^2_1$\\
  \hline
 
  $v_2$&$  0$&$  0$&$ 0$&$    0$&$ 0$&$ 0$&$-1$ & $ 1/b_3$\\
  \hline
 
     \end{tabular}
  \end{center}
\caption{Charges of the perfect matchings under the global symmetry of the $\cE_2$ model. Here $s_\alpha$ are the fugacities of the R-charges, $x$ is the weight of the $SU(2)$ symmetry, $q_1, q_2, b_1, b_2$ and $b_3$ are, respectively, the charges under the mesonic abelian symmetries $U(1)_1$, $U(1)_2$ and under the three baryonic $U(1)_{B_1}, U(1)_{B_2}$ and $U(1)_{B_3}$.}
\label{t:chargefano275}
\end{table}

\begin{table}[h]
 \begin{center}  
  \begin{tabular}{|c||c|}
  \hline
  \; Quiver fields & R-charge\\
  \hline  \hline 
  $X^1_{12}, X^2_{12}, X^1_{34}, X^2_{34}$ & 0.347 \\
  \hline
  $X^1_{23}$ & 0.468 \\
  \hline
  $X^2_{23}$ & 0.643 \\
  \hline
  $X_{41}$ & 0.663 \\
  \hline
  $X_{45}$ & 0.604 \\
  \hline
  $X_{51}$ & 0.234 \\
  \hline
  \end{tabular}
  \end{center}
\caption{R-charges of the quiver fields for the $\cE_2$ model.}
\label{t:Rquivfano275}
\end{table}

\paragraph{The Hilbert series.} The Hilbert series of the Master space can be obtained by integrating that of the space of perfect matchings over the $z_i$ fugacities:
{\footnotesize
\bea
g^{\firr{}}  (s_{\alpha}, x, q_1, q_2, b_1, b_2, b_3; \cE_2) &=& \left( \prod^3_{i=1} \oint \limits_{|z_i| =1} {\frac{\ud z_i}{2 \pi i z_i}} \right)\frac{1}{\left(1- s_1 x z_1\right)\left(1-\frac{s_1 z_1}{x}\right)\left(1 - s_2 q_1 z_2\right)\left(1-\frac{s_3 z_2 z_3}{q_1}\right)}\nn \\
&&\times \frac{1}{\left(1- \frac{s_4 q_2}{z_3}\right)\left(1-\frac{s_5 b_1}{q_2 z_2 z_3}\right)\left(1- \frac{s_6 b_1 b_2}{z_2}\right)\left(1-\frac{1}{z_1 b_2}\right)\left(1-\frac{b_3}{z_1 b^2_1}\right)\left(1-\frac{z_3}{b_3}\right)}\nn \\
&=& \frac{\left(1- \frac{s^2_1 b_3}{b^2_1 b_2}\right)}{\left(1-\frac{s_1 x}{b_2}\right)\left(1-\frac{s_1}{x b_2}\right)\left(1-	\frac{s_1 x b_3}{b^2_1}\right)\left(1-\frac{s_1 b_3}{x b^2_1}\right)\left(1-\frac{s_4 q_2}{b_3}\right)\left(1- \frac{s_2 s_5 q_1 b_1}{q_2 b_3}\right)}\nn \\
&&\times \frac{\left(1-\frac{s_2 s_3 s_4 s_5 s_6 b^2_1 b_2}{b_3}\right)}{\left(1-\frac{s_3 s_5 b_1}{q_1 q_2}\right)\left(1-\frac{s_3 s_4 s_6 q_2 b_1 b_2}{q_1}\right)\left(1- s_2 s_6 q_1 b_1 b_2\right)}~.
\label{e:HSMasterfano275}
\eea}
Integrating the Hilbert series of the Master space over the three baryonic fugacities, the Hilbert series of the mesonic moduli space is
{\footnotesize
\bea
\gm (s_{\alpha}, x, q_1, q_2; \cE_2) &=& \left( \prod^3_{j=1} \oint \limits_{|b_j| =1} {\frac{\ud b_j}{2 \pi i b_j}} \right) g^{\firr{}}  (s_{\alpha}, x, q_1, q_2, b_1, b_2, b_3;  \cE_2)\nn \\
&=& \frac{P\left(s_{\alpha}, x, q_1, q_2; \cE_2\right)}{\left(1- s^3_1 s^2_2 s_4 s^2_6 x^3 q^2_1 q_2\right)\left(1-\frac{s^3_1 s^2_2 s_4 s^2_6 q^2_1 q_2}{x^3}\right)\left(1-\frac{s^3_1 s^2_3 s^3_4 s^2_6 x^3 q^3_2}{q^2_1}\right)\left(1-\frac{s^3_1 s^2_3 s^3_4 s^2_6 q^3_2}{x^3 q^2_1}\right)}\nn \\
&\times& \frac{1}{\left(1-\frac{s^2_1 s^2_2 s_5 s_6 x^2 q^2_1}{q_2}\right)\left(1-\frac{s^2_1 s^2_2 s_5 s_6 q^2_1}{x^2 q_2}\right)\left(1-\frac{s_1 s^2_3 s_4 s^2_5 x}{q^2_1 q_2}\right)\left(1-\frac{s_1 s^2_3 s_4 s^2_5}{x q^2_1 q_2}\right)}\nn \\
&\times& \frac{1}{\left(1-\frac{s_1 s_2 s_3 s^2_5 x}{q^2_2}\right)\left(1-\frac{s_1 s_2 s_3 s^2_5}{x q^2_2}\right)},
\label{e:HSmesfano275}
\eea}
where $P\left(s_{\alpha}, x, q_1, q_2; \cE_2\right)$ is polynomial which is not reported here. The plethystic logarithm of (\ref{e:HSmesfano275}) can be written as:
\bea
\PL[\gm (t_1,t_2,t_3, x, q_1, q_2; \cE_2)] &=& [3] \left(\frac{q^2_1 q_2 t_2 t^2_3}{t^2_1}+\frac{q^2_2 t^2_2 t^2_3}{t^3_1}+\frac{q^3_2 t^3_2 t^2_3}{q^2_1 t^4_1}\right) + [2] \frac{t_2 t_3}{t_1}\nn \\
&&+ [2]\left(\frac{q^2_1 t_3}{q_2} + \frac{q_2 t^2_2 t_3}{q^2_1 t^2_1}\right) + [1]\left(\frac{t_1}{q^2_2}+\frac{t_2}{q^2_1 q_2}\right)\nn\\
&&- O(t_1)O(t_2)O(t_3)
\eea
Therefore, the generators of the mesonic moduli space are:
\bea
\begin{array}{llll}
 p_i p_j p_k r^2_1 r_3 r^2_5 r'^2_5 v_1 v_2, \quad & p_i p_j p_k r_1 r_2 r^2_3 r^2_5 r'^2_5 v_1 v_2, \quad  & p_i p_j p_k r^2_2 r^3_3 r^2_5 r'^2_5 v_1 v_2, \quad &p_i p_j r_1 r_2 r_3 r_4 r_5 r'_5 v_1 v_2,\nn \\
 p_i p_j r^2_1 r_4 r_5 r'_5 v_1 v_2, \quad & p_i p_j r^2_2 r^2_3 r_4 r_5 r'_5 v_1 v_2, \quad  & p_i r_1 r_2 r^2_4 v_1 v_2, \quad &  p_i r^2_2 r_3 r^2_4 v_1 v_2~.
 \end{array}
\eea
with $i,j,k=1,2$. The R-charges of these generators are listed in Table \ref{t:Rgenfanomms275}.

\begin{table}[h]
 \begin{center}  
  \begin{tabular}{|c||c|}
  \hline
  \; Generators & R-charge \\
  \hline  \hline 
  $p_i p_j p_k r^2_1 r_3 r^2_5 r'^2_5 v_1 v_2$ & 2.211 \\
  \hline
  $p_i p_j p_k r_1 r_2 r^2_3 r^2_5 r'^2_5 v_1 v_2$ & 2.386\\
  \hline
  $p_i p_j p_k r^2_2 r^3_3 r^2_5 r'^2_5 v_1 v_2$ &  2.561\\
  \hline
  $p_i p_j r_1 r_2 r_3 r_4 r_5 r'_5 v_1 v_2$ & 2\\
  \hline
  $p_i p_j r^2_1 r_4 r_5 r'_5 v_1 v_2$ & 1.825\\
  \hline
  $p_i p_j r^2_2 r^2_3 r_4 r_5 r'_5 v_1 v_2$ &  2.175\\
  \hline
  $p_i r_1 r_2 r^2_4 v_1 v_2$ & 1.614\\
  \hline
  $p_i r^2_2 r_3 r^2_4 v_1 v_2$ & 1.789\\
  \hline
  \end{tabular}
  \end{center}
\caption{R-charges of the generators of the mesonic moduli space for the $\cE_2$ model.}
\label{t:Rgenfanomms275}
\end{table}

\begin{figure}[ht]
\begin{center}
\includegraphics[totalheight=4cm]{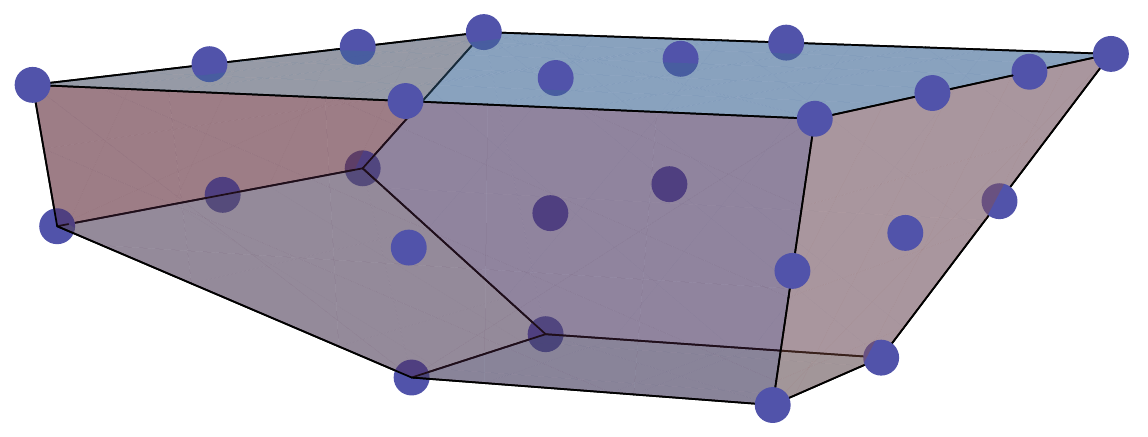}
\caption{The lattice of generators of the $\cE_2$ theory.}
  \label{f:latd2}
  \end{center}
\end{figure}

\section{$\cE_3$ (Toric Fano 266): $dP_2 \times \BP^1$}
This model has 5 gauge groups and bi-fundamental fields $X_{34}^i$, $X_{12}^i$, $X_{23}^i$, $X_{41}$, $X_{51}$, $X_{45}$ (with $i=1,2$). The tiling and quiver of this model coincide with those in Figure \ref{t:fano275tileandquiver}.

\comment{The tiling of this model is a `double bonding' of Phase I of the $\BF_0$ theory.
\begin{figure}[ht]
 \centerline{  \epsfxsize = 5cm \epsfbox{QuiverFano266.pdf}\hskip 8mm \epsfxsize = 6.5cm \epsfbox{fano266tilinggg.pdf}}
 \caption{(i) Quiver diagram of the $\cE_3$  model. (ii) Tiling of the $\cE_3$  model.}
  \label{f:fano266tileandquiver}
\end{figure} }

The superpotential is the same as (\ref{e:spotfano275})
Let us choose that CS levels to be $\vec{k} = (1,1,-1,0,-1)$.

\comment{
\begin{figure}[ht]
\begin{center}
\includegraphics[totalheight=7.5cm]{fano266tilingfd.pdf}
  \caption{The fundamental domain of the tiling for the $\cE_3$.}
  \label{t:fano266tileandquiver}
  \end{center}
 \end{figure} 
}

\paragraph{The Kasteleyn matrix.}   A parametrization of the Chern-Simons levels in terms of the integers $n^i_{jk}$ and $n_{jk}$ is given by:
\bea
\begin{array}{ll}
\text{Gauge group 1:} \qquad  k_1 &=   n^{1}_{12} + n^{2}_{12} - n_{41} - n_{51} ~, \nn \\
\text{Gauge group 2:} \qquad  k_2 &=   n^{1}_{23} + n^{2}_{23} - n^{1}_{12} - n^{2}_{12} ~, \nn \\
\text{Gauge group 3:} \qquad  k_3 &=   n^{1}_{34} + n^{2}_{34} - n^{1}_{23} - n^{2}_{23} ~,   \nn \\
\text{Gauge group 4:} \qquad  k_4 &=   n_{41} + n_{45} - n^{1}_{34} - n^{2}_{34} ~, \nn \\
\text{Gauge group 5:} \qquad  k_5 &=   n_{51} - n_{45} ~.
\label{e:kafano266}
\end{array}
\eea

Let us choose 
\bea
n^{2}_{23} =-n_{51} =   1,~ n^i_{jk}=n_{jk}=0 \; \text{otherwise}~.
\eea
The Kasteleyn matrix can now be constructed. The fundamental domain contains two black 
nodes and two white nodes and, therefore, the Kasteleyn matrix is a $2 \times 2$ matrix\footnote{The weight assignment is different from that chosen in \ref{e:kastfano275}. This will make the non-abelian part of the mesonic symmetry more evident in the $G_K$ matrix}: 

\bea
K= \left(
\begin{array}{c|cc}
& w_1 & w_2 \\
\hline
b_1 & z^{n^{1}_{34}} + \frac{1}{x} z^{n^{2}_{12}} &\ z^{n^{1}_{23}} + y z^{n_{45}} +  y z^{n_{51}}     \\
b_2 & z^{n^{2}_{23}} + \frac{1}{y} z^{n_{41}} &\ z^{n^{2}_{34}} + x z^{n^{1}_{12}}
\end{array}
\right)~.
\label{e:kastfano266}
\eea
The permanent of this matrix is given by
\bea
\perm(K) &=&  x {z^{(n^{1}_{12} + n^{1}_{34})}} +  x^{-1} z^{(n^{2}_{12} + n^{2}_{34})}+  y z^{(n^{2}_{23} + n_{51})}+  y^{-1} {z^{(n^1_{23} + n_{41})}}\nn \\
&+&  y z^{(n^2_{23} + n_{45})}+  z^{(n^{1}_{23} + n^{2}_{23})}+  z^{(n_{41} + n_{51})}+  z^{(n^{1}_{12} + n^{2}_{12})}\nn \\
&+& z^{(n^{1}_{34} + n^{2}_{34})}+  z^{(n_{45} + n_{41})}\nn \\
&=& x + x^{-1} +y + y^{-1}+ y z+  z + z^{-1}+ 3\qquad \nn \\
&&\text{(for $n^{2}_{23} =-n_{51} =   1,~ n^i_{jk}=n_{jk}=0\; \text{otherwise}$)} ~. \nn \\ 
\label{e:charpolyfano266}
\eea
The perfect matchings can be written in terms of the chiral fields as
\bea 
&&   p_1 = \left\{X^1_{12}, X^1_{34}\right\}, \;\; p_2 = \left\{X^2_{12}, X^2_{34}\right\}, \;\; q_1 = \left\{X^2_{23},X_{51}\right\}, \;\; r_1 = \left\{X_{41},X^1_{23}\right\}, \nn \\
&&   u_1 = \left\{X_{45}, X^2_{23}\right\}, \;\; q_2 = \left\{X^1_{23}, X^2_{23}\right\}, \;\; r_2 = \left\{X_{41},X_{51}\right\}, \;\; v_1 = \left\{X^1_{12},X^2_{12}\right\},\nn \\
&&  v_2 = \left\{X^1_{34}, X^2_{34}\right\}, \;\;v_3 = \left\{X_{41}, X_{45}\right\}\ . \qquad
\eea
Note that the perfect matchings $v_1, v_2$ and $v_3$ correspond to the internal point in the toric diagram.
In turn, the chiral fields can be written as products of perfect matchings as follows:
\bea
\begin{array}{lllll}
X^1_{12} = p_1 v_1, \quad  & X^2_{12} = p_2 v_1, \quad &  X^1_{34} = p_1 v_2, \quad & X^2_{34} = p_2 v_2, \quad & X_{45} = u_1 v_3,\nn \\
X^1_{23} = r_1 q_2, \quad &  X^2_{23} = q_1 u_1 q_2, \quad &  X_{41} = r_1 r_2 v_3, \quad & X_{51} = q_1 r_2~. 
\end{array}
\eea

\paragraph{The perfect matching.}  From \eref{e:charpolyfano266}, The relationship between the quiver fields and the perfect matchings is encoded in the $P$ matrix:
\beq
P=\left(\begin{array} {c|cccccccccc}
&p_1 & p_2 & q_1 & r_1 & u_1 & q_2 & r_2 & v_1 & v_2 & v_3\\
\hline
X^1_{12}  &1&0&0&0&0&0&0&1&0&0\\
X^2_{12}  &0&1&0&0&0&0&0&1&0&0\\
X^1_{34}  &1&0&0&0&0&0&0&0&1&0\\
X^2_{34}  &0&1&0&0&0&0&0&0&1&0\\
X_{45}    &0&0&0&0&1&0&0&0&0&1\\
X^1_{23}  &0&0&0&1&0&1&0&0&0&0\\
X^2_{23}  &0&0&1&0&1&1&0&0&0&0\\
X_{41}    &0&0&0&1&0&0&1&0&0&1\\
X_{51}    &0&0&1&0&0&0&1&0&0&0
\end{array}\right)~.
\label{e:pfano266}
\eeq
Basis vectors of the null space of $P$ are given in the rows of the charge matrix: 
\bea
Q_F= \left(
\begin{array}{cccccccccc}
 1 & 1 & 0 & 0 & 0 & 0 & 0 &-1 &-1 & 0 \\
 0 & 0 & 1 & 1 & 0 &-1 &-1 & 0 & 0 & 0 \\
 0 & 0 & 0 & 1 & 1 &-1 & 0 & 0 & 0 & -1
\end{array}
\right)~. \label{e:qffano266}
\eea
The relations between the perfect matchings are therefore given by
\bea \label{relperm266}
p_1 + p_2 - v_1 - v_2 &=& 0~, \nn \\
q_1 - q_2 + r_1 - r_2 &=& 0~, \nn \\
q_2 - r_1 - u_1 + v_3 &=& 0~.
\eea
Since the coherent component of the Master space is generated by the perfect matchings (subject to the relations \eref{relperm266}), it follows that 
\bea
\firr{} = \BC^{10}//Q_F~.
\eea

\paragraph{The toric diagram.} Two methods of constructing the toric diagram are demonstrated.
\begin{itemize}
 \item {\bf The charge matrices.}  Since this model has 5 gauge groups, there are 3 baryonic symmetries coming from the D-terms.  The charges of the perfect matchings are collected in the $Q_D$ matrix: 
\bea
Q_D = \left( \begin{array}{cccccccccc}
 0 & 0 & 0 & 0 & 0 & 1 & 1 &-2 & 0 & 0 \\
 0 & 0 & 0 & 0 & 0 & 0 & 0 & 1 & 0 &-1 \\
 0 & 0 & 0 & 0 & 0 & 0 & 0 & 0 & 1 &-1
 \end{array} \right) ~.
 \label{e:qdfano266}
\eea
Combining \eref{e:qffano266} and \eref{e:qdfano266}, the total charge matrix is given by
\bea
Q_t =  { \Blue Q_F \choose \Green Q_D \Black } = \left( \begin{array}{cccccccccc} \Blue
 1 & 1 & 0 & 0 & 0 & 0 & 0 &-1 &-1 & 0 \\
 0 & 0 & 1 & 1 & 0 &-1 &-1 & 0 & 0 & 0 \\
 0 & 0 & 0 & 1 & 1 &-1 & 0 & 0 & 0 & -1\\ \Green
 0 & 0 & 0 & 0 & 0 & 1 & 1 &-2 & 0 & 0 \\
 0 & 0 & 0 & 0 & 0 & 0 & 0 & 1 & 0 &-1 \\
 0 & 0 & 0 & 0 & 0 & 0 & 0 & 0 & 1 &-1\Black
 \end{array} \right) ~.
 \label{e:qfanddfano266}
\eea
The kernel of the total charge matrix gives the matrix $G_t$ which, after the removal of its first row, gives the coordinates of the points of the toric diagram in its columns:
\bea
G'_t = \left(
\begin{array}{cccccccccc}
 1 &-1 & 0 & 0 & 0 & 0 & 0 & 0 & 0 & 0 \\
 0 & 0 & 1 &-1 & 1 & 0 & 0 & 0 & 0 & 0 \\
 0 & 0 & 0 & 0 & 1 & 1 &-1 & 0 & 0 & 0
\end{array}
\right)~.
\eea
The toric diagram is shown in \fref{f:fano266toric}. Note that the 6 blue points form the toric diagram of the $dP_2$, and the 2 black points together with the internal point form the toric diagram of $\BP^1$. Thus, the mesonic moduli space of this theory is $dP_2\times \BP^1$.
\begin{figure}[ht]
\begin{center}
\includegraphics[totalheight=4.5cm]{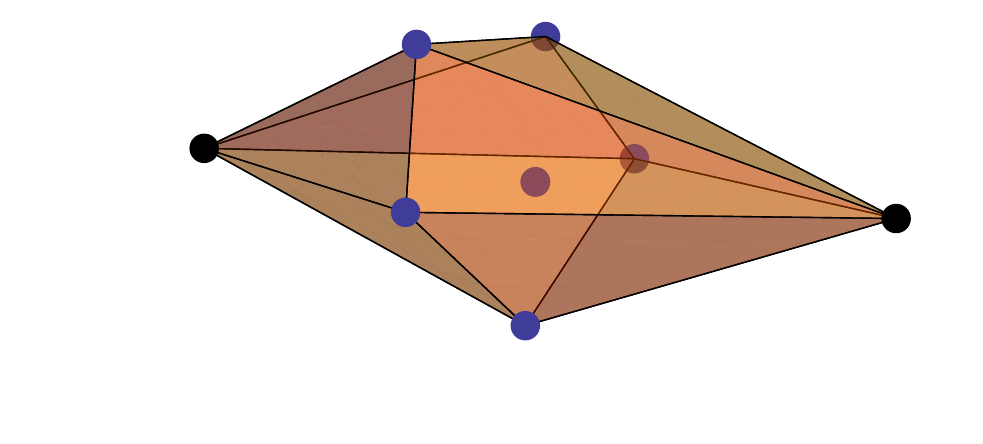}
  \caption{The toric diagram of the $\cE_3$ model.}
 \label{f:fano266toric}
 \end{center}
 \end{figure} 

\item {\bf The Kasteleyn matrix.} The powers of $x, y, z$ in each term of the permanent of the Kasteleyn matrix give the coordinates of each point in the toric diagram. The coordinates of each point in the toric diagram form columns of the following $G_K$ matrix: 
\bea
G_K = \left(
\begin{array}{cccccccccc}
 1 &-1 & 0 & 0 & 0 & 0 & 0 & 0 & 0 & 0 \\
 0 & 0 & 1 &-1 & 1 & 0 & 0 & 0 & 0 & 0 \\
 0 & 0 & 0 & 0 & 1 & 1 &-1 & 0 & 0 & 0
\end{array}
\right) = G'_t ~.
\eea
The first row of the Kasteleyn matrix contains the weights of the fundamental representation of $SU(2)$. Therefore, the mesonic moduli space contains one $SU(2)$ as the only non-abelian factor.
\end{itemize}

\paragraph{The baryonic charges.} Since the toric diagram has 7 external points, this model has precisely $7-4 = 3$ baryonic symmetries, which will be denoted by $U(1)_{B_1}, U (1)_{B_2}$ and $U(1)_{B_3}$. It can be seen that these symmetries arise from D-terms from the discussion above. Therefore, the baryonic charges of the perfect matchings are given by the rows of the $Q_D$ matrix.

\paragraph{The global symmetry.} The $Q_t$ matrix has one pair of repeated columns, thus confirming that the mesonic symmetry of this model is $SU(2) \times U(1)^3$.  From this it is also possible to see that $p_1$ and $p_2$ transform as a doublet under this $SU(2)$ symmetry.  Since the perfect matchings $v_1, v_2$ and $v_3$ correspond to the internal point of the toric diagram, they carry zero R-charge.
In order to compute the R-charge of each perfect matching, it is necessary to solve a volume minimisation problem first. Let us assign the R-charge fugacity $s_1$ to the perfect matchings $p_1$ and $p_2$ and R-charge fugacities $s_2, s_3, s_4, s_5$ and $s_6$ to the perfect matchings $q_1,r_1,u_1,q_2$ and $r_2$ respectively.  The Hilbert series of the mesonic moduli space is
\bea
\gm (s_{\alpha}; \cE_3) &=& \left( \prod^{3}_{i=1} \oint \limits_{|z_i|=1} {\frac{\ud z_i}{2 \pi i z_i}} \prod^{3}_{j=1} \oint \limits_{|b_j|=1} {\frac{\ud b_j}{2 \pi i b_j}} \right) \frac{1}{\left(1- s_1 z_1\right)^2\left(1- s_2 z_2\right)\left(1-s_3 z_2 z_3 \right)}\nn \\
&\times& \frac{1}{\left(1- s_4 z_3\right)\left(1- \frac{s_5 b_1}{z_2 z_3}\right)\left(1-\frac{s_6 b_1}{z_2}\right)\left(1-\frac{b_2}{b^2_1 z_1}\right)\left(1-\frac{b_3}{z_1}\right)\left(1-\frac{1}{b_2 b_3 z_3}\right)}~.\nn \\
\label{e:HSmesvolfano266}
\eea
Since there are 3 factors of $U(1)$ in the mesonic symmetry, the integral (\ref{e:HSmesvolfano266}) depends only on three combinations of perfect matchings. Choosing: 
\bea
t_1 = s^2_1 s^2_2 s_4 s^2_6, \quad t_2 = s^2_1 s_2 s_3 s^2_6, \quad t_3 = s^2_1 s^2_3 s_5 s_6~,
\eea
the mesonic Hilbert series can then be written as
{\small
\bea
\gm (t_1, t_2, t_3; \cE_3) &=& \frac{P\left(t_1, t_2, t_3; \cE_3\right)}{\left(1-t_1\right)^2\left(1-t_2\right)^2\left(1-t_3\right)^2\left(1-\frac{t_1 t^2_3}{t^2_2}\right)^2\left(1-\frac{t^3_1 t^2_3}{t^4_2}\right)^2}~, \quad
\eea}
where:
{\scriptsize
\bea
P \left(t_1, t_2, t_3; \cE_3\right) &=& 1+t_1+t_2-3 t_1 t_2+t_3-9 t_1 t_3+5 t_1^2 t_3-t_1^3 t_3-3 t_2 t_3+8 t_1 t_2 t_3-t_1^2 t_2 t_3+5 t_1 t_3^2- 5 t_1^2 t_3^2+2 t_1^3 t_3^2- \nn \\
&& t_1^2 t_2 t_3^2-t_1 t_3^3+2 t_1^2 t_3^3-t_1^3 t_3^3-t_1 t_2 t_3^2+\frac{3 t_1^2 t_3}{t_2^2}- \frac{t_1^3 t_3}{t_2^2}+\frac{3 t_1 t_3}{t_2}- \frac{7 t_1^2 t_3}{t_2}+\frac{2 t_1^3 t_3}{t_2}+ \frac{t_1^3 t_3^2}{t_2^4}-\frac{3 t_1^4 t_3^2}{t_2^4}+ \nn \\
&& \frac{3 t_1^2 t_3^2}{t_2^3}-\frac{9 t_1^3 t_3^2}{t_2^3}+\frac{8 t_1^4 t_3^2}{t_2^3}+\frac{t_1 t_3^2}{t_2^2}-\frac{15 t_1^2 t_3^2}{t_2^2}+\frac{7 t_1^3 t_3^2}{t_2^2}-\frac{t_1^4 t_3^2}{t_2^2}-\frac{9 t_1 t_3^2}{t_2}+\frac{22 t_1^2 t_3^2}{t_2}-\frac{5 t_1^3 t_3^2}{t_2}-\frac{t_1^5 t_3^3}{t_2^6}-\frac{t_1^6 t_3^3}{t_2^6}-\nn \\
&& \frac{t_1^4 t_3^3}{t_2^5}+\frac{t_1^5 t_3^3}{t_2^5}+\frac{2 t_1^6 t_3^3}{t_2^5}-\frac{9 t_1^3 t_3^3}{t_2^4}+\frac{9 t_1^4 t_3^3}{t_2^4}+ \frac{t_1^5 t_3^3}{t_2^4}-\frac{t_1^6 t_3^3}{t_2^4}-\frac{7 t_1^2 t_3^3}{t_2^3}+\frac{21 t_1^3 t_3^3}{t_2^3}-\frac{13 t_1^4 t_3^3}{t_2^3}-\frac{t_1^5 t_3^3}{t_2^3}-\frac{3 t_1 t_3^3}{t_2^2}+\nn \\
&&\frac{22 t_1^2 t_3^3}{t_2^2}-\frac{10 t_1^3 t_3^3}{t_2^2}+\frac{t_1^4 t_3^3}{t_2^2}+\frac{8 t_1 t_3^3}{t_2}-\frac{21 t_1^2 t_3^3}{t_2}+\frac{3 t_1^3 t_3^3}{t_2}-\frac{t_1^5 t_3^4}{t_2^7}+\frac{2 t_1^6 t_3^4}{t_2^7}-\frac{t_1^7 t_3^4}{t_2^7}-\frac{3 t_1^4 t_3^4}{t_2^6}+\frac{9 t_1^5 t_3^4}{t_2^6}+\frac{3 t_1^6 t_3^4}{t_2^6}-\nn \\
&& \frac{t_1^7 t_3^4}{t_2^6}-\frac{t_1^3 t_3^4}{t_2^5}+  \frac{9 t_1^4 t_3^4}{t_2^5}-\frac{11 t_1^5 t_3^4}{t_2^5}-\frac{5 t_1^6 t_3^4}{t_2^5}+\frac{7 t_1^3 t_3^4}{t_2^4}+\frac{12 t_1^4 t_3^4}{t_2^4}-\frac{13 t_1^5 t_3^4}{t_2^4}+\frac{2 t_1^6 t_3^4}{t_2^4}+\frac{5 t_1^2 t_3^4}{t_2^3}-\frac{10 t_1^3 t_3^4}{t_2^3}-\nn \\
&& \frac{11 t_1^4 t_3^4}{t_2^3}+ \frac{8 t_1^5 t_3^4}{t_2^3}-\frac{5 t_1^2 t_3^4}{t_2^2}+\frac{t_1^4 t_3^4}{t_2^2}+\frac{2 t_1^2 t_3^4}{t_2}+\frac{2 t_1^3 t_3^4}{t_2}+\frac{2 t_1^6 t_3^5}{t_2^8}+\frac{2 t_1^7 t_3^5}{t_2^8}+\frac{t_1^5 t_3^5}{t_2^7}-\frac{5 t_1^7 t_3^5}{t_2^7}+\frac{8 t_1^4 t_3^5}{t_2^6}-\frac{11 t_1^5 t_3^5}{t_2^6}- \nn \\
&&\frac{10 t_1^6 t_3^5}{t_2^6}+\frac{5 t_1^7 t_3^5}{t_2^6}+\frac{2 t_1^3 t_3^5}{t_2^5}-\frac{13 t_1^4 t_3^5}{t_2^5}+\frac{12 t_1^5 t_3^5}{t_2^5}+\frac{7 t_1^6 t_3^5}{t_2^5}-\frac{5 t_1^3 t_3^5}{t_2^4}-\frac{11 t_1^4 t_3^5}{t_2^4}+\frac{9 t_1^5 t_3^5}{t_2^4}-\frac{t_1^6 t_3^5}{t_2^4}-\frac{t_1^2 t_3^5}{t_2^3}+\nn \\
&& \frac{3 t_1^3 t_3^5}{t_2^3}+\frac{9 t_1^4 t_3^5}{t_2^3}-\frac{3 t_1^5 t_3^5}{t_2^3}-\frac{t_1^2 t_3^5}{t_2^2}+\frac{2 t_1^3 t_3^5}{t_2^2}-\frac{t_1^4 t_3^5}{t_2^2}-\frac{t_1^6 t_3^6}{t_2^9}+\frac{2 t_1^7 t_3^6}{t_2^9}-\frac{t_1^8 t_3^6}{t_2^9}+\frac{3 t_1^6 t_3^6}{t_2^8}-\frac{21 t_1^7 t_3^6}{t_2^8}+\frac{8 t_1^8 t_3^6}{t_2^8}+\nn \\
&&\frac{t_1^5 t_3^6}{t_2^7}-\frac{10 t_1^6 t_3^6}{t_2^7}+\frac{22 t_1^7 t_3^6}{t_2^7}-\frac{3 t_1^8 t_3^6}{t_2^7}-\frac{t_1^4 t_3^6}{t_2^6}-\frac{13 t_1^5 t_3^6}{t_2^6}+\frac{21 t_1^6 t_3^6}{t_2^6}-\frac{7 t_1^7 t_3^6}{t_2^6}-\frac{t_1^3 t_3^6}{t_2^5}+\frac{t_1^4 t_3^6}{t_2^5}+\frac{9 t_1^5 t_3^6}{t_2^5}-\frac{9 t_1^6 t_3^6}{t_2^5}+\nn \\
&& \frac{2 t_1^3 t_3^6}{t_2^4}+\frac{t_1^4 t_3^6}{t_2^4}-\frac{t_1^5 t_3^6}{t_2^4}-\frac{t_1^3 t_3^6}{t_2^3}-\frac{t_1^4 t_3^6}{t_2^3}-\frac{t_1^7 t_3^7}{t_2^{10}}-\frac{t_1^8 t_3^7}{t_2^{10}}+\frac{2 t_1^6 t_3^7}{t_2^9}- \frac{5 t_1^7 t_3^7}{t_2^9}+\frac{5 t_1^8 t_3^7}{t_2^9}-\frac{5 t_1^6 t_3^7}{t_2^8}+\frac{22 t_1^7 t_3^7}{t_2^8}-\nn \\
&& \frac{9 t_1^8 t_3^7}{t_2^8}-\frac{t_1^5 t_3^7}{t_2^7}+\frac{7 t_1^6 t_3^7}{t_2^7}-\frac{15 t_1^7 t_3^7}{t_2^7}+\frac{t_1^8 t_3^7}{t_2^7}+\frac{8 t_1^5 t_3^7}{t_2^6}-\frac{9 t_1^6 t_3^7}{t_2^6}+ \frac{3 t_1^7 t_3^7}{t_2^6}-\frac{3 t_1^5 t_3^7}{t_2^5}+\frac{t_1^6 t_3^7}{t_2^5}-\frac{t_1^7 t_3^8}{t_2^{10}}+\frac{8 t_1^8 t_3^8}{t_2^{10}}-\nn \\
&&\frac{3 t_1^9 t_3^8}{t_2^{10}}-\frac{t_1^6 t_3^8}{t_2^9}+\frac{5 t_1^7 t_3^8}{t_2^9}-\frac{9 t_1^8 t_3^8}{t_2^9}+\frac{t_1^9 t_3^8}{t_2^9}+\frac{2 t_1^6 t_3^8}{t_2^8}-\frac{7 t_1^7 t_3^8}{t_2^8}+\frac{3 t_1^8 t_3^8}{t_2^8}-\frac{t_1^6 t_3^8}{t_2^7}+\frac{3 t_1^7 t_3^8}{t_2^7}-\frac{3 t_1^8 t_3^9}{t_2^{10}}+\frac{t_1^9 t_3^9}{t_2^{10}}+\nn \\
&&\frac{t_1^8 t_3^9}{t_2^9}+\frac{t_1^9 t_3^9}{t_2^9}~.
\eea}
The superpotential, which has R-charge fugacity $t_1 t_3 / t_2$, must have R-charge equal to 2. Therefore, it must be imposed that
\bea
R_1 - R_2 + R_3 = 2,
\eea
where $R_i$ is the R-charge corresponding to the fugacity $t_i$.
The volume of the $\cE_3$ is given by
{\small
\bea
\lim_{\mu \rightarrow 0} \mu^4 \gm (e^{-\mu R_1}, e^{-\mu R_2}, e^{-\mu (2 + R_2 - R_1)}; \cE_3)= \frac{p\left(R_1, R_2; \cE_3\right)}{R^2_1 R^2_2\left(R_1 - 4\right)^2\left(4 + R_1 - 2R_2\right)^2\left(2 + R_2 - R_1\right)^2}~,\nn \\
\eea}
where
\bea
p\left(R_1, R_2; \cE_3\right) &=& 1024 R_1 - 512 R_1^2 - 128 R_1^3 + 64 R_1^4 + 4 R_1^5 - 2 R_1^6 + 1024 R_2 - 512 R_1 R_2+\nn \\
&& 640 R_1^2 R_2 - 128 R_1^3 R_2 - 44 R_1^4 R_2 + 10 R_1^5 R_2 + 512 R_2^2 - 320 R_1^2 R_2^2 +\nn \\
&& 160 R_1^3 R_2^2 - 26 R_1^4 R_2^2 - 256 R_2^3 + 384 R_1 R_2^3 - 160 R_1^2 R_2^3 + 32 R_1^3 R_2^3-\nn \\
&& 128 R_2^4 + 64 R_1 R_2^4 - 16 R_1^2 R_2^4~.
\eea
This function has a minimum at
\bea
R_1 = 2 , \qquad R_2 = R_3 \approx 1.824 ~.
\eea
The R-charge of the perfect matching corresponding to the divisor $D_\alpha$ is 
\bea
\lim_{\mu\rightarrow0}\frac{1}{\mu} \left[ \frac{g(D_\alpha; e^{- \mu R_1}, e^{- \mu R_2 }, e^{- \mu R_3 }; \cE_3) }{\gm(e^{-\mu R_1}, e^{- \mu R_2 }, e^{- \mu R_3 };\cE_3)}- 1 \right]~,
\eea
where $g(D_\alpha; e^{- \mu R_1}, e^{- \mu R_2 }, e^{- \mu R_3 }; \cE_3)$ is the Molien-Weyl integral with the insertion of the inverse of the weight corresponding to the divisor $D_\alpha$. The results are presented in Table \ref{t:chargefano266}.

The assignment of charges under the remaining abelian symmetries can be done by requiring that the superpotential is not charged under them and that the charge vectors are linearly independent. The assignments are shown in Table \ref{t:chargefano266}.

\begin{table}[h!]
 \begin{center}  
  \begin{tabular}{|c||c|c|c|c|c|c|c|c|}
  \hline
  \;& $SU(2)_1$&$U(1)_1$&$U(1)_2$&$U(1)_R$&$U(1)_{B_1}$&$U(1)_{B_2}$&$U(1)_{B_3}$&fugacity\\
  \hline\hline  
   
  $p_1$&$  1$&$  0$&$ 0$&$0.334$&$ 0$&$ 0$&$ 0$ & $s_1 x$\\
  \hline
  
  $p_2$&$ -1$&$  0$&$ 0$&$0.334$&$ 0$&$ 0$&$ 0$ & $s_1 / x$\\
  \hline  
  
  $q_1$&$  0$&$  1$&$ 0$&$0.226$&$ 0$&$ 0$&$ 0$ & $s_2 q_1$\\
  \hline
  
  $q_2$&$  0$&$ -1$&$ 0$&$0.226$&$ 1$&$ 0$&$ 0$ & $s_2 b_1/q_1$\\
  \hline
   
  $r_1$&$  0$&$  0$&$ 1$&$0.310$&$ 0$&$ 0$&$ 0$ & $s_3 q_2$\\
  \hline
      
  $r_2$&$  0$&$  0$&$ 1$&$0.310$&$ 1$&$ 0$&$ 0$ & $s_3 b_1  q_2 $\\
  \hline
  
  $u_1$&$  0$&$  0$&$-2$&$0.260$&$ 0$&$ 0$&$ 0$ & $s_4 / q^2_2$\\
  \hline

  $v_1$&$  0$&$  0$&$ 0$&$    0$&$-2$&$ 1$&$ 0$ & $ b_2 / b^2_1$\\
  \hline

  $v_2$&$  0$&$  0$&$ 0$&$    0$&$ 0$&$ 0$&$ 1$ & $  b_3 $\\
  \hline
 
  $v_3$&$  0$&$  0$&$ 0$&$    0$&$ 0$&$-1$&$-1$ & $ 1 / (b_2 b_3)$\\
  \hline
 
     \end{tabular}
  \end{center}
\caption{Charges of the perfect matchings under the global symmetry of the $\cE_3$ model. Here $s_\alpha$ are the fugacities of the R-charges, $x$ is the weight of the $SU(2)$ symmetry, $q_1, q_2, b_1, b_2$ and $b_3$ are, respectively, the charges under the mesonic abelian symmetries $U(1)_1$, $U(1)_2$ and under the three baryonic $U(1)_{B_1}, U(1)_{B_2}$ and $U(1)_{B_3}$.}
\label{t:chargefano266}
\end{table}

\begin{table}[h]
 \begin{center}  
  \begin{tabular}{|c||c|}
  \hline
  \; Quiver fields &R-charge\\
  \hline  \hline 
  $ X^i_{12}, X^i_{34}$ & 0.334 \\
  \hline
  $ X_{51}, X^1_{23}$ & 0.536 \\
  \hline
  $ X^2_{23}$ & 0.712 \\
  \hline
  $ X_{45}$ & 0.26 \\
  \hline
  $ X_{41}$ & 0.62 \\
  \hline
  \end{tabular}
  \end{center}
\caption{R-charges of the quiver fields for the $\cE_3$ model.}
\label{t:Rquivfano266}
\end{table}

\paragraph{The Hilbert series.} The Hilbert series of the Master space can be computed by integrating the Hilbert series of the space of perfect matching over the $z_i$ fugacities:
{\small
\bea
g^{\firr{}}  (s_{\alpha}, x, q_1, q_2, b_1, b_2, b_3; \cE_3) &=& \left( \prod^3_{i=1} \oint \limits_{|z_i| =1} {\frac{\ud z_i}{2 \pi i z_i}} \right) \frac{1}{\left(1- s_1 x z_1\right)\left(1-\frac{s_1 z_1}{x}\right)\left(1- s_2 q_1 z_2\right)\left(1-\frac{s_2 b_1}{q_1 z_2 z_3}\right)}\nn \\
&&\times \frac{1}{\left(1- s_3 q_2 z_2 z_3\right)\left(1-\frac{s_3 q_2 b_1}{z_2}\right)\left(1- \frac{s_4 z_3}{q^2_2} \right)\left(1-\frac{b_2}{b^2_1 z_1}\right)\left(1-\frac{b_3}{z_1}\right)\left(1-\frac{1}{b_2 b_3 z_3}\right)}\nn \\
&=& \frac{\left(1- \frac{s^2_2 s^2_3 s_4 b^2_1}{b_2 b_3}\right)}{\left(1- s_2 s_3 q_1 q_2 b_1\right)\left(1-\frac{s_2 s_3 q_2 b_1}{q_1}\right)\left(1-\frac{s^2_2 s_4 b_1}{q^2_2}\right)\left(1-\frac{s_4}{q^2_2 b_2 b_3}\right)\left(1-\frac{s^2_3 q^2_2 b_1}{b_2 b_3}\right)}\nn \\
&&\times \frac{\left(1-\frac{s^2_1 b_2 b_3}{b^2_1}\right)}{\left(1- s_1 x b_3\right)\left(1-\frac{s_1 b_3}{x}\right)\left(1-\frac{s_1 x b_2}{b^2_1}\right)\left(1-\frac{s_1 b_2}{x b^2_1}\right)}~.
\label{e:HSMasterfano266}
\eea}
Integrating the Hilbert series of the Master space over the baryonic fugacities yields the Hilbert series of the mesonic moduli space:
{\small
\bea
\gm (s_{\alpha}, x, q_1, q_2; \cE_3) &=& \left( \prod^3_{j=1} \oint \limits_{|b_j| =1} {\frac{\ud b_j}{2 \pi i b_j}}  \right) g^{\firr{}}  (s_{\alpha}, x, q_1, q_2, b_1, b_2, b_3; \cE_3)\nn \\
&=& \frac{P\left(s_{\alpha}, x, q_1, q_2; \cE_3\right)}
{\left(1-s^2_1 s^2_2 s^2_3 s_4 x^2 q^2_1\right)\left(1-\frac{s^2_1 s^2_2 s^2_3 s_4 q^2_1}{x^2}\right)\left(1-\frac{s^2_1 s^2_2 s^2_3 s_4 x^2}{q^2_1}\right)\left(1-\frac{s^2_1 s^2_2 s^2_3 s_4}{x^2 q^2_1}\right)}\nn \\
&\times& \frac{1}{\left(1-s^2_1 s_2 s^3_3 x^2 q_1 q^3_2\right)\left(1-\frac{s^2_1 s_2 s^3_3 q_1 q^3_2}{x^2}\right)\left(1-\frac{s^2_1 s_2 s^3_3 q^3_2 x^2}{q_1}\right)\left(1-\frac{s^2_1 s_2 s^3_3 q^3_2}{x^2 q_1}\right)}\nn \\
&\times& \frac{1}{\left(1-\frac{s^2_1 s^4_2 s^3_4 x^2}{q^6_2}\right)\left(1-\frac{s^2_1 s^4_2 s^3_4}{x^2 q^6_2}\right)}~,
\label{e:HSmesfano266}
\eea}
where $P\left(s_{\alpha}, x, q_1, q_2; \cE_3\right)$ is a polynomial which is not reported here. The plethystic logarithm of (\ref{e:HSmesfano266}) can be written as:
\bea
\PL[\gm (t_{\alpha}, x, q_1, q_2; \cE_3)] &=& [2]\left(q_1 + \frac{1}{q_1}\right)\left(q^3_2 t_2 + \frac{t^2_1}{q^3_2 t_2}\right) + [2]\left(q^2_1 + 1 + \frac{1}{q^2_1}\right) t_1\nn \\
&&+  [2]\frac{t^3_1}{q^6_2 t^2_2}  - O(t^2_1)~. \nn \\
\label{e:PLfano266}
\eea
From (\ref{e:PLfano266}), it can be seen that the abelian symmetry $U(1)_1$ is enhanced to $SU(2)$ in the mesonic moduli space.

\paragraph{The generators.}
The generators of the mesonic moduli space are
\bea
&& p_i p_j q_k r_k r^2_l v_1 v_2 v_3, \quad p_i p_j q^2_k q_l r_l u^2_1 v_1 v_2 v_3, \quad p_i p_j q^2_k r^2_l u_1 v_1 v_2 v_3, \nn \\
&&  p_i p_j q_1 q_2 r_1 r_2 u_1 v_1 v_2 v_3,\quad p_i p_j q^2_1 q^2_2 u^3_1 v_1 v_2 v_3.
\eea
with $i,j,k,l=1,2$ and $k\neq l$.
The R-charges of the generators are listed in Table \ref{t:Rgenfano266}.

\begin{table}[h]
 \begin{center}  
  \begin{tabular}{|c||c|}
  \hline
  \; Generators & R-charge \\
  \hline  \hline 
  $p_i p_j q_k r_k r^2_l v_1 v_2 v_3$ & 1.824 \\
  \hline
  $p_i p_j q^2_k q_l r_l u^2_1 v_1 v_2 v_3$ & 2.176 \\
  \hline
  $p_i p_j q^2_k r^2_l u_1 v_1 v_2 v_3$ & 2 \\
  \hline
  $p_i p_j q_1 q_2 r_1 r_2 u_1 v_1 v_2 v_3$ & 2 \\
  \hline
  $p_i p_j q^2_1 q^2_2 u^3_1 v_1 v_2 v_3$ & 2.352 \\
  \hline
  \end{tabular}
  \end{center}
\caption{R-charges of the generators of the mesonic moduli space for the $\cE_3$ theory.}
\label{t:Rgenfano266}
\end{table}

\begin{figure}[ht]
\begin{center}
\includegraphics[totalheight=4cm]{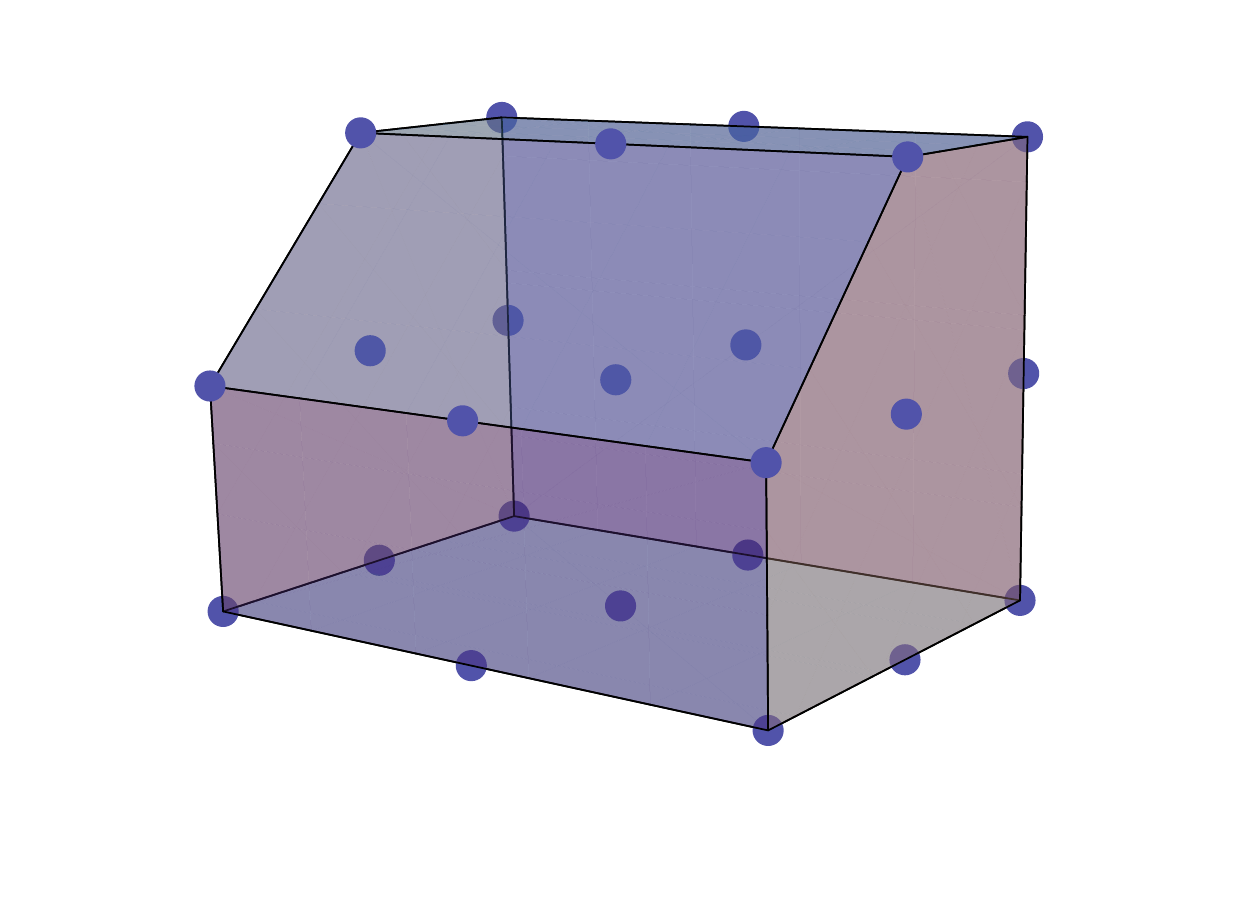}
\caption{The lattice of generators of the $\cE_3$ model.}
  \label{f:latd3}
  \end{center}
\end{figure}

\section{$\cE_4$ (Toric Fano 271): $dP_2$ bundle over $\BP^1$}
This theory has 9 chiral fields: $X^i_{12}$, $X^i_{23}$, $X^i_{41}$ (with $i=1,2$), $X_{35}$, $X_{54}$ and $X_{34}$. The quiver diagram and the tiling coincide with those of Figure \ref{t:fano275tileandquiver}. The superpotential can be read from (\ref{e:spotfano275}). However, for this model let us choose the CS levels to be $\vec{k}=(1,-1,0,-1,1)$.


\comment{
\begin{figure}[ht]
\begin{center}
   \includegraphics[totalheight=7cm]{fano271tilingfd.pdf}
 \caption{The fundamental domain of the tiling for the $\cE_4$ model.}
  \label{f:fdtoricfano271}
\end{center}
\end{figure}
}

\paragraph{The Kasteleyn matrix.} The Chern-Simons levels for this model can be written as:
\bea
\begin{array}{ll}
\text{Gauge group 1:} \qquad  k_1 &=   n^{1}_{12} + n^{2}_{12} - n_{41} - n_{51} ~, \nn \\
\text{Gauge group 2:} \qquad  k_2 &=   n^{1}_{23} + n^{2}_{23} - n^{1}_{12} - n^{2}_{12} ~, \nn \\
\text{Gauge group 3:} \qquad  k_3 &=   n^{1}_{34} + n^{2}_{34} - n^{1}_{23} - n^{2}_{23} ~,   \nn \\
\text{Gauge group 4:} \qquad  k_4 &=   n_{41} + n_{45} - n^{1}_{34} - n^{2}_{34} ~, \nn \\
\text{Gauge group 5:} \qquad  k_5 &=   n_{51} - n_{45} ~,
\label{e:kafano271}
\end{array}
\eea

with the choice:
\bea
n^2_{12}=-n_{45}= 1,\qquad n^i_{jk}=n_{jk}=0 \;\;\text{otherwise}~.
\eea
The fundamental domain of the tiling contains two white nodes and two black nodes, thus the Kasteleyn matrix is a $2 \times 2$ matrix and can be written as:
\bea
K =   \left(
\begin{array}{c|cc}
& w_1 & w_2\\
\hline
b_1 & z^{n^{1}_{34}} + x z^{n^{2}_{12}} &\ z^{n^{1}_{23}} + y z^{n_{45}} + y z^{n_{51}}   \\
b_2 & z^{n^{2}_{23}} + \frac{1}{y} z^{n_{41}} &\ z^{n^{2}_{34}} + \frac{1}{x} z^{n^{1}_{12}}
\end{array}
\right) ~.
\label{e:kastfano271}
\eea
The permanent of the Kasteleyn matrix is a generating function for the perfect matchings and can be written as:
\bea
\mathrm{perm}(K) &=&  x z^{(n^2_{12} + n^2_{34})}+  x^{-1} z^{(n^1_{12} + n^1_{34})}+  z^{(n^1_{12} + n^2_{12})}+  z^{(n_{41} + n_{45})}\nn \\
&+&  y z^{(n^2_{23} + n_{45})}+  y z^{(n_{51} + n^2_{23})} +  y^{-1} z^{(n_{41} + n^1_{23})}+  z^{(n_{51} + n_{41})}\nn \\
&+&  z^{(n^1_{23} + n^2_{23})}+  z^{(n^1_{34} + n^2_{34})}\nn \\
&=&  x + x^{-1} z+  z+  z^{-1} + y z^{-1} + y + y^{-1} + 3 \nn \\
&&\text{(for $n^2_{12}=-n_{45}= 1,\qquad n^i_{jk}=n_{jk}=0 \; \text{otherwise}$)}.
\label{e:charpolyfano271}
\eea
The perfect matchings can be written in terms of the chiral fields as:
\bea 
&& p_1 = \left\{X^2_{12}, X^2_{34}\right\}, \;\; p_2 = \left\{X^1_{12}, X^1_{34}\right\}, \;\; r_1 = \left\{X^1_{12}, X^2_{12}\right\}, \;\; r_2 = \left\{X_{41}, X_{45}\right\},\nn \\ 
&& r_3 = \left\{X^2_{23}, X_{45}\right\}, \;\; r_4 = \left\{X_{51}, X^2_{23}\right\}, \;\; r_5 = \left\{X_{41}, X^1_{23}\right\}, \;\; v_1 = \left\{X_{51}, X_{41}\right\},\nn \\
&& v_2 = \left\{X^1_{23}, X^2_{23}\right\}, \;\; v_3 = \left\{X^1_{34}, X^2_{34}\right\}\ . \qquad
\eea
In turn, the chiral fields can be written as products of perfect matchings as:
\bea
\begin{array}{lll}
X^2_{34} = p_1 v_3, \quad & X^1_{34} = p_2 v_3, \quad & X^1_{23} = r_5 v_2, \nn \\
X_{51} = r_4 v_1, \quad &  X^2_{23} = r_3 r_4 v_2, \quad  & X_{45} = r_2 r_3,\nn \\
X_{41} = r_2 r_5 v_1, \quad &  X^2_{12} = p_1 r_1 ,   \quad  & X^1_{12} = p_2 r_1 ~.
\end{array}
\eea
These pieces of information can be collected in the $P$ matrix:
\beq
P=\left(\begin{array} {c|cccccccccc}
  \;& p_1 & p_2 & r_1 & r_2 & r_3 & r_4 & r_5 & v_1 & v_2 & v_3\\
  \hline 
X^{1}_{34}  & 1&0&0&0&0&0&0&0&0&1\\
X^{2}_{34}  & 0&1&0&0&0&0&0&0&0&1\\
X^{1}_{23}  & 0&0&0&0&0&0&1&0&1&0\\
X_{51}      & 0&0&0&0&0&1&0&1&0&0\\
X^{2}_{23}  & 0&0&0&0&1&1&0&0&1&0\\
X_{45}      & 0&0&0&1&1&0&0&0&0&0\\
X_{41}      & 0&0&0&1&0&0&1&1&0&0\\
X^{1}_{12}  & 1&0&1&0&0&0&0&0&0&0\\
X^{2}_{12}  & 0&1&1&0&0&0&0&0&0&0\\
  \end{array}
\right).
\label{e:pmatrifano271}
\eeq
The kernel of the $P$ matrix is given by:
\be
Q_F =   \left(
\begin{array}{cccccccccc}
 1 & 1 &-1 & 0 &  0 & 0 &  0 &  0 &  0 & -1\\
 0 & 0 & 0 & 1 & -1 & 0 & -1 &  0 &  1 &  0\\
 0 & 0 & 0 & 0 &  0 & 1 &  1 & -1 & -1 &  0
\end{array}
\right)~.  
\label{e:qffano271}
\ee
Thus, the relations among the perfect matchings are:
\bea
p_1 + p_2 - r_1 - v_3 &=& 0~,\nn \\
r_2 - r_3 - r_5 + v_2 &=& 0~, \nn \\
r_4 + r_5 - v_1 - v_2 &=& 0~.
\label{e:relpmfano271}
\eea

\paragraph{The toric diagram.} The toric diagram is constructed using two different methods.
\begin{itemize}
\item {\bf The Kasteleyn matrix.} The powers of $x$, $y$ and $z$ in each of the terms in \eref{e:charpolyfano271} are the coordinates of the toric diagram in the following matrix:
\bea
\left(
\begin{array}{cccccccccc}
  1 & -1 &  0 &  0 & 0 & 0 &  0 & 0 & 0 & 0 \\
  0 &  0 &  0 &  0 & 1 & 1 & -1 & 0 & 0 & 0 \\
  0 &  1 &  1 & -1 &-1 & 0 &  0 & 0 & 0 & 0 
\end{array}
\right)~.
\label{e:Gkfano271}
\eea
Multiplying this matrix on the left by {\footnotesize $\left( \begin{array}{ccc} 1&0&0\\0&0&1\\0&1&0 \end{array} \right) \in GL(3, \BZ)$} gives us
\bea
G_K = \left(
\begin{array}{cccccccccc}
  1 & -1 &  0 &  0 & 0 & 0 &  0 & 0 & 0 & 0 \\
  0 &  1 &  1 & -1 &-1 & 0 &  0 & 0 & 0 & 0 \\
  0 &  0 &  0 &  0 & 1 & 1 & -1 & 0 & 0 & 0 
\end{array}
\right)~.
\label{e:Gkpfano271}
\eea
The first row of \eqref{e:Gkpfano271} contains the weights of the fundamental representation of $SU(2)$. Therefore the mesonic symmetry contains one $SU(2)$ as the non-abelian factor.

\item {\bf The charge matrices.} Since the model has 5 gauge groups, there are 3 baryonic symmetries coming from the D-terms. The charges of the perfect matchings are collected in the columns of the following matrix:
\be
Q_D =   \left(
\begin{array}{cccccccccc}
0 & 0 & 1 & 1 & 0 & 0 & 0 &  0 & 0 & -2\\
0 & 0 & 0 & 0 & 0 & 0 & 0 &  1 &-1 &  0\\
0 & 0 & 0 & 0 & 0 & 0 & 0 &  0 & 1 & -1
\end{array}
\right) 
\label{e:qdfano271}
\ee
The total charge matrix $Q_t$ is then given by:
\be
Q_t = { \Blue Q_F \choose \Green Q_D \Black } =   \left( 
\begin{array}{cccccccccc} \Blue
1 & 1 &-1 & 0 &  0 & 0 &  0 &  0 &  0 & -1\\
0 & 0 & 0 & 1 & -1 & 0 & -1 &  0 &  1 &  0\\
0 & 0 & 0 & 0 &  0 & 1 &  1 & -1 & -1 &  0\\ \Green
0 & 0 & 1 & 1 & 0 & 0 & 0 &  0 & 0 & -2\\
0 & 0 & 0 & 0 & 0 & 0 & 0 &  1 &-1 &  0\\
0 & 0 & 0 & 0 & 0 & 0 & 0 &  0 & 1 & -1\Black
\end{array}
\right)~. 
\label{e:qtfano271}
\ee
The kernel of this matrix gives the $G_t$ matrix which, after the removal of its first row, yields the $G'_t$ matrix, whose columns gives the coordinates of the toric diagram:
\bea
G'_t = \left(
\begin{array}{cccccccccc}
  1 & -1 &  0 &  0 & 0 & 0 &  0 & 0 & 0 & 0 \\
  0 &  1 &  1 & -1 &-1 & 0 &  0 & 0 & 0 & 0 \\
  0 &  0 &  0 &  0 & 1 & 1 & -1 & 0 & 0 & 0 
\end{array}
\right)= G_K~. \label{e:toricdiafano271}
\eea
	
The toric diagram constructed from (\ref{e:toricdiafano271}) is drawn in Figure \ref{f:tdtoricfano271}.
\begin{figure}[ht]
\begin{center}
\includegraphics[totalheight=3.0cm]{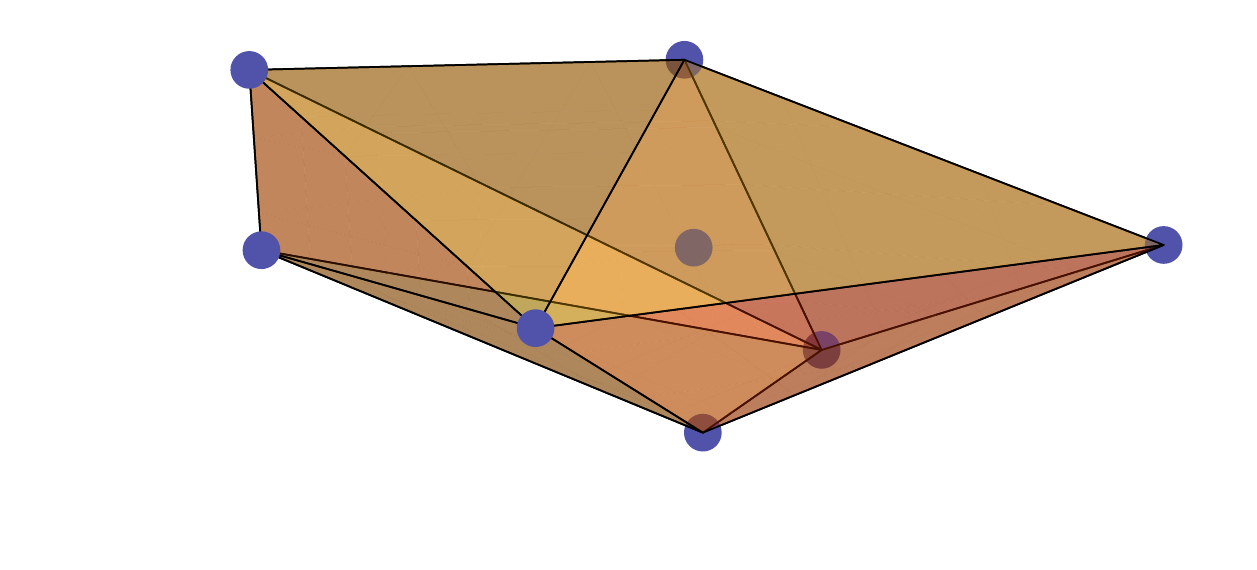}
\caption{The toric diagram of $\cE_4$.}
\label{f:tdtoricfano271}
\end{center}
\end{figure}

\end{itemize}

\paragraph{The baryonic charges.} From Figure \ref{f:tdtoricfano271} it can be seen that the toric diagram of this model contains 7 external points. Hence, the total number of baryonic symmetries is 3. These come from the D-terms and, henceforth, they will be referred to as $U(1)_{B_1}, U(1)_{B_2}$ and $U(1)_{B_3}$.

\paragraph{The global symmetry.} The $Q_t$ matrix has only one pair of repeated columns, confirming that the mesonic symmetry of this model has one $SU(2)$ as the non-abelian factor.  Since the mesonic symmetry has total rank of is 4, it can be identified with $SU(2)\times U(1)^3$, where one of the abelian factors corresponds to the R-symmetry. The perfect matchings $p_1$ and $p_2$ transform as a doublet under the $SU(2)$. The perfect matchings $v_1, v_2$ and $v_3$ correspond to internal points in the toric diagram and so they carry zero R-charges. 

In order to determine the R-charge of a perfect matching, it is necessary to solve a volume minimisation problem first.  Let us assign the R-charge fugacity $s_1$ to the perfect matchings $p_1$ and $p_2$ (note that the non-abelian symmetry does not play any role in the volume minimisation, so $p_1$ and $p_2$ carry the same R-charges), and R-charge fugacities $s_2, s_3, s_4, s_5$ and $s_6$ to the perfect matchings $r_1, r_2, r_3, r_4$ and $r_5$ respectively. The Hilbert series of the mesonic moduli space is given by:
\bea
\gm (s_{\alpha}; \cE_4) &=& \left( \prod^3_{i=1}\oint\limits_{|b_i|=1}  \frac{\ud b_i}{2\pi i b_i} \prod^3_{j=1} \oint\limits_{|z_j|=1} \frac{\ud z_j}{2 \pi i z_j} \right) \frac{1}{\left(1- s_1 z_1\right)^2\left(1- \frac{s_2 b_1}{z_1}\right)\left(1-s_3 b_1 z_2\right)\left(1- \frac{s_4}{z_2}\right)}\nn \\
&& \times \frac{1}{\left(1- s_5 z_3\right)\left(1-\frac{s_6 z_3}{z_2}\right)\left(1-\frac{ b_2}{z_3}\right)\left(1-\frac{ b_3 z_2}{b_2 z_3}\right)\left(1-\frac{1}{b^2_1 b_3 z_1}\right)}~.
\label{e:HSmesvolfano271}
\eea
Since there are three $U(1)$ factors in the mesonic symmetry, the integral \eref{e:HSmesvolfano271} depends only on 3 combinations of $s_\alpha$'s. In particular, choosing:
\bea
t^2_1 = s^2_1 s_2 s_3 s^2_6, \quad t^2_2 = s_1 s^2_3 s^3_4 s^2_5, \quad t^2_3 = s_1 s^2_3 s_4 s^2_6~,
\eea
the mesonic Hilbert series can be written as:
\bea
\gm (t_1, t_2, t_3; \cE_4) &=& \frac{P\left(t_1, t_2, t_3; \cE_4\right)}{\left(1- t^2_1\right)^2\left(1-t^2_2\right)^2\left(1-t^2_3\right)^2\left(1-\frac{t^4_1 t_2}{t^3_3}\right)^2\left(1-\frac{t^4_1  t^2_2}{t^4_3}\right)^2}~, \qquad \quad 
\eea
where:
{\scriptsize
\bea
P\left(t_1, t_2, t_3; \cE_4\right) &=& 1+t_1^2-8 t_1^2 t_2^2+8 t_1^4 t_2^2-6 t_1^6 t_2^2+2 t_1^8 t_2^2+4 t_1^2 t_2^4-3 t_1^4 t_2^4+t_1^6 t_2^4+2 t_2 t_3-6 t_1^2 t_2 t_3+ 6 t_1^4 t_2 t_3-2 t_1^6 t_2 t_3-\nn \\
&& t_2^3 t_3+3 t_1^2 t_2^3 t_3-3 t_1^4 t_2^3 t_3+t_1^6 t_2^3 t_3-2 t_1^2 t_3^2-t_2^2 t_3^2+6 t_1^2 t_2^2 t_3^2-t_1^4 t_2^2 t_3^2-2 t_1^2 t_2^4 t_3^2-t_2 t_3^3+2 t_1^2 t_2 t_3^3-\nn \\
&& t_1^4 t_2 t_3^3-\frac{t_1^{14} t_2^8}{t_3^{12}}+\frac{2 t_1^{16} t_2^8}{t_3^{12}}- \frac{t_1^{18} t_2^8}{t_3^{12}}-\frac{2 t_1^{16} t_2^5}{t_3^{11}}-\frac{t_1^{14} t_2^7}{t_3^{11}}+\frac{6 t_1^{16} t_2^7}{t_3^{11}}-\frac{t_1^{18} t_2^7}{t_3^{11}}-\frac{2 t_1^{16} t_2^9}{t_3^{11}}+\frac{t_1^{12} t_2^6}{t_3^{10}}-\frac{3 t_1^{14} t_2^6}{t_3^{10}}+\nn \\
&&\frac{3 t_1^{16} t_2^6}{t_3^{10}}-\frac{t_1^{18} t_2^6}{t_3^{10}}-\frac{2 t_1^{12} t_2^8}{t_3^{10}}+\frac{6 t_1^{14} t_2^8}{t_3^{10}}-\frac{6 t_1^{16} t_2^8}{t_3^{10}}+\frac{2 t_1^{18} t_2^8}{t_3^{10}}+\frac{t_1^{12} t_2^5}{t_3^9}-\frac{3 t_1^{14} t_2^5}{t_3^9}+\frac{4 t_1^{16} t_2^5}{t_3^9}+\frac{2 t_1^{10} t_2^7}{t_3^9}-\frac{6 t_1^{12} t_2^7}{t_3^9}+\nn \\
&&\frac{8 t_1^{14} t_2^7}{t_3^9}-\frac{8 t_1^{16} t_2^7}{t_3^9}+\frac{t_1^{16} t_2^9}{t_3^9}+\frac{t_1^{18} t_2^9}{t_3^9}+\frac{3 t_1^{10} t_2^4}{t_3^8}-\frac{t_1^{12} t_2^4}{t_3^8}-\frac{12 t_1^{12} t_2^6}{t_3^8}+ \frac{10 t_1^{14} t_2^6}{t_3^8}-\frac{2 t_1^{16} t_2^6}{t_3^8}+\frac{8 t_1^{12} t_2^8}{t_3^8}-\frac{9 t_1^{14} t_2^8}{t_3^8}+\nn \\
&&\frac{3 t_1^{16} t_2^8}{t_3^8}-\frac{5 t_1^8 t_2^3}{t_3^7}+\frac{10 t_1^{10} t_2^3}{t_3^7}-\frac{t_1^{12} t_2^3}{t_3^7}+\frac{8 t_1^8 t_2^5}{t_3^7}-\frac{10 t_1^{10} t_2^5}{t_3^7}-\frac{14 t_1^{12} t_2^5}{t_3^7}+ \frac{10 t_1^{14} t_2^5}{t_3^7}-  \frac{2 t_1^{16} t_2^5}{t_3^7}-\frac{t_1^8 t_2^7}{t_3^7}-\frac{t_1^6 t_2^6}{t_3^6}-\nn \\
&&\frac{5 t_1^{10} t_2^7}{t_3^7}+\frac{20 t_1^{12} t_2^7}{t_3^7}-\frac{13 t_1^{14} t_2^7}{t_3^7}+\frac{3 t_1^{16} t_2^7}{t_3^7}-\frac{4 t_1^{12} t_2^4}{t_3^6}+\frac{2 t_1^{14} t_2^4}{t_3^6}+\frac{5 t_1^8 t_2^6}{t_3^6}-\frac{10 t_1^{10} t_2^6}{t_3^6}+ \frac{17 t_1^{12} t_2^6}{t_3^6}-\frac{7 t_1^{14} t_2^6}{t_3^6}-\nn \\
&&\frac{4 t_1^{12} t_2^8}{t_3^6}+\frac{2 t_1^{14} t_2^8}{t_3^6}-\frac{6 t_1^6 t_2^3}{t_3^5}+\frac{20 t_1^8 t_2^3}{t_3^5}-\frac{20 t_1^{10} t_2^3}{t_3^5}+\frac{2 t_1^{12} t_2^3}{t_3^5}+\frac{3 t_1^6 t_2^5}{t_3^5}-\frac{15 t_1^8 t_2^5}{t_3^5}+\frac{14 t_1^{10} t_2^5}{t_3^5}+\frac{9 t_1^{12} t_2^5}{t_3^5}- \frac{3 t_1^{14} t_2^5}{t_3^5}+\nn \\
&&\frac{4 t_1^{10} t_2^7}{t_3^5}-\frac{10 t_1^{12} t_2^7}{t_3^5}+\frac{2 t_1^{14} t_2^7}{t_3^5}+\frac{2 t_1^4 t_2^2}{t_3^4}-\frac{10 t_1^6 t_2^2}{t_3^4}+\frac{4 t_1^8 t_2^2}{t_3^4}-\frac{3 t_1^4 t_2^4}{t_3^4}+\frac{9 t_1^6 t_2^4}{t_3^4}+\frac{14 t_1^8 t_2^4}{t_3^4}-\frac{15 t_1^{10} t_2^4}{t_3^4}+\frac{3 t_1^{12} t_2^4}{t_3^4}+\nn \\
&&\frac{2 t_1^6 t_2^6}{t_3^4}-\frac{20 t_1^8 t_2^6}{t_3^4}+\frac{20 t_1^{10} t_2^6}{t_3^4}-\frac{6 t_1^{12} t_2^6}{t_3^4}+\frac{2 t_1^4 t_2}{t_3^3}-\frac{4 t_1^6 t_2}{t_3^3}-\frac{7 t_1^4 t_2^3}{t_3^3}+\frac{17 t_1^6 t_2^3}{t_3^3}-\frac{10 t_1^8 t_2^3}{t_3^3}+\frac{5 t_1^{10} t_2^3}{t_3^3}-\frac{t_1^{12} t_2^3}{t_3^3}+\nn \\
&&\frac{2 t_1^4 t_2^5}{t_3^3}- \frac{4 t_1^6 t_2^5}{t_3^3}+\frac{3 t_1^2 t_2^2}{t_3^2}-\frac{13 t_1^4 t_2^2}{t_3^2}+\frac{20 t_1^6 t_2^2}{t_3^2}-\frac{5 t_1^8 t_2^2}{t_3^2}-\frac{t_1^{10} t_2^2}{t_3^2}-\frac{2 t_1^2 t_2^4}{t_3^2}+\frac{10 t_1^4 t_2^4}{t_3^2}-\frac{14 t_1^6 t_2^4}{t_3^2}-\frac{10 t_1^8 t_2^4}{t_3^2}+\nn \\
&& \frac{8 t_1^{10} t_2^4}{t_3^2}-\frac{t_1^6 t_2^6}{t_3^2}+ \frac{10 t_1^8 t_2^6}{t_3^2}- \frac{5 t_1^{10} t_2^6}{t_3^2}+\frac{3 t_1^2 t_2}{t_3}-\frac{9 t_1^4 t_2}{t_3}+\frac{8 t_1^6 t_2}{t_3}-\frac{2 t_1^2 t_2^3}{t_3}+\frac{10 t_1^4 t_2^3}{t_3}-\frac{12 t_1^6 t_2^3}{t_3}-\frac{t_1^6 t_2^5}{t_3}+\frac{3 t_1^8 t_2^5}{t_3}\nn \\~.
\eea}
Let $R_i$ be the R-charges corresponding to the fugacity $t_i$.  Since the superpotential, which has fugacity $t^2_1 t_2 / t_3$, carries R-charge 2, it follows that:
\bea
2R_1 + R_2 - R_3 = 2~.
\label{e:cyfano271}
\eea
The volume of $\cE_4$ is given by:
{\footnotesize
\bea
\lim_{\mu \rightarrow 0} \mu^4 \gm (e^{-\mu R_1}, e^{-\mu R_2}, e^{-\mu(2R_1 + R_2 - 2)}; \cE_4)= \frac{p\left(R_1, R_2; \cE_4\right)}{16 R^2_1 R^2_2(3 - R_1 - R_2)^2(4 - 2R_1 - R_2)^2(2 - 2R_1 - R_2)^2},\nn \\
\label{e:volfano271}
\eea}
where:
{\small
\bea
p\left(R_1, R_2; \cE_4\right) &=& 624 R_1^3-288 R_1^2 - 464 R_1^4 + 144 R_1^5 - 16 R_1^6 + 336 R_1^2 R_2- 448 R_1^3 R_2 +\nn \\
&& 208 R_1^4 R_2 - 32 R_1^5 R_2 -  192 R_2^2 + 400 R_1 R_2^2 - 392 R_1^2 R_2^2 + 192 R_1^3 R_2^2 -\nn \\
&& 40 R_1^4 R_2^2 + 256 R_2^3 - 376 R_1 R_2^3 + 192 R_1^2 R_2^3 - 40 R_1^3 R_2^3 - 124 R_2^4 +\nn \\
&& 117 R_1 R_2^4 - 29 R_1^2 R_2^4 + 26 R_2^5 - 12 R_1 R_2^5 - 2 R_2^6 ~.
\eea
This function has a minimum at:
\bea
R_1 \approx 0.916, \qquad R_2 \approx 1.065, \qquad R_3\approx 0.897.
\eea
The R-charge of the perfect matching corresponding to the divisor $D_\alpha$ is:
\bea
\lim_{\mu\rightarrow 0}\frac{1}{\mu} \left[ \frac{g(D_\alpha; e^{- \mu R_1}, e^{- \mu R_2 }, e^{- \mu R_3 }; \cE_4) }{\gm(e^{-\mu R_1}, e^{- \mu R_2 }, e^{- \mu R_3 };\cE_4)}- 1 \right]~,
\eea
where $g(D_\alpha; e^{- \mu R_1}, e^{- \mu R_2 }, e^{- \mu R_3 }; \cE_4)$ is the Molien-Weyl integral with the insertion of the inverse of the weight corresponding to the divisor $D_\alpha$. The results of the computations are reported in Table \ref{t:chargefano271}.
The assignment of charges under the remaining abelian symmetries can be done by requiring that the superpotential is not charged under them and that the charge vectors are linearly independent. The assignments are shown in Table \ref{t:chargefano271}.
\begin{table}[h!]
 \begin{center}  
  \begin{tabular}{|c||c|c|c|c|c|c|c|c|}
  \hline
  \;& $SU(2)$&$U(1)_1$&$U(1)_2$&$U(1)_R$&$U(1)_{B_1}$&$U(1)_{B_2}$&$U(1)_{B_3}$&fugacity\\
  \hline\hline  
   
  $p_1$&$  1$&$  0$&$ 0$&$0.357$&$ 0$&$ 0$ &$ 0$ & $s_1 x $\\
  \hline
  
  $p_2$&$ -1$&$  0$&$ 0$&$0.357$&$ 0$&$ 0$ &$ 0$ & $s_1 / x $\\
  \hline 
   
  $r_1$&$  0$&$  1$&$ 0$&$0.221$&$ 1$&$ 0$ &$ 0$ & $s_2 q_1 b_1$\\
  \hline
   
  $r_2$&$  0$&$ -1$&$ 0$&$0.258$&$ 1$&$ 0$ &$ 0$ & $s_3 b_1 / q_1$\\
  \hline
  
  $r_3$&$  0$&$  0$&$ 1$&$0.282$&$ 0$&$ 0$ &$ 0$ & $s_4 q_2$\\
  \hline
 
  $r_4$&$  0$&$  0$&$-1$&$0.206$&$ 0$&$ 0$ &$ 0$ & $s_5 / q_2 $\\
  \hline
  
  $r_5$&$  0$&$  0$&$ 0$&$0.319$&$ 0$&$ 0$ &$ 0$ & $s_6 $\\
  \hline
  
  $v_1$&$  0$&$  0$&$ 0$&$    0$&$ 0$&$ 1$ &$ 0$ & $ b_2 $\\
  \hline
  
  $v_2$&$  0$&$  0$&$ 0$&$    0$&$ 0$&$-1$ &$ 1$ & $ b_3 / b_2 $\\
  \hline
 
  $v_3$&$  0$&$  0$&$ 0$&$    0$&$-2$&$ 0$ &$-1$ & $1 / (b^2_1 b_3)$\\
  \hline
  
   \end{tabular}
  \end{center}
\caption{Charges of the perfect matchings under the global symmetry of the $\cE_4$ model. Here $s_\alpha$ is the fugacity of the R-charge, $x$ is the weight of the $SU(2)$ symmetry, $q_1, q_2, b_1, b_2$ and $b_3$ are, respectively, the charges under the mesonic abelian symmetries $U(1)_1, U(1)_2$ and of the three baryonic $U(1)_{B_1},U(1)_{B_2}$ and $U(1)_{B_3}$.}
\label{t:chargefano271}
\end{table}

\begin{table}[h]
 \begin{center}  
  \begin{tabular}{|c||c|}
  \hline
  \; Quiver fields & R-charge\\
  \hline  \hline 
  $ X^i_{34}$ & 0.357 \\
  \hline
  $ X^i_{12}$ & 0.578 \\
  \hline
  $ X^1_{23} $ &  0.319\\
  \hline
  $ X^2_{23} $ & 0.488 \\
  \hline
  $ X_{51} $ & 0.206 \\
  \hline
  $ X_{41} $ & 0.577 \\
  \hline
  $ X_{45} $ & 0.540 \\
  \hline
  \end{tabular}
  \end{center}
\caption{R-charges of the quiver fields for the $\cE_4$ Model.}
\label{t:Rfieldfano271}
\end{table}

\paragraph{The Hilbert series.}  The Hilbert series of the Master space can be obtained by integrating that of the space of perfect matchings over the $z_i$ fugacities:
{\footnotesize
\bea
g^{\firr{}} (s_{\alpha}, x, q_1, q_2, b_i; \cE_4) &=& \left( \prod^3_{j=1} \oint\limits_{|z_j|=1}  \frac{\ud z_j}{2 \pi i z_j} \right)\frac{1}{\left(1- s_1 x z_1\right)\left(1- \frac{s_1 z_1}{x}\right)\left(1- \frac{s_2 q_1 b_1}{z_1}\right)\left(1-\frac{s_3 b_1 z_2}{q_1}\right)}\nn \\
&&\times \frac{1}{\left(1- \frac{s_4 q_2}{z_2}\right)\left(1- \frac{s_5 z_3}{q_2}\right)\left(1-\frac{s_6 z_3}{z_2}\right)\left(1-\frac{b_2}{z_3}\right)\left(1-\frac{b_3 z_2}{b_2 z_3}\right)\left(1-\frac{1}{b^2_1 b_3 z_1}\right)}\nn \\
&=& \frac{\left(1-\frac{s^2_1 s_2 q_1}{b_1 b_3}\right)}{\left(1- \frac{s_1 x}{b^2_1 b_3}\right)\left(1-\frac{s_1}{x b^2_1 b_3}\right)\left(1- s_1 s_2 x q_1 b_1\right)\left(1-\frac{s_1 s_2 q_1 b_1}{x}\right)\left(1-\frac{s_6 b_3}{b_2}\right)}\nn \\
&&\times \frac{\left(1-\frac{s_3 s_4 s_5 s_6 b_1 b_3}{q_1}\right)}{\left(1- \frac{s_5 b_2}{q_2}\right)\left(1-\frac{s_3 s_4 q_2 b_1}{q_1}\right)\left(1-\frac{s_4 s_5 b_3}{b_2}\right)\left(1-\frac{s_3 s_6 b_1 b_2}{q_1}\right)}~.
\label{e:HSMasterfano271}
\eea}
The Hilbert series of the mesonic moduli space can be obtained by integrating over the three baryonic fugacities:
{\footnotesize
\bea
\gm (s_{\alpha},x,q_1,q_2; \cE_4) &=& \left( \prod^3_{i=1} \oint\limits_{|b_i|=1} \frac{\ud b_i}{2 \pi i b_i} \right)g^{\firr{}} (s_{\alpha}, x, q_1, q_2, b_i; \cE_4)\nn \\
&=& \frac{P\left(s_{\alpha},x,q_1,q_2; \cE_4\right)}{\left(1- \frac{s^3_1 s^2_2 s_4 s^2_5 x^3 q^2_1}{q_2}\right)\left(1-\frac{s^3_1 s^2_2 s_4 s^2_5 q^2_1}{x^3 q_2}\right)\left(1-\frac{s^3_1 s^2_2 s_5 s_6 q^2_1 x^3}{q_2}\right)\left(1-\frac{s^3_1 s^2_2 s_5 s_6 q^2_1}{x^3 q_2} \right)}\nn \\
&&\times \frac{1}{\left(1-s^2_1 s_2 s_3 s^2_6 x^2\right)\left(1-\frac{s^2_1 s_2 s_3 s^2_6 }{x^2}\right)\left(1 - \frac{s_1 s^2_3 s^3_4 s^2_5 x q_2}{q^2_1}\right)\left(1-\frac{s_1 s^2_3 s^3_4 s^2_5 q_2}{x q^2_1}\right)}\nn \\
&&\times \frac{1}{\left(1-\frac{s_1 s^2_3 s_4 s^2_6 x q_2}{q^2_1}\right)\left(1-\frac{s_1 s^2_3 s_4 s^2_6 q_2}{x q^2_1}\right)}~,
\label{e:HSmesonic}
\eea}
where $P\left(s_{\alpha},x,q_1,q_2; \cE_4\right)$ is a polynomial which is not reported here.
The plethystic logarithm of the Hilbert series above can be written as:
\bea
\PL[\gm (t_{\alpha},x,q_1,q_2; \cE_4)] &=& [3]\frac{q^2_1}{q_2}\left(\frac{t^4_1 t^2_2}{t^4_3} + \frac{t^4_1 t_2}{t^3_3}\right) + [2]\left(\frac{t^2_2 t^2_2}{t^2_3} + \frac{t^2_1 t_2}{t_3} +  t^2_1 \right)\nn \\
&+& [1]\frac{q_2}{q^2_1}\left(t^2_2 + t_2 t_3 + t^2_3\right) - O(t^2_1)O(t^2_2).
\label{e:plfano271}
\eea
Thus, the generators of the mesonic moduli space are
\bea
\begin{array}{lll}
 p_i p_j p_k r^2_1 r_3  r^2_4 v_1 v_2 v_3, \quad & p_i p_j p_k r^2_1 r_4 r_5 v_1 v_2 v_3, \quad &p_i p_j r_1 r_2 r^2_3 r^2_4 v_1 v_2 v_3,  \nn \\
  p_i p_j r_1 r_2 r_3 r_4 r_5 v_1 v_2 v_3, \quad & p_i p_j r_1 r_2 r^2_5 v_1 v_2 v_3, \quad & p_i r_1 r^2_2 r^2_3 r_4 r_5 v_1 v_2 v_3, \nn \\
p_i r^2_2 r^3_3 r^2_4 v_1 v_2 v_3, \qquad & p_i r^2_2 r_3 r^2_5 v_1 v_2 v_3~. & 
\label{e:genfano271}
\end{array}
\eea
with $i,j,k=1,2$. The R-charges of the generators are presented in \tref{t:Rgenfano271}. The lattice of generators is drawn in \fref{f:late4}.

\begin{table}[h]
 \begin{center}  
  \begin{tabular}{|c||c|}
  \hline
  \; Generators &$U(1)_R$\\
  \hline  \hline 
  $p_i p_j p_k r^2_1 r_3  r^2_4 v_1 v_2 v_3$ & 2.207 \\
  \hline
  $p_i p_j p_k r^2_1 r_4 r_5 v_1 v_2 v_3$ & 2.038\\
  \hline
  $p_i p_j r_1 r_2 r^2_3 r^2_4 v_1 v_2 v_3$ & 2.169 \\
  \hline
  $p_i p_j r_1 r_2 r_3 r_4 r_5 v_1 v_2 v_3$ & 2 \\
  \hline
  $p_i p_j r_1 r_2 r^2_5 v_1 v_2 v_3$ & 1.831 \\
  \hline
  $p_i r_1 r^2_2 r^2_3 r_4 r_5 v_1 v_2 v_3$ & 2.131 \\
  \hline
  $p_i r^2_2 r^3_3 r^2_4 v_1 v_2 v_3$ & 1.962 \\
  \hline
  $p_i r^2_2 r_3 r^2_5 v_1 v_2 v_3$ & 1.793 \\
  \hline
  \end{tabular}
  \end{center}
\caption{R-charges of the generators of the mesonic moduli space for the $\cE_4$ Model.}
\label{t:Rgenfano271}
\end{table}

\begin{figure}[ht]
\begin{center}
\includegraphics[totalheight=3.8cm]{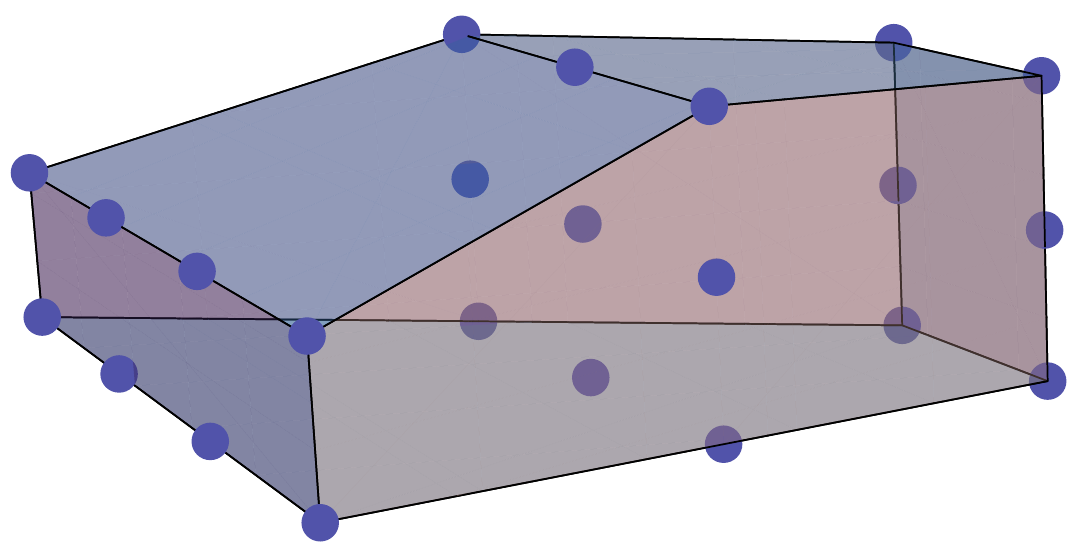}
\caption{The lattice of generators of the $\cE_4$ theory.}
  \label{f:late4}
\end{center}
\end{figure}

\section{$\cF_2$ (Toric Fano 369): $dP_3$ bundle over $\BP^1$}
This theory has 6 gauge groups and chiral fields $X^{i}_{23}, X^{i}_{31}, X^{i}_{42}$ (with $i=1,2$), $X_{12}, X_{34}, X_{26},X_{63}, X_{15}$ and $X_{54}$. The quiver diagram and the tiling of this model are presented in Figure \ref{f:tqfano369}.  Note that this tiling is actually that of $dP_1$ with 2 double bonds.  The superpotential of this model can be read off from the tiling as:
\bea
W = \tr \left[\epsilon_{ij} \left(X_{12}X^i_{23}X^j_{31} + X_{34}X^i_{42}X^j_{23} + X_{26}X_{63}X^{i}_{31}X_{15}X_{54}X^j_{42} \right) \right].
\label{spfano369}
\eea
The CS levels are $\vec{k}=(0,-1,0,-1,1,1)$.
\begin{figure}[ht]
\begin{center}
\hskip -1.2cm
\includegraphics[totalheight=3cm]{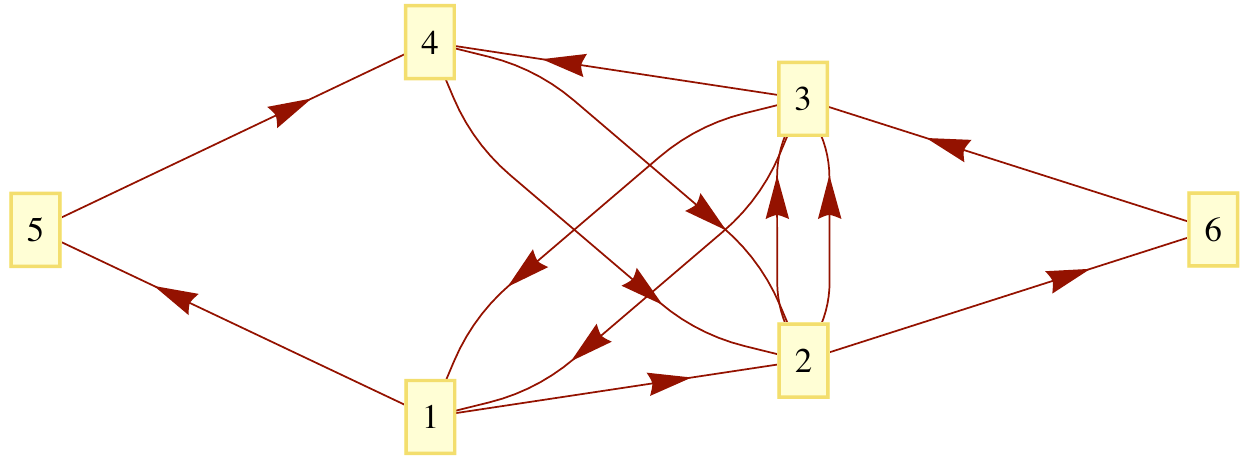}
\hskip 0.3cm
\includegraphics[totalheight=4.5cm]{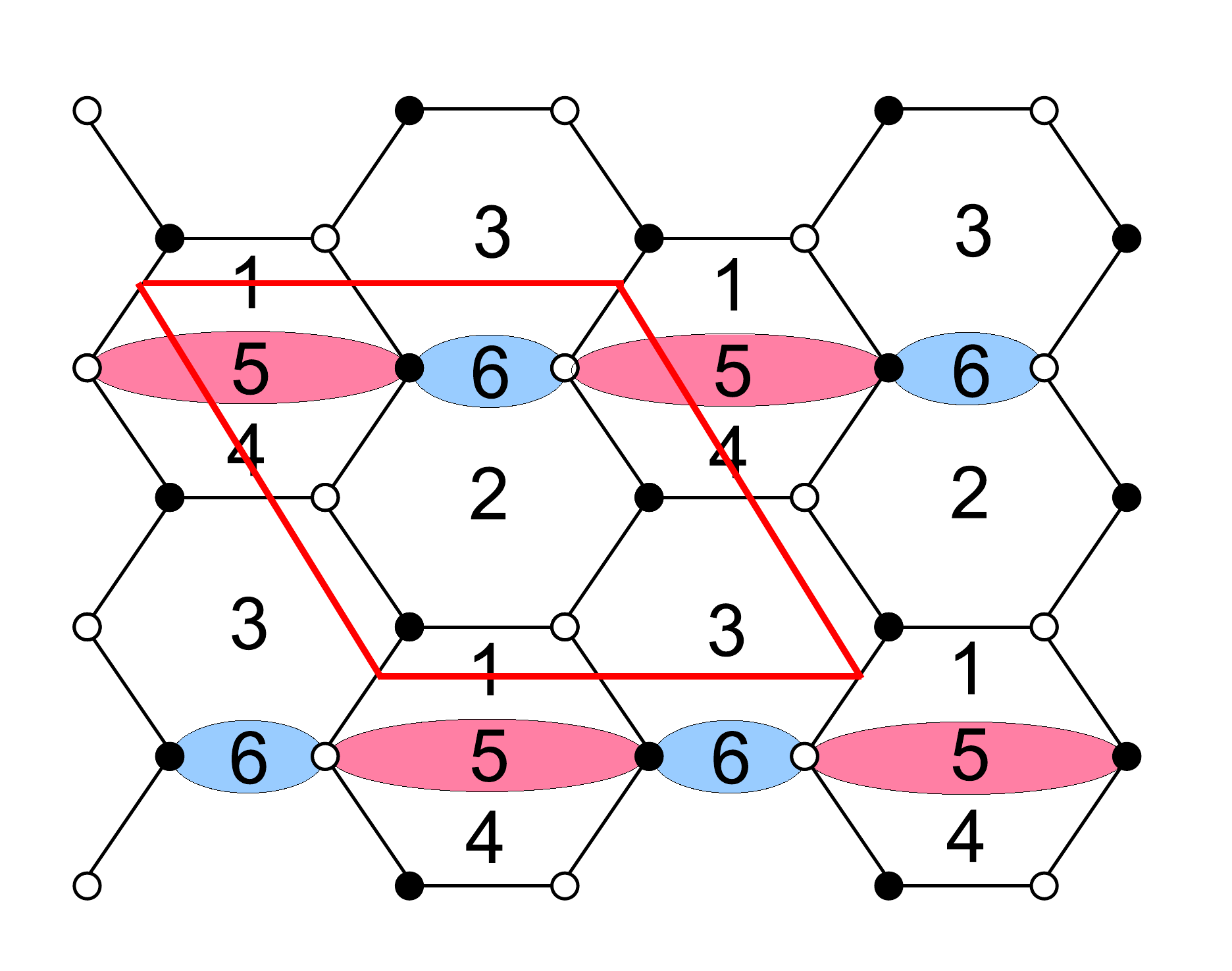}
\caption{(i) Quiver of the $\cF_2$ model. \quad (ii) Tiling of the $\cF_2$ model.}
  \label{f:tqfano369}
\end{center}
\end{figure}

\paragraph{The Kasteleyn matrix.} The Chern-Simons levels can be parametrized in terms of the integers $n^i_{jk}$ or $n_{jk}$ as follows:
\bea
\begin{array}{ll}
\text{Gauge group 1~:} \qquad k_1 &=   n_{12} + n_{15} - n^{1}_{31} - n^{2}_{31}  ~,  \nn \\
\text{Gauge group 2~:} \qquad k_2 &=   n_{26} + n^{1}_{23} + n^{2}_{23} - n^{1}_{42} - n^{2}_{42} - n_{12}  ~,  \nn \\
\text{Gauge group 3~:} \qquad k_3 &=   n_{34} + n^{1}_{31} + n^{2}_{31} - n^{1}_{23} - n^{2}_{23} - n_{63} ~,  \nn \\
\text{Gauge group 4~:} \qquad k_4 &=   n^{1}_{42} + n^{2}_{42} - n_{34} - n_{54} ~,  \nn \\
\text{Gauge group 5~:} \qquad k_5 &=   n_{54} - n_{15} ~,  \nn \\
\text{Gauge group 6~:} \qquad k_6 &=   n_{63} - n_{26} ~.
\label{e:kafano369}
\end{array}
\eea  
Let us choose
\bea
n_{54}=-n_{26}= 1,\qquad n^i_{jk}=n_{jk}=0 \;\;\text{otherwise}~.
\eea
Since the fundamental domain contains 3 pairs of black and white nodes, the Kasteleyn matrix of this model is a $3\times 3$ matrix:
\bea
K =   \left(
\begin{array}{c|ccc}
& b_1 & b_2 & b_3\\
\hline
w_1 & z^{n_{12}} & \frac{y}{x} z^{n^2_{31}} & z^{n^1_{23}} \\
w_2 & z^{n^2_{23}} & z^{n^1_{42}} & \frac{1}{y} z^{n_{34}} \\
w_3 & x z^{n^1_{31}} \quad & z^{n_{63}} + z^{n_{26}} + y z^{n_{15}} + y z^{n_{54}} \quad & z^{n^2_{42}} \end{array}
\right) ~.
\label{e:kastfano369}
\eea
The permanent of the Kasteleyn matrix is
\bea
\mathrm{perm}(K) &=&  x z^{(n^1_{23} + n^1_{42} + n^1_{31})} +  x^{-1} y z^{(n^2_{23} + n^2_{42} + n^2_{31})}+   y  z^{(n^1_{23} + n^2_{23} + n_{15})}\nn \\
&+&  y^{-1} z^{(n_{63} + n_{12} + n_{34})} +  y z^{(n^1_{23} + n^2_{23} + n_{54})} +  y^{-1} z^{(n_{26} + n_{12} + n_{34})} \nn \\
&+& z^{(n^1_{23} + n^2_{23} + n_{26} )}+ z^{(n_{54} + n_{12} + n_{34})}+ z^{(n^1_{23} + n^2_{23} + n_{63})} \nn \\
&+&  z^{(n_{15} + n_{12} + n_{34})} + z^{(n^1_{42} + n^2_{42} + n_{12})}+  z^{(n^1_{31} + n^2_{31} + n_{34})}\nn \\
&=& x + x^{-1} y + y + y^{-1} + y z + y^{-1} z^{-1} + z^{-1} + z + 4\nn \\
&& \; \text{(for $n_{54}=-n_{26}= 1,\qquad n^i_{jk}=n_{jk}=0 \; \text{otherwise}$)}~.
\label{e:charpolyfano369}
\eea
Therefore, the perfect matchings can be written in terms of the chiral fields as:
\bea 
&& p_1 = \left\{X^1_{23}, X^1_{42}, X^1_{31}\right\}, \;\; p_2 = \left\{X^2_{23}, X^2_{42}, X^2_{31}\right\}, \;\;  u_1 = \left\{X^1_{23}, X^2_{23}, X_{15}\right\}, \nn \\  
&& u_2 = \left\{X_{12}, X_{34}, X_{63}\right\}, \;\; q_1 = \left\{X^1_{23}, X^2_{23}, X_{54}\right\}, \;\; r_1 = \left\{X_{12}, X_{34}, X_{26}\right\}, \nn \\  
&& q_2 = \left\{X^1_{23}, X^2_{23}, X_{26}\right\}, \;\; r_2 = \left\{X_{12}, X_{34}, X_{54}\right\}, \;\; v_1 = \left\{X^1_{23}, X^2_{23}, X_{63}\right\}, \nn \\  
&&  v_2 = \left\{X_{12}, X_{34}, X_{15}\right\}, \;\; v_3 = \left\{X_{12}, X^1_{42}, X^2_{42}\right\}, \;\; v_4 = \left\{X_{34}, X^1_{31}, X^2_{31}\right\}\ . \qquad
\eea
In turn, chiral fields can be written as products of perfect matchings:
\bea
\begin{array}{lll}
X^1_{23} = p_1 u_1 q_1 q_2 v_1, \quad & X^2_{23} = p_2 u_1 q_1 q_2 v_1, \quad & X_{12} = u_2 r_1 r_2 v_2 v_3~,\nn \\
X_{34} = u_2 r_1 r_2 v_2 v_4, \quad & X^1_{42} = p_1 v_3, \quad & X^2_{42} = p_2 v_3~,\nn \\
X^1_{31} = p_1 v_4, \quad & X^2_{31} = p_2 v_4,   \quad & X_{15} = u_1 v_2~, \nn \\
X_{63} = u_2 v_1,   \quad  & X_{54} = q_1 r_2,   \quad &  X_{26} = r_1 q_2~.
\end{array}
\eea
These pieces of information can be collected in the following $P$ matrix:
\beq
P=\left(\begin{array} {c|cccccccccccc}
  \;& p_1 & p_2 & u_1 & u_2 & q_1 & r_1 & q_2 & r_2 & v_1 & v_2 & v_3 & v_4\\
  \hline 
  X^{1}_{23}& 1&0&1&0&1&0&1&0&1&0&0&0\\
  X^{2}_{23}& 0&1&1&0&1&0&1&0&1&0&0&0\\
  X_{12}    & 0&0&0&1&0&1&0&1&0&1&1&0\\
  X_{34}    & 0&0&0&1&0&1&0&1&0&1&0&1\\
  X^{1}_{42}& 1&0&0&0&0&0&0&0&0&0&1&0\\
  X^{2}_{42}& 0&1&0&0&0&0&0&0&0&0&1&0\\
  X^{1}_{31}& 1&0&0&0&0&0&0&0&0&0&0&1\\
  X^{2}_{31}& 0&1&0&0&0&0&0&0&0&0&0&1\\
  X_{15}    & 0&0&1&0&0&0&0&0&0&1&0&0\\
  X_{63}    & 0&0&0&1&0&0&0&0&1&0&0&0\\
  X_{54}    & 0&0&0&0&1&0&0&1&0&0&0&0\\
  X_{26}    & 0&0&0&0&0&1&1&0&0&0&0&0
  \end{array}
\right).
\label{e:pmatrifano369}
\eeq
The $Q_F$ matrix, defined as the kernel of the $P$ matrix, can be written as:
\be
Q_F =   \left(
\begin{array}{cccccccccccc}
1 & 1 & 0 & 0 & 0 & 1 &-1 & 0 & 0 & 0 &-1 &-1\\
0 & 0 & 1 & 0 & 0 & 1 &-1 & 0 & 0 &-1 & 0 & 0\\
0 & 0 & 0 & 1 & 0 &-1 & 1 & 0 &-1 & 0 & 0 & 0\\
0 & 0 & 0 & 0 & 1 & 1 &-1 &-1 & 0 & 0 & 0 & 0
\end{array}
\right)~.  
\label{e:qffano369}
\ee
Therefore, the relations between the perfect matchings are:
\bea
  p_1 + p_2 - q_2 + r_1 - v_3 - v_4 &=& 0~, \nn \\
  q_2 - r_1 - u_1 + v_2 &=& 0~, \nn \\
  q_2 - r_1 + u_2 - v_1 &=& 0~, \nn \\
  q_1 - q_2 + r_1 - r_2 &=& 0~. 
\label{e:relpmfano369}
\eea

\paragraph{The toric diagram.}  The toric diagram for this model is constructed using two different methods.
\begin{itemize}
\item {\bf The Kasteleyn matrix.} The powers of $x, y$ and $z$ of each term in \eref{e:charpolyfano369} give the coordinates of the toric diagram:
\bea
G_K = \left(
\begin{array}{cccccccccccc}
  1 & -1 & 0 & 0 & 0 &  0 & 0 & 0 & 0 & 0 & 0 & 0 \\
  0 &  1 & 1 &-1 & 1 & -1 & 0 & 0 & 0 & 0 & 0 & 0 \\
  0 &  0 & 0 & 0 & 1 & -1 &-1 & 1 & 0 & 0 & 0 & 0 
\end{array}
\right)~.
\label{e:Gkfano369}
\eea
The first row contains the powers of the weights of the fundamental representation of $SU(2)$. Thus, the mesonic symmetry of this model contains one $SU(2)$ as the non-abelian factor.

\item {\bf The charge matrices.}
Since this model has 6 gauge groups, the total number of baryonic symmetries is 4. The charges of the perfect matchings under these four symmetries are collected in the columns of the $Q_D$ matrix:
\be
Q_D =   \left(
\begin{array}{cccccccccccc}
0 & 0 & 0 & 0 & 0 & 0 & 1 & 1 & 0 & 0 &-2 & 0\\
0 & 0 & 0 & 0 & 0 & 0 & 0 & 0 & 1 & 0 & 0 &-1\\
0 & 0 & 0 & 0 & 0 & 0 & 0 & 0 & 0 & 1 & 0 &-1\\
0 & 0 & 0 & 0 & 0 & 0 & 0 & 0 & 0 & 0 & 1 &-1
\end{array}
\right). 
\label{e:qdfano369}
\ee
Combining the $Q_F$ and $Q_D$ matrices, the total charge matrix $Q_t$ can be written as:
\be
Q_t = { \Blue Q_F \choose \Green Q_D \Black } =   \left( 
\begin{array}{cccccccccccc} \Blue
1 & 1 & 0 & 0 & 0 & 1 &-1 & 0 & 0 & 0 &-1 &-1\\
0 & 0 & 1 & 0 & 0 & 1 &-1 & 0 & 0 &-1 & 0 & 0\\
0 & 0 & 0 & 1 & 0 &-1 & 1 & 0 &-1 & 0 & 0 & 0\\
0 & 0 & 0 & 0 & 1 & 1 &-1 &-1 & 0 & 0 & 0 & 0\\ \Green
0 & 0 & 0 & 0 & 0 & 0 & 1 & 1 & 0 & 0 &-2 & 0\\
0 & 0 & 0 & 0 & 0 & 0 & 0 & 0 & 1 & 0 & 0 &-1\\
0 & 0 & 0 & 0 & 0 & 0 & 0 & 0 & 0 & 1 & 0 &-1\\
0 & 0 & 0 & 0 & 0 & 0 & 0 & 0 & 0 & 0 & 1 &-1\Black
\end{array}
\right). 
\label{e:qtfano369}
\ee
The kernel of the $Q_t$ matrix, after the removal of the first trivial row, contains the coordinates of the toric diagram in its columns:
\bea
G'_t = \left(
\begin{array}{cccccccccccc}
  1 & -1 & 0 & 0 & 0 &  0 & 0 & 0 & 0 & 0 & 0 & 0 \\
  0 &  1 & 1 &-1 & 1 & -1 & 0 & 0 & 0 & 0 & 0 & 0 \\
  0 &  0 & 0 & 0 & 1 & -1 &-1 & 1 & 0 & 0 & 0 & 0 
\end{array}
\right) = G_K~. \label{e:toricdiafano369}
\eea
The toric diagram constructed from (\ref{e:toricdiafano369}) is presented in Figure \ref{f:tdtoricfano369}.
\begin{figure}[ht]
\begin{center}
  \includegraphics[totalheight=2.8cm]{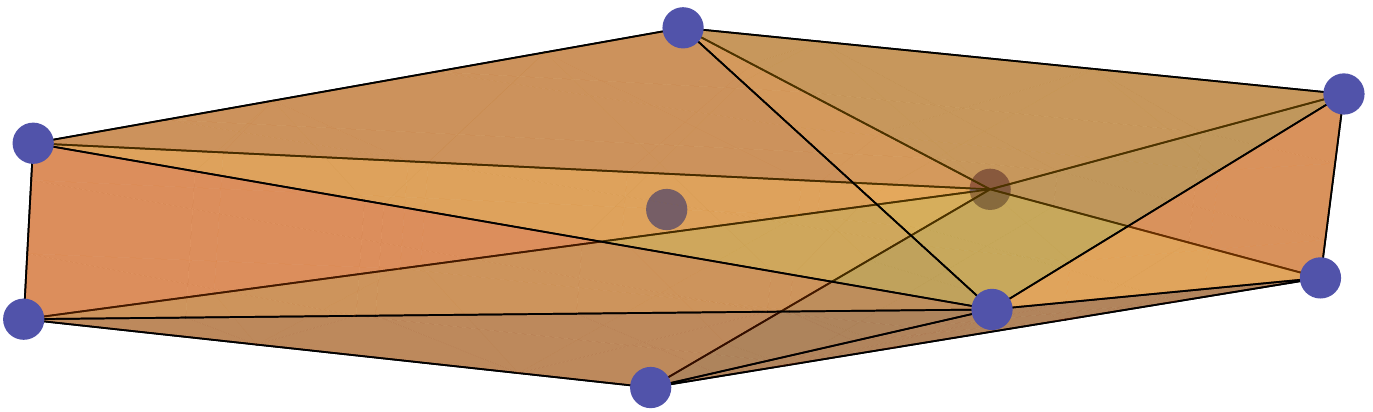}
 \caption{The toric diagram of $\cF_2$.}
  \label{f:tdtoricfano369}
\end{center}
\end{figure}
\end{itemize}

\paragraph{The baryonic charges.} The toric diagram of this model contains 8 external points. It follows that the number of baryonic symmetries of the gauge theory is $8-4=4$. These symmetries come from the D-terms and will be denoted as $U(1)_{B_1}, \ldots, U(1)_{B_4}$.

\paragraph{The global symmetry.} The $Q_t$ matrix contains only one pair of repeated columns, confirming that the mesonic symmetry contains one $SU(2)$ as the only non-abelian factor.  Since the mesonic symmetry has rank 4, it can be identified precisely with $SU(2)\times U(1)^3$, where one of the three abelian factors corresponds to the R-symmetry. The perfect matchings $p_1$ and $p_2$ transform as a doublet under the non-abelian mesonic symmetry. The perfect matchings $v_1, v_2, v_3$ and $v_4$ correspond to the internal point in the toric diagram and, therefore, they carry zero R-charges.

Since there are three $U(1)$ factors in the mesonic symmetry, a volume minimisation problem needs to be solved in order to assign the R-charges to the perfect matchings.
Let us assign the R-charge fugacity $s_1$ to the perfect matchings $p_1$ and $p_2$ (note that the non-abelian symmetry does not play any role in the volume minimisation, so $p_1$ and $p_2$ carry the same R-charges), and the R-charge fugacities $s_2,s_3, s_4, s_5, s_6$ and $s_{7}$ to the perfect matchings $u_1, u_2, q_1, r_1, q_2$ and $r_2$, respectively. The Hilbert series of the mesonic moduli space is:
{\footnotesize
\bea
\gm (s_{\alpha}; \cF_2) &=& \left( \prod_{i=1}^4 \oint \limits_{|b_i|=1} \frac{\ud b_i}{2\pi i b_i}\prod_{j=1}^4 \oint \limits_{|z_j|=1} \frac{\ud z_j}{2\pi i z_j} \right) \frac{1}{\left(1-s_1 z_1\right)^2\left(1- s_2 z_2\right)\left(1- s_3 z_3\right)\left(1- s_4 z_4\right)}\nn \\
&&\times \frac{1}{\left(1-\frac{s_5 z_1 z_2 z_4}{z_3}\right)\left(1-\frac{s_6 b_1 z_3}{z_1 z_2 z_4}\right)\left(1-\frac{s_7 b_1}{z_4}\right)\left(1-\frac{b_2}{z_3}\right)\left(1-\frac{ b_3}{z_2}\right)\left(1-\frac{ b_4}{b^2_1 z_1}\right)\left(1-\frac{1}{b_2 b_3 b_4 z_1}\right)}\nn \\~.
\label{e:hsvolfano369}
\eea}
Since there are three factors of $U(1)$ in the mesonic symmetry, the integral (\ref{e:hsvolfano369}) depends only on three combinations of fugacities. By choosing:
\bea
 t_1 = s^3_1 s^2_2 s_4 s_5 s_6^2, \quad t_2 = s^2_1 s_2 s_3 s^2_5 s^2_6, \quad t_3 = s^3_1 s^2_2 s^2_4 s_6 s_7~, 
\eea
we have
{\footnotesize
\bea
\gm (t_1, t_2, t_3; \cF_2) &=& \frac{P(t_1,t_2,t_3; \cF_2)}{\left(1-t_1\right)^2\left(1-t_2\right)^2\left(1-t_3\right)^2\left(1-\frac{t^2_2 t_3}{t^2_1}\right)^2\left(1-\frac{t_2 t^2_3}{t^2_1}\right)^2\left(1-\frac{t^2_2 t^2_3}{t^3_1}\right)^2}~, \qquad \quad
\label{e:hsvolfano369T}
\eea}
where: 
{\scriptsize
\bea
P \left(t_1,t_2,t_3; \cF_2 \right) &=& 1 + 2 t_1 + t_2 - 4 t_1 t_2 + 2 t_3 - 5 t_1 t_3 - 10 t_2 t_3 + 10 t_1 t_2 t_3 + 4 t_2^2 t_3 - t_1 t_2^2 t_3 - t_2^2 t_3^2  + 3 t_2 t_3^2-2 t_1 t_2^2 t_3^2 +\nn \\
&&  \frac{3 t_2 t_3}{t_1}  - \frac{9 t_2^2 t_3}{t_1} - \frac{2 t_2^3 t_3}{t_1^2} + \frac{8 t_2^3 t_3}{t_1} + \frac{t_2 t_3^2}{t_1^2} - \frac{10 t_2 t_3^2}{t_1} - \frac{19 t_2^2 t_3^2}{t_1^2} + \frac{28 t_2^2 t_3^2}{t_1}  - \frac{8 t_2^3 t_3^2}{t_1^3} + \frac{28 t_2^3 t_3^2}{t_1^2} -\frac{20 t_2^3 t_3^2}{t_1} +\nn \\
&&  \frac{6 t_2^4 t_3^2}{t_1^3} - \frac{5 t_2^4 t_3^2}{t_1^2} + \frac{2 t_2^4 t_3^2}{t_1} - \frac{2 t_2^5 t_3^2}{t_1^3} - \frac{t_2^5 t_3^2}{t_1^2} - \frac{4 t_2 t_3^3}{t_1^2} + \frac{10 t_2 t_3^3}{t_1} - \frac{9 t_2^2 t_3^3}{t_1^3} + \frac{28 t_2^2 t_3^3}{t_1^2} - \frac{19 t_2^2 t_3^3}{t_1} -\frac{8 t_2^3 t_3^3}{t_1^4} +\nn \\
&&  \frac{39 t_2^3 t_3^3}{t_1^3} -\frac{24 t_2^3 t_3^3}{t_1^2} - \frac{t_2^3 t_3^3}{t_1} - \frac{t_2^4 t_3^3}{t_1^5} + \frac{28 t_2^4 t_3^3}{t_1^4} - \frac{30 t_2^4 t_3^3}{t_1^3} - \frac{4 t_2^4 t_3^3}{t_1^2} + \frac{4 t_2^4 t_3^3}{t_1} + \frac{6 t_2^5 t_3^3}{t_1^5} - \frac{22 t_2^5 t_3^3}{t_1^4} + \frac{5 t_2^5 t_3^3}{t_1^3} + \nn \\
&&\frac{2 t_2^5 t_3^3}{t_1^2} - \frac{t_2^6 t_3^3}{t_1^5}+ \frac{2 t_2^6 t_3^3}{t_1^4} - \frac{t_2^6 t_3^3}{t_1^3} + \frac{4 t_2^2 t_3^4}{t_1^3} - \frac{t_2^2 t_3^4}{t_1^2} - \frac{2 t_2^3 t_3^4}{t_1^5} + \frac{28 t_2^3 t_3^4}{t_1^4} - \frac{24 t_2^3 t_3^4}{t_1^3} - \frac{9 t_2^3 t_3^4}{t_1^2} + \frac{4 t_2^3 t_3^4}{t_1} + \frac{28 t_2^4 t_3^4}{t_1^5} -\nn \\
&& \frac{66 t_2^4 t_3^4}{t_1^4} + \frac{18 t_2^4 t_3^4}{t_1^3}+ \frac{11 t_2^4 t_3^4}{t_1^2} + \frac{7 t_2^5 t_3^4}{t_1^6} - \frac{36 t_2^5 t_3^4}{t_1^5} + \frac{27 t_2^5 t_3^4}{t_1^4} + \frac{6 t_2^5 t_3^4}{t_1^3} - \frac{t_2^5 t_3^4}{t_1^2} - \frac{9 t_2^6 t_3^4}{t_1^6} + \frac{6 t_2^6 t_3^4}{t_1^5} + \frac{5 t_2^6 t_3^4}{t_1^4} - \nn \\
&&\frac{2 t_2^6 t_3^4}{t_1^3} + \frac{4 t_2^7 t_3^4}{t_1^6} + \frac{2 t_2^7 t_3^4}{t_1^5}-\frac{t_2^2 t_3^5}{t_1^3} - \frac{2 t_2^2 t_3^5}{t_1^2} + \frac{8 t_2^3 t_3^5}{t_1^5} - \frac{20 t_2^3 t_3^5}{t_1^4} - \frac{t_2^3 t_3^5}{t_1^3} + \frac{4 t_2^3 t_3^5}{t_1^2} + \frac{6 t_2^4 t_3^5}{t_1^6} - \frac{30 t_2^4 t_3^5}{t_1^5} + \frac{18 t_2^4 t_3^5}{t_1^4} +\nn \\
&& \frac{11 t_2^4 t_3^5}{t_1^3} - \frac{2 t_2^4 t_3^5}{t_1^2} + \frac{6 t_2^5 t_3^5}{t_1^7}-\frac{36 t_2^5 t_3^5}{t_1^6} - \frac{6 t_2^5 t_3^5}{t_1^5} + \frac{36 t_2^5 t_3^5}{t_1^4} - \frac{9 t_2^5 t_3^5}{t_1^3} - \frac{23 t_2^6 t_3^5}{t_1^7} + \frac{36 t_2^6 t_3^5}{t_1^6} + \frac{27 t_2^6 t_3^5}{t_1^5} - \frac{22 t_2^6 t_3^5}{t_1^4} -\nn \\
&& \frac{2 t_2^7 t_3^5}{t_1^8} + \frac{11 t_2^7 t_3^5}{t_1^7} - \frac{4 t_2^7 t_3^5}{t_1^6}- \frac{5 t_2^7 t_3^5}{t_1^5} - \frac{5 t_2^4 t_3^6}{t_1^6} - \frac{4 t_2^4 t_3^6}{t_1^5} + \frac{11 t_2^4 t_3^6}{t_1^4} - \frac{2 t_2^4 t_3^6}{t_1^3} - \frac{22 t_2^5 t_3^6}{t_1^7} + \frac{27 t_2^5 t_3^6}{t_1^6} +\frac{36 t_2^5 t_3^6}{t_1^5} -\nn \\
&&  \frac{23 t_2^5 t_3^6}{t_1^4} - \frac{9 t_2^6 t_3^6}{t_1^8} + \frac{36 t_2^6 t_3^6}{t_1^7} - \frac{6 t_2^6 t_3^6}{t_1^6} - \frac{36 t_2^6 t_3^6}{t_1^5} + \frac{6 t_2^6 t_3^6}{t_1^4} - \frac{2 t_2^7 t_3^6}{t_1^9} + \frac{11 t_2^7 t_3^6}{t_1^8} + \frac{18 t_2^7 t_3^6}{t_1^7} - \frac{30 t_2^7 t_3^6}{t_1^6} +\frac{6 t_2^7 t_3^6}{t_1^5} +\nn \\
&&  \frac{4 t_2^8 t_3^6}{t_1^9} - \frac{t_2^8 t_3^6}{t_1^8} - \frac{20 t_2^8 t_3^6}{t_1^7} + \frac{8 t_2^8 t_3^6}{t_1^6} - \frac{2 t_2^9 t_3^6}{t_1^9}- \frac{t_2^9 t_3^6}{t_1^8} + \frac{2 t_2^4 t_3^7}{t_1^6} + \frac{4 t_2^4 t_3^7}{t_1^5} - \frac{2 t_2^5 t_3^7}{t_1^8} + \frac{5 t_2^5 t_3^7}{t_1^7} + \frac{6 t_2^5 t_3^7}{t_1^6} -\frac{9 t_2^5 t_3^7}{t_1^5} - \nn \\
&& \frac{t_2^6 t_3^7}{t_1^9} + \frac{6 t_2^6 t_3^7}{t_1^8} + \frac{27 t_2^6 t_3^7}{t_1^7} - \frac{36 t_2^6 t_3^7}{t_1^6} + \frac{7 t_2^6 t_3^7}{t_1^5}+\frac{11 t_2^7 t_3^7}{t_1^9} + \frac{18 t_2^7 t_3^7}{t_1^8} - \frac{66 t_2^7 t_3^7}{t_1^7} + \frac{28 t_2^7 t_3^7}{t_1^6} + \frac{4 t_2^8 t_3^7}{t_1^{10}} - \frac{9 t_2^8 t_3^7}{t_1^9} -\nn \\
&& \frac{24 t_2^8 t_3^7}{t_1^8} + \frac{28 t_2^8 t_3^7}{t_1^7} - \frac{2 t_2^8 t_3^7}{t_1^6} - \frac{t_2^9 t_3^7}{t_1^9} + \frac{4 t_2^9 t_3^7}{t_1^8} - \frac{t_2^5 t_3^8}{t_1^8}+ \frac{2 t_2^5 t_3^8}{t_1^7}- \frac{t_2^5 t_3^8}{t_1^6} + \frac{2 t_2^6 t_3^8}{t_1^9} + \frac{5 t_2^6 t_3^8}{t_1^8} - \frac{22 t_2^6 t_3^8}{t_1^7} +\frac{6 t_2^6 t_3^8}{t_1^6} +\nn \\
&&  \frac{4 t_2^7 t_3^8}{t_1^{10}} - \frac{4 t_2^7 t_3^8}{t_1^9} - \frac{30 t_2^7 t_3^8}{t_1^8} + \frac{28 t_2^7 t_3^8}{t_1^7} - \frac{t_2^7 t_3^8}{t_1^6} - \frac{t_2^8 t_3^8}{t_1^{10}}- \frac{24 t_2^8 t_3^8}{t_1^9} + \frac{39 t_2^8 t_3^8}{t_1^8} - \frac{8 t_2^8 t_3^8}{t_1^7} - \frac{19 t_2^9 t_3^8}{t_1^{10}} +\frac{28 t_2^9 t_3^8}{t_1^9} - \nn \\
&& \frac{9 t_2^9 t_3^8}{t_1^8} + \frac{10 t_2^{10} t_3^8}{t_1^{10}} - \frac{4 t_2^{10} t_3^8}{t_1^9} - \frac{t_2^6 t_3^9}{t_1^9} - \frac{2 t_2^6 t_3^9}{t_1^8} + \frac{2 t_2^7 t_3^9}{t_1^{10}} - \frac{5 t_2^7 t_3^9}{t_1^9}+ \frac{6 t_2^7 t_3^9}{t_1^8} - \frac{20 t_2^8 t_3^9}{t_1^{10}} + \frac{28 t_2^8 t_3^9}{t_1^9} - \frac{8 t_2^8 t_3^9}{t_1^8} -\nn \\
&& \frac{2 t_2^9 t_3^9}{t_1^{12}} - \frac{t_2^9 t_3^9}{t_1^{11}} + \frac{28 t_2^9 t_3^9}{t_1^{10}} - \frac{19 t_2^9 t_3^9}{t_1^9} + \frac{3 t_2^{10} t_3^9}{t_1^{11}} - \frac{10 t_2^{10} t_3^9}{t_1^{10}} + \frac{t_2^{10} t_3^9}{t_1^9} + \frac{8 t_2^8 t_3^{10}}{t_1^{10}}- \frac{2 t_2^8 t_3^{10}}{t_1^9} - \frac{t_2^9 t_3^{10}}{t_1^{12}} + \frac{4 t_2^9 t_3^{10}}{t_1^{11}} -\nn \\
&& \frac{9 t_2^9 t_3^{10}}{t_1^{10}} + \frac{10 t_2^{10} t_3^{10}}{t_1^{12}} - \frac{10 t_2^{10} t_3^{10}}{t_1^{11}} + \frac{3 t_2^{10} t_3^{10}}{t_1^{10}} - \frac{5 t_2^{11} t_3^{10}}{t_1^{12}} + \frac{2 t_2^{11} t_3^{10}}{t_1^{11}} - \frac{4 t_2^{10} t_3^{11}}{t_1^{12}} + \frac{t_2^{10} t_3^{11}}{t_1^{11}}+ \frac{2 t_2^{11} t_3^{11}}{t_1^{12}} +\frac{t_2^{11} t_3^{11}}{t_1^{11}}~.\nn \\
\eea}
Let $R_i$ be the charges corresponding to the fugacity $t_i$.  Since the superpotential, which has fugacity $t_2 t_3 / t_1$, carries R-charge 2, it must be imposed that:
\bea
R_2 + R_3 - R_1 = 2.
\label{e:cyfano369}
\eea
Thus, the volume of $\cF_2$ is given by:
{\small
\bea
\lim_{\mu \rightarrow 0} \mu^4 \gm (e^{-\mu R_1}, e^{-\mu R_2}, e^{-\mu (2 + R_1 - R_2)};\cF_2)= \frac{p\left(R_1, R_2; \cF_2 \right)}{R_1^2 R_2^2(2 - R_1 + R_2)^2 (2 + R_1 - R_2)^2 (4 - R_1)^2 (4 - R_2)^2 }~, \nn \\
\label{e:volfano369}
\eea}
where:
{\small
\bea
p\left(R_1, R_2; \cF_2 \right) &=& 	4096 R_1 - 1792 R_1^3 + 128 R_1^4 + 192 R_1^5 - 32 R_1^6 + 6144 R_2 - 3072 R_1 R_2\nn \\
&& + 896 R_1^2 R_2 + 1280 R_1^3 R_2 - 480 R_1^4 R_2 - 32 R_1^5 R_2 + 16 R_1^6 R_2 + 4352 R_1 R_2^2 \nn \\
&& - 3680 R_1^2 R_2^2 + 352 R_1^3 R_2^2 + 184 R_1^4 R_2^2 - 16 R_1^5 R_2^2 - 4 R_1^6 R_2^2 - 2688 R_2^3 \nn \\
&& + 1472 R_1 R_2^3 + 880 R_1^2 R_2^3 - 336 R_1^3 R_2^3 - 12 R_1^4 R_2^3 + 12 R_1^5 R_2^3 + 192 R_2^4 \nn \\
&&- 1056 R_1 R_2^4 + 232 R_1^2 R_2^4 + 40 R_1^3 R_2^4 - 16 R_1^4 R_2^4 + 288 R_2^5 + 16 R_1 R_2^5 \nn \\
&& - 52 R_1^2 R_2^5 + 12 R_1^3 R_2^5 - 48 R_2^6 + 24 R_1 R_2^6 - 4 R_1^2 R_2^6~.
\label{e:numvolfano369}
\eea}
This function has a minimum at:
\bea
R_1 = R_3 \approx 2.210, \qquad R_2 = 2~.
\eea
The R-charge of the perfect matching corresponding to the divisor $D_\alpha$ is given by:
\bea
\lim_{\mu\rightarrow0}\frac{1}{\mu} \left[ \frac{g(D_\alpha; e^{- \mu R_1}, e^{- \mu R_2 }, e^{- \mu R_3 }; \cF_2) }{\gm(e^{-\mu R_1}, e^{- \mu R_2 }, e^{- \mu R_3 };\cF_2)}- 1 \right]~,
\eea
where $g(D_\alpha; e^{- \mu R_1}, e^{- \mu R_2 }, e^{- \mu R_3 }; \cF_2)$ is the Molien-Weyl integral with the insertion of the inverse of the weight corresponding to the divisor $D_\alpha$. The results are shown in Table \ref{t:chargefano369}.  The assignment of charges under the remaining abelian symmetries can be done by requiring that the superpotential is not charged under them and that the charge vectors are linearly independent. The assignments are shown in Table \ref{t:chargefano369}.

\begin{table}[h!]
 \begin{center}  
  \begin{tabular}{|c||c|c|c|c|c|c|c|c|c|}
  \hline
  \;& $SU(2)$&$U(1)_1$&$U(1)_2$&$U(1)_R$&$U(1)_{B_1}$&$U(1)_{B_2}$&$U(1)_{B_3}$&$U(1)_{B_4}$&fugacity\\
  \hline\hline  
   
  $p_1$&$  1$&$  0$&$ 0$&$0.350$&$ 0$&$ 0$ &$ 0$ &$ 0$ & $s_1 x $\\
  \hline
  
  $p_2$&$ -1$&$  0$&$ 0$&$0.350$&$ 0$&$ 0$ &$ 0$ &$ 0$ & $s_1 / x $\\
  \hline  
  
  $q_1$&$  0$&$  1$&$ 0$&$0.199$&$ 0$&$ 0$ &$ 0$ &$ 0$ & $s_2 q_1 $\\
  \hline
  
  $q_2$&$  0$&$ -1$&$ 0$&$0.199$&$ 1$&$ 0$ &$ 0$ &$ 0$ & $s_2 b_1/ q_1 $\\
  \hline
   
  $r_1$&$  0$&$  0$&$ 1$&$0.244$&$ 0$&$ 0$ &$ 0$ &$ 0$ & $s_3 q_2 $\\
  \hline
   
  $r_2$&$  0$&$  0$&$ 1$&$0.244$&$ 1$&$ 0$ &$ 0$ &$ 0$ & $s_3 b_1 q_2 $\\
  \hline
  
  $u_1$&$  0$&$  0$&$-1$&$0.160$&$ 0$&$ 0$ &$ 0$ &$ 0$ & $s_4 / q_2$\\
  \hline
  
  $u_2$&$  0$&$  0$&$-1$&$0.254$&$ 0$&$ 0$ &$ 0$ &$ 0$ & $s_5 / q_2$\\
  \hline
 
  $v_1$&$  0$&$  0$&$ 0$&$    0$&$ 0$&$ 1$ &$ 0$ &$ 0$ & $ b_2 $\\
  \hline

  $v_2$&$  0$&$  0$&$ 0$&$    0$&$ 0$&$ 0$ &$ 1$ &$ 0$ & $ b_3 $\\
  \hline
  
  $v_3$&$  0$&$  0$&$ 0$&$    0$&$-2$&$ 0$ &$ 0$ &$ 1$ & $ b_4 / b^2_1 $\\
  \hline
 
  $v_4$&$  0$&$  0$&$ 0$&$    0$&$ 0$&$-1$ &$-1$ &$-1$ & $1 / (b_2 b_3 b_4) $\\
  \hline
  
   \end{tabular}
  \end{center}
\caption{Charges of the perfect matchings under the global symmetry of the $\cF_2$ model. Here $s_\alpha$ is the fugacity of the R-charge, $x$ is the weight of the $SU(2)$ symmetry, $q_1, q_2, b_1, b_2, b_3$ and $b_4$ are, respectively, the charges under the mesonic abelian symmetries $U(1)_1, U(1)_2$ and of the three baryonic $U(1)_{B_1},U(1)_{B_2}, U(1)_{B_3}$ and $U(1)_{B_4}$.}
\label{t:chargefano369}
\end{table}

\begin{table}[h]
 \begin{center}  
  \begin{tabular}{|c||c|}
  \hline
  \; Quiver fields & R-charge\\
  \hline  \hline 
  $ X^{i}_{23}$ & 0.908 \\
  \hline
  $ X_{12},  X_{34} $ & 0.742 \\
  \hline
  $ X^{i}_{31}, X^i_{42}$ & 0.350 \\
  \hline
  $ X_{54}, X_{26} $ & 0.443 \\
  \hline
  $ X_{15} $ & 0.160 \\
  \hline
  $ X_{63} $ & 0.254 \\
  \hline
  \end{tabular}
  \end{center}
\caption{R-charges of the quiver fields for the $\cF_2$ model.}
\label{t:Rquivfano369}
\end{table}

\paragraph{The Hilbert series.}
The Hilbert series of the Master space can be obtained by integrating that of the space of perfect matchings over the $z_i$ fugacities:
{\footnotesize
\bea
g^{\firr{}}  (s_{\alpha}, x, q_i, b_j; \cF_2) &=& \left( \prod_{i=1}^4 \oint \limits_{|z_i|=1} \frac{\ud z_i}{2\pi i z_i} \right) \frac{1}{\left(1- s_1 x z_1\right)\left(1-\frac{s_1 z_1}{x}\right)\left(1- s_2 q_1 z_4\right)\left(1- \frac{s_2 z_3 b_1}{q_1 z_1 z_2 z_4}\right)}\nn \\
&\times& \frac{1}{\left(1- \frac{s_3 q_2 z_1 z_2 z_4}{z_3}\right)\left(1- \frac{s_3 q_2 b_1}{z_4}\right)\left(1- \frac{s_4 z_2}{q_2}\right)\left(1- \frac{s_5 z_3}{q_2}\right)\left(1-\frac{b_2}{z_3}\right)\left(1-\frac{b_3}{z_2}\right)}\nn \\
&\times& \frac{1}{\left(1-\frac{b_4}{z_1 b^2_1}\right)\left(1-\frac{1}{z_1 b_2 b_3 b_4}\right)}\nn \\
&=& \frac{\cP(s_{\alpha}, x, q_i, b_j; \cF_2)}{\left(1- \frac{s_1 s^2_2 s_4 x b_1 b_2}{q_2}\right)\left(1- \frac{s_1 s^2_2 s_4 b_1 b_2}{x q_2}\right)\left(1-\frac{s_1 x b_4}{b^2_1}\right)\left(1-\frac{s_1 b_4}{x b^2_1}\right)\left(1-\frac{s_1 x}{b_2 b_3 b_4}\right)}\nn \\
&\times& \frac{1}{\left(1-\frac{s_1}{x b_2 b_3 b_4}\right)\left(1- s_2 s_3 q_1 q_2 b_1\right)\left(1-\frac{s_2 s_3 q_2 b_1}{q_1}\right)\left(1-\frac{s^2_3 s_5 b_1 q_2}{b_2 b_4}\right)\left(1-\frac{s^2_3 s_5 b_3 b_4}{b_1}\right)}\nn \\
&\times& \frac{1}{\left(1-\frac{s_4 b_3}{q_2}\right)\left(1- \frac{s_5 b_2}{q_2}\right)}~, \nn \\
\label{e:HSmasterfano369}
\eea}
where $\cP\left(s_{\alpha}, x, q_1, q_2, b_j; \cF_2\right)$ is a polynomial which is not reported here. The Hilbert series of the mesonic moduli space can be obtained by integrating over the three baryonic fugacities, 
{\footnotesize
\bea
\gm(s_{\alpha}, x, q_1, q_2; \cF_2) &=& \left( \prod^4_{i=1} \oint\limits_{|b_j|=1} \frac{\ud b_j}{2 \pi i b_j} \right) g^{\firr{}} (s_{\alpha}, x, q_i, b_j; \cF_2)\nn \\
&=& \frac{P(s_{\alpha}, x, q_1, q_2; \cF_2)}{\left(1-\frac{s^3_1 s^3_2 s_3 s^2_4 x^3 q_1}{q_2}\right)\left(1-\frac{s^3_1 s^3_2 s_3 s^2_4 q_1}{x^3 q_2}\right)\left(1-\frac{s^3_1 s^3_2 s_3 s^2_4 x^3}{q_
1 q_2}\right)\left(1-\frac{s^3_1 s^3_2 s_3 s^2_4}{x^3 q_1 q_2}\right)} \nn \\
&\times& \frac{1}{\left(1-s^2_1 s^2_2 s^2_3 s_4 s_5 x^2 q^2_1\right)\left(1-\frac{s^2_1 s^2_2 s^2_3 s_4 s_5 q^2_1}{x^2}\right)\left(1-\frac{s^2_1 s^2_2 s^2_3 s_4 s_5 x^2}{q^2_1}\right)\left(1-\frac{s^2_1 s^2_2 s^2_3 s_4 s_5}{x^2 q^2_1}\right)}\nn \\
&\times& \frac{1}{\left(1-s_1 s_2 s^3_3 s^2_5 x q_1 q_2\right)\left(1-\frac{s_1 s_2 s^3_3 s^2_5 q_1 q_2}{x}\right)\left(1-\frac{s_1 s_2 s^3_3 s^2_5 x q_2}{q_1}\right)\left(1-\frac{s_1 s_2 s^3_3 s^2_5 q_2}{x q_1}\right)}~. \nn \\
\label{e:HSmesfano369}
\eea}
where $P(s_{\alpha}, x, q_1, q_2; \cF_2)$ is a polynomial that is not reported here.
The plethystic logarithm of the Hilbert series of the mesonic moduli space is:
\bea
\PL[\gm(t_{\alpha}, x, q_1, q_2; \cF_2)] &=& [3]\left(q_1+\frac{1}{q_1}\right)\frac{t_1}{q_2} + [2]\left(q^2_1 + 1 + \frac{1}{q^2_1}\right)t_2 + [1]\left(q_1+\frac{1}{q_1}\right) \frac{q_2 t^2_2}{t_1} - O(t^2_2) \nn \\
\label{e:plfano369}
\eea
The plethystic logarithm shows that in the mesonic moduli space the abelian symmetry $U(1)_1$ is enhanced to $SU(2)$.

\paragraph{The generators.} From the positive terms of the plethystic logarithm, it can be deduced that the generators of the mesonic moduli space are:
\bea
&& p_i p_j p_k q_l^2 q_m r_m u^2_1 v_1 v_2 v_3 v_4, \quad p_i p_j q^2_l r^2_m u_1 u_2 v_1 v_2 v_3 v_4,\nn \\
&& p_i p_j q_1 q_2 r_1 r_2 u_1 u_2 v_1 v_2 v_3 v_4, \quad p_i q_l r_l r^2_m u^2_2 v_1 v_2 v_3 v_4,
\label{e:genfano369}
\eea
with $i,j,k,l,m=1,2$ with $l\neq m$.
The R-charges of the generators are listed in Table \ref{t:Rgenfano369}.
\begin{table}[h]
 \begin{center}  
  \begin{tabular}{|c||c|}
  \hline
  \; Generators& R-charge \\
  \hline  \hline 
  $p_i p_j p_k q_l^2 q_m r_m u^2_1 v_1 v_2 v_3 v_4$ & 2.211\\
  \hline
  $p_i p_j q^2_l r^2_m u_1 u_2 v_1 v_2 v_3 v_4$ & 2\\
  \hline
  $p_i p_j q_1 q_2 r_1 r_2 u_1 u_2 v_1 v_2 v_3 v_4$ & 2\\
  \hline
  $p_i q_l r_l r^2_m u^2_2 v_1 v_2 v_3 v_4$ & 1.789\\
  \hline
  \end{tabular}
  \end{center}
\caption{R-charges of the generators of the mesonic moduli space for the $\cF_2$ Model.}
\label{t:Rgenfano369}
\end{table}

\begin{figure}[ht]
\begin{center}
\includegraphics[totalheight=4cm]{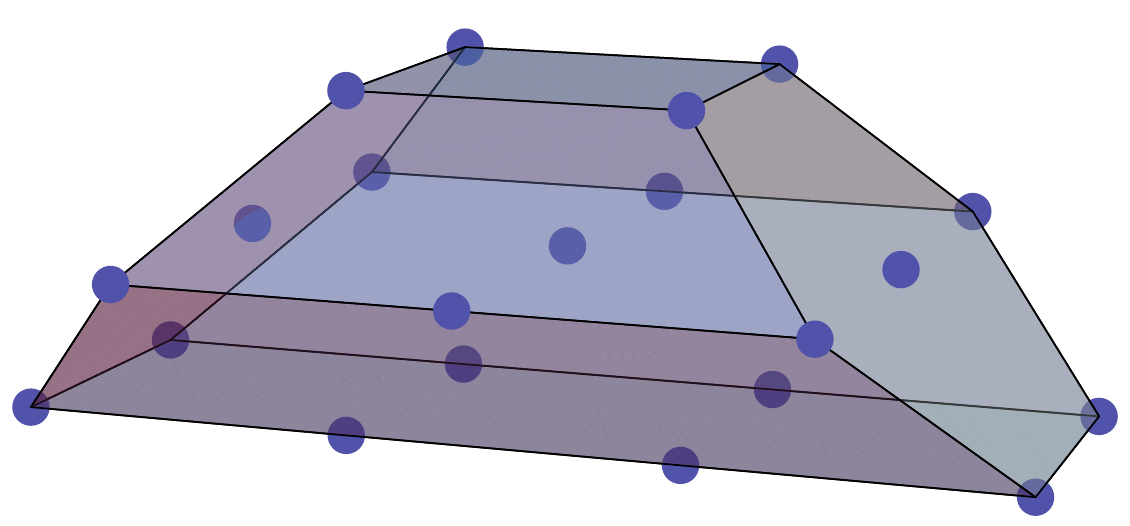}
\caption{The lattice of generators of the $\cF_2$ theory.}
  \label{f:latd4}
  \end{center}
\end{figure}

\section{$\cF_1$ (Toric Fano 324): $dP_3 \times \BP^1$}
The model has 6 gauge groups and 10 chiral fields: $X^i_{12}, X^i_{23}, X^i_{34}$ (with $i=1,2$), $X_{46},X_{61}, X_{45}$ and $X_{51}$. The quiver diagram and tiling are presented in Figure \ref{f:fano324tileandquiver}.
The superpotential can be read off from the tiling as
\bea
W = \tr \left[ \epsilon_{ij}  \left(X^i_{12}X^1_{23}X^j_{34}X_{45}X_{51} - X^j_{12}X^2_{23}X^i_{34}X_{46}X_{61}\right) \right]~.
\label{e:spotfano324}
\eea
Let us choose the CS levels to be $\vec{k}=(0,0,0,0,-1,1)$.
\begin{figure}[ht]
\begin{center}
\vskip 0.5cm
\hskip -6cm
\includegraphics[totalheight=3.4cm]{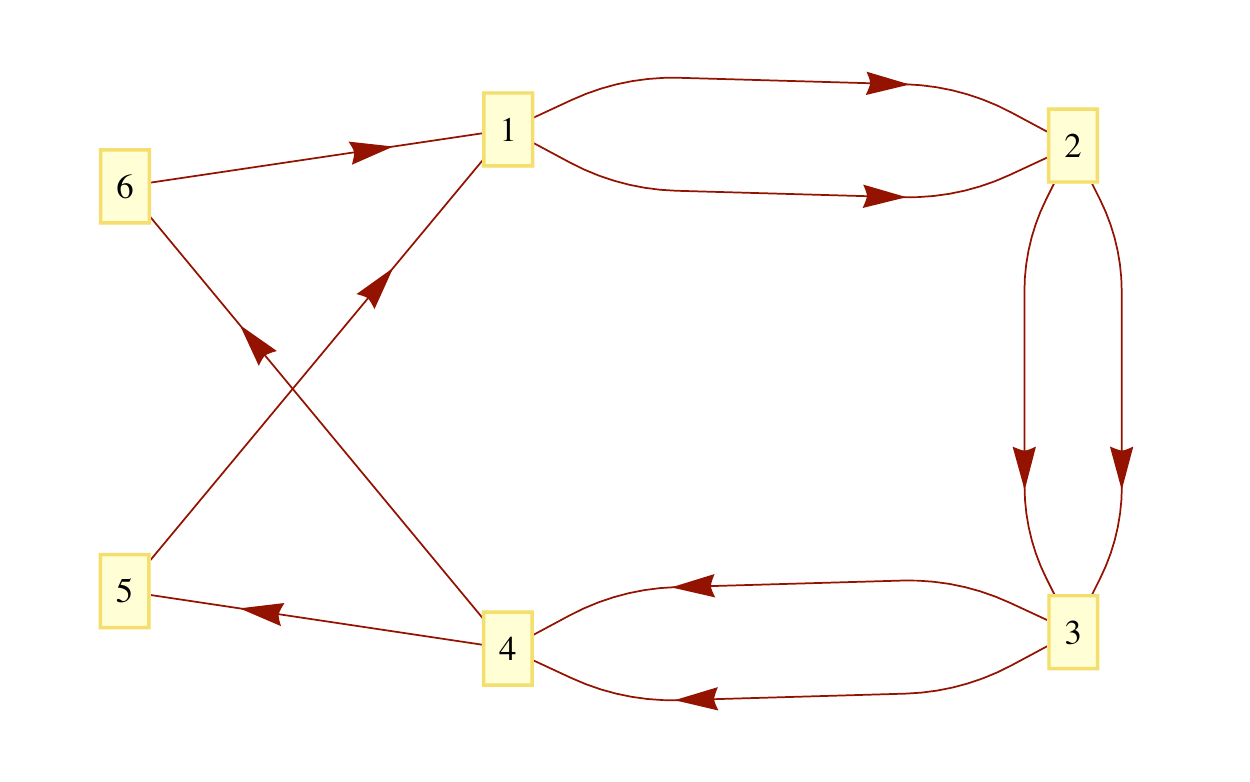}
\vskip -4 cm
\hskip 8 cm
\includegraphics[totalheight=4.5cm]{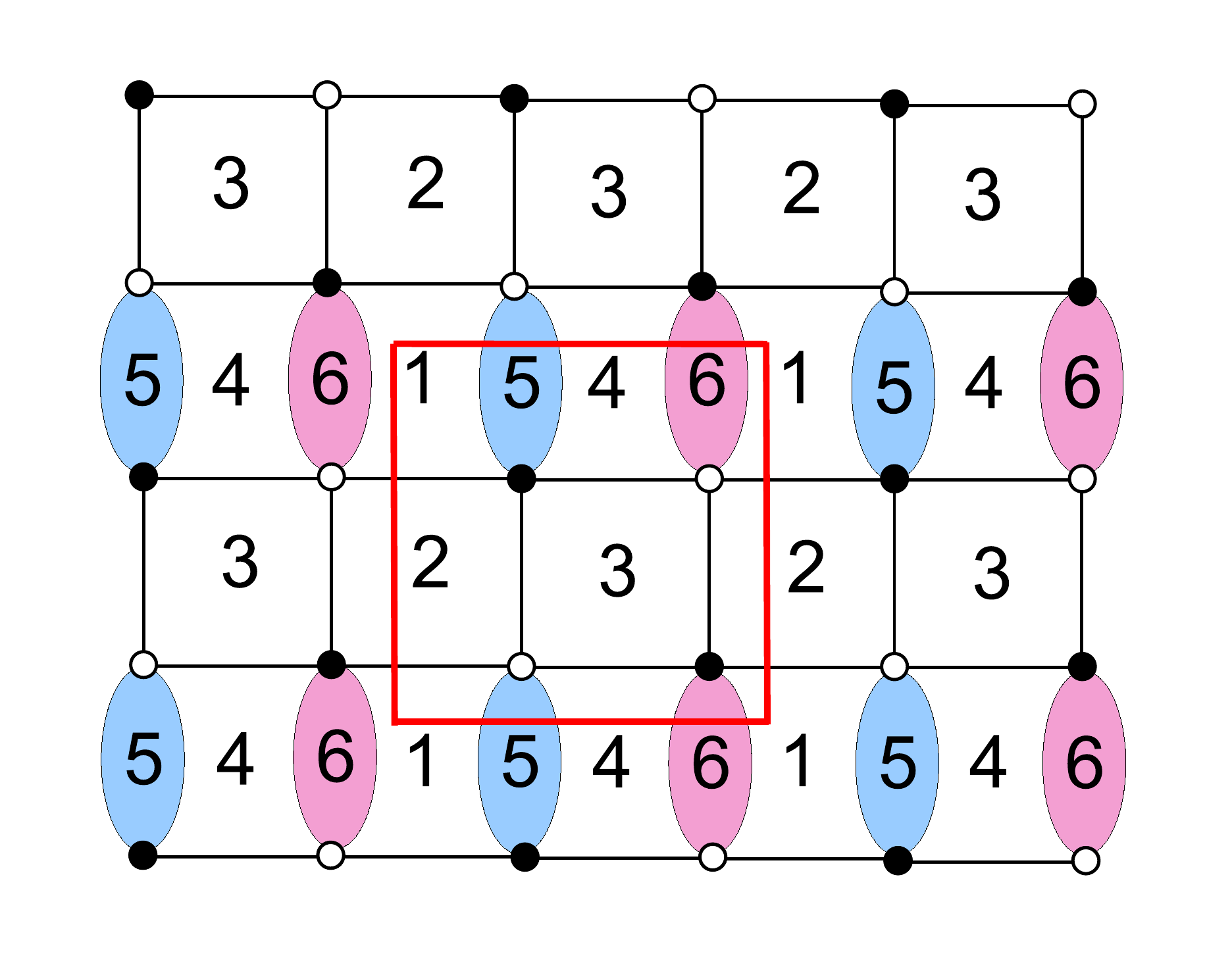}
 \caption{(i) Quiver of the $\cF_1$  model. \qquad (ii) Tiling of the $\cF_1$  model.}
  \label{f:fano324tileandquiver}
  \end{center}
\end{figure} 

\comment{
\begin{figure}[ht]
\begin{center}
\vskip -1cm
\includegraphics[totalheight=8cm]{fano324tilingfd.pdf}
  \caption{The fundamental domain of the tiling for the $\cF_1$ model.}
  \label{f:fano324funddom}
\end{center}
 \end{figure} 
}

\paragraph{The Kasteleyn matrix.} The Chern-Simons levels of this model can be written in terms of the integers $n^i_{jk}$ and $n_{jk}$ as:
\bea
\begin{array}{ll}
\text{Gauge group 1:} \qquad  k_1 &=   n^{1}_{12} + n^{2}_{12} - n_{51} - n_{61} ~, \nn \\
\text{Gauge group 2:} \qquad  k_2 &=   n^{1}_{23} + n^{2}_{23} - n^{1}_{12} - n^{2}_{12} ~, \nn \\
\text{Gauge group 3:} \qquad  k_3 &=   n^{1}_{34} + n^{2}_{34} - n^{1}_{23} - n^{2}_{23} ~, \nn \\
\text{Gauge group 3:} \qquad  k_4 &=   n_{45} + n_{46} - n^{1}_{34} - n^{2}_{34} ~, \nn \\
\text{Gauge group 4:} \qquad  k_5 &=   n_{51} - n_{45} ~, \nn \\
\text{Gauge group 5:} \qquad  k_6 &=   n_{61} - n_{46} ~.
\label{e:kafano324}
\end{array}
\eea
Let us choose 
\bea
n_{45} =-n_{46} =   1,~ n^i_{jk}=n_{jk}=0 \; \text{otherwise}~.
\eea
The fundamental domain contains two pairs of white and black nodes, and so the Kasteleyn matrix is a $2 \times 2$ matrix:
\bea
K= \left(
\begin{array}{c|cc}
& w_1 & w_2 \\
\hline
b_1 & z^{n^{2}_{34}} + x z^{n^{1}_{12}} &\ z^{n^{2}_{23}} + \frac{1}{y} z^{n_{46}}+ \frac{1}{y} z^{n_{61}}   \\
b_2 & z^{n^{1}_{23}} + y z^{n_{45}} + y z^{n_{51}} &\ z^{n^{1}_{34}} + \frac{1}{x} z^{n^{2}_{12}}  
\end{array}
\right)~.
\label{e:kastfano324}
\eea
The permanent of the Kasteleyn matrix can be written as
\bea
\perm~K &=&  x z^{(n^1_{12} + n^1_{34})} +  x^{-1} z^{(n^2_{12} + n^2_{34})} +  y z^{(n_{51} + n^2_{23})}+  y^{-1} z^{(n_{61} + n^1_{23})} \nn \\ 
&+&  y z^{(n_{45} + n^2_{23})} +  y^{-1} z^{(n^1_{23} + n_{46})} +  z^{(n_{51} + n_{46})}+  z^{(n_{61} + n_{45})}\nn \\  
&+& z^{(n^1_{12} + n^2_{12})} + z^{(n^1_{34} + n^2_{34})} +  z^{(n_{61} + n_{51})} + z^{(n^1_{23} + n^2_{23})} + z^{(n_{46} + n_{45} )}\nn \\
&=& x + x^{-1} + y + y^{-1} + y z+ y^{-1} z^{-1} + z^{-1} + z + 5\nn \\
&& \text{(for $n_{45} =-n_{46} =   1,~ n^i_{jk}=n_{jk}=0 \; \text{otherwise}$)} ~. 
\label{e:charpolyfano324}
\eea
The perfect matchings can then be written in terms of the chiral fields as:
\bea 
&&p_1 = \left\{X^1_{12}, X^1_{34}\right\}, \;\; p_2 = \left\{X^2_{12}, X^2_{34}\right\}, \;\; q_1 = \left\{X_{51}, X^2_{23}\right\}, \;\; q_2 = \left\{X_{61},X^1_{23}\right\}, \nn \\
&&   r_1 = \left\{X_{45}, X^2_{23}\right\}, \;\; r_2 = \left\{X^1_{23},X_{46}\right\}, \;\; u_1 = \left\{X_{51},X_{46}\right\}, \;\; u_2 = \left\{X_{61},X_{45}\right\},\nn \\
&&   v_1 = \left\{X^1_{12}, X^2_{12}\right\}, \;\; v_2 = \left\{X^1_{34},X^2_{34}\right\}, \;\; v_3 = \left\{X_{61},X_{51}\right\}, \;\; v_4 = \left\{X^1_{23},X^2_{23}\right\},\nn \\
&&   v_5 = \left\{X_{46},X_{45}\right\}~.
\eea
Note that the perfect matchings $v_1, v_2, v_3, v_4$ and $v_5$ correspond to internal points in the toric diagram.
In turn, chiral fields can be written as products of perfect matchings:
\bea
\begin{array}{llll}
X^1_{12} = p_1 v_1, \quad & X^2_{12} = p_2 v_1, \quad & X^1_{34} = p_1 v_2, \quad & X^2_{34} = p_2 v_2,\nn \\
X_{51} = q_1 u_1 v_3, \quad & X_{61} = q_2 u_2 v_3, \quad &  X^2_{23} = q_1 r_1 v_4, \quad & X^1_{23} = q_2 r_2 v_4,\nn \\
 X_{46} = r_2 u_1 v_5 , \quad & X_{45} = r_1 u_2 v_5~. & &
\end{array}
\eea
These pieces of information are summarized in the following $P$ matrix:
\beq
P=\left(\begin{array} {c|ccccccccccccc}
&p_1 & p_2 & q_1 & q_2 & r_1 & r_2 & u_1 & u_2 & v_1 & v_2 & v_3 & v_4 & v_5\\
\hline
X^1_{12}  &1&0&0&0&0&0&0&0&1&0&0&0&0\\
X^2_{12}  &0&1&0&0&0&0&0&0&1&0&0&0&0\\
X^1_{34}  &1&0&0&0&0&0&0&0&0&1&0&0&0\\
X^2_{34}  &0&1&0&0&0&0&0&0&0&1&0&0&0\\
X_{51}    &0&0&1&0&0&0&1&0&0&0&1&0&0\\
X_{61}    &0&0&0&1&0&0&0&1&0&0&1&0&0\\
X^2_{23}  &0&0&1&0&1&0&0&0&0&0&0&1&0\\
X^1_{23}  &0&0&0&1&0&1&0&0&0&0&0&1&0\\
X_{46}    &0&0&0&0&0&1&1&0&0&0&0&0&1\\
X_{45}    &0&0&0&0&1&0&0&1&0&0&0&0&1
\end{array}\right)~.
\label{e:pfano324}
\eeq
The kernel of $P$ is given by:
\bea
Q_F= \left(
\begin{array}{ccccccccccccc}
 1 & 1 & 0 & 0 & 0 & 0 & 0 & 0 &-1 &-1 & 0 & 0 & 0\\
 0 & 0 & 1 & 0 &-1 & 0 &-1 & 0 & 0 & 0 & 0 & 0 & 1\\
 0 & 0 & 0 & 1 & 0 &-1 & 1 & 0 & 0 & 0 &-1 & 0 & 0\\
 0 & 0 & 0 & 0 & 1 & 1 & 0 & 0 & 0 & 0 & 0 &-1 &-1\\
 0 & 0 & 0 & 0 & 0 & 0 & 1 & 1 & 0 & 0 &-1 & 0 &-1
 \end{array}
\right)~. \label{e:qffano324}
\eea
Therefore, the relations between the perfect matchings are given by:
\bea \label{relperm324}
p_1 + p_2 - v_1 - v_2 &=& 0~, \nn \\
q_1 - r_1 - u_1 + v_5 &=& 0~, \nn \\
q_2 - r_2 + u_1 - v_3 &=& 0~, \nn \\
r_1 + r_2 - v_4 - v_5 &=& 0~, \nn \\
u_1 + u_2 - v_3 - v_5 &=& 0~.
\eea

\paragraph{The toric diagram.} The toric diagram of this model is constructed using two different methods.
\begin{itemize}
\item {\bf The Kasteleyn matrix.} The powers of $x$, $y$ and $z$ in each of the terms of \eref{e:charpolyfano324} give the coordinates of the toric diagram and can be collected in the columns of the $G_K$ matrix:
\bea
G_K = \left(
\begin{array}{ccccccccccccc}
  1 & -1 & 0 &  0 & 0 & 0 &  0 &  0 & 0 & 0 & 0 & 0 & 0 \\
  0 &  0 & 1 & -1 & 1 &-1 &  0 &  0 & 0 & 0 & 0 & 0 & 0 \\
  0 &  0 & 0 &  0 & 1 &-1 & -1 &  1 & 0 & 0 & 0 & 0 & 0
\end{array}
\right)~.
\label{e:Gkfano324}
\eea
The first row contains the weights of the fundamental representation of $SU(2)$. Thus, the mesonic symmetry contains one $SU(2)$ as the only non-abelian symmetry. 

\item {\bf The charge matrices.} Since the model has 6 gauge groups, there are 4 baryonic symmetries that come from the D-terms. The charges of the perfect matchings under these symmetries are collected in the columns of the following matrix:
\be
Q_D =   \left(
\begin{array}{ccccccccccccc}
0 & 0 & 0 & 0 & 0 & 0 & 0 &  0 & 1 & 0 &-1 & 0 & 0\\
0 & 0 & 0 & 0 & 0 & 0 & 0 &  0 & 0 & 1 & 0 &-1 & 0\\
0 & 0 & 0 & 0 & 0 & 0 & 0 &  0 & 0 & 0 & 1 & 0 &-1\\
0 & 0 & 0 & 0 & 0 & 0 & 0 &  0 & 0 & 0 & 0 & 1 &-1
\end{array}
\right). 
\label{e:qdfano324}
\ee
Combining the matrices (\ref{e:qffano324}) and (\ref{e:qdfano324}) gives the total charge matrix $Q_t$:
\be
Q_t = { \Blue Q_F \choose \Green Q_D \Black } =   \left( 
\begin{array}{ccccccccccccc} \Blue
1 & 1 & 0 & 0 & 0 & 0 & 0 & 0 &-1 &-1 & 0 & 0 & 0\\
0 & 0 & 1 & 0 &-1 & 0 &-1 & 0 & 0 & 0 & 0 & 0 & 1\\
0 & 0 & 0 & 1 & 0 &-1 & 1 & 0 & 0 & 0 &-1 & 0 & 0\\
0 & 0 & 0 & 0 & 1 & 1 & 0 & 0 & 0 & 0 & 0 &-1 &-1\\
0 & 0 & 0 & 0 & 0 & 0 & 1 & 1 & 0 & 0 &-1 & 0 &-1\\ \Green
0 & 0 & 0 & 0 & 0 & 0 & 0 &  0 & 1 & 0 &-1 & 0 & 0\\
0 & 0 & 0 & 0 & 0 & 0 & 0 &  0 & 0 & 1 & 0 &-1 & 0\\
0 & 0 & 0 & 0 & 0 & 0 & 0 &  0 & 0 & 0 & 1 & 0 &-1\\
0 & 0 & 0 & 0 & 0 & 0 & 0 &  0 & 0 & 0 & 0 & 1 &-1 \Black
\end{array}
\right). 
\label{e:qtfano324}
\ee
The kernel of this matrix is the $G_t$ matrix. The removal of the first row gives the $G'_t$ matrix, whose columns contain the coordinates of the toric diagram:
\bea
G'_t = \left(
\begin{array}{ccccccccccccc}
  1 & -1 & 0 &  0 & 0 & 0 &  0 &  0 & 0 & 0 & 0 & 0 & 0 \\
  0 &  0 & 1 & -1 & 1 &-1 &  0 &  0 & 0 & 0 & 0 & 0 & 0 \\
  0 &  0 & 0 &  0 & 1 &-1 & -1 &  1 & 0 & 0 & 0 & 0 & 0
\end{array}
\right) = G_K~. \label{e:toricdiafano324}
\eea
The toric diagram constructed from (\ref{e:toricdiafano324}) is drawn in Figure \ref{f:tdtoricfano324}.  Note that the 7 blue points form the toric diagram of $dP_3$, and the 2 black points together with the blue internal point form the toric diagram of $\BP^1$. Therefore, the mesonic moduli space of this theory is $dP_3 \times \BP^1$.

\begin{figure}[ht]
\begin{center}
  \includegraphics[totalheight=3.0cm]{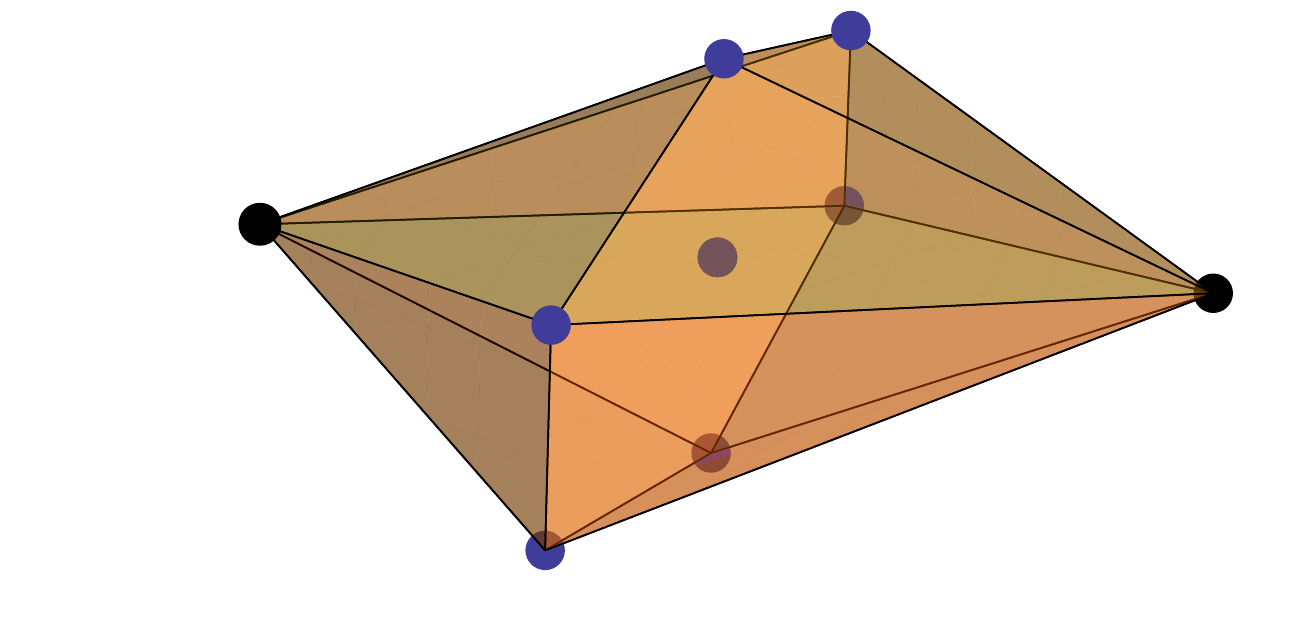}
 \caption{The toric diagram of $\cF_1$.}
  \label{f:tdtoricfano324}
\end{center}
\end{figure}
\end{itemize}

\paragraph{The baryonic charges.} The toric diagram contains 8 external points and, therefore, the model has $8-4=4$ baryonic charges, which are denoted by $U(1)_{B_1}, U(1)_{B_2}, U(1)_{B_3}$ and $U(1)_{B_4}$. All these come from the D-terms and the baryonic charges of the perfect matchings are encoded in the rows of the $Q_D$ matrix  (\ref{e:qdfano324}).

\paragraph{The global symmetry.} The $Q_t$ matrix contains a single pair of repeated columns, confirming that the mesonic symmetry contains one $SU(2)$ as the only non-abelian factor.  Since the mesonic symmetry must have total rank 4, it follows that it can be identified with $SU(2)\times U(1)^3$, where one of the three abelian factors corresponds to the R-symmetry. The perfect matchings $p_1$ and $p_2$ transform as a doublet under the non-abelian factor. Also, the perfect matchings $v_1,\ldots,v_5$ correspond to internal points in the toric diagram and so they carry zero R-charges.

Since there are three $U(1)$ factors in the mesonic symmetry, a volume minimisation problem needs to be solved in order to assign the R-charges to the perfect matchings. Let us assign the R-charge fugacities $s_1$ to both $p_1$ and $p_2$ (note that the non-abelian symmetry does not play any role in the volume minimisation, so $p_1$ and $p_2$ carry the same R-charges), and the R-charge fugacities $s_2, s_3, s_4, s_5, s_6$ and $s_7$ to the perfect matchings $q_1, q_2, r_1, r_2, u_1$ and $u_2$ respectively. Therefore, the Hilbert series of the mesonic moduli space is:
{\small
\bea
\gm (t_{\alpha}; \cF_1) &=& \left( \prod^4_{i=1}  \oint \limits_{|b_i|=1} \frac{\ud b_i}{2\pi i b_i}  \prod^5_{j=1} \oint \limits_{|z_j|=1} \frac{\ud z_j}{2 \pi i z_j} \right) \frac{1}{\left(1- s_1 z_1\right)^2\left(1- s_2 z_2\right)\left(1-s_3 z_3\right)\left(1-\frac{s_4 z_4}{z_2}\right)}\nn \\
&&\times \frac{1}{\left(1-\frac{s_5 z_4}{z_3}\right)\left(1-\frac{s_6 z_3 z_5}{z_2}\right)\left(1-s_7 z_5\right)\left(1-\frac{b_1}{z_1}\right)\left(1-\frac{b_2}{z_1}\right)\left(1-\frac{b_3}{b_1 z_3 z_5}\right)}\nn \\
&&\times \frac{1}{\left(1-\frac{ b_4}{b_2 z_4}\right)\left(1-\frac{ z_2}{b_3 b_4 z_4 z_5}\right)}~.
\label{e:HSmesvolfano324}
\eea}
Since there are three $U(1)$ factors in the mesonic symmetry, the integral (\ref{e:HSmesvolfano324}) depends only on three combinations of fugacities. Defining:
\bea
t^2_1 = s^2_1 s^2_2 s_4 s_5 s_6^2 , \quad t^2_2 = s^2_1 s_2 s_3 s^2_5 s^2_6, \quad t^2_3 = s^2_1 s^2_3 s_4 s_5 s^2_7~,
\eea
gives us
{\footnotesize
\bea
\gm (t_1, t_2, t_3; \cF_1) &=& \frac{P(t_1, t_2, t_3; \cF_1)}{\left(1-t^2_1\right)^2\left(1-t^2_2\right)^2\left(1-t^2_3\right)^2\left(1-\frac{t^2_1 t^2_3}{t^2_2}\right)^2\left(1-\frac{t^3_1 t_3}{t^2_2}\right)^2\left(1-\frac{t^2_2 t_3}{t_1}\right)^2}~, \qquad \quad
\label{e:HSmesvolTfano324}
\eea}
where: 
{\scriptsize
\bea
P(t_1, t_2, t_3; \cF_1) &=& 1 + t_1^2 + t_2^2 - 3 t_1^2 t_2^2 + 3 t_1 t_3 - 9 t_1^3 t_3 + 8 t_1^5 t_3 - 9 t_1 t_2^2 t_3 + 5 t_1^3 t_2^2 t_3- t_1^5 t_2^2 t_3+ 8 t_1 t_2^4 t_3-t_1^3 t_2^4 t_3 + t_3^2 - \nn \\
&& 19 t_1^2 t_3^2 + 28 t_1^4 t_3^2 - 5 t_1^6 t_3^2 - t_1^8 t_3^2- 9 t_2^2 t_3^2 + 28 t_1^2 t_2^2 t_3^2 - 21 t_1^4 t_2^2 t_3^2 + 2 t_1^6 t_2^2 t_3^2 + 5 t_2^4 t_3^2-5 t_1^2 t_2^4 t_3^2 + 2 t_1^4 t_2^4 t_3^2 -\nn \\
&&  t_2^6 t_3^2 - t_1^2 t_2^6 t_3^2 - 9 t_1 t_3^3 + 39 t_1^3 t_3^3- 30 t_1^5 t_3^3+ 5 t_1^7 t_3^3 - t_1^9 t_3^3 + 28 t_1 t_2^2 t_3^3- 30 t_1^3 t_2^2 t_3^3+t_1^5 t_2^2 t_3^3 + 2 t_1^7 t_2^2 t_3^3  -\nn \\
&&  21 t_1 t_2^4 t_3^3 + 5 t_1^3 t_2^4 t_3^3 + 2 t_1^5 t_2^4 t_3^3  + 2 t_1 t_2^6 t_3^3 - t_1^3 t_2^6 t_3^3 + 28 t_1^2 t_3^4 - 66 t_1^4 t_3^4 + 27 t_1^6 t_3^4+ 5 t_1^8 t_3^4 + 5 t_2^2 t_3^4 - \nn \\
&&30 t_1^2 t_2^2 t_3^4 + 27 t_1^4 t_2^2 t_3^4 + t_1^6 t_2^2 t_3^4 - t_1^8 t_2^2 t_3^4 - 5 t_2^4 t_3^4 + t_1^2 t_2^4 t_3^4 + 5 t_1^4 t_2^4 t_3^4 - t_1^6 t_2^4 t_3^4 +2 t_2^6 t_3^4 + 2 t_1^2 t_2^6 t_3^4 + 8 t_1 t_3^5 -\nn \\
&&  30 t_1^3 t_3^5 - 6 t_1^5 t_3^5 + 27 t_1^7 t_3^5 - 5 t_1^9 t_3^5 - 21 t_1 t_2^2 t_3^5 + 27 t_1^3 t_2^2 t_3^5 + 27 t_1^5 t_2^2 t_3^5-21 t_1^7 t_2^2 t_3^5  + 5 t_1 t_2^4 t_3^5 + t_1^3 t_2^4 t_3^5 -\nn \\
&&  5 t_1^5 t_2^4 t_3^5 + 2 t_1 t_2^6 t_3^5 - t_1^3 t_2^6 t_3^5 - 5 t_1^2 t_3^6 + 27 t_1^4 t_3^6 - 6 t_1^6 t_3^6 - 30 t_1^8 t_3^6 + 8 t_1^{10} t_3^6  - t_2^2 t_3^6 + t_1^2 t_2^2 t_3^6 + 27 t_1^4 t_2^2 t_3^6 - \nn \\
&&30 t_1^6 t_2^2 t_3^6 + 5 t_1^8 t_2^2 t_3^6 + 2 t_2^4 t_3^6 + 5 t_1^2 t_2^4 t_3^6 - 21 t_1^4 t_2^4 t_3^6 + 8 t_1^6 t_2^4 t_3^6 -t_2^6 t_3^6 - t_1^2 t_2^6 t_3^6 + 5 t_1^3 t_3^7 + 27 t_1^5 t_3^7 -66 t_1^7 t_3^7 +\nn \\
&&   28 t_1^9 t_3^7  + 2 t_1 t_2^2 t_3^7 + t_1^3 t_2^2 t_3^7 - 30 t_1^5 t_2^2 t_3^7 + 28 t_1^7 t_2^2 t_3^7-3 t_1^9 t_2^2 t_3^7 + 2 t_1 t_2^4 t_3^7 - 5 t_1^3 t_2^4 t_3^7+ 5 t_1^5 t_2^4 t_3^7 -t_1^2 t_3^8 + \nn \\
&&5 t_1^4 t_3^8 -  30 t_1^6 t_3^8 + 39 t_1^8 t_3^8 - 9 t_1^{10} t_3^8  + 2 t_1^2 t_2^2 t_3^8 -21 t_1^4 t_2^2 t_3^8 + 28 t_1^6 t_2^2 t_3^8 - 9 t_1^8 t_2^2 t_3^8 - t_1^2 t_2^4 t_3^8 +  8 t_1^4 t_2^4 t_3^8-\nn \\
&& 3 t_1^6 t_2^4 t_3^8 - t_1^3 t_3^9 - 5 t_1^5 t_3^9 + 28 t_1^7 t_3^9 - 19 t_1^9 t_3^9 +t_1^{11} t_3^9 - t_1^3 t_2^2 t_3^9 + 5 t_1^5 t_2^2 t_3^9 - 9 t_1^7 t_2^2 t_3^9 + t_1^9 t_2^2 t_3^9 + 8 t_1^6 t_3^{10} -\nn \\
&&  9 t_1^8 t_3^{10} + 3 t_1^{10} t_3^{10} - 3 t_1^6 t_2^2 t_3^{10} + t_1^8 t_2^2 t_3^{10} +t_1^9 t_3^{11} + t_1^{11} t_3^{11} + \frac{t_1^3 t_3}{t_2^2} - \frac{3 t_1^5 t_3}{t_2^2} + \frac{t_2^2 t_3}{t_1} - \frac{3 t_2^4 t_3}{t_1}  +\frac{t_1^2 t_3^2}{t_2^2} - \nn \\
&&  \frac{9 t_1^4 t_3^2}{t_2^2} + \frac{ 5 t_1^6 t_3^2}{t_2^2} - \frac{t_1^8 t_3^2}{t_2^2} - \frac{3 t_1^5 t_3^3}{t_2^4} + \frac{8 t_1^7 t_3^3}{t_2^4} -\frac{t_1^9 t_3^3}{t_2^4} - \frac{9 t_1^3 t_3^3}{t_2^2} + \frac{28 t_1^5 t_3^3}{t_2^2} - \frac{21 t_1^7 t_3^3}{t_2^2} + \frac{2 t_1^9 t_3^3}{t_2^2} - \frac{3 t_2^2 t_3^3}{t_1}+ \nn \\
&& \frac{8 t_2^4 t_3^3}{t_1}- \frac{t_2^6 t_3^3}{t_1} + \frac{5 t_1^6 t_3^4}{t_2^4} - \frac{5 t_1^8 t_3^4}{t_2^4} + \frac{2 t_1^{10} t_3^4}{t_2^4}-\frac{3 t_1^2 t_3^4}{t_2^2} + \frac{28 t_1^4 t_3^4}{t_2^2} - \frac{30 t_1^6 t_3^4}{t_2^2} + \frac{t_1^8 t_3^4}{t_2^2} + \frac{2 t_1^{10} t_3^4}{t_2^2}- \frac{t_1^9 t_3^5}{t_2^6} -\nn \\
&&  \frac{t_1^{11} t_3^5}{t_2^6} + \frac{8 t_1^5 t_3^5}{t_2^4} - \frac{21 t_1^7 t_3^5}{t_2^4} + \frac{5 t_1^9 t_3^5}{t_2^4} + \frac{2 t_1^{11} t_3^5}{t_2^4}+\frac{5 t_1^3 t_3^5}{t_2^2} - \frac{30 t_1^5 t_3^5}{t_2^2}+ \frac{27 t_1^7 t_3^5}{t_2^2} + \frac{t_1^9 t_3^5}{t_2^2} - \frac{t_1^{11} t_3^5}{t_2^2}- \frac{t_2^4 t_3^5}{t_1} -\nn \\
&&   \frac{t_2^6 t_3^5}{t_1} - \frac{t_1^8 t_3^6}{t_2^6} +\frac{2 t_1^{10} t_3^6}{t_2^6} - \frac{t_1^{12} t_3^6}{t_2^6} - \frac{5 t_1^6 t_3^6}{t_2^4}+ \frac{t_1^8 t_3^6}{t_2^4} +\frac{5 t_1^{10} t_3^6}{t_2^4} - \frac{t_1^{12} t_3^6}{t_2^4} - \frac{21 t_1^4 t_3^6}{t_2^2}+ \frac{27 t_1^6 t_3^6}{t_2^2} + \frac{27 t_1^8 t_3^6}{t_2^2} - \nn \\
&&  \frac{21 t_1^{10} t_3^6}{t_2^2}+\frac{2 t_1^9 t_3^7}{t_2^6} + \frac{2 t_1^{11} t_3^7}{t_2^6} - \frac{t_1^5 t_3^7}{t_2^4} + \frac{5 t_1^7 t_3^7}{t_2^4} + \frac{t_1^9 t_3^7}{t_2^4} -\frac{5 t_1^{11} t_3^7}{t_2^4}- \frac{t_1^3 t_3^7}{t_2^2} + \frac{t_1^5 t_3^7}{t_2^2} + \frac{27 t_1^7 t_3^7}{t_2^2} - \frac{30 t_1^9 t_3^7}{t_2^2} + \nn \\
&& \frac{5 t_1^{11} t_3^7}{t_2^2}- \frac{t_1^8 t_3^8}{t_2^6} + \frac{2 t_1^{10} t_3^8}{t_2^6} - \frac{t_1^{12} t_3^8}{t_2^6} + \frac{2 t_1^6 t_3^8}{t_2^4} + \frac{5 t_1^8 t_3^8}{t_2^4} -\frac{21 t_1^{10} t_3^8}{t_2^4}+ \frac{8 t_1^{12} t_3^8}{t_2^4} + \frac{2 t_1^4 t_3^8}{t_2^2} + \frac{t_1^6 t_3^8}{t_2^2} - \frac{30 t_1^8 t_3^8}{t_2^2} +\nn \\
&&  \frac{28 t_1^{10} t_3^8}{t_2^2} - \frac{3 t_1^{12} t_3^8}{t_2^2} - \frac{t_1^9 t_3^9}{t_2^6} - \frac{t_1^{11} t_3^9}{t_2^6} + \frac{2 t_1^7 t_3^9}{t_2^4} - \frac{5 t_1^9 t_3^9}{t_2^4} +\frac{5 t_1^{11} t_3^9}{t_2^4} + \frac{2 t_1^5 t_3^9}{t_2^2} - \frac{21 t_1^7 t_3^9}{t_2^2}+ \frac{28 t_1^9 t_3^9}{t_2^2} - \frac{9 t_1^{11} t_3^9}{t_2^2} - \nn \\
&& \frac{t_1^8 t_3^{10}}{t_2^4} + \frac{8 t_1^{10} t_3^{10}}{t_2^4} - \frac{3 t_1^{12} t_3^{10}}{t_2^4}- \frac{t_1^6 t_3^{10}}{t_2^2} + \frac{5 t_1^8 t_3^{10}}{t_2^2} - \frac{9 t_1^{10} t_3^{10}}{t_2^2} + \frac{t_1^{12} t_3^{10}}{t_2^2}- \frac{3 t_1^9 t_3^{11}}{t_2^2} + \frac{t_1^{11} t_3^{11}}{t_2^2}~.
\eea}
Let $R_i$ be the R-charge corresponding to the fugacity $t_i$. Since the superpotential, which has fugacity $t_1 t_3$, carries R-charge 2, it follows that:
\bea
R_1 + R_3 &=& 2
\label{e:cyfano324}
\eea
The volume of $\cF_1$ is given by:
{\footnotesize
\bea
\lim_{\mu \rightarrow 0} \mu^4 \gm (e^{-\mu R_1}, e^{-\mu R_2}, e^{-\mu(2 - R_1)}; \cF_1) &=& \frac{p\left(R_1, R_2; \cF_1 \right)}{4 R_1^2 R_2^2(2 - R_1)^2(2 - R_2)^2(1+R_1-R_2)^2(1-R_1+R_2)^2}~, \nn \\
\label{e:volfano324}
\eea}
where:
{\footnotesize
\bea
p\left(R_1, R_2; \cF_1 \right) &=& 8 R_1 - 14 R_1^3 + 2 R_1^4 + 6 R_1^5 - 2 R_1^6 + 8 R_2 - 8 R_1 R_2+ 18 R_1^2 R_2 + 16 R_1^3 R_2 - 20 R_1^4 R_2\nn \\
&& + 2 R_1^6 R_2 + 18 R_1 R_2^2 - 48 R_1^2 R_2^2 + 17 R_1^3 R_2^2 + 9 R_1^4 R_2^2 - 3 R_1^5 R_2^2 - R_1^6 R_2^2 - 14 R_2^3\nn \\
&& + 16 R_1 R_2^3 + 17 R_1^2 R_2^3 - 15 R_1^3 R_2^3+ R_1^4 R_2^3 + 3 R_1^5 R_2^3 + 2 R_2^4 - 20 R_1 R_2^4 + 9 R_1^2 R_2^4\nn \\
&&  + R_1^3 R_2^4 - 4 R_1^4 R_2^4 + 6 R_2^5 - 3 R_1^2 R_2^5 + 3 R_1^3 R_2^5 - 2 R_2^6 + 2 R_1 R_2^6 - R_1^2 R_2^6~.
\eea}
This function has a minimum at: 
\bea
R_1 = R_2 = R_3 = 1.
\eea
The R-charge of the perfect matching corresponding to the divisor $D_\alpha$ is given by:
\bea
\lim_{\mu\rightarrow0}\frac{1}{\mu} \left[ \frac{g(D_\alpha; e^{- \mu R_1}, e^{- \mu R_2 }, e^{- \mu R_3 }; \cF_1) }{\gm(e^{-\mu R_1}, e^{- \mu R_2 }, e^{- \mu R_3 };\cF_1)}- 1 \right]~,
\eea
where $g(D_\alpha; e^{- \mu R_1}, e^{- \mu R_2 }, e^{- \mu R_3 }; \cF_1)$ is the Molien-Weyl integral with the insertion of the inverse of the weight corresponding to the divisor $D_\alpha$. The results are shown in Table \ref{t:chargefano324}.  The assignment of charges under the remaining abelian symmetries can be done by requiring that the superpotential is not charged under them and that the charge vectors are linearly independent.  The charge assignments are listed in Table \ref{t:chargefano324}.

\begin{table}[h!]
 \begin{center}  
  \begin{tabular}{|c||c|c|c|c|c|c|c|c|c|}
  \hline
  \;& $SU(2)$&$U(1)_1$&$U(1)_2$&$U(1)_R$&$U(1)_{B_1}$&$U(1)_{B_2}$&$U(1)_{B_3}$&$U(1)_{B_4}$&fugacity\\
  \hline\hline  
   
  $p_1$&$  1$&$  0$&$ 0$&$1/3$&$ 0$&$ 0$ &$ 0$ &$ 0$  & $t^3 x $\\
  \hline
  
  $p_2$&$ -1$&$  0$&$ 0$&$1/3$&$ 0$&$ 0$ &$ 0$ &$ 0$  & $t^3 / x $\\
  \hline  
  
  $q_1$&$  0$&$  1$&$ 0$&$2/9$&$ 0$&$ 0$ &$ 0$ &$ 0$  & $t^2 q_1 $\\
  \hline
  
  $q_2$&$  0$&$ -1$&$ 0$&$2/9$&$ 0$&$ 0$ &$ 0$ &$ 0$  & $t^2 / q_1 $\\
  \hline
   
  $r_1$&$  0$&$  0$&$ 1$&$2/9$&$ 0$&$ 0$ &$ 0$ &$ 0$  & $t^2 q_2$\\
  \hline
  
  $r_2$&$  0$&$  0$&$-1$&$2/9$&$ 0$&$ 0$ &$ 0$ &$ 0$  & $t^2 / q_2$\\
  \hline
  
  $u_1$&$  0$&$  0$&$ 0$&$2/9$&$ 0$&$ 0$ &$ 0$ &$ 0$  & $t^2$\\
  \hline
  
  $u_2$&$  0$&$  0$&$ 0$&$2/9$&$ 0$&$ 0$ &$ 0$ &$ 0$  & $t^2$\\
  \hline

  $v_1$&$  0$&$  0$&$ 0$&$    0$&$ 1$&$ 0$ &$ 0$ &$ 0$  & $ b_1$\\
  \hline
  
  $v_2$&$  0$&$  0$&$ 0$&$    0$&$ 0$&$ 1$ &$ 0$ &$ 0$  & $ b_2$\\
  \hline
 
  $v_3$&$  0$&$  0$&$ 0$&$    0$&$-1$&$ 0$ &$ 1$ &$ 0$  & $ b_3 / b_1$\\
  \hline

  $v_4$&$  0$&$  0$&$ 0$&$    0$&$ 0$&$-1$ &$ 0$ &$ 1$  & $ b_4 / b_2 $\\
  \hline
 
  $v_5$&$  0$&$  0$&$ 0$&$    0$&$ 0$&$ 0$ &$-1$ &$-1$  & $1 / (b_3 b_4)$\\
  \hline 
   \end{tabular}
  \end{center}
\caption{Charges of the perfect matchings under the global symmetry of the $\cF_1$ model. Here $t$ is the fugacity of the R-charge, $x$ is the weight of the $SU(2)$ symmetry, $q_1, q_2, b_1, b_2, b_3$ and $b_4$ are, respectively, the charges under the mesonic abelian symmetries $U(1)_1, U(1)_2$ and of the three baryonic $U(1)_{B_1},U(1)_{B_2}, U(1)_{B_3}$ and $U(1)_{B_4}$.}
\label{t:chargefano324}
\end{table}

\begin{table}[h]
 \begin{center}  
  \begin{tabular}{|c||c|}
  \hline
  \; Quiver fields & R-charge\\
  \hline  \hline 
  $ X^i_{12}, X^i_{34}$ & 1/3 \\
  \hline
  $ X^i_{23}, X_{51}, X_{61}, X_{45}, X_{46}$ & 4/9 \\
  \hline
  \end{tabular}
  \end{center}
\caption{R-charges of the quiver fields for the $\cF_1$ model.}
\label{t:Rquivefano324}
\end{table}

\paragraph{The Hilbert series.} 
The Hilbert series of the Master space can be obtained by integrating that of the space of perfect matchings the fugacities $z_1,\ldots,z_5$:
{\footnotesize
\bea
g^{\firr{}} (t, x, q_1, q_2, b_i; \cF_1) &=& \left( \prod^5_{j=1} \oint\limits_{|z_j|=1} \frac{\ud z_j}{2 \pi i z_j}\right) 
\frac{1}{\left(1- t^3 x z_1\right)\left(1- \frac{t^3 z_1}{x}\right)\left(1- t^2 q_1 z_2\right)\left(1- \frac{t^2 z_3}{q_1}\right)\left(1- \frac{t^2 q_2 z_4}{z_2}\right)}\nn \\
&&\times \frac{1}{\left(1-\frac{t^2 z_4}{q_2 z_3}\right)\left(1-\frac{t^2 z_3 z_5}{z_2}\right)\left(1 - t^2 z_5\right)\left(1-\frac{b_1}{z_1}\right)\left(1-\frac{b_2}{z_1}\right)\left(1-\frac{b_3}{z_3 z_5 b_1}\right)}\nn \\
&&\times \frac{1}{\left(1-\frac{b_4}{z_4 b_2}\right)\left(1-\frac{z_2}{b_3 b_4 z_4 z_5}\right)} \nn \\
&=& \frac{\left(1- t^6 b_1 b_2\right)}{\left(1- t^3 x b_2\right)\left(1-\frac{t^3 b_2}{x}\right)\left(1- t^3 x b_1\right)\left(1-\frac{t^3 b_1}{x}\right)\left(1- \frac{t^4 q_2}{b_3 b_4}\right)\left(1-\frac{t^4}{q_2 b_3 b_4}\right)}\nn \\
&&\times \frac{\left(1-\frac{t^{12}}{b_1 b_2}\right)}{\left(1-\frac{t^4 q_1 b_3}{b_1}\right)\left(1-\frac{t^4 b_3}{q_1 b_1}\right)\left(1-\frac{t^4 q_1 q_2 b_4}{b_2}\right)\left(1-\frac{t^4}{q_1 q_2 b_3 b_4}\right)}~.
\label{e:HSMasterfano324}
\eea}
The fully unrefined Hilbert series of the Master space can be written as:
\bea
g^{\firr{}} (t, 1, 1, 1, 1; \cF_1) &=& \frac{(1-t^6)(1-t^{12})}{(1-t^3)^4(1-t^4)^6}
\eea
The Hilbert series of the mesonic moduli space is obtained by integrating Hilbert series of the Master space over the baryonic fugacities:
\bea
\gm(t, x, q_1, q_2; \cF_1) &=& \left( \prod^4_{i=1} \oint\limits_{|b_i|=1} \frac{\ud b_i}{2 \pi i b_i} \right) g^{\firr{}} (t, x, q_1, q_2, b_i; \cF_1)\nn \\
&=& \frac{P(t, x, q_1, q_2; \cF_1)}{\left(1-t^{18} x^2 q^2_1\right)\left(1-\frac{t^{18} q^2_1}{x^2}\right)\left(1- \frac{t^{18} x^2}{q^2_1}\right)\left(1-\frac{t^{18}}{x^2 q^2_1}\right)}\nn \\
&&\times \frac{1}{\left(1-t^{18} x^2 q^2_2\right)\left(1-\frac{t^{18} q^2_2}{x^2}\right)\left(1-\frac{t^{18} x^2}{q^2_2}\right)\left(1-\frac{t^{18}}{x^2 q^2_2}\right)}\nn \\
&&\times \frac{1}{\left(1- t^{18} x^2 q^2_1 q^2_2\right)\left(1-\frac{t^{18} q^2_1 q^2_2}{x^2}\right)\left(1-\frac{t^{18} x^2}{q^2_1 q^2_2}\right)\left(1-\frac{t^{18}}{x^2 q^2_1 q^2_2}\right)}\nn \\
\label{e:HSmesfano324}
\eea
The fully unrefined version of the Hilbert series of the mesonic moduli space can be written as:
\bea
\gm(t, 1, 1, 1; \cF_1) &=& \frac{1 + 17 t^{18} + 17 t^{36} +  t^{54}}{(1-t^{18})^4}
\eea
The plethystic logarithm of the Hilbert series of the mesonic moduli space is:
\bea
\PL[\gm(t, x, q_1, q_2; \cF_1)] &=& [2]\left(q^2_1 + q^2_2 + q^2_1 q^2_2 + 1 +\frac{1}{q^2_1 q^2_2} + \frac{1}{q^2_2} + \frac{1}{q^2_1}\right)t^{18} -  O(t^{36}).\nn \\~
\label{e:plfano324}
\eea

\paragraph{The generators.} The generators of the mesonic moduli space arem/
\bea
&& p_i p_j q^2_k r^2_k u_1 u_2 v_1 v_2 v_3 v_4 v_5, \qquad  p_i p_j q^2_k r_1 r_2 u^2_k v_1 v_2 v_3 v_4 v_5,\nn \\
&& p_i p_j q_1 q_2 r^2_k u^2_l v_1 v_2 v_3 v_4 v_5, \qquad  p_i p_j q_1 q_2 r_1 r_2 u_1 u_2 v_1 v_2 v_3 v_4 v_5,\
\label{e:genfano324}
\eea
with $i,j,k,l=1,2$, and $k \neq l$. All the generators of the mesonic moduli space have R-charge 2. The lattice of generators is drawn in Figure \ref{f:latf1}.

\begin{figure}[ht]
\begin{center}
\includegraphics[totalheight=4cm]{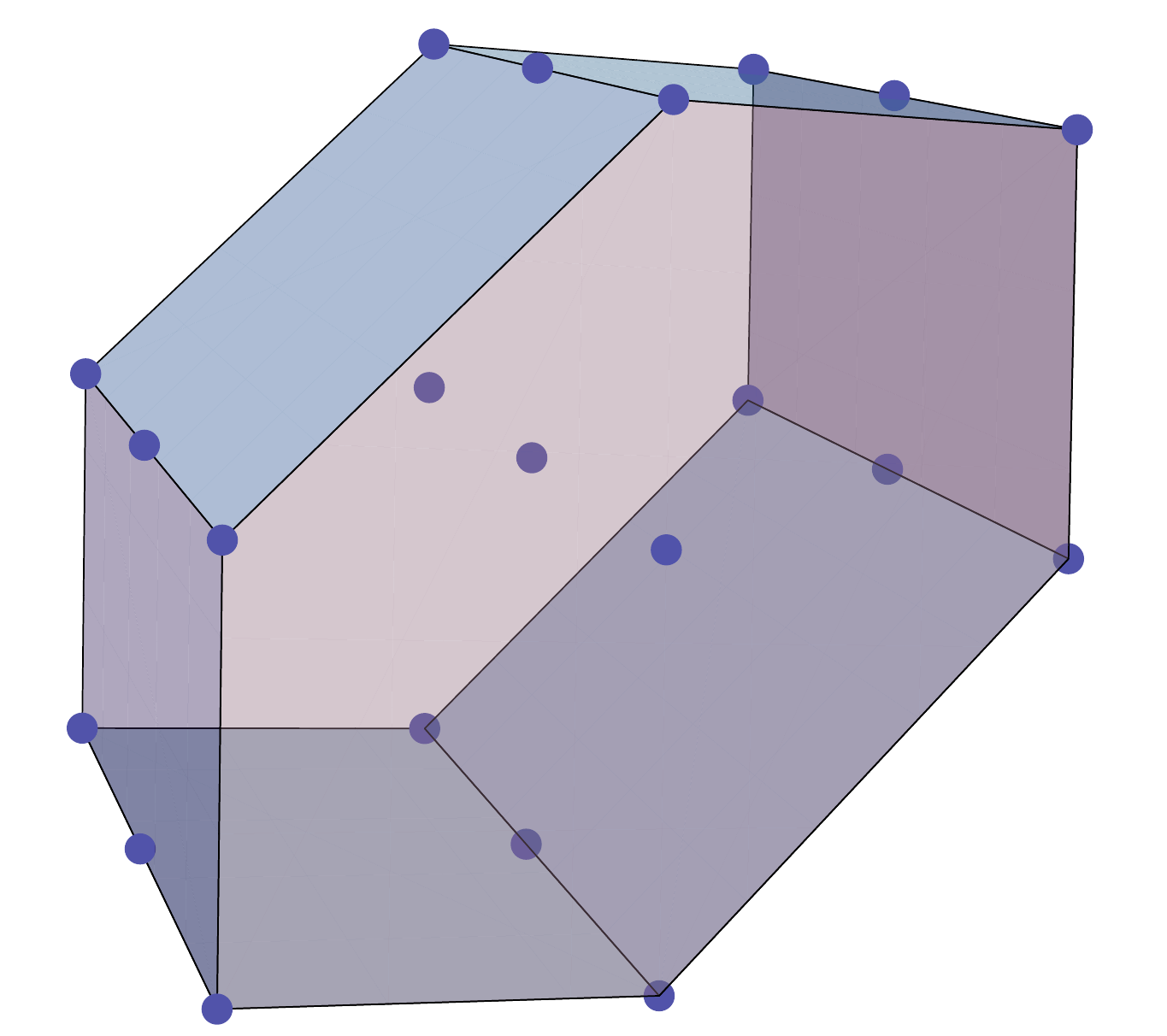}
\caption{The lattice of generators of the $\cF_1$ theory.}
  \label{f:latf1}
  \end{center}
\end{figure}

\section{Discussion and Outlook}

In this paper 14 of the fano 3-folds and their corresponding tilings and quiver CS theories are analysed in detail. Some of them, such as the $M^{1,1,1}$ and $Q^{1,1,1}/\BZ_2$ theories, have already been studied in the literature \cite{Martelli:2008si,Hanany:2008cd,Hanany:2008fj,phase,Hanany:2009vx,Davey:2009qx,Davey:2009bp,Davey:2009et,Franco:2009sp,Amariti:2009rb,Petrini:2009ur,Fabbri:1999hw}, but to our knowledge all of the other models that are discussed here were not known before. 

Starting from tilings the forward algorithm has been used to determine the Hilbert series, the generators of the mesonic moduli space and the spectrum of scaling dimensions of the chiral fields of each of the theories. It is our hope that this investigation will prove to be an important source of models in the examination of CS gauge theories living on M2-branes.

Indeed this work demonstrates the strength of the forward algorithm - a detailed analysis of the structure of a CS gauge theory can be carried out by a small number of relatively simple computations. However it should be noted that the limitations of the forward algorithm when studying 2+1 dimensional Chern-Simons theories are also apparent.
In spite of a long and exhaustive study of all the tilings with less than 10 nodes, it has not been possible to identify tiling that can correspond to $\BP^3, \cB_1, \cB_2$ and $\cB_3$ models. In particular we believe that the current forward algorithm is incapable of dealing with a tiling that has $\BP^3$ as its mesonic moduli space. It might be that there cannot exist a consistent CS gauge theory on M2-branes probing certain toric CY 4-folds. Another possibility is that such theories do not admit a brane tiling description.

In the study of $(3+1)$-dimensional gauge theories living on D3-branes, it is possible to use the so called `inverse algorithm' which given a toric CY 3-fold allows the construction of the brane tiling of a gauge theory which has the CY as its mesonic moduli space. Unfortunately an inverse algorithm that relates toric CY 4-folds to $(2+1)$-dimensional CS theories is not known. Further investigation of such an algorithm and also the improvement of the current forward algorithm is of great importance and should be studied in the future.

\section{Acknowledgements}
J.D. would like to thank STFC for his studentship. N.~M.~ is grateful to the following institutes and collaborators for their very kind hospitality during the completion of this script: Max-Planck-Institut f\"ur Physik (Werner-Heisenberg-Institut), Rudolf Peierls Centre for Theoretical Physics (University of Oxford), Department of Applied Mathematics and Theoretical Physics (University of Cambridge), String Theory Group at the Universiteit van  Amsterdam, Frederik Beaujean, Francis Dolan, Yang-Hui He, Sven Krippendorf and Alexander Shannon.  He also thanks Aroonroj Mekareeya for his help in graphic drawing as well as his family for the warm encouragement and support. This research is supported by the DPST project, the Royal Thai Government. G.T. wants to thank the Fondazione Angelo Della Riccia for financial suppport. He is extremely grateful to Stanford University, UC Berkeley and UC Davis for their kind hospitality during the preparation of this work and would like to thank Alberto Zaffaroni, Mina Aganagic, Yu Nakayama, Alessio Marrani, Rak-Kyeong Seong and Kevin Schaeffer for precious and enlightening discussions. He would like to dedicate this paper to the loving memory of professor Carlo Alberto Quaranta, whose great sense of duty, endless dedication and enormous skills have made him an invaluable example to look at and learn from through all these years.

%
%

\end{document}